%% file: main.tex
\documentclass[twocolumn, twocolappendix, tighten]{aastex63}
\usepackage{graphicx}
\usepackage{url}
\usepackage{epstopdf}
\usepackage{color}
\usepackage{gensymb}
\usepackage{multirow}
\usepackage{ragged2e}
\usepackage{url}
\usepackage{amsmath} 
\usepackage{booktabs}
\usepackage{afterpage}
\usepackage{mathtools}
\usepackage{tabularx}
\usepackage{bold-extra}
\usepackage{xspace}
\usepackage{relsize}
\usepackage[utf8]{inputenc}
\usepackage{braket}
\usepackage{bbm}

\usepackage{eht}

\usepackage[T1]{fontenc} 

\usepackage[encapsulated]{CJK}
\usepackage{ucs}
\newcommand{\cntext}[1]{\begin{CJK}{UTF8}{gbsn}#1\end{CJK}}

\graphicspath{{compressed_figures/}}
\DeclareGraphicsExtensions{.pdf,.png,.jpeg}

\newcommand{\uv}{$(u,v)$\xspace}
\newcommand{\sdiam}{$\sim$50\,$\mu$as\xspace}

\usepackage{amsmath}
\DeclareMathOperator*{\median}{median}

\newcolumntype{P}[1]{>{\centering\arraybackslash}p{#1}}

\begin{document}

\title{First Sagittarius A$^{*}$ Event Horizon Telescope Results III: Imaging of the Galactic Center Supermassive Black Hole}
\shorttitle{Event-horizon-scale Imaging of \sgra with the Event Horizon Telescope}
\shortauthors{The EHT Collaboration et al.}

\input{GAL-only}

\collaboration{0}{The Event Horizon Telescope Collaboration}

\input{abstract}

\nocite{PaperI}
\nocite{PaperII}
\nocite{PaperIII}
\nocite{PaperIV}
\nocite{PaperV}
\nocite{PaperVI}

\input{intro}
\input{observations}

\input{pre-imaging}

\input{imaging_methods}
\input{synthetic_data}

\input{parameter_survey}

\input{topset_images}

\input{image_analysis}
\input{best_times}

\input{summary}

\facility{EHT}.
\software{ \difmap \citep{Shepherd_1995,Shepherd_1997,Shepherd_2011}, \ehtim \citep{Chael_2016,Chael_2018_Imaging}, ehtplot, \smili \citep{Akiyama_2017a,Akiyama_2017b}, \themis \citep{Broderick_2020a}, \vida \citep{Tiede_2020}, numpy \citep{numpy}, scipy \citep{scipy}, matplotlib \citep{matplotlib}, pandas \citep{pandas1, pandas2}, astropy \citep{astropy1, astropy2}}

\input{Acks_PaperIII_may9}

\begin{appendix}
\input{appendix_imaging_methods}

\input{appendix_scattering}

\input{appendix_first_imaging}

\input{appendix_pipelines}

\input{appendix_topsetselection}

\input{appendix_clustering}

\input{appendix_smili_dynamical}
\input{appendix_bestbetMAD_from_theory}
\input{appendix_ringfits}
\input{appendix_besttimes}

\end{appendix}

\bibliographystyle{yahapj}
\bibliography{SgrA_Imaging, EHTCPapers, software}

\allauthors

\end{document}

%% file: GAL-only.tex
\author[0000-0002-9475-4254]{Kazunori Akiyama}
\affiliation{Massachusetts Institute of Technology Haystack Observatory, 99 Millstone Road, Westford, MA 01886, USA}
\affiliation{National Astronomical Observatory of Japan, 2-21-1 Osawa, Mitaka, Tokyo 181-8588, Japan}
\affiliation{Black Hole Initiative at Harvard University, 20 Garden Street, Cambridge, MA 02138, USA}

\author[0000-0002-9371-1033]{Antxon Alberdi}
\affiliation{Instituto de Astrof\'{\i}sica de Andaluc\'{\i}a-CSIC, Glorieta de la Astronom\'{\i}a s/n, E-18008 Granada, Spain}

\author{Walter Alef}
\affiliation{Max-Planck-Institut f\"ur Radioastronomie, Auf dem H\"ugel 69, D-53121 Bonn, Germany}

\author[0000-0001-6993-1696]{Juan Carlos Algaba}
\affiliation{Department of Physics, Faculty of Science, Universiti Malaya, 50603 Kuala Lumpur, Malaysia}

\author[0000-0003-3457-7660]{Richard Anantua}
\affiliation{Black Hole Initiative at Harvard University, 20 Garden Street, Cambridge, MA 02138, USA}
\affiliation{Center for Astrophysics $|$ Harvard \& Smithsonian, 60 Garden Street, Cambridge, MA 02138, USA}
\affiliation{Department of Physics \& Astronomy, The University of Texas at San Antonio, One UTSA Circle, San Antonio, TX 78249, USA}

\author[0000-0001-6988-8763]{Keiichi Asada}
\affiliation{Institute of Astronomy and Astrophysics, Academia Sinica, 11F of Astronomy-Mathematics Building, AS/NTU No. 1, Sec. 4, Roosevelt Rd, Taipei 10617, Taiwan, R.O.C.}

\author[0000-0002-2200-5393]{Rebecca Azulay}
\affiliation{Departament d'Astronomia i Astrof\'{\i}sica, Universitat de Val\`encia, C. Dr. Moliner 50, E-46100 Burjassot, Val\`encia, Spain}
\affiliation{Observatori Astronòmic, Universitat de Val\`encia, C. Catedr\'atico Jos\'e Beltr\'an 2, E-46980 Paterna, Val\`encia, Spain}
\affiliation{Max-Planck-Institut f\"ur Radioastronomie, Auf dem H\"ugel 69, D-53121 Bonn, Germany}

\author[0000-0002-7722-8412]{Uwe Bach}
\affiliation{Max-Planck-Institut f\"ur Radioastronomie, Auf dem H\"ugel 69, D-53121 Bonn, Germany}

\author[0000-0003-3090-3975]{Anne-Kathrin Baczko}
\affiliation{Max-Planck-Institut f\"ur Radioastronomie, Auf dem H\"ugel 69, D-53121 Bonn, Germany}

\author{David Ball}
\affiliation{Steward Observatory and Department of Astronomy, University of Arizona, 933 N. Cherry Ave., Tucson, AZ 85721, USA}

% ...

%% file: abstract.tex
\begin{abstract}
We present the first event-horizon-scale images and spatiotemporal analysis of \sgra taken with the Event Horizon Telescope in 2017 April at a wavelength of 1.3\,mm. Imaging of \sgra has been conducted through surveys over a wide range of imaging assumptions using the classical CLEAN algorithm, regularized maximum likelihood methods, and a Bayesian posterior sampling method. Different prescriptions have been used to account for scattering effects by the interstellar medium towards the Galactic Center. Mitigation of the rapid intra-day variability that characterizes \sgra has been carried out through the addition of a ``variability noise budget" in the observed visibilities, facilitating the reconstruction of static full-track images. Our static reconstructions of \sgra can be clustered into four representative morphologies that correspond to ring images with three different azimuthal brightness distributions, and a small cluster that contains diverse non-ring morphologies. 
Based on our extensive analysis of the effects of sparse $(u,v)$-coverage, source variability and interstellar scattering, as well as studies of simulated visibility data, we conclude that the Event Horizon Telescope \sgra data show compelling evidence for an image that is dominated by a bright ring of emission with a ring diameter of \sdiam, consistent with the expected ``shadow" of a $4\times10^6 M_\odot$ black hole in the Galactic Center located at a distance of 8\,kpc.
\end{abstract}

%% file: intro.tex
\section{Introduction}
\label{sec:introduction}

At the center of our galaxy is the nearest candidate supermassive black hole (SMBH), Sagittarius A* (\sgra). We present here the first horizon-scale images of it. Compared with M87$^{*}$, \sgra is more challenging to image mainly due to its rapid variability and the distortions of the intervening scattering medium. We develop methods to characterize and mitigate these two factors in order to reconstruct images that take us a step closer to establishing that \sgra is a black hole.  This paper is the third in the Event Horizon Telescope (EHT)'s series of six \sgra articles \citep[][hereafter Papers I, II, III, IV, V, and VI]{PaperI,PaperII,PaperIII,PaperIV,PaperV,PaperVI}.

Since the first discovery of \sgra as a compact radio source with interferometric observations at centimeter (cm) wavelengths \citep{Balick_Brown_1974}, there have been many studies of this closest SMBH to Earth. 
Of particular importance is the study of stellar dynamics showing that the position of \sgra in the Galactic Center coincides with the center of gravity of a dense cluster of young and old stars \citep{Eckart_1997,Ghez_1998,Menten_1997,Reid_2004,Reid_2009}. 
Moreover, the gravitational potential is dominated by a compact object of mass of $4\times10^6 M_\odot$ contained within 120\,AU of \sgra, at a distance of 8\,kpc from Earth \citep{Ghez_2008,Gillessen_2009,Gillessen_2017,Gravity_2018,Gravity_2019,Do_2019}. 
Based on these facts, together with the continuous variability on characteristic time scales from minutes to hours, especially at near-infrared (NIR) wavelengths on angular scales of typically 150 micro-arcseconds ($\mu$as) \citep{Gravity_2018b,Gravity_2020}, the likely scenario of \sgra is that this compact object is a SMBH. 
The combination of mass and proximity make \sgra the black hole subtending the largest angle on the sky with a Schwarzschild radius of 0.08\,AU$\,\sim10\,\mu$as and an expected ``shadow'' angular size of $\sim 50\,\mu$as. 
\sgra was thus identified early on as a primary target for imaging a black hole ``shadow'' \citep{Falcke_2000} predicted by Einstein's theory of general relativity \citep{Hilbert_1917, von-Laue_1921, Bardeen_1973, Luminet_1979}.
Similar calculations put the shadow angular size of M87$^{*}$ at $\sim40\,\mu$as ($6.5\times10^9 M_\odot$, 16.4\,Mpc from Earth) confirmed by the EHT through imaging and analysis \citep[hereafter M87$^*$ Papers I-VI]{M87PaperI,M87PaperII,M87PaperIII,M87PaperIV,M87PaperV,M87PaperVI}.

In very-long-baseline interferometry (VLBI) observations at cm wavelengths, the structure of \sgra is unresolved and dominated by scatter broadening caused by the ionized interstellar medium (ISM; see, e.g., \citealt{Rickett_1990, Narayan_1992}). 
As a result, the measured sizes are proportional to $\lambda^2$, where $\lambda$ is the observing wavelength 
\citep{Davies_1976}, with an asymmetric Gaussian shape elongated toward the east-west direction (i.e., stronger angular broadening; \citealt{Lo_1985,Lo_1998,Alberdi_1993,Krichbaum_1993,Frail_1994,Bower_1998}).
For several decades, many VLBI observations have attempted to reach smaller angular (and spatial) scales. These studies
found that the observed size at millimeter (mm) wavelengths deviates from the $\lambda^2$-relation, indicating a larger intrinsic size than expected from scatter broadening of an intrinsically unresolved source \citep[e.g.,][]{Krichbaum_1998, Lo_1998, Doeleman_2008, Falcke_2009}. 

After constraining the scattering effects (see \autoref{sec:pre_scatter}), the intrinsic structure of \sgra at long radio wavelengths can be modelled with a single nearly isotropic Gaussian source \citep{Bower_2004, Shen_2005, Lu_2011a, Bower_2014b, Johnson_2018, Issaoun_2021, Cho_2022}.
Its size and orientation on the sky have remained fairly constant over days to years timescales \citep[e.g.,][]{Alberdi_1993, Marcaide_1999, Lu_2011}, but a marginal variation has also been suggested (\citealt{Bower_2004, Akiyama_2013}). 
At observing wavelengths $>1\,$mm, some evidence for structure beyond a single Gaussian model has also been reported. While this is likely attributed to refractive scattering sub-structure (i.e., not intrinsic) at cm-wavelengths \citep[e.g.,][]{Gwinn_2014}, its cause is still unclear at mm-wavelengths. %
For instance, at 7\,mm, \citet{rauch_2016} reported a short-lived secondary component that could possibly be related to a preceding NIR flare. However, the detection of non-zero closure phases was only marginal, and does not exclude a realization of thermal or other systematic errors. 
At 3.5\,mm, several studies have found slight non-zero closure phases \citep{Ortiz_2016, Brinkerink_2016} and asymmetric non-Gaussian structure along the minor axis \citep{Issaoun_2019,Issaoun_2021}, but its physical origin remained non-conclusive: it could be due either to scattering or an intrinsic asymmetry of \sgra. 

Multi-wavelength observations of \sgra show an inverted spectral energy distribution rising with frequency in the radio due to synchrotron emission, with spectral break at THz frequencies (submillimeter wavelengths), where the accretion flow becomes optically thin \citep{Falcke_1998,Bower_2015_spectrum,Bower_2019}. Its bolometric luminosity was measured to be $\sim 5 \times 10^{35}\,{\rm erg\,s}^{-1}$, or $10^{-9}L_{\rm Edd}$ \citep{Genzel_2010,Bower_2019}. A more detailed description of the spectral properties of \sgra is presented in \citetalias{PaperII}. %
At an observing frequency of 230\,GHz (1.3\,mm wavelength), the accretion flow is expected to be sufficiently optically thin to detect the black hole shadow in \sgra with an Earth-sized interferometric array, such as the Event Horizon Telescope (EHT)~(\citealt{Falcke_2000,Doeleman_2009,Broderick_2016}; \citetalias{M87PaperII}). 

\sgra additionally exhibits variability across the entire electromagnetic spectrum \citep{Genzel_2003,Ghez_2004,Neilsen_2013,Neilsen_2015,Bower_2015_spectrum,Boyce_2019}, with frequent flaring in the radio, infrared and X-ray regimes. Variability and motion can be observed within a single observing night, with variability timescales of the order of seconds to hours, characteristic dynamic timescales for a $4\times 10^6 M_\odot$ black hole \citep{Baganoff_2003,Marrone_2006,Meyer_2008,Hora_2014,Dexter_2014,Witzel_2018,Bower_2018,Bower_2019}. In the radio and submillimeter, \sgra is constantly varying, with a variability level of $<10\%$ during quiescence \citep[\citealt{Macquart_2006}; \citetalias{PaperII};][]{Wielgus_2021}. A detailed description of the multi-wavelength properties of \sgra is presented in \citetalias{PaperII}.

At 1.3\,mm, \sgra was detected for the first time with VLBI on a single baseline with the IRAM 30-m telescope and the Plateau de Bure Interferometer in 1995 \citep{Krichbaum_1998}. In 2007, the first successful observations with the early EHT array, consisted of the Arizona Radio Observatory Sub-Millimeter Telescope (SMT) in Arizona, the James Clerk Maxwell Telescope (JCMT) in \hawaii and the Combined Array for Research in Millimeter-wave Astronomy (CARMA) in California, offered the first line of evidence that the source size at 1.3\,mm is comparable to the expected size of the shadow of a SMBH with the mass and position of \sgra \citep{Doeleman_2008,Fish_2011}. In 2013, 1.3\,mm observations were carried out with an early subset of the present EHT array: five stations at four geographical sites (Arizona, California, \hawaii, and Chile). An early processing of the US-only data resulted in a first measurement of relatively high linear polarization on the $50-100\,\mu$as scale by \citet{Johnson_2015}, and the detection of non-zero closure phases by \citet{Fish_2016}, indicative of asymmetric source structure on the Arizona--California--\hawaii triangle. A final processing of the data with the addition of Atacama Pathfinder Experiment telescope (APEX) in Chile, by \citet{Lu_2018}, which provides a resolution of $\sim 30\,\mu$as (3 Schwarzschild Radii for the estimated black hole mass) in the north-south direction, revealed the presence of compact structure residing within the scale of $50\,\mu$as and confirmed the previously reported asymmetry (non-zero closure phase) by \citet{Fish_2016}. A subsequent expansion effort of the EHT to increase array sensitivity and imaging ability culminated in the April 2017 EHT observing campaign \citepalias{M87PaperII}.

The 2017 EHT observing campaign was scheduled over a 12-day time window in April 2017, to minimize weather impact. The two primary targets, M87$^{*}$ (at the center of the giant elliptical galaxy M87) and \sgra, were observed for four and five nights, respectively. With a similar mass-to-distance ratio, the two targets are expected to exhibit similar angular sizes on the sky. Because M87$^{*}$ is about three orders of magnitude more massive compared to \sgra, its dynamical timescale is much longer, allowing for the straightforward use of standard aperture synthesis VLBI techniques over each observing night. The effects of scattering toward the M87 galaxy are also minimal. These factors render M87$^{*}$ the optimal first imaging target for the EHT. Based on the observations of M87$^{*}$ the EHT Collaboration presented first direct images of a SMBH, showing a bright ring-like structure surrounding a central dark circular area \citepalias{M87PaperI,M87PaperII,M87PaperIII,M87PaperIV,M87PaperV,M87PaperVI}.
Under the stellar dynamics mass measurement prior \citep{Gebhardt_2011} and with a scaling based on numerical simulations of the accretion flow \citepalias{M87PaperV}, these images confirm the predictions of general relativity about the diameter of a black hole shadow \citepalias{M87PaperVI}. Following these results, the analysis of the linear polarization observations of M87$^{*}$ produced the first polarized images of the M87$^{*}$ black hole and inferred a magnetic field strength and geometry in the immediate vicinity of the SMBH~\citep[hereafter M87$^*$ Papers VII and VIII]{M87PaperVII,M87PaperVIII}. The technique and workflow developments for the analysis of M87$^{*}$ data serve as the basis for \sgra analysis, although new significant developments were introduced to address the challenges of interstellar scattering and short-timescale variability.

In this paper, we present the first imaging results of \sgra with the EHT for the 2017 April 6 and 7 observations. 
In \autoref{sec::Observations}, we describe the EHT observations of \sgra in 2017 and their properties. 
In \autoref{sec::PreImaging_Considerations}, we estimate additional properties of two major effects anticipated for \sgra, interstellar scattering and intrinsic intra-day brightness variations, from non-imaging analysis to aid the imaging process. In \autoref{sec::Background}, we provide a brief review of the employed imaging techniques. 
In \autoref{sec:synthetic_data}, we describe the process for synthetic data generation for the imaging parameter surveys outlined in \autoref{sec:survey}. 
These surveys provide a set of imaging parameters that are used to produce \sgra images. 
In \autoref{sec:sgra_images}, we describe the resulting \sgra images for four different imaging pipelines and assess their properties and uncertainties. 
In \autoref{sec:image_analysis}, we extract source parameters from \sgra images via ring parameter fitting.
In \autoref{sec::dynamic}, we utilize dynamical imaging and geometric modeling techniques to explore and characterize potential azimuthal time variations in the data. We summarize our results in \autoref{sec::summary}. 

%% file: observations.tex
\section{Observations and Data Processing}\label{sec:observations}
\label{sec::Observations}
In this section we describe the EHT observations of \sgra performed in 2017 April (\autoref{sub:eht-obs}), the data reduction (\autoref{sub:data-red}), and overall data properties (\autoref{sub:data-prop}). A description of interferometric measurements and associated data products is provided in \citetalias{M87PaperIV} for reference.

\subsection{EHT Observations}
\label{sub:eht-obs}

The EHT observed \sgra with eight stations at six geographic sites on 2017 April 5, 6, 7, 10, and 11. 
The participating radio observatories are the phased Atacama Large Millimeter/submillimeter Array (ALMA) and APEX  in the Atacama Desert in Chile, the JCMT and the phased Submillimeter Array (SMA) on Maunakea in \hawaii, the SMT on Mt.\ Graham in Arizona, the IRAM 30-m (PV) telescope on Pico Veleta in Spain, the Large Millimeter Telescope Alfonso Serrano (LMT) on the Sierra Negra in Mexico, and the South Pole Telescope (SPT) in Antarctica. The observations of \sgra were interleaved with two AGN calibrator sources, the quasars NRAO\,530 and J1924-2914. Scientific analysis of the observations of calibrators will be presented in future publications (Jorstad et al. in prep., Issaoun et al. in prep.). The geocentric coordinates for each of the telescopes are presented in Table 2 of \citetalias{M87PaperII}.

The 2017 VLBI data were recorded in two polarizations and two frequency bands at a total data rate of 32 Gbps (for 2 bit sampling). All sites recorded two 2\,GHz wide frequency windows centered at 227.1 and 229.1 GHz, low and high band respectively. 
An extensive description of  the EHT array setup, equipment, and station upgrades 
leading up to the 2017 observations is provided in \citetalias{M87PaperII}. 
All sites except ALMA and JCMT recorded dual circular polarization (RCP and LCP). ALMA recorded dual linear polarization, subsequently converted to a circular basis via the CASA-based software package {\tt PolConvert} \citep{Marti_2016,Matthews_2018,Goddi_2019}, and JCMT recorded a single circular polarization (the recorded polarization component varied from day to day). Since JCMT recorded a single circular polarization, baselines to JCMT use the available parallel-hand component ($RR$ or $LL$) visibilities to approximate Stokes~$\mathcal{I}$ (``pseudo $\mathcal{I}$'', $\mathcal{I} \equiv RR \equiv LL$). This is consistent with the Stokes~$\mathcal{V}\equiv0$ assumption taken for the data calibration, justified by the expected $|\mathcal{I}| \gg |\mathcal{V}|$ relation \citep{Munoz2012,Goddi2021}. A small but detectable amount of intrinsic circular source polarization is present in our observations, which we account for in the systematic error analysis \citepalias{PaperII}.

\subsection{Data Reduction}
\label{sub:data-red}

Following the correlation of the data recorded at different sites, instrumental bandpass effects and phase turbulence introduced by the Earth's atmosphere were corrected using established fringe fitting algorithms \citepalias{M87PaperIII}. We use two independent software packages, the CASA-based \citep{McMullin_2007} rPICARD pipeline \citep{Janssen_2019} and the HOPS-based \citep{Whitney_2004} EHT-HOPS pipeline \citep{Blackburn_2019}.
The mitigation of the atmospheric phase variation allows for coherent averaging of the data in order to build up signal-to-noise ratio (\textit{S/N}) without substantial losses from decoherence.
Instrumental RCP/LCP phase and delay offsets were corrected by referencing fringe solutions to ALMA, calibrated with \texttt{PolConvert} \citep{Marti_2016}. %
The assumption of Stokes $\mathcal{V}=0$ on VLBI baselines is taken for the RCP/LCP gain calibration. Following the band-averaging in frequency, data were amplitude-calibrated using station-specific measurements of the system-equivalent flux density and time-averaged in 10\,s segments. Stations with an intra-site partner (i.e., ALMA, APEX, SMA, and JCMT) were subsequently ``network calibrated'' \citepalias[][\citealt{ Blackburn_2019}]{M87PaperIII} to further improve the amplitude calibration accuracy and stability via constraints among redundant baselines. Polarimetric leakage is not corrected for this Stokes $\mathcal{I}$ analysis, but rather included as a source of systematic uncertainty in the parallel-hand visibilities \citepalias{PaperII}.

The %
data processing pipeline has been slightly updated with respect to the one described in \citetalias{M87PaperIII}. Some notable changes include: a recorrelation of the data following setting changes at ALMA and more accurate sky coordinates of \sgra; updated amplitude calibration (most notably for LMT and SMA) using more accurate measurements of the telescope aperture efficiency, found to be variable across the campaign; stronger polarimetric calibration assumptions ($\mathcal{V}\!=\!0$); time-variable network calibration of \sgra using ALMA and SMA connected-element light curves \citep{Wielgus_2021}; %
and a time dependent transfer of the antennas gains to the visibility amplitudes, following the analysis of the data from the calibrators.

After the calibration using data reduction pipelines described by \citetalias{M87PaperIII} and \citetalias{PaperII}, additional steps were taken to mitigate specific data issues related to poorly constrained LMT gains and JCMT coherence losses. Following the source size constraints derived in the Section 5.1.5 of \citetalias{PaperII}, LMT amplitude gains have been pre-corrected assuming the 60\,$\mu$as source size seen by the baselines shorter than $2$\,G$\lambda$ (only SMT-LMT). 
Visibility phases on JCMT baselines were stabilized by calibrating phase on an intra-site JCMT-SMA baseline to zero degrees, in agreement with the unresolved point source visibility phases seen, for similar baseline lengths, in the intra-ALMA observations. A detailed description of the theoretical background from visibilities to images is presented in \cite{TMS}, \citetalias{M87PaperIV}, and \cite{2020ApJ...894...31B}, as well as in the Appendix of \citetalias{PaperIV}.

\begin{figure*}[t]
\centering
\includegraphics[trim={0.1cm 0.0cm 0.2cm 0.cm},clip,width=1.0\linewidth]{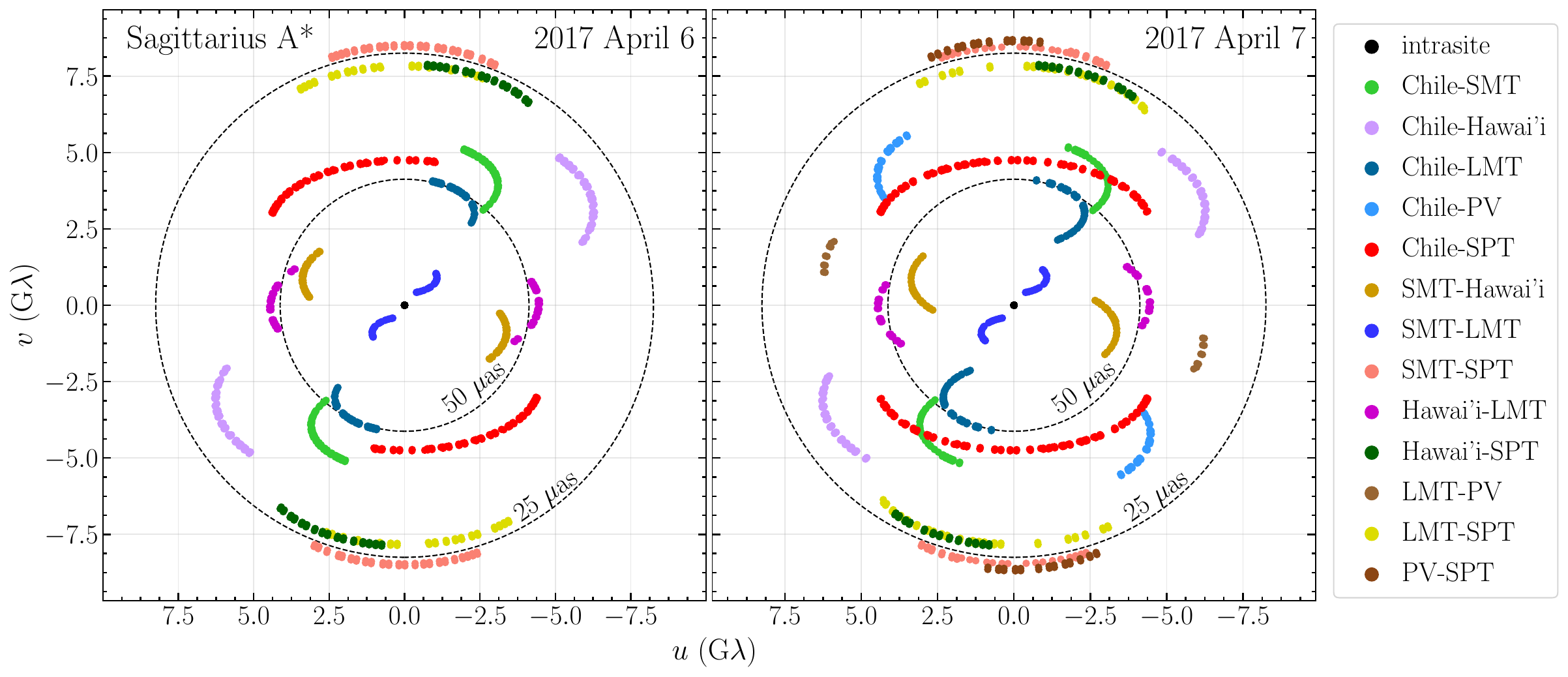}
\caption{$(u,v)$-coverage of the EHT observations of \sgra on 2017 April 6 and 7, from the HOPS dataset. Each point represents scan-averaged data, both bands are shown. ``Chile'' represents the stations ALMA and APEX. ``\hawaii'' represents the stations SMA and JCMT. Dashed line circles indicate the fringe spacing of 50\,$\mu$as and 25\,$\mu$as.
}
\label{fig:uv_coverage}
\end{figure*}

\subsection{Data Properties}
\label{sub:data-prop}
\subsubsection{General Aspects of \sgra Data}
The highly sensitive phased ALMA array participated in three of the five observing days, 2017 April 6, 7, and 11. April 7 is the only day that additionally includes PV observations of \sgra, and is therefore the day with the longest observation duration, largest number of detections, and the best overall $(u,v)$-coverage, as shown in Fig.~\ref{fig:uv_coverage}. On April 11 an X-ray flare was reported shortly before the start of the EHT observations \citepalias{PaperII}. Strongly enhanced flux density variability is seen in the light curves on that day
\citep{Wielgus_2021}, possibly posing difficulties for the static imaging of the April 11 data set. These constraints motivate utilizing the less variable 2017 April 7 data as the primary data set for static image reconstruction, with the April 6 observations as a secondary validation data set. Analysis of the remaining 2017 EHT observations of \sgra will be presented elsewhere.

In Fig.~\ref{fig:uv_coverage}, the $(u,v)$-coverage on 2017 April 7 is shown to be asymmetric, with the longest baselines along the north-south direction.
The shortest baselines in the EHT are intra-site and sensitive to arcsecond-scale structure (i.e., the SMA and JCMT are separated by 0.16\, km; ALMA and APEX are separated by 2.6\,km). In contrast, the longest baselines are sensitive to microarcsecond-scale structure (see Tab.~\ref{tab:resolution}). The $\sim8.7$\,G$\lambda$ detections on PV--SPT and SMT--SPT baselines are among the longest published projected baseline lengths obtained with ground-based VLBI, alongside the recent EHT 3C\,279 results of \citet{Kim_2020}, slightly longer than the longest baselines in the EHT observations of M87* \citepalias[8.3\,G$\lambda$;][]{M87PaperIV}. \sgra was detected on all baselines between stations with mutual visibility, leading to the April 7 $(u,v)$-coverage approaching the best one theoretically possible with the EHT 2017 array. 
\autoref{tab:resolution} shows the angular resolutions derived from \uv-coverage on both April 6 and 7 data.

\begin{table}[t]
    \centering
    \begin{tabular}{@{}lccc}
        \toprule
         & ${\rm FWHM}_{\rm maj}$ & ${\rm FWHM}_{\rm min}$ & P.A. \\ 
                 & ($\mu$as)              & ($\mu$as)              & ($^\circ$)\\ \midrule
        \multicolumn{4}{@{}l}{\noindent Minimum Fringe Spacing $1/|\mathbf{u}|_{\rm max}$ (All Baselines)}\\ 
        \quad April 6 & 24.2 & --- & --- \\[0.5ex] 
        \quad April 7 & 23.7 & --- & --- \\[0.5ex]
        \multicolumn{4}{@{}l}{\noindent Minimum Fringe Spacing (ALMA Baselines)}\\ 
        \quad April 6-7 & 28.6 & --- & --- \\[0.5ex]        
        \multicolumn{4}{@{}l}{\noindent CLEAN Beam (Uniform Weighting)}\\ 
        \quad April 6  & 24.8 & 15.3 & 67.0\\
        \quad April 7 & 23.0 & 15.3 & 66.6\\[0.5ex]  
        \multicolumn{4}{@{}l}{\noindent CLEAN Restoring Beam (Used in This Paper)}\\ 
        \quad April 6-7 & 20 & 20 & --- \\
         \bottomrule
    \end{tabular}
    \caption{Metrics of EHT angular resolution for the 2017 observations of \sgra for the 229.1\,GHz band. 
    In order to avoid asymmetries introduced by restoring beams, and to homogenize the images among epochs, we adopt a circular Gaussian restoring beam with $20\,\mu{\rm as}$ FWHM for all CLEAN reconstructions.
    }
    \label{tab:resolution}
\end{table}

\begin{figure}[t!]
\centering
\includegraphics[trim={0cm 0.0cm 0.0cm 0.cm},clip,width=1.0\linewidth]{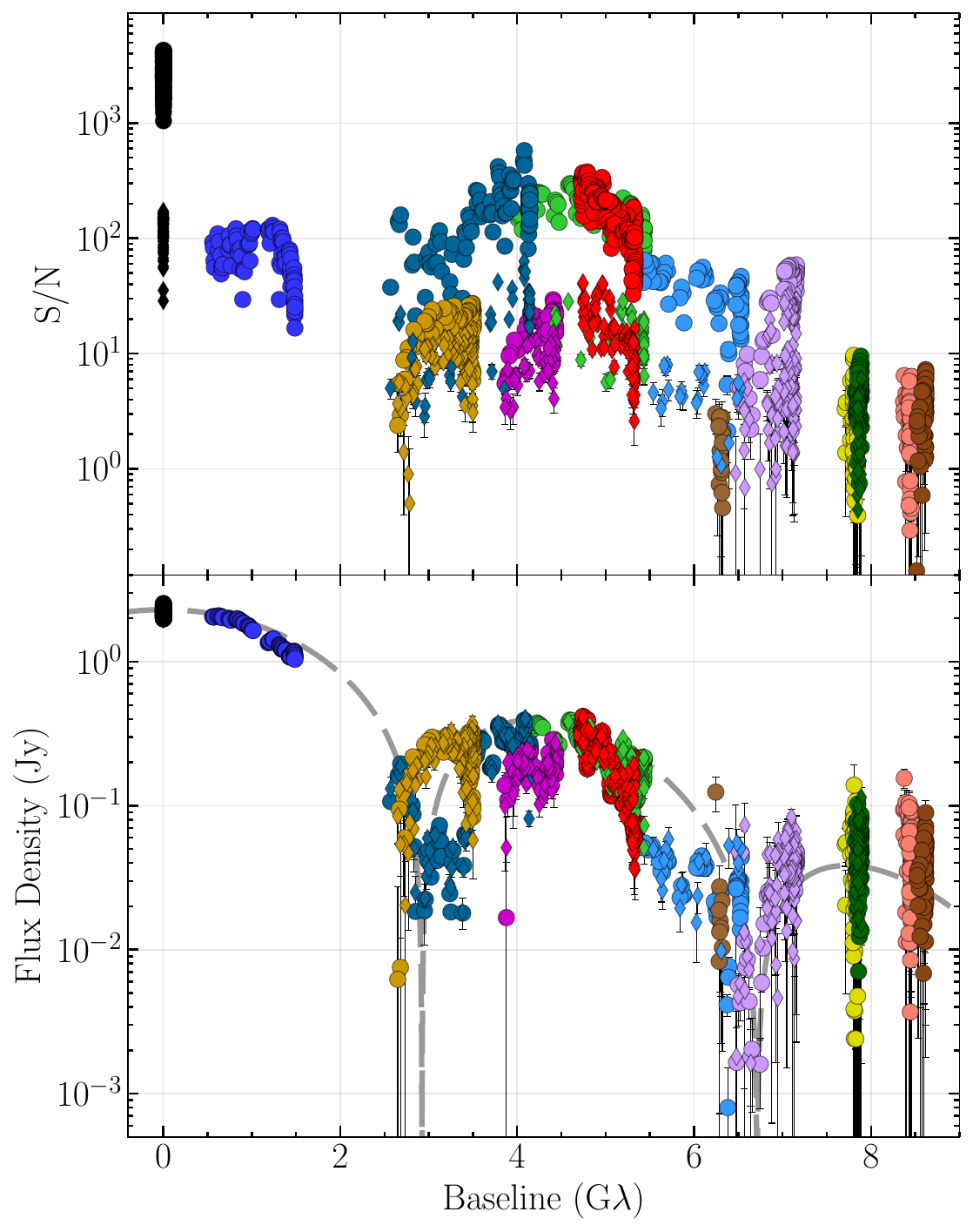}
\caption{Top: signal to noise ratio (\textit{S/N}) as a function of the $(u,v)$-distance (projected baseline length). Low band HOPS data recorded on 2017 April 7, averaged in 120\,s segments, are shown. The data points are color-coded with the baseline, following Fig.~\ref{fig:uv_coverage}. Bottom: fully calibrated visibility amplitude data. 
A model corresponding to a thin ring with a 54\,$\mu$as diameter, blurred with a 23\,$\mu$as FWHM circular Gaussian is overplotted for a reference (dashed curve).}
\label{fig:proc_stages}
\end{figure}

In the top panel of Fig.~\ref{fig:proc_stages} we show in the signal-to-noise ratio of the \sgra observations as a function of projected baseline length, for the coherent averaging time of 120\,s. The split in \textit{S/N} distributions at various projected baseline lengths is due to the difference in sensitivity for the co-located Chile sites ALMA and APEX, with ALMA baselines yielding detections stronger by about an order of magnitude. In the bottom panel, we show the visibility amplitude (correlated flux density in units of Jansky) for \sgra as a function of projected baseline length after applying full data calibration. 

\begin{figure}[t!]
\centering
\includegraphics[trim={0cm 0.0cm 0.0cm 0.cm},clip,width=1.0\linewidth]{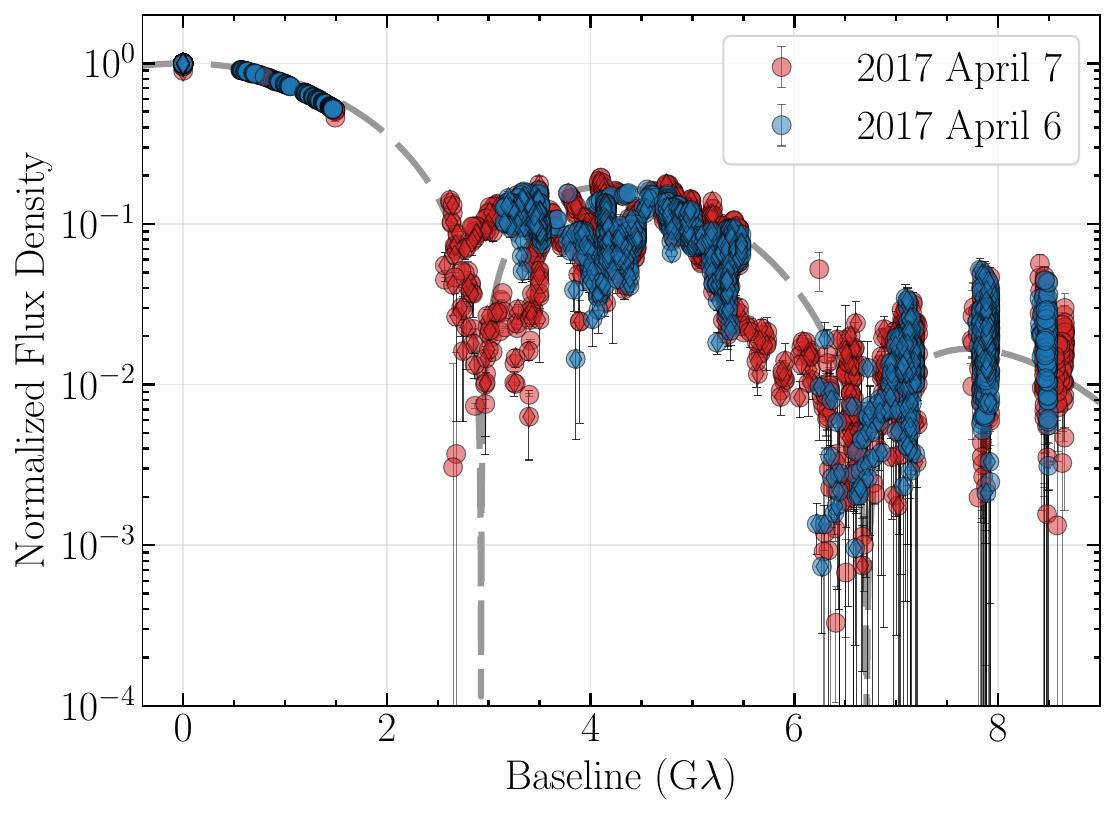}
\caption{Comparison of the light-curve-normalized flux density measurements on 2017 April 6 and 7 in the fully calibrated HOPS data, averaged in 120\,s segments and between low and high bands, are shown.
}
\label{fig:two_days_amps}
\end{figure}

The fully-calibrated visibility amplitudes exhibit a prominent secondary peak between two local minima. %
The first minimum is located at $\sim$3.0\,G$\lambda$ and is probed by the Chile--LMT north-south baselines on 2017 April 7. On April 6 recording started about 2 h later, thus missing the relevant detections, as shown in Fig.~\ref{fig:two_days_amps}. The second minimum appears at $\sim$6.5\,G$\lambda$, probed by the Chile--\hawaii baselines on both 2017 April 6 and 7. Overall amplitude structure of the source appears to be consistent across both days, which is particularly well visible in the fully-calibrated, light-curve-normalized data sets shown in Fig.~\ref{fig:two_days_amps}, as the light-curve-normalization procedure strongly suppresses the large-scale source intrinsic variability \citep{Broderick2022}.
The observed local visibility amplitude minima can be associated with the nulls of the Bessel function $J_0$, corresponding to the Fourier transform of an infinitely thin ring. For a ring that is 54\,$\mu$as in diameter we would obtain local amplitude minima at 2.92\,G$\lambda$ and 6.71\,G$\lambda$.
This is illustrated in the bottom panel of Fig.~\ref{fig:proc_stages}, where an analytic Fourier transform of an infinitely thin ring blurred with a 23\,$\mu$as full width at half maximum (FWHM) Gaussian kernel is shown with dashed lines\footnote{We note that plotting the visibilities from a thin ring over the measured visibility amplitudes is meant only to guide the eye in observing the double null structure. Other geometries, such as a disk model, can also align with the double null structure seen in visibility amplitudes (see \autoref{fig:synthetic_summary}). Detailed fitting of different simple geometries to the visibilities is performed in \citetalias{PaperIV}.}. While a blurred ring model roughly captures the dependence of visibility amplitudes on projected baseline length, there is also a clear indication of the source asymmetry, manifesting as amplitude differences between the Chile--LMT and SMT--\hawaii baselines at the first minimum, probing the same range in projected baseline length ($\sim$2.5-3.5\,G$\lambda$) in orthogonal directions. There is also a deficit of flux density with respect to the simple ring model at projected baseline lengths of $\sim$ 5-6\,G$\lambda$.

\begin{figure*}[t!]
\centering
\includegraphics[trim={0cm 0.0cm 0.0cm 0.cm},clip,width=1.0\linewidth]{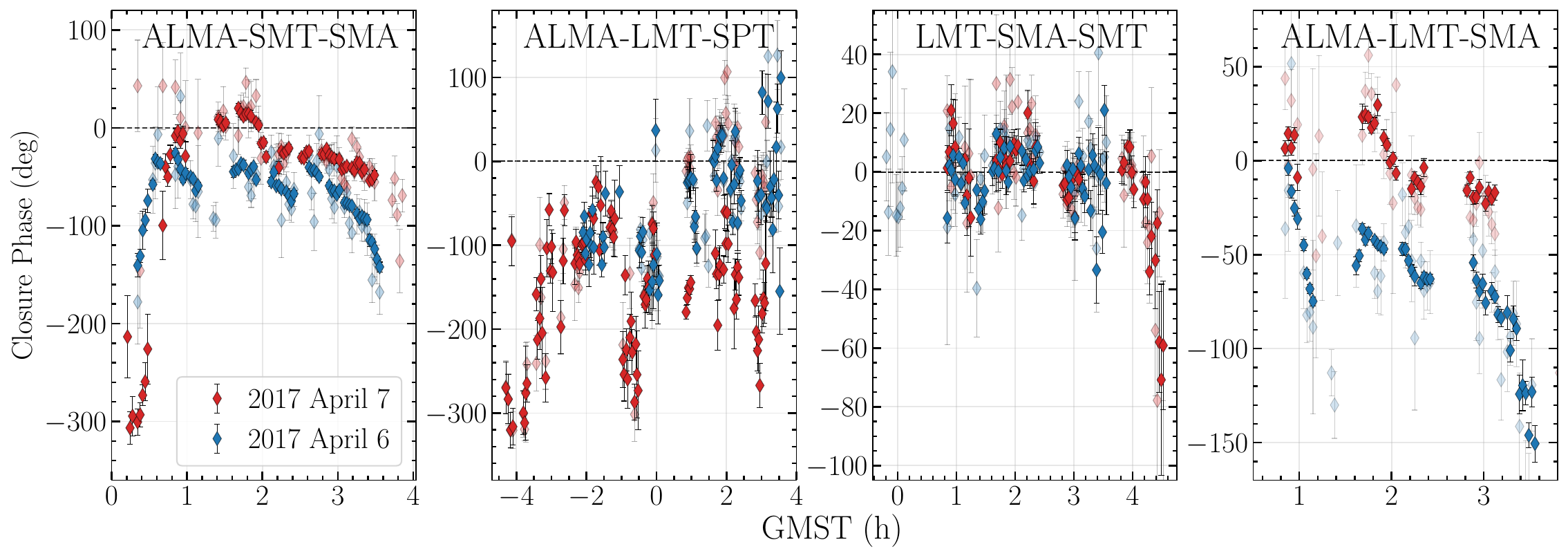}
\caption{Examples of closure phases of \sgra, observed on 2017 April 6 and 7. Semi-transparent points correspond to measurements on redundant APEX/JCMT triangles. Data reduced with the HOPS pipeline, integrated in 120\,s segments, averaged between both bands, are shown. No significant differences are seen with respect to the CASA-based calibration.
}
\label{fig:two_days_cphases}
\end{figure*}

Finally, we detect very clear and unambiguous non-zero closure phases, that are indicative of source asymmetry. With multiple independent triangles and high \textit{S/N}, these data sets offer insight into the source phase structure greatly surpassing that of any previous mm-wavelength observations \citep{Fish_2016,Lu_2018}. In Fig.~\ref{fig:two_days_cphases}, we show examples of closure phases on several triangles exhibiting various degrees of probed asymmetry and inter/intra-day time variability. Closure phases on ALMA-SMT-SMA and ALMA-LMT-SMA triangles immediately show inter-day variability of the source structure. In the case of \sgra, intrinsic source variability is expected also on timescales as short as minutes, adding to the closure phase intra-day variability caused by the non-trivial average structure of the source (see Section \ref{sec:PSD_noise}). Very long baselines, such as LMT-SPT in the ALMA-LMT-SPT triangle, are additionally affected by the presence of refractive scattering structure (see Section \ref{sec:pre_scatter}).

\subsubsection{Station Gain Uncertainties and Non-closing Errors}

Typically, the antenna gains and sensitivities as a function of elevation are derived using polynomial fits to opacity-corrected antenna temperature measurements from quasars and solar system objects, tracked over a wide range of elevations. Residual errors in the characterization of these antenna gains lead to corruptions in the flux calibration of the visibilities. Quantifying these effects enables us to disentangle astrophysical variability of \sgra from apparent flux variations caused by the imperfect calibration. In this work, we mitigate the systematic gain errors in the \sgra data sets based on the analysis of the calibrator sources (J1924-2914 and NRAO\,530), that remained stationary in their source structure and flux density on relevant timescales. 
The detailed procedure to estimate the antenna gains from the two calibrators are described in section 5.1.3 in \citetalias{PaperII}. In particular, it is shown that the {\it a priori} gains are 5-15\% for all baselines, except intrasite baselines ($\sim 1\%$) and those including the LMT ($\sim 35\%$).

In addition to the thermal noise and the antenna gain uncertainties, we also estimate the non-closing errors in the data, based on deviations of the trivial closure quantities from zero, as well as from the observed inconsistencies in the distributions of closure quantities between bands and polarizations \citepalias{M87PaperIII, PaperII}. These non-closing errors are expected to arise from the presence of a small circular polarization component, as well as from uncorrected polarimetric leakages, and other systematic errors, such as residual bandpass effects. For \sgra, the non-closing errors are estimated to be 2\degree ~in closure phase and 8\% in log closure amplitude \citepalias{PaperII}. Assuming the errors are baseline-independent, these translate to 1\degree~systematic non-closing uncertainties in visibility phases and 4\% systematic non-closing uncertainties in visibility amplitudes, on the top of the uncertainties related to the amplitude gains calibration. We found that the $RR-LL$ discrepancies in closure quantities are more significant for \sgra than in the case of the calibrators. This hints at an intrinsic source property, possibly a contribution from a small circular polarization component. This is consistent with \citet{Goddi2021}, reporting $\sim\,1\%$ circular polarization in the simultaneous ALMA-only data.

%% file: pre-imaging.tex
\section{Mitigation of Scattering and Time Variability}
\label{sec::PreImaging_Considerations}
The imaging of Sgr A* at 230 GHz with the EHT is challenged by two important effects: interstellar scattering and short-timescale variability.
In this section, we introduce strategies for mitigating the effects of scattering (Section~\ref{sec:pre_scatter}) and intrinsic variability (Section~\ref{sec:PSD_noise}) adopted in this work.

\subsection{Effects of Interstellar Scattering} \label{sec:pre_scatter}
Fluctuations in the tenuous plasma's electron density along the line of sight causes scattering of the radio waves from \sgra. The scattering properties of \sgra can be well described by a single, thin, phase changing screen $\phi(\textbf{r})$, where $\textbf{r}$ is a two-dimensional vector transverse to the line of sight. 
The electron density fluctuation on the phase screen is typically characterized by a single power-law spectral shape between the outer ($r_{\rm out}$) and inner ($r_{\rm in}$) scale as $Q(\mathbf{q})\propto |\mathbf{q}|^{-\beta}$, where $\mathbf{q}$ is the wavevector of the propagating radio wave and a Kolmogorov spectrum of density fluctuations gives $\beta = 11/3$ \citep{Goldreich_Sridhar_1995}.
The statistical effects of the scattering can then be related to a spatial structure function
$D_{\phi}(\mathbf{r}) = \left< [ \phi (\mathbf{r}+\mathbf{r_0})- \phi (\mathbf{r_0}) ]^2 \right>_{\mathbf{r_0}} \propto \lambda ^2$, where $\left< \cdot \right>_{\mathbf{r_0}}$ denotes the ensemble average over $\mathbf{r_0}$. 

The interstellar scattering of radio waves from \sgra is in the regime of strong scattering, where scintillation is dominated by two distinct effects, {\it diffraction} and {\it refraction}, attributed to widely separated scales \citep[see][]{Narayan_1992,Johnson_Gwinn_2015}. Diffractive scintillation  arises from fluctuations on the scale of the phase coherence (or diffractive scale) given by $D_\phi (\textbf{r}) \sim 1$. It causes rapid temporal variations on a time scale much shorter than 1 s for \sgra, which is also much shorter than the integration time of radio observations. As a result, radio observations measure ensemble averages of the diffractively scattered structure, appearing as the intrinsic structure blurred with the scattering kernel (see \autoref{sec:pre_scatter_diff}). 

Refractive scintillation 
arises from fluctuations on the scale of the scattering kernel much larger than the phase coherence length in the strong scattering regime. For \sgra, the refractive scintillation causes temporal variations of the source images over a time scale of $\sim 1$ d at 1.3\,mm \citep[e.g.][]{Johnson_2018} -- longer than the typical length of radio observations including our EHT observations. Consequently, a single realization of refractive scintillation will be observed by the EHT over an observing run; this will appear as an angular-broadened (i.e., diffractively-scattered) source structure with compact substructure caused by refractive scintillation \citep{Narayan_Nityananda_1986, Johnson_Gwinn_2015, Johnson_Narayan_2016}.

A brief introduction of the expected scattering properties in the EHT 2017 data is described in Section 5.1 of \citetalias{PaperII}. Here we describe the scattering mitigation strategy for the effects of angular broadening by diffractive scattering (\autoref{sec:pre_scatter_diff}) and substructure induced by refractive scattering (\autoref{sec:pre_scatter_ref}). 
To describe scattering effects on \sgra, we use a theoretical framework of these scattering effects developed by \citet{Psaltis_2018}, whose model parameters have been observationally studied by \citet{Johnson_2018}, \citet{Issaoun_2019,Issaoun_2021} and \citet{Cho_2022}. 
For general background and reviews on interstellar scattering, see \citet{Rickett_1990}, \citet{Narayan_1992}, or \citet{TMS}. 

\subsubsection{Mitigation of Angular Broadening}\label{sec:pre_scatter_diff}
Angular broadening is described by a convolution of an unscattered image with a scattering kernel, or equivalently by a multiplication of unscattering interferometric visibilities by the appropriate Fourier-conjugate kernel. 
The Fourier-conjugate kernel is given by $\exp \left[ -\frac{1}{2}D_{\phi}{(\mathbf{b}/(1+\mathcal{M}))} \right]$, where $\mathbf{b}$ is the baseline vector of the interferometer and $\mathcal{M}$ is the magnification of the scattering screen given by the ratio of the observer-to-screen distance to the screen-to-source distance. 
Interferometric measurements of \sgra with the EHT at the observing wavelength of 1.3\,mm are primarily obtained on long baselines of $|\textbf{b}| \gsim (1+\mathcal{M})r_{\rm in}$, or equivalently on angular scales of $\theta \lesssim \lambda / (1+\mathcal{M}) r_{\rm in}$, where $r _{\rm in}$ is the inner scale of the fluctuations. 
In this regime, the angular broadening is affected by the power-law density fluctuations on scales between the inner and outer scales, giving the phase structure function of $D_{\phi} (\mathbf{r}) \propto \lambda ^2 |\mathbf{r}|^{-\alpha}$, where $\alpha = \beta - 2$.

\begin{figure}[t]   %
    \centering
    \includegraphics[width=\columnwidth]{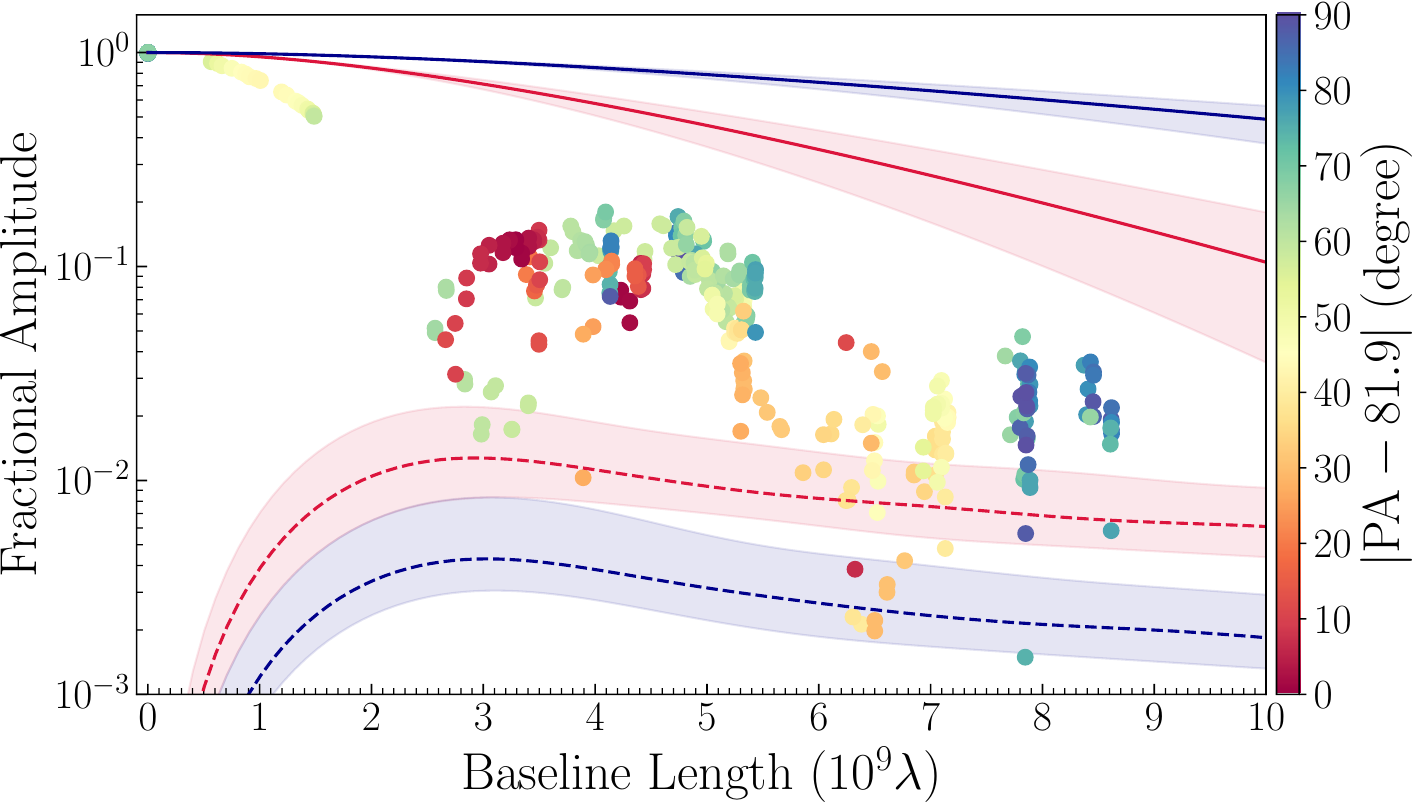}
    \caption{
    Projection of the diffractive scattering kernel (solid lines) and flux-normalized refractive noise amplitudes (dashed lines) at 1.3~mm based on the scattering model of \citet{Johnson_2018}, overlaid on light-curve-normalized \sgra data (low band on April 7). The red and blue lines correspond to the major and minor axes of the scattering, respectively. The associated shaded area indicates the 3$\sigma$ uncertainty of the scattering model in \citet{Johnson_2018}.
    \sgra data are colored by their position angle difference from the major axis of the scattering kernel --- as the points change from red to blue the \uv coordinates move from being closer to the major to minor scattering axis.
    Regardless of the position angle, \sgra amplitudes appear to more rapidly decrease than the scattering kernel, indicating that the intrinsic structure is well resolved against the diffractive angular broadening effects. 
    Additionally, most \sgra amplitudes are above the refractive floor. Thus, refractive effects should only dominate for a small amount of data above $\sim 6\,\mathrm{G}\lambda$.
    }
    \label{fig:scatt_kernel}
\end{figure}

In Figure \ref{fig:scatt_kernel}, we show the scattering kernel in the visibility domains based on the scattering parameters in \citet{Johnson_2018}. 
\citet{Johnson_2018} imply a near-Kolmogorov power-law spectral index $\beta \sim 3.38$ (or $\alpha \sim 1.38$), providing a non-Gaussian kernel more compact than the conventional Gaussian kernel.
Consequently, the angular broadening effect, i.e. multiplication of the intrinsic visibilities with the Fourier-conjugate kernel of scattering, causes a slight decrease in visibility amplitudes and therefore also the \textit{S/N} by a factor of a few at maximum. 

Angular broadening provides deterministic and multiplicative effects on the observed visibility. Therefore, it is invertible --- they can be mitigated by dividing the observed visibility and associated uncertainties by the diffractive kernel visibility \citep[often called {\it deblurring};][]{Fish_2014}. 
However, the actual interferometric measurements of \sgra with the EHT have contributions from substructure that arises from the refractive scattering, often referred as the ``refractive noise.'' 
The refractive noise is stochastic and additive, and therefore not invertible \citep{Johnson_Narayan_2016}.
In fact, since refractive effects are included in the diffractively blurred image, the refractive noise will be amplified by simply deblurring with the scattering kernel, and likely create artifacts in the reconstructions if not accounted for \citep{Johnson_2016}.
To avoid effects from refractive noise we expand the noise budgets of the visibility data prior to deblurring, as described in \autoref{sec:pre_scatter_ref}.

\subsubsection{Mitigation of Refractive Scattering}
\label{sec:pre_scatter_ref}

The contribution of the refractive noise to the observed visibility is anticipated to be not dominant, except for a small fraction of data beyond $\sim 6\,G\lambda$ (\autoref{fig:scatt_kernel})\footnote{The refractive noise in \autoref{fig:scatt_kernel} is estimated for a circular Gaussian source with an intrinsic FWHM of 45\,${\rm \mu}$as under the same condition of interstellar scattering as constrained for \sgra. The size of the Gaussian is broadly consistent with the equivalent second moments of the geometric models in \autoref{sec:synthetic_data} that share similar visibility amplitude profiles with \sgra data.}. %
Signature of the refractive noise, namely the long, flat tail of the visibility amplitude at long baselines found in recent longer wavelength observations of \sgra~\citep{Johnson_Gwinn_2015, Johnson_2018, Issaoun_2019, Issaoun_2021, Cho_2022} is not clearly seen in the EHT data. 
In this EHT regime, where the refractive substructure is not unambiguously constrained from data, it is challenging to apply complex strategies that account for the stochastic properties or explicitly recover the refractive screen \citep[e.g.][]{Johnson_2016, Johnson_2018, Issaoun_2019, Broderick_2020a}.

We instead mitigate the effect of refractive substructure by introducing error budget models that approximate the anticipated refractive noise. The visibility error budget is increased based on these models prior to the mitigation of angular broadening via deblurring (\autoref{sec:pre_scatter_diff}). 
In this work, we consider four base models that approximate the refractive noise budgets:
i. {\tt Const}: a constant noise floor (e.g., $10\,$mJy) for all baselines motivated by the fact that the refractive noise has a mostly flat profile as a function of the baseline length (see \autoref{fig:scatt_kernel}), 
ii. {\tt J18model1}: \uv dependent noise floor based on the scattering model and parameters described in \citet{Psaltis_2018} and \citet{Johnson_2018}. Since the scattered image is not unique, we have simulated hundreds of scattering realizations and generated corresponding synthetic data that matches the \uv-coverage of the actual April 7 observation of \sgra. The refractive noise values are then computed by taking the standard deviation of the complex visibilities across different realizations. Since the refractive noise is also dependent on the intrinsic source structure, in this case we consider a circular Gaussian model with the 2nd moment that matches the pre-imaging size constraints (see~\citetalias{PaperII}),
iii. {\tt J18model2}: Same as {\tt J18model1} but using the average refractive noise value of seven geometric models as possible intrinsic source structures (see Section~\ref{sec:synthetic_data}), 
iv. considering not only the standard deviation of the refractive effects, but their correlations via a covariance matrix as well. %

\begin{figure}
    \centering
    \includegraphics[width=.99\linewidth]{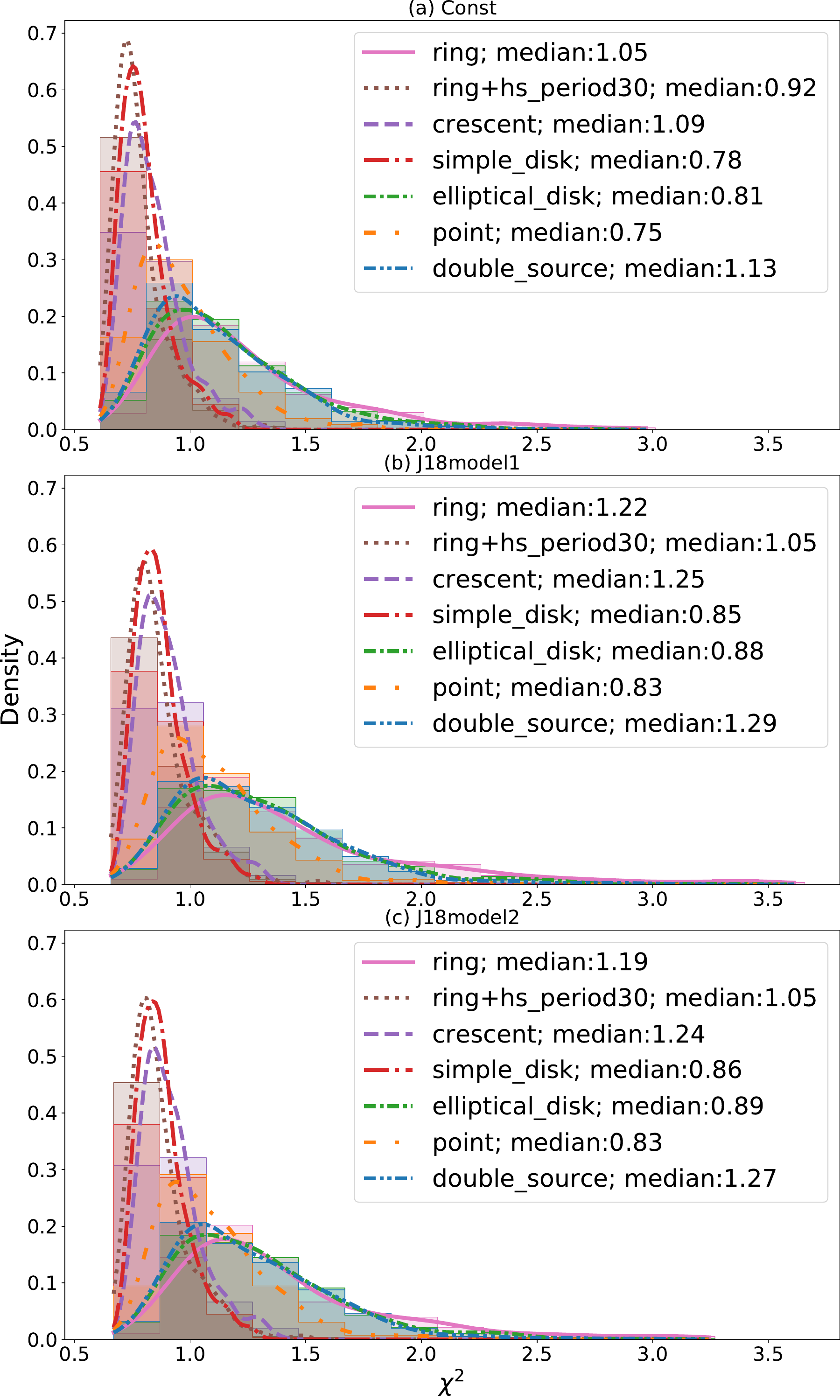}
    \caption{Reduced $\chi^2$ distributions and their density function (in solid lines) for all geometric models of intrinsic source structure using the three different refractive noise models (\texttt{Const}, \texttt{J18model1} and \texttt{J18model2}). For all noise models, the majority of the $\chi^2$s falls in the range between 0.5 and 2.0 (ideally $\chi^2=1.0$), which indicates the ensemble average images provide reasonable fits to the simulated data after taking into account one of the proposed refractive noise budgets.} %
    \label{fig:chisq_metrics}
\end{figure}

Note that using all the information encoded in the covariance matrix (not only in the variance of the refractive noise variables) will provide a better approximation of the refractive noise. However, the short time cadence and the redundant baselines in our data make this covariance matrix non-invertible, and thus difficult to use in imaging. For the remaining three refractive noise models, we compute the complex visibility $\chi^2$ for a suite of synthetic data (corrupted only by thermal noise along with scattering effects) based on seven geometric models of the intrinsic source structure (Section~\ref{sec:synthetic_data}), 
\begin{equation}
    \chi^{2}\left(\sigma_\mathrm{ref,i}\right) = \frac{1}{2N_\mathrm{vis}}\sum_\mathrm{i=1}^{N_\mathrm{vis}}\frac{\lvert V_\mathrm{i} - V_\mathrm{ea}\rvert^2}{\sigma_\mathrm{th,i}^{2} + \sigma_\mathrm{ref,i}^{2}},
    \label{eq:ref_chisq}
\end{equation}
where $V_i$ is the data visibilities for each scattering realization, $V_{ea}$ is the ensemble averaged visibilities (i.e., corresponding to the image experiencing only diffractive scattering), $\sigma_\mathrm{th,i}$ is the thermal noise, and $\sigma_\mathrm{ref,i}$ is the corresponding refractive noise budget for each strategy. 
The $\chi^2$ metric provides us with a statistic on how well the ensemble average image represents the synthetic data after taking into account the different modeled refractive noise budgets.
Figure~\ref{fig:chisq_metrics} shows a comparison of the different $\chi^2$ distributions for 400 realizations of synthetic data for every scattering mitigation strategy. 
For {\tt J18model1} we have derived a scaling factor to make the median of $\chi^2$ of all models equal to 1 in order to overcome the dependence of the refractive noise level on the intrinsic source structure (see Appendix~\ref{appendix:ref_scatt}).

Figure~\ref{fig:chisq_metrics} demonstrates that the ({\tt Const}, {\tt J18model1}, and {\tt J18model2}) refractive noise models result in reasonable $\chi^2$s for the simulated scattering realizations we tested, with a majority of the $\chi^2$ values below 3.0 (ideally $\chi^2=1$). For this reason, in the rest of the paper we focus on the simpler \texttt{Const} and {\tt J18model1} strategies.  %

\subsection{Effects of Time Variability}
\label{sec:PSD_noise}

With a gravitational timescale of only $G M/c^3 \approx$ 20 s, \sgra is expected to be able to exhibit substantial changes in its emission structure on timescales of a few minutes or less.  A single multi-hour observing track is thus sufficiently long for \sgra to significantly alter its appearance, potentially hundreds of times.  Such structural variability violates a core assumption of Earth-rotation aperture synthesis -- namely, that the source structure must remain static throughout the observation -- and necessitates modifications to standard imaging practices.

\subsubsection{Evidence for Source Variability in the Data}

The EHT \sgra data contain unambiguous signatures of an evolving emission structure.  On the largest spatial scales, the light curve varies at the $\sim$10\% level on timescales of $\sim$hours \citep{Wielgus_2021}.  Variations in excess of those expected from thermal noise are seen on timescales as short as $\sim$1 min.  Over timescales typical of observation scans, $\sim$10~min, the degree of variation is on the order of 5\% \citep{Wielgus_2021}. %

Direct evidence for short-timescale structural variations may be found in the evolution of closure quantities. Closure phases measured on certain triangles of baselines (e.g., ALMA-SMT-SMA) exhibit significant differences between April 6 and 7 (see \autoref{fig:two_days_cphases}).  The variations seen in the closure phases measured on multiple triangles show significant excesses, relative to thermal noise levels, as captured using the $\mathcal{Q}$-metric statistic (\citealt{Roelofs_2017}, \citetalias{PaperII}).

Non-parametric estimates of the degree of variability as a function of baseline length may be generated by inspecting the visibility amplitudes directly.  This is made possible by two fortuitous facts: first, the existence of crossing tracks, and second, that \sgra was observed on multiple days.  As a result, many presumably independent realizations of the source structure may be compared.
Practically, this is obtained by collecting visibility amplitudes in \uv bins, linearly detrending to remove 
the contribution from the static component of the image, and computing the mean and variance of the residuals.  We average these estimates azimuthally to improve the significance of variability detection. For details on the procedure and validation examples, we direct the reader to \citet{Georgiev_2021} and \citetalias{PaperIV}.

In the case of \sgra, this non-parametric estimate produces a clear detection of variability that is significantly in excess of the expected thermal noise \citepalias{PaperIV}. 
Within the range of baseline lengths over which meaningful estimates can be produced, roughly $2~{\rm G}\lambda<|u|<6~{\rm G}\lambda$, the observed excess variability is broadly consistent with that anticipated by GRMHD simulations, both in magnitude and dependence upon baseline length \citepalias{PaperIV,PaperV}.

\subsubsection{Strategies for Imaging Variable Data}
\label{sec:stratvariabiledata}

Strategies for imaging in the face of source variability can be classified into one of three general categories:%

\begin{enumerate}
    \item Variability reconstruction, or ``dynamic imaging,'' in which the evolution of the source emission structure is explicitly recovered during the imaging process.  The output of this strategy is a movie of the source emission structure. We refer the reader to \autoref{sec:dynamicimagingmethods} for more discussion on methods for dynamic imaging.
    \item Variability circumvention, or ``snapshot imaging,'' in which standard image reconstruction is performed on segments of data (``snapshots'') that are sufficiently short that the source may be approximated as static across them.  The output of this strategy is a time series of static images.
    \item Variability mitigation, or ``variability noise modeling,'' in which the impact of structural changes in the visibilities is absorbed into an appropriately inflated error budget.  The output of this strategy is a single static image of the source, indicative of the time-averaged image over this observation period.
\end{enumerate}

In practice, for segments of data short enough that \sgra may be reasonably approximated as static, the \uv-coverage of the EHT is insufficient to support reliable snapshot image reconstruction (though more restrictive parameterizations of the source structure, such as permitted using geometric modeling, can still be applied; see Section~\ref{sec::dynamic} and \citetalias{PaperIV}).  However, because dynamic imaging enforces a degree of temporal continuity, it is able to leverage the information provided by densely-covered intervals of time to augment the lack of information available during intervals of sparser coverage. Dynamic imaging can thus be thought of as a generalization of both standard (static) imaging and snapshot imaging, with the former being equivalent to dynamic imaging with maximal temporal continuity enforcement and the latter being equivalent to dynamic imaging with no temporal continuity enforcement at all. Because dynamic imaging falls in between these two extremes, it can potentially recover reliable source structure in regions where the data are both too variable for standard imaging and too sparse for snapshot imaging.  Our efforts to perform dynamic imaging in the most densely \uv-covered regions of data are described in \autoref{sec::dynamic}.

The third strategy listed above -- the variability noise modeling approach -- permits static images to be reconstructed even from time-variable data.  Depending on the specifics of the implementation, the recovered image captures some representation of ``typical'' source structure.  Imaging with variability noise modeling requires that the error budget of the data be first inflated in a way to capture the statistics -- or ``noise'' -- of the source variability.  The specific form of the noise model we use in this work is a broken power law, for which the variance $\sigma^2_\text{var}$, as a function of baseline length $|u|$, takes the form
\begin{equation}
\sigma_{\text{var}}^2(|u|) = a^2 
\left(\frac{|u|}{4~\text{G}\lambda}\right)^c
\frac{1 + \left[ (4~\text{G}\lambda)/u_0 \right]^{b+c}}{1 + \left( |u| / u_0 \right)^{b+c}}.
 \label{eq:PSD_noise}
\end{equation}
Here, $|u| \equiv \sqrt{u^2+v^2}$ is the dimensionless length of the baseline located at \uv, $u_0$ is the baseline length corresponding to the break in the power law, $a$ is the variability noise amplitude at a baseline length of 4\,G$\lambda$, and $b$ and $c$ are the long- and short-baseline power-law indices, respectively. 
 \autoref{eq:PSD_noise} represents the variance that is associated with structural variability after removing the mean and normalizing by the light curve; see \citetalias{PaperIV} for details.

By adding the variability noise given by \autoref{eq:PSD_noise} in quadrature to the uncertainty of every visibility data point, the image becomes constrained to fit each data point to only within the tolerance permitted by the expected source variability.  This parameterized variability noise model is generic and can explain well a wide range of source evolution, including complicated physical GRMHD simulations of \sgra \citep{Georgiev_2021}.

\citetalias{PaperIV} presents a non-parametric analysis of \sgra's variability, which is further inspected to provide ranges of broken-power-law model parameters that fit \sgra data \citep[see][for details]{Georgiev_2021}.
As explained in \citetalias{PaperIV}, given the baseline coverage of the 2017 \sgra campaign, little traction is found on the location of the break, $u_0$, and the short-baseline power law, $c$. %
However, the amplitude $a$ is well constrained with an interquartile range from 1.9\% to 2.1\%. %
Similarly, the long-baseline power law, $b$, is modestly constrained, with interquartile range 2.2 to 3.2. %
These interquatile ranges are used to provide approximate priors on the variability noise that should be considered during static imaging reconstruction; values employed are listed in \autoref{tab:premodeling} for \sgra as well as for a number of synthetic data sets described in \autoref{sec:synthetic_data}.
In the case of the later, theoretical considerations imply that under very general conditions, $c\approx2$, and thus we adopt a general prior of $[1.5,2.5]$.
For the CLEAN (\autoref{sec:cleanmethods}) and regularized maximum likelihood (RML) (\autoref{subsec:rml_static}) imaging surveys described in Section~\ref{sec:survey}, variability noise models in the identified \sgra range are added to the visibility noise budget before static imaging (for both synthetic and real \sgra data). %
For the \themis imaging method (\autoref{sec:Themis}), the parameters of the noise model are fit simultaneously with the image structure, subject to the data-set-specific values in \autoref{tab:premodeling} being used to define a uniform prior over each data set.

\begin{deluxetable}{lccc}
\tablecaption{Variability Noise Model Ranges used for Static Imaging  \label{tab:premodeling}}
\tablehead{
\colhead{Source\tablenotemark{\dag}} & 
\colhead{$a$ (\%)} &
\colhead{$b$\tablenotemark{\ddag}} & 
\colhead{$u_0$\tablenotemark{\ddag} (${\rm G}\lambda$)}
}
\startdata
          \sgra & $[1.9,2.1]$ & $[2.2,3.2]$ & $[0.37,1.45]$ \\
\hline
           Ring & $[0.7,0.8]$ & $[1.9,3.0]$ & $[0.18,1.06]$ \\
        Ring+hs & $[0.6,0.7]$ & $[3.0,4.1]$ & $[0.33,0.90]$ \\
       Crescent & $[0.9,1.1]$ & $[2.4,3.4]$ & $[0.29,1.15]$ \\
    Simple Disk & $[0.6,0.7]$ & $[2.3,3.3]$ & $[0.18,0.82]$ \\
Elliptical Disk & $[0.6,0.7]$ & $[2.1,3.0]$ & $[0.16,0.77]$ \\
          Point & $[0.8,0.9]$ & $[1.8,3.1]$ & $[0.32,2.67]$ \\
         Double & $[0.8,0.9]$ & $[1.9,2.9]$ & $[0.19,1.09]$ \\
          GRMHD & $[2.4,2.7]$ & $[2.6,3.8]$ & $[0.56,1.78]$
\enddata
\tablenotetext{\dag}{All sources include high- and low-bands on observation days April 5, 6, 7, and 10.}
\tablenotetext{\ddag}{Interquartile (25\%-75\% percentile) ranges based on non-parametric analysis of suprathermal fluctuations in the visibility amplitudes on a per-scan basis.}
\end{deluxetable}

%% file: imaging_methods.tex
\section{Imaging Methods for \sgra}
\label{sec::Background}

Recovering an image of \sgra from interferometric measurements amounts to solving an inverse problem. This inverse problem is challenging because of four primary reasons: 1) the interferometer incompletely samples the visibility domain, 2) there is significant structured noise included in the visibility measurements, 3) the source structure is evolving over the duration of the observation, and 4) the source is both diffractively and refractively scattered. The methods used in \citetalias{M87PaperIV} to recover an image of \m87 from interferometric measurements had to address challenges 1 and 2 above; challenges 3 and 4 are unique to the rapidly evolving \sgra source, which we observe through the interstellar medium. Strategies to mitigate the effects of scattering and time variability are discussed in detail in Section~\ref{sec::PreImaging_Considerations}. In this section we assume that the data have already been modified by the appropriate descattering strategy and variability noise budget prior to imaging\footnote{In \themis static imaging and dynamic imaging the variability noise is not included before imaging. In \themis the variability noise budget is estimated along with the image. In dynamic imaging variability noise is not included.}.

To choose among the possible \sgra images, additional information, assumptions, or constraints must be included when solving the inverse problem. We broadly categorize imaging algorithms into three methodologies: CLEAN, regularized maximum likelihood, and Bayesian posterior sampling. We summarize these approaches, but refer the reader to  \citetalias{M87PaperIV} for a more complete discussion of static imaging methods for EHT data. Additionally, we introduce the idea of dynamic imaging, which aims to reconstruct a movie rather than a single static image over an observation.

\subsection{CLEAN Static Imaging}
\label{sec:cleanmethods}

Traditionally, radio interferometric images have been made using non-linear deconvolution algorithms of the CLEAN family \citep[e.g.,][]{Hogbom_1974, Schwarz:1978, Clark_1980, Schwab:1984, Cornwell:1999a, Cornwell:2008}. These algorithms iteratively deconvolve the effects of the limited sampling of the $(u,v)$ plane, i.e., the interferometer's point source response (also known as \emph{dirty beam}) from the inverse Fourier transform of the visibilities (\emph{dirty image}). 

The classical CLEAN algorithm assumes that the sky brightness distribution can be represented as a collection of point sources. The imaging process consists of rounds of locating the brightness peak in the dirty image, generating a point source (\emph{CLEAN component}) at this location with an intensity of some fraction of the map peak, and either convolving the CLEAN component with the dirty beam and subtracting it from the dirty image \citep{Hogbom_1974,Clark_1980} or subtracting the CLEAN components directly from the ungridded visibilities \citep{Schwab:1984}. This is continued until some specified cleaning stopping criterion is reached. One can supplement the process by restricting the area in which the peaks are searched (so-called \emph{CLEAN windows}). This limits the parameter space in fitting and is especially important for data with sparse $(u,v)$ sampling. The final image is made by convolving the obtained set of CLEAN components with a Gaussian restoring beam to smooth out the higher spatial frequencies and adding the last residual image to represent the remaining noise.

After image deconvolution, further improving of the image quality can be achieved using \textit{self-calibration}, which uses the current image estimate to apply a correction to amplitude and phase information.
Self-calibration is usually applied as part of an iterative process following each CLEAN iteration.

In this work we implement the CLEAN method using the \difmap pipeline described in Section~\ref{sec:survey} and Appendix~\ref{appendix:difmap-pipeline}.

\subsection{RML Static Imaging}

\label{subsec:rml_static}

The general approach in Regularized Maximum Likelihood static imaging methods is to find an image, $\hat{I}$, that minimizes a specified objective function. As described further in \citetalias{M87PaperIV}, by using $\chi^2\left(I, V\right)$ as a measure of the inconsistency of the image, $I$, with the measurements, $V$, we can specify the objective function: 
\begin{align}
 \label{eq::objfunc}
 J(I) = \sum_{\mathclap{\text{data terms}}} \alpha_{\textrm{D}} \chi^2_{\textrm{D}}\left(I, V\right) - \sum_{\mathclap{\text{regularizers}}} \beta_{\textrm{R}} S_{\textrm{R}}\left(I\right).
\end{align}
In this expression, the $\chi^2_{\textrm{D}}$'s are goodness-of-fit functions corresponding to the data product $D$, and the $S_{\textrm{R}}$'s are regularization terms corresponding to the regularizer $R$. 
Maximum entropy~\citep{Narayan_1986, Chael_2016}, total variation, and sparsity priors~\citep{Wiaux_2009a,Wiaux_2009b,Honma_2014,Akiyama_2017a} have all been used to define $S_{\textrm{R}}(I)$ and have been demonstrated in the interferometic imaging of \m87~\citepalias{M87PaperIV}.
The $\chi^2_{\textrm{D}}(I, V)$ and $S_{\textrm{R}}(I)$ terms often
have different preferences for the “best” image, and compete
against each other in minimizing $J(I)$. Their relative impact in
this minimization process is specified with the hyperparameters $\beta_{\textrm{R}}$. 

This expression can be interpreted probabilistically when
$\exp \left(-\sum \alpha_{\textrm{D}} \chi^2_{\textrm{D}}\left(I, V\right) \right) \propto p(V|I)$ and $\exp \left( \sum \beta_{\textrm{R}} S_{\textrm{R}}\left(I\right) \right) \propto p(I)$. %
In this case, minimizing the cost function $J(I)$ is equivalent to maximizing the log-posterior $\log p(I|V)$. 
Not all regularizer cost functions $S_{\textrm{R}}$ correspond to a formal probability distribution. Nonetheless, while not all RML methods have a probabilistic interpretation, their formulation leads to a similar optimization setup. %

For the EHT, RML methods have an advantage of being able to naturally constrain closure data products that are insensitive to atmospheric noise that corrupts EHT visibilities~\citep{Bouman_2016,Chael_2016,Akiyama_2017a,Chael_2018_Imaging}.
In this work we implement RML methods using the \ehtim and \smili pipelines described in Section~\ref{sec:survey} and Appendices~\ref{appendix:ehtim-pipeline} and~\ref{appendix:smili-pipeline}.

\subsection{Bayesian Full Posterior Static Imaging} \label{sec:Themis}

A fully Bayesian approach to imaging is a natural extension of the RML approach to image reconstruction.
The primary output of Bayesian methods is an image posterior, i.e., not only a single ``best fit'' image but the family of images that are consistent with the underlying visibility data.  In this way, the  Bayesian image posterior encapsulates both the typical image reconstruction and its aleatoric (e.g., statistical) uncertainty, permitting quantitative analyses of the robustness of image features, array calibration quantities, and the relationships between each (see, e.g., \citealt{Broderick_2020b, 2019Arras, sun2021, DMC}; \citetalias{M87PaperVII}).

We employ the general modeling framework \themis, developed specifically to compare parameterized models with the VLBI data produced by the EHT \citep{Broderick_2020a}.  The image model consists of three conceptually distinct components: a description of the brightness distribution on the sky, the variable complex gains at each station, and the additional ``noise'' associated with intra-day structural variability in the source.  
Scan-specific complex station gains and variability ``noise'' parameters are recovered and marginalized over simultaneously with image exploration.
Details on the model construction, adopted priors, sampling methods, and fidelity criteria are collected in \autoref{sec:themisdescription}, and are only briefly summarized here.

We make use of the adaptive splined raster models within \themis, consisting of a set of brightness control points that may vary in brightness on an adjustable rectilinear grid.  In practice, only a handful of resolution elements are required (see \autoref{subsec:themis_survey} and \autoref{sec:themis_survey_details}), and full images are produced via an approximate cubic spline
\citep{Broderick_2020b}.
The dimensions of the raster, $N_x$ and $N_y$, are selected based upon the Bayesian evidence as discussed further in \autoref{subsec:themis_survey}.

The combined parameter space, comprised of the brightness control points, raster size and orientation, complex station gains, and noise model parameters are sampled via a parallely tempered, Hamiltonian Monte Carlo scheme, producing a chain of candidate images and ancillary quantities that are distributed according to their posterior probability, $p(I|V)$. In practice, the sampler must explore the parameter space sufficiently to produce an accurate reproduction of the posterior, often referred to as ``convergence,'' which we assess via standard convergence criteria.
A fully converged Markov chain will have identified all available image modes that can be captured by the specified image representation, and assessed their relative likelihoods.

\subsection{Dynamic Imaging}
\label{sec:dynamicimagingmethods}

As discussed in Section~\ref{sec:stratvariabiledata}, the quickly evolving structure of \sgra poses significant challenges in reconstructing an image. Imaging techniques traditionally rely on Earth rotation aperture synthesis, which are based on the fundamental assumption that the target being imaged remains stationary during the whole duration of the observation. This is no longer valid when the target source is expected to exhibit significant structural changes in time scales smaller than the observing run; thus for static imaging we must incorporate an inflated ``variability noise budget" to capture the ``typical" source structure (refer to Section~\ref{sec:stratvariabiledata}). If we instead wish to capture the evolving structure of \sgra we can attempt to recover a full movie from the data, rather than just a single image.

Extensions of the CLEAN approach have been proposed to address time-variable sources~\cite{stewart2011multiple, 2012SPIEcleanvariable, Farah_2021}. In~\cite{miller2019rapidly} evolution of the microquasar V404 Cygni was reconstructed using model fitting in \difmap. 
\cite{2019Arras} developed a variational inference approach for dynamic imaging that was used to simultaneously reconstruct images of M87$^{*}$ over four nights from EHT 2017 data.
In this work we focus on methods that explicitly incorporate temporal regularization to allow for recovery of evolving sources with complex spatial structure in the presence of especially sparse \uv-coverage.

\subsubsection{RML Dynamic Imaging}
\label{sec:rmldynamicimaging}

Extending the RML approach from static to dynamic imaging is simple conceptually. 
Rather than solving for a single image, $\hat{I}$, our new goal is to solve for a series of $K$ images $\hat{\{I_k\}}$. Each of these images correspond with small segments of data, which have been divided to have a time duration similar to the  expected time variability of the target (typically tens of minutes for \sgra). Since the \uv-coverage of each data segment is severely limited, we must include an additional term that regularizes the images ${\{I_k\}}$ in time rather than just space. A general prescription in terms of the temporal regularizer $S_{\textrm{Q}}$ can be written mathematically as:  
\begin{align}
 \label{eq::objfunc}
\notag J(\{I_k\}) = & \sum_{\mathclap{\text{data terms}}} \alpha_{\textrm{D}} \chi^2_{\textrm{D}}\left(\{I_k\}, V\right) \\ &- \sum_{\mathclap{\text{spatial reg.}}} \beta_{\textrm{R}} S_{\textrm{R}}\left(\{I_k\}\right) - \sum_{\mathclap{\text{temporal reg.}}} \beta_{\textrm{Q}} S_{\textrm{Q}}\left(\{I_k\}\right).
\end{align}
The additional temporal regularization terms, $S_{\textrm{Q}}$, encourages smooth evolution of the target over the full observation. Descriptions of temporal regularizers and their application to EHT data are described in~\cite{johnson2017dynamical}.

In Appendix~\ref{sec:smili_dynam_survey} the RML Dynamic Imaging method is used to explore the structure of \sgra over the course of a night independent of the variability noise model introduced in Section~\ref{sec:stratvariabiledata}.

\subsubsection{\texttt{StarWarps} Dynamic Imaging}
\label{subsec:starwarps}
\texttt{StarWarps} is a dynamic imaging method based on a probabilistic graphical model~\citep{Bouman_2018}. 
Similar to RML Dynamic Imaging, \texttt{StarWarps} makes use of temporal regularization to solve for the frames of a movie $\hat{\{I_k\}}$ over an observation rather than a static image.
In contrast to RML, \texttt{StarWarps} independently solves for the marginal posterior of each frame conditioned on all measurements in time; the reconstructed movie $\hat{\{I_k\}}$ is the mean of each marginal distribution. 
The advantage of \texttt{StarWarps} with respect to RML is that, when using a linearized measurement model, \texttt{StarWarps} can solve for the frames of a video $\hat{\{I_k\}}$ with exact inference -- resulting in a better behaved optimization problem that is less likely to get stuck in local minima when compared to RML Dynamic Imaging.

\texttt{StarWarps} defines a dynamic imaging model for
observed data using the following potential functions:
\begin{align}
    \psi_{V_k|I_k} &= \mathcal{N}_{V_k}(f_k(I_k), \sigma_k^2) \\
    \psi_{I_k} &= \mathcal{N}_{I_k}(\mu, \Lambda) \\
    \psi_{I_k | I_{k-1}} &= \mathcal{N}_{I_k}(I_{k-1}, \beta_{Q}^{-1} \mathbbm{1}) \label{eq:starwarpsreg},
\end{align}
for a normal distribution $\mathcal{N}(m,C)$ with mean $m$ and matrix covariance or scalar standard deviation $C$. 
Each set of observed
data $V_k$ taken at time $k$ is related to the underlying image, $I_k$, through the measurement model,
$f_k(I_k)$ (e.g., visibility model, closure phase model). Spatial regularization is imposed through the second potential; $I_k$ is encouraged to be a sample from a multivariate Gaussian distribution with mean $\mu$ and covariance $\Lambda$. 
In this work, we define $\Lambda$ to encourage spatial smoothness with a spectral distribution profile $(u^2 + v^2)^{-a/2}$ controlled by hyperparameter $a$, as described in detail in~\cite{Bouman_2018}.
The third potential describes how images evolve over time; as $\beta_{Q}$ increases the temporal regularization increases and visa versa.
Although more complex evolution models are described in~\cite{Bouman_2018}, in this paper we restrict ourselves to a simplified evolution model that encourages only small changes between adjacent frames $I_{k-1}$ and $I_k$.

The joint probability distribution of this dynamic model can be written as:
\begin{align}
    p(\{ I_k \}, \{ V_k \} ) = p(I_1) \prod_{k=1}^K p(V_k|I_k) \prod_{k=2}^K p(I_k | I_{k-1})
\end{align}
where $p(I_1) = \psi_{I_1}$, $p(V_k|I_k) = \psi_{V_k|I_k}$, and $p(V_k|I_k) \propto \psi_{I_k} \psi_{I_k|I_{k-1}}$. In the case of a linear measurement model, $f(I)$, (e.g., complex visibility model) the expected value of every $I_k$ conditioned on all data $V=\{ V_k \}$ can be solved in closed-form efficiently using the Elimination Algorithm. However, in the case of complex gain errors the measurement model is no longer linear. By linearizing the model we can solve in closed form for a linearized solution, $\hat{\{I_k\}}$. We then iterate between linearizing the measurement model around our current solution and solving the linearized solution in closed-form until convergence.

The \texttt{StarWarps} method is used in Section~\ref{sec::dynamic} alongside snapshot geometric modeling methods to help analyze the short-timescale variations of \sgra over a $\sim100$ min region of time on April 6 and 7. 

%% file: synthetic_data.tex
\newcommand{\fr}[1]{\textcolor{blue}{FR: #1}}
\newcommand{\al}[1]{\textcolor{red}{AL: #1}}

\section{Synthetic Data}
\label{sec:synthetic_data}

\begin{figure*}[t]
    \centering
    \includegraphics[width=1.\textwidth]{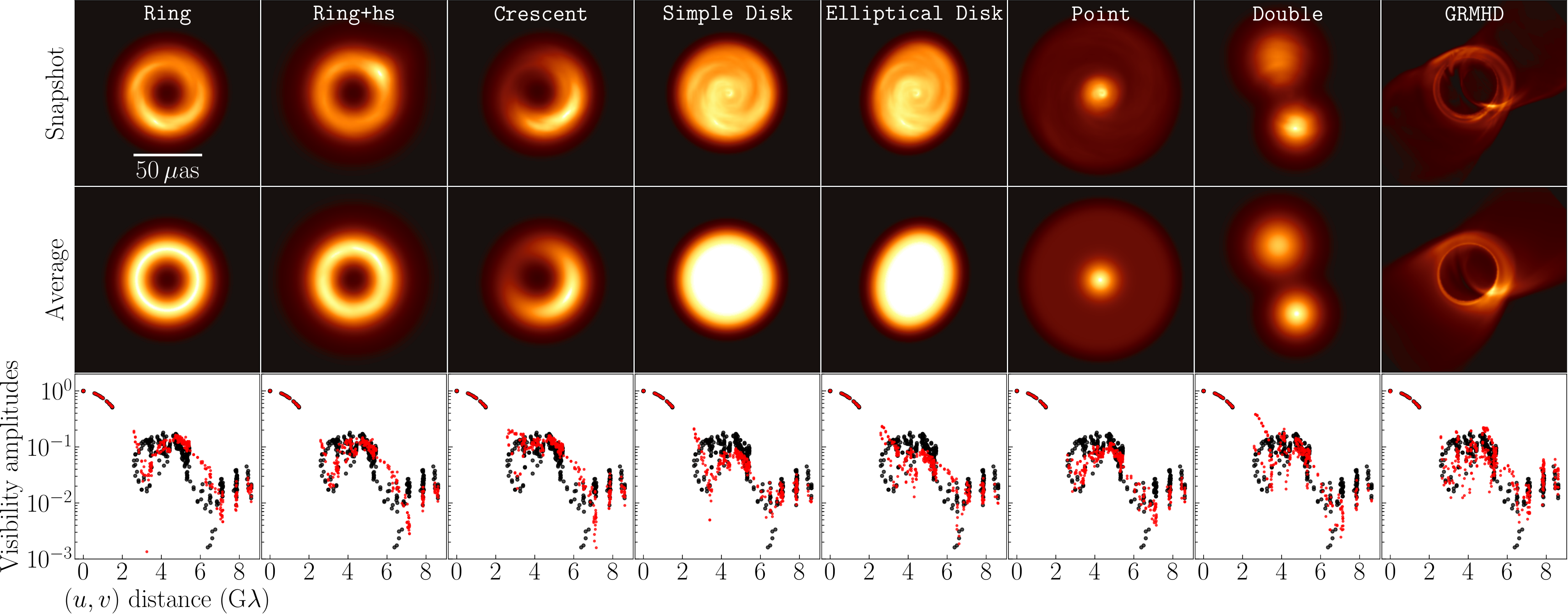}
    \caption{Eight synthetic models and corresponding visibility amplitudes. From left to right, we show the seven geometric models and the single GRMHD model. \textit{Top panels}: a single frame of the eight synthetic movies highlighting the effect of temporal variability. 
    \textit{Middle panels}: time-averaged images illustrating the static component of the source structure. \textit{Bottom panels}: a comparison of the simulated visibility amplitudes (red) and real \sgra measurements (black) as a function of projected baseline length. 
    }
    \label{fig:synthetic_summary}
\end{figure*}

While imaging is a powerful tool to identify the source morphology without a specific source model, reconstructed images obtained with the techniques described in \autoref{sec::Background} are sensitive to hyper-parameter and optimization choices (in this paper, often referred to simply as parameter choices). For instance, in RML imaging methods, a common design choice is the type of regularizers and how much weight to assign the regularization terms relative to data fitting terms. In CLEAN, common design choices include the location of CLEAN windows and the initial model used for self-calibration. Reconstructed images can be sensitive to these choices, especially when the data constraints are severely limited, as is the case in the sparse EHT measurements of \sgra. 

In the second half of 2019, images of \sgra were initially reconstructed by five teams that worked independently of each other to identify the morphology of \sgra through imaging. 
As summarized in \autoref{sec:first_images}, the five independent teams identified a \sdiam feature, but with a significant uncertainty in the detailed morphology. While many of the images contained a ring structure, some of the teams obtained non-ring images that also reasonably fit the data. Furthermore, the flux distribution around the recovered rings showed large variation across different reconstructions. 
These initial images motivated a series of tests presented in this paper to systematically study the possible underlying source structure of \sgra. 

To systematically explore and evaluate the imaging algorithms design choices and their effects on the resulting image reconstructions, we generated a series of synthetic data sets. The synthetic data were carefully constructed to match properties of \sgra  EHT measurements. The use of synthetic data enables quantitative evaluation of image reconstruction by comparison to the known ground truth. This in turn enables evaluation of the design choices and imaging algorithms performance (\autoref{sec::Background}).
As summarized in \autoref{fig:synthetic_summary} two sets of time-variable synthetic data were generated for slightly different purposes. The first set are the geometric models (\autoref{sec:synthetic_data:geometric}), which were used to both assess the capability of identifying and distinguishing different morphologies as well as to select optimal design choices and parameters (for RML and CLEAN) that perform well across the entire data set. The second data set is the GRMHD model which was used to evaluate imaging performance on physically motivated models of \sgra (see \autoref{sec:survey}).

Data sets were generated using \ehtim's simulation library with a \uv-coverage identical to \sgra measurements. Prior to the synthetic observations, all movies were scattered based on the best-fit model of \citet{Johnson_2018} (see \autoref{sec:pre_scatter} for details). 
The observed visibilities were further corrupted by thermal noise, amplitude gains, and polarization leakage, consistent with \sgra data \citepalias{PaperII}.
Atmospheric phase fluctuations were simulated by randomizing the visibility phase gains on a scan-by-scan basis.

\subsection{Geometric Models}
\label{sec:synthetic_data:geometric}
To assess the capability of identifying source morphology, seven geometric models were used to generate synthetic data (Figure \ref{fig:synthetic_summary}). 
As described in \autoref{subsec:geometric_model_envelope}, the time-averaged morphology of these models was motivated by the first imaging results (\autoref{sec:first_images}). Furthermore, the geometric model parameters were adjusted and selected to be qualitatively consistent with \sgra measurements. 
To assess the effects of temporal variability on the reconstructed images, a dynamic component is added to the time-averaged models. The static geometric models are modulated by an evolution generated statistical model with parameters optimized to match metrics seen in \sgra data. 

\subsubsection{Time-averaged Morphology}
\label{subsec:geometric_model_envelope}
We use the following three ring models motivated by the morphology identified in many ``first images" presented in \autoref{sec:first_images}: symmetric and asymmetric ring models (henceforth \texttt{Ring} and \texttt{Crescent}, respectively) and a symmetric ring model with a bright hot spot that rotates in the counter-clock-wise direction with a period of 30 min (henceforth \texttt{Ring+hs}). The first two models are designed to test whether our imaging methods can identify a symmetric vs. asymmetric ring, while the latter hot spot model tests the effects of a fast-moving localized emission on the reconstructed images.
Besides the ring models, we use four non-ring images.
To assess the robustness of the central depression seen in ring reconstructions, we adopt a uniform circular and a elliptical disk model (henceforth \texttt{Simple Disk} and \texttt{Elliptical Disk}, respectively).
Finally, motivated by the non-ring images recovered in \autoref{sec:first_images}, we adopt a point-source and double point-source model (henceforth \texttt{Point} and \texttt{Double}, respectively). 

The parameters (e.g., diameter, width) of each geometric model are selected to be broadly consistent with representative properties of \sgra's deblurred visibility amplitudes. We use the following four criteria: 
(1) the first null traced by Chile-LMT baselines is located at the baseline length of $3.25-3.65$\,G$\lambda$ and position angle of $\sim 50^\circ$, with an amplitude of $\sim 0.1$\,Jy; (2) the peak of the visibility amplitudes between the first two nulls has $\sim 0.3$\,Jy; (3) the second null traced by Chile-\hawaii and/or Chile-PV baselines is located at the \uv-distance of 6\,G$\lambda$ with an amplitude less than 0.1\,Jy; and (4) for asymmetric models, visibility amplitudes on Chile-LMT baselines are $\sim 1.5$ times larger than on SMT-\hawaii baselines. 
Figure \ref{fig:synthetic_summary} shows the comparison of visibility amplitudes between \sgra data on April 7 and corresponding synthetic data (after adding temporal variability -- see \autoref{subsec:geometric_model_inoisy}), demonstrating qualitative agreement between the synthetic data and \sgra visibility amplitudes. 

\subsubsection{Characterization of the Time Variability}
\label{subsec:geometric_model_inoisy}
\begin{figure}[t!]
\centering
\includegraphics[width=\linewidth]{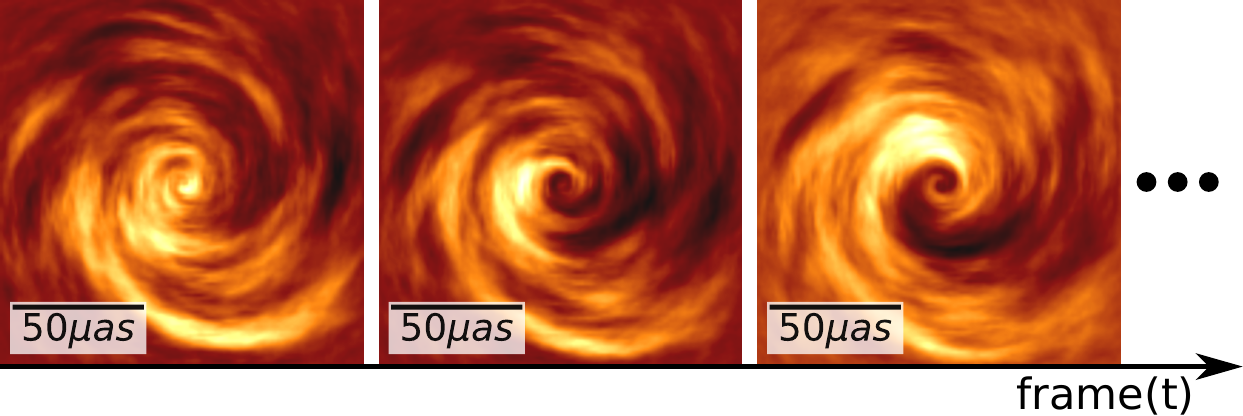}
\caption{A sequence of three frames of statistical evolution $\exp\left\{\rho(t,x,y)\right\}$ that was sampled using \texttt{inoisy}. The random field was generated with correlations that mimic a disk rotating clockwise. Here the image sequence corresponds to $\sim 10$ min of observation time.
}
\label{fig:inoisy}
\end{figure}
To mimic the temporal variability of \sgra, the geometric models, denoted by $I_{\rm geo}(x, y)$, are modulated by a temporal evolution sampled from a statistical model: \texttt{inoisy} \citep{Lee_2021}. This model enables sampling random spatio-temporal fields, $\rho(t,x,y)$, according to specified local correlations. \citet{Lee_2021} and \citet{levis2021inference} showed that \texttt{inoisy} is able to generate random fields that capture the statistical properties of accretion disks (see \autoref{fig:inoisy}). Using this model we modulate the static geometric models according to
\begin{equation}
I (t, x, y) = I_{\rm geo}(x, y) \exp\left\{\rho(t,x,y)\right\}.
\end{equation}
\autoref{fig:synthetic_summary} shows both the time-average and a single snapshot of each model highlighting the effect of temporal evolution.

The \texttt{inoisy} model parameters were selected to generate a similar degree of time-variability as \sgra measurements. We impose the following conditions to match metrics of temporal variations; (1) the mean of each movie's total flux is 2.3\,Jy, consistent with the ALMA light curve on April 7 (\autoref{sec::Observations}; \citealt{Wielgus_2021}); (2) the standard deviation of the total flux is in the range of 0.09-0.28\,Jy, as seen in the ALMA and SMA light curves and intra-site baselines; (3) the $\mathcal{Q}$-metric \citep{Roelofs_2017} of the intrinsic closure phase variability is comparable with \sgra data on all triangles. 

\begin{figure*}[t]
\centering
\includegraphics[width=\linewidth]{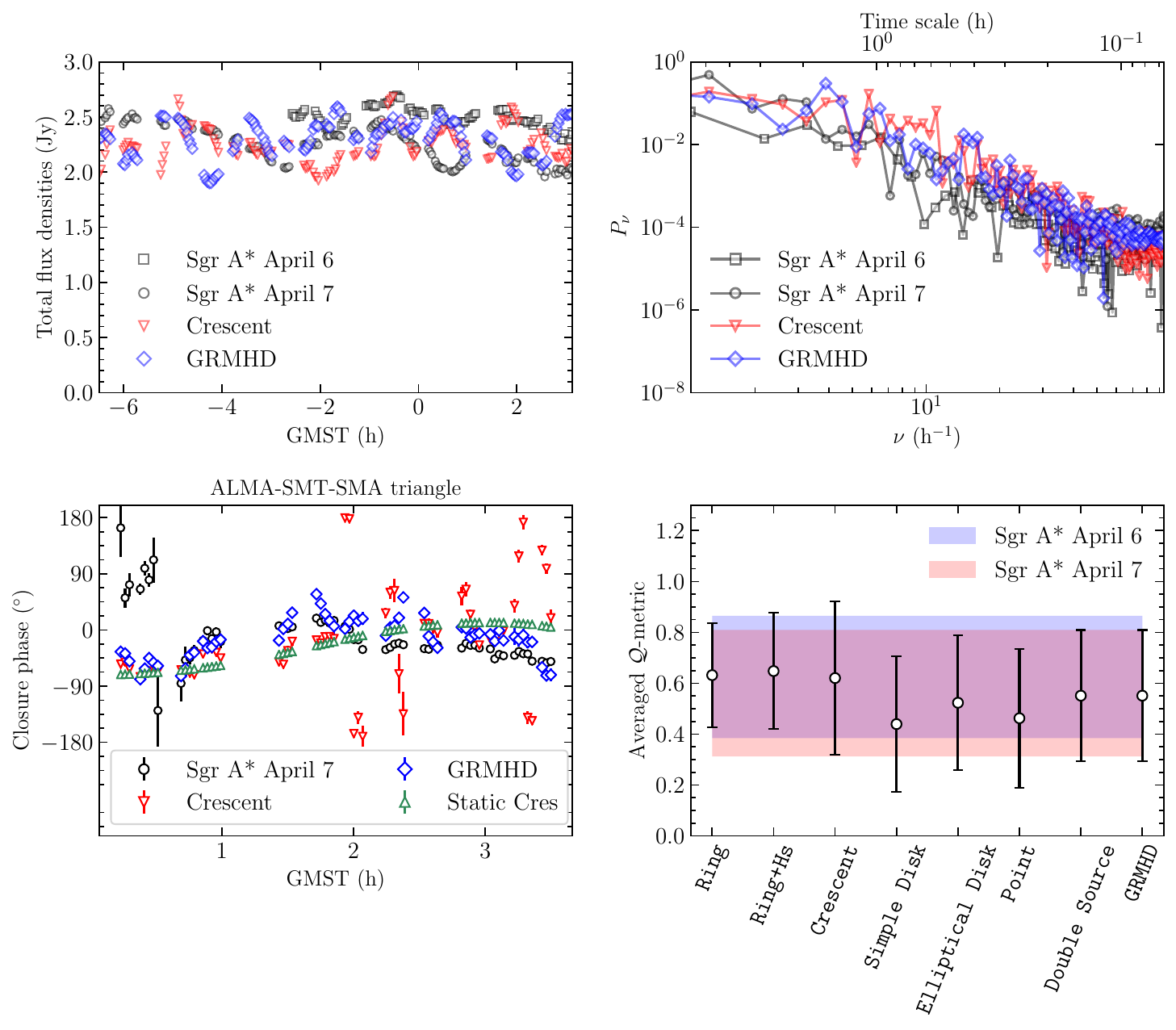}
\caption{
Temporal variability in \sgra and synthetic data. (top left) Light curves of ALMA \sgra data (gray points), the crescent model (red points) and the GRMHD model (blue points). (top right) Power spectrum density distributions of the light curves in the same color conventions. (bottom left) Closure phases on ALMA-SMT-SMA triangle from the April 7 \sgra data (gray), the crescent model (red), its time-averaged static image (green), and movie of the GRMHD model (blue). (bottom right) $\mathcal{Q}$-metrics of closure phases from \sgra and synthetic data. We show the mean and standard deviation of $\mathcal{Q}$-metrics over all independent triangles for each data set, overlaid by two shaded areas indicating the corresponding ranges for \sgra data on April 6 and 7.
}
\label{fig:synthetic_data_variation}
\end{figure*}

\autoref{fig:synthetic_data_variation} shows temporal variations in the total flux density and closure phases with comparison to \sgra data. The synthetic data variability is able to capture the real data light curve variablity. Moreover, the power spectrum density distributions of the light curves from synthetic models are broadly consistent with \sgra data. For closure phases, \texttt{inoisy} produces data with visible time variability seen in high \textit{S/N} triangles, such as the ALMA-SMT-LMT triangle. These synthetic movies also roughly match the \sgra variability amplitudes, averaged over all triangles, as evaluated using the $\mathcal{Q}$-metric.
We note that while these synthetic data are in good agreement of the above aspects of the EHT data, their variability amplitudes in Fourier domain are slightly less than \sgra data (see \autoref{tab:premodeling} in \autoref{sec:PSD_noise}).

\subsection{GRMHD Model}
\label{sec:synthetic_data:grmhd}

In addition to the geometric models, we also generated synthetic data from GRMHD simulations to evaluate the performance of our imaging procedures on more complicated physically-motivated models of \sgra.
These GRMHD models are selected from the simulation library presented in \citetalias{PaperV}, and are in general agreement with \sgra data (\citetalias{PaperV} section~3.1.2).
Section~\ref{subsec:synthetic_data_images_grmhd} shows the result of applying our imaging procedure to a weakly magnetized ``standard and normal evolution'' (SANE) model, with dimensionless spin $a_*=-0.94$, electron temperature ratio $R_{\mathrm{high}}=160$, and viewing inclination $i=50^{\circ}$.
Although failed in other constraints (see \citetalias{PaperV} appendix~A), this model satisfies the same criteria used for selection of the geometric models as seen by the resulting visibility amplitudes (\autoref{subsec:geometric_model_envelope}) and temporal variability (see \autoref{subsec:geometric_model_inoisy}), as shown in \autoref{fig:synthetic_data_variation}.
\autoref{fig:synthetic_summary}  shows a single snapshot frame of the GRMHD movie along with a time-averaged structure. The GRMHD movie frames contain a sharp photon ring with a faint emission broadly extended over $\sim 100\,\mu$as. Some of the frames contain notable spiral arm features surrounding the photon ring and extending beyond the compact structure (refer to the SANE frames shown in Figure~\ref{fig:grmhd_pa}). This spiral arm feature is smoothed out by averaging over the observational time.

In addition, \autoref{sec:appendix_best_grmhd} shows the result of applying the same imaging procedure on a strongly magnetized ``magnetically arrested disk'' (MADs) model with positive spin, which passes more observational constraints and is in the ``best bet region'' considered by \citetalias{PaperV}. By using these two GRMHD models, generated with different physical parameters, we demonstrate that our imaging procedure and the resulting performance is robust against the details of GRMHD models.

%% file: parameter_survey.tex
\section{Imaging Surveys with Synthetic Data}
\label{sec:survey}
We conducted surveys over a wide range of imaging assumptions with four scripted imaging pipelines using RML, CLEAN and a Bayesian posterior sampling method. 
The surveys were performed on the synthetic data sets presented in \autoref{sec:synthetic_data} as well as on the real \sgra data. 
Reconstructing synthetic data with exactly the same procedure used on \sgra allows us to assess our ability to identify the true underlying time-averaged morphology.
Both synthetic and real \sgra data sets were pre-processed with a common pipeline described in \autoref{subsec:pre-im_pipeline}. 
We describe the RML and CLEAN imaging parameter surveys in \autoref{subsec:rml_clean_imaging_surveys} and imaging with a Bayesian posterior sampling method in \autoref{subsec:themis_survey}. Images of synthetic data from imaging surveys are described in \autoref{subsec:synthetic_data_images}.
We present images of the actual \sgra data in \autoref{sec:sgra_images}.

\subsection{Common Pre-imaging Processing}
\label{subsec:pre-im_pipeline}
To reconstruct a time-averaged image of \sgra, each pipeline used the original calibrated data sets described in \autoref{sec:observations} and/or data sets further normalized by the 
time-dependent total flux density of \sgra.  %
Within each imaging pipeline, data sets were first time-averaged to enhance the \textit{S/N} of visibilities; each pipeline adopted a single integration time or explored multiple choices of the integration time (see Tables \ref{table:topset-difmap}-\ref{table:topset-smili}). After time averaging, fractional errors of 0\%, 2\,\% or 5\,\% were added to the visibility error budget in quadrature to account for the non-closing systematic errors (refer to \citetalias{PaperII}). 

As described in Sections~\ref{sec:pre_scatter} and~\ref{sec:PSD_noise}, we employ additional strategies to mitigate extrinsic scattering and intra-day variations. To assess these proposed strategies, and our ability to account for these two effects in the imaging process, we incorporate parameterized error budgets and systematically explore the various assumptions on these two effects in the RML and CLEAN surveys discussed in Section~\ref{subsec:rml_clean_imaging_surveys}. In total, we potentially include up to three additional noise budgets that account for 1) systematic error, 2) interstellar scattering, and 3) time variability. 

The second budget accounts for the substructure arising from refractive scattering. We added the anticipated refractive noise level for the observed (i.e. interstellar-scattered) visibilities using two base models described in \autoref{sec:pre_scatter}. The first model we explore is the \texttt{Const} model, which adds a constant noise budget across all visibilities prior to deblurring;
we examined two noise levels: 0.4\,\% and 0.9\,\% of the total flux density at each time segment (i.e., \texttt{Const} and \texttt{2$\times$Const}). The second model explored is the {\tt J18model1 (J18)} model, which is no longer constant in the \uv-space;
we adopt two scaling factors of 1.0 and 2.0 for this noise floor (i.e, \texttt{J18model1 (J18)} and \texttt{2$\times$J18model1 (2$\times$J18}). The two noise levels adopted in each model reasonably cover differences in the noise levels caused by different potential intrinsic structures (see \autoref{sec:pre_scatter}). After including one of the above budgets for the refractive noise, visibilities were divided by the diffractive scattering kernel based on the J18 model to mitigate diffractive scattering. In addition to the above four sets of the scattering mitigation schemes, we attempted imaging without any form of scattering mitigation to probe the interstellar-scattered source structure of \sgra (referred to as {\it on-sky} images).

The third noise budget explored accounts for the structural deviations from the time-averaged morphology due to the intra-day variations. 
We further inflated the visibility error budget %
using the variability noise model described by \autoref{eq:PSD_noise} in \autoref{sec:PSD_noise}. 
This budget was added in quadrature to the visibility noise budget, after being normalized by the time-dependent total flux density.
We systematically explored various sets of parameters in \autoref{eq:PSD_noise} including the variability rms level at 4\,G$\lambda$ ($a$), the break location ($u_0$), and the variability power-law spectra index at long baselines ($b$). Similar to scattering, we also attempted reconstructions without this error budget (i.e., assuming no intra-day variation in data).

\subsection{RML and CLEAN Imaging Parameter Surveys}
\label{subsec:rml_clean_imaging_surveys}
In a manner similar to previous EHT imaging of \m87 \citepalias{M87PaperIV,M87PaperVII}, we explore how recovered images are influenced by different imaging and optimization choices.
In particular, we objectively evaluate each set of imaging parameters in scripted RML and CLEAN imaging pipelines using synthetic data with known ground truth images. Each parameter survey leads to a {\it Top Set} of parameters: parameter combinations that each produce acceptable images on our entire suite of synthetic data. 
The distribution of \sgra images recovered with the Top Set parameter combinations reflects our uncertainty due to modeling and optimization choices made in imaging; thus, it is different from a Bayesian posterior and instead attempts to characterize what is sometimes referred to as epistemic uncertainty.

\begin{table}[t]
\footnotesize
\tabcolsep=0.05cm
\centering
\caption{Parameters in the \difmap Pipeline Top Set 
}
\label{table:topset-difmap}
\begin{tabularx}{\columnwidth}{lccccccc} 
\hline
\hline
\multicolumn{8}{c}{April~7 (8400 Param.\ Combinations; 1626 in Top Set) } \\
\hline
\textbf{Systematic} & \multicolumn{2}{c}{\textbf{0}} & \multicolumn{2}{c}{\textbf{0.02}} & \multicolumn{3}{c}{\textbf{0.05}} \\
\textbf{error} & \multicolumn{2}{c}{25.6\%} & \multicolumn{2}{c}{36.8\%} & \multicolumn{3}{c}{37.5\%} \\
\hline
\textbf{$\bf Ref\ Type$} & \textbf{No} & \textbf{Const} & \multicolumn{2}{c}{\textbf{2$\times$Const}} & \textbf{J18} & \multicolumn{2}{c}{\textbf{2$\times$J18}} \\
& 14.9\% & 20.7\% & \multicolumn{2}{c}{21.3\%} & 22.1\% & \multicolumn{2}{c}{20.9\%} \\
\hline
\textbf{$\bf a_{\bf \rm psd}$} & \textbf{No} & \multicolumn{2}{c}{\textbf{0.015}} & \multicolumn{2}{c}{\textbf{0.02}} & \multicolumn{2}{c}{\textbf{0.025}} \\
& 5.1\% & \multicolumn{2}{c}{28.5\%} & \multicolumn{2}{c}{32.1\%} & \multicolumn{2}{c}{34.3\%} \\
\hline
\textbf{$\bf b_{\bf \rm psd}$} & \textbf{No} & \multicolumn{2}{c}{\textbf{1}} & \multicolumn{2}{c}{\textbf{3}} & \multicolumn{2}{c}{\textbf{5}} \\
& 5.1\% & \multicolumn{2}{c}{20.2\%} & \multicolumn{2}{c}{35.5\%} & \multicolumn{2}{c}{39.2\%} \\
\hline
\textbf{$|u|_{0}$} & \multicolumn{3}{c}{\textbf{No}} & \multicolumn{4}{c}{\textbf{2}} \\
& \multicolumn{3}{c}{5.1\%} & \multicolumn{4}{c}{94.9\%} \\
\hline
\textbf{Time average} & \multicolumn{3}{c}{\textbf{10}} & \multicolumn{4}{c}{\textbf{60}}\\
(sec) & \multicolumn{3}{c}{45.0\%} & \multicolumn{4}{c}{55.0\%} \\
\hline
\textbf{ALMA weight} & \multicolumn{3}{c}{\textbf{0.1}} & \multicolumn{4}{c}{\textbf{0.5}}\\
& \multicolumn{3}{c}{41.1\%} & \multicolumn{4}{c}{58.9\%} \\
\hline
\textbf{UV weight} & \multicolumn{3}{c}{\textbf{0}} & \multicolumn{4}{c}{\textbf{2}} \\
& \multicolumn{3}{c}{54.7\%} & \multicolumn{4}{c}{45.3\%} \\
\hline
\textbf{Mask Diameter} & \textbf{80} & \textbf{85} & \textbf{90} & \textbf{95} & \textbf{100} & \textbf{105} & \textbf{110} \\
(\uas) & 0.2\% & 2.5\% & 22.3\% & 25.0\% & 21.3\% & 20.0\% & 8.7\%\\
\hline
\end{tabularx}
\\ \vspace{0.3cm}
\raggedright{\textbf{Note. }
In each row, the upper line with bold text shows the surveyed parameter value corresponding to the parameter of left column, while the lower line shows the number fraction of each value in Top Set. The total number of surveyed parameter combinations and Top Set are shown in the first row. } 
\end{table}

\begin{table}[t]
\footnotesize
\tabcolsep=0.2cm
\centering
\caption{
Parameters in the \ehtim Pipeline Top Set 
}
\label{table:topset-ehtim}
\begin{tabularx}{\columnwidth}{lccccc}
\hline
\hline
\multicolumn{6}{c}{April~7 (112320 Param.\ Combinations; 5594 in Top Set) } \\
\hline
\textbf{Systematic} & \multicolumn{2}{c}{\textbf{0}} & \textbf{0.02} & \multicolumn{2}{c}{\textbf{0.05}} \\
\textbf{error}& \multicolumn{2}{c}{21.4\%} & 36.7\% & \multicolumn{2}{c}{41.8\%} \\
\hline
\textbf{$\bf Ref\ Type$} & \textbf{No} & \textbf{Const} & \textbf{2$\times$Const} & \textbf{J18} & \textbf{2$\times$J18} \\
& 27.0\% & 23.9\% & 20.6\% & 16.4\% & 12.1\% \\
\hline
\textbf{$\bf a_{\bf \rm psd}$} & \textbf{No} & \multicolumn{2}{c}{\textbf{0.015}} & \textbf{0.02} & \textbf{0.025} \\
& 11.4\% & \multicolumn{2}{c}{40.6\%} & 26.6\% & 21.4\% \\
\hline
\textbf{$\bf b_{\bf \rm psd}$} & \textbf{No} & \textbf{1} & \textbf{2} & \textbf{3} & \textbf{5} \\
& 11.4\% & 24.8\% & 20.4\% & 21.5\% & 21.8\%\\
\hline
\textbf{$|u|_{0}$} & \multicolumn{3}{c}{\textbf{No}} & \multicolumn{2}{c}{\textbf{2}} \\
& \multicolumn{3}{c}{11.4\%} & \multicolumn{2}{c}{88.6\%} \\
\hline
\textbf{TV} & $\bf 0$ & \multicolumn{2}{c}{\textbf{0.01}} & \textbf{0.1} & \textbf{1} \\
& 13.2\% & \multicolumn{2}{c}{16.0\%} & 36.1\% & 34.7\% \\
\hline
\textbf{TSV} & $\bf 0$ & \multicolumn{2}{c}{\textbf{0.01}} & \textbf{0.1} & \textbf{1} \\
& 29.5\% & \multicolumn{2}{c}{32.3\%} & 26.6\% & 11.5\% \\
\hline
\textbf{Prior size} & \multicolumn{2}{c}{\textbf{70}} & \textbf{80} & \multicolumn{2}{c}{\textbf{90}} \\
(\uas) & \multicolumn{2}{c}{33.9\%} & 34.7\% & \multicolumn{2}{c}{31.4\%} \\
\hline
\textbf{MEM} & $\bf 0$ & \multicolumn{2}{c}{\textbf{0.01}} & \textbf{0.1} & \textbf{1} \\
& 7.8\% & \multicolumn{2}{c}{18.4\%} & 54.3\% & 19.6\% \\
\hline
\textbf{Amplitude} & \multicolumn{2}{c}{\textbf{0}} & \textbf{0.1} & \multicolumn{2}{c}{\textbf{1}} \\
\textbf{weight} & \multicolumn{2}{c}{0.9\%} & 23.3\% & \multicolumn{2}{c}{75.8\%} \\
\hline
\end{tabularx}
\\ \vspace{0.3cm}
\raggedright{\textbf{Note. } %
Same as \autoref{table:topset-difmap}} 
\end{table}

\begin{table}[t]
\footnotesize
\tabcolsep=0.2cm
\centering
\caption{Parameters in the \smili Pipeline Top Set %
}
\label{table:topset-smili}
\begin{tabularx}{\columnwidth}{lccccc} 
\hline
\hline
\multicolumn{6}{c}{April~7 (54000 Param.\ Combinations; 2763 in Top Set) } \\
\hline
\textbf{Systematic} & \multicolumn{2}{c}{\textbf{0}} & \textbf{0.02} & \multicolumn{2}{c}{\textbf{0.05}} \\
\textbf{error}& \multicolumn{2}{c}{33.9\%} & 33.3\% & \multicolumn{2}{c}{32.8\%} \\
\hline
\textbf{$\bf Ref\ Type$} & \textbf{No} & \textbf{Const} & \textbf{2$\times$Const} & \textbf{J18} & \textbf{2$\times$J18} \\
& 15.7\% & 22.1\% & 18.5\% & 21.4\% & 22.3\% \\
\hline
\textbf{$\bf a_{\bf \rm psd}$} & \textbf{No} & \multicolumn{2}{c}{\textbf{0.015}} & \textbf{0.02} & \textbf{0.025}\\
& 7.5\% & \multicolumn{2}{c}{37.2\%} & 26.7\% & 28.6\%\\
\hline
\textbf{$\bf b_{\bf \rm psd}$} & \textbf{No} & \textbf{1} & \textbf{2} & \textbf{3} & \textbf{5}\\
& 7.5\% & 16.8\% & 27.7\% & 31.9\% & 16.1\%\\
\hline
\textbf{$|u|_{0}$} & \multicolumn{2}{c}{\textbf{No}} & \textbf{1} & \multicolumn{2}{c}{\textbf{2}} \\
& \multicolumn{2}{c}{7.5\%} & 40.7\% & \multicolumn{2}{c}{51.8\%} \\
\hline
\textbf{TV} & $\bf 10^2$ & \multicolumn{2}{c}{$\bf 10^3$} & $\bf 10^4$ & $\bf 10^5$\\
& 8.6\% & \multicolumn{2}{c}{46.7\%} & 44.4\% & 0.3\%\\
\hline
\textbf{TSV} & $\bf 10^2$ & \multicolumn{2}{c}{$\bf 10^3$} & $\bf 10^4$ & $\bf 10^5$\\
& 38.7\% & \multicolumn{2}{c}{53.3\%} & 8.0\% & 0.0\%\\
\hline
\textbf{Prior size} & \multicolumn{2}{c}{\textbf{140}} & \textbf{160} & \multicolumn{2}{c}{\textbf{180}} \\
(\uas) & \multicolumn{2}{c}{33.0\%} & 33.6\% & \multicolumn{2}{c}{33.3\%} \\
\hline
\textbf{$\bf \ell_1$} & \multicolumn{2}{c}{\textbf{0.1}} & \textbf{1} & \multicolumn{2}{c}{$\bf 10$} \\
& \multicolumn{2}{c}{44.8\%} & 54.8\% & \multicolumn{2}{c}{0.3\%} \\
\hline
\end{tabularx}
\\ \vspace{0.3cm}
\raggedright{\textbf{Note. }
Same as \autoref{table:topset-difmap}} 
\end{table}

\subsubsection{Imaging Pipelines}
Similar to previous EHT work \citepalias{M87PaperIV, M87PaperVII}, we designed three scripted imaging pipelines utilizing the \difmap, \ehtim and \smili software packages. After completing the common pre-imaging processing of data (Section~\ref{subsec:pre-im_pipeline}), each pipeline reconstructs images using a broad parameter space (e.g., weights for the regularization functions, mask sizes, station gain constraints, variability noise budget parameters, etc.). We describe each pipeline in detail in \autoref{appendix:difmap-pipeline}, \autoref{appendix:ehtim-pipeline}, and \autoref{appendix:smili-pipeline}.

Each pipeline explored on the order of $10^{3-5}$ parameter combinations, as summarised in Table \ref{table:topset-difmap}, \ref{table:topset-ehtim}, and \ref{table:topset-smili} for \difmap, \ehtim, and \smili, respectively. 
Each pipeline has some unique choices that are fixed (e.g. the pixel size, or the convergence criterion) and surveyed (e.g., the regularizer weights), while some parameters are commonly explored (e.g. parameters for the scattering and intra-day variations in \autoref{subsec:pre-im_pipeline}).

While all imaging pipelines adopt the common pre-processing of data described in \autoref{subsec:pre-im_pipeline}, there are some differences in data processing. 
For instance, the noise budgets for refractive scattering and intra-day variability %
are updated during self-calibration rounds in \smili. 
RML imaging pipelines (\ehtim and \smili) adopt the same prior and initial images across all synthetic and real data sets. The \difmap pipeline uniformly explores a library of initial models for a first phase self-calibration, selecting the one that provides the best fit to the closure phases after a first run of cleaning (see \autoref{appendix:difmap-pipeline}). All three pipelines use combined low- and high-band data for imaging without any data flagging (including the intra-site baselines). 

\subsubsection{Top Sets of Imaging Parameters via Surveys on Synthetic Data}
\label{subsubsec:Top Sets}
\noindent Large imaging surveys on synthetic data facilitate the evaluation of different potential parameter combinations. %
Following \citetalias{M87PaperIV}, the principal output from each parameter survey is a Top Set: a set of parameter combinations that produce acceptable images on the suite of synthetic data presented in Section~\ref{sec:synthetic_data}.

\begin{figure}
    \centering
    \includegraphics[width=\columnwidth]{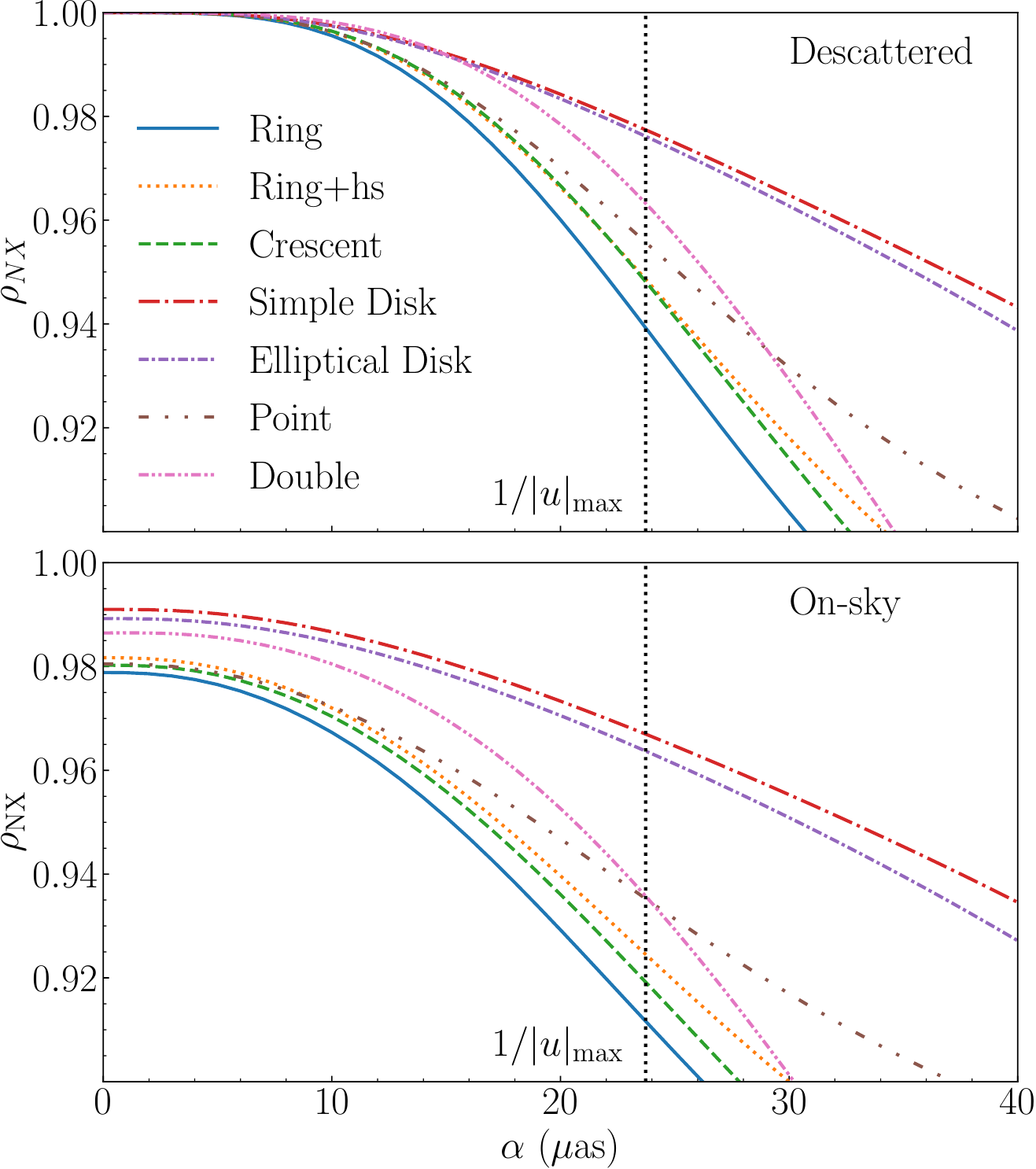}
    \caption{Normalized cross correlation, $\rho_{\rm NX}$, between time-averaged ground truth images and their corresponding blurred images, as a function of the blurring kernel size. These curves are shown for both the intrinsic images (top panels) and the scattered (i.e., on-sky) images (bottom panels). The dashed black line indicates the angular resolution equivalent to the maximum fringe spacing of \sgra observations, 24\,\uas; the corresponding value of $\rho_{\rm NX}$ at 24\,\uas is used to define a threshold that selects which \difmap, \ehtim and \smili imaging parameter combination are applied to \sgra data.}
    \label{fig:Nxcorr-cut}
\end{figure}

The fidelity of synthetic image reproduction is measured using the normalized cross-correlation between the reconstructed images and ground truth images.
We define the normalized cross-correlation of two images $X$ and $Y$ made of $N$ pixels as 
\begin{equation}
\rho_{\rm NX} = \frac{1}{N} \sum_i \frac {(X_i-\braket{X}) (Y_i-\braket{Y})}{\sigma_X \sigma_Y}.
\end{equation}
Here $X_i$ and $Y_i$ denote the image intensity at $i$-th pixel, $\braket{X}$ and $\braket{Y}$ denote the mean pixel values of the images, and $\sigma_X$ and $\sigma_Y$ are the standard deviations of pixel values.
The position offset between the frames is corrected by shifting one frame relative to the other along R.A. and decl. and identifying the shift coordinate that corresponds with the largest $\rho_{\rm NX}$.

In order to recover a Top Set, a threshold for $\rho_{\rm NX}$ is defined for each synthetic data set. Imaging parameter combinations that recover images that score above that thresholds for all synthetic data set are selected as Top Sets.
The threshold values of $\rho_{\rm NX}$ are determined in a manner similar to \citetalias{M87PaperIV}. For each ``ground truth" image (obtained by time averaging the synthetic movie), we evaluate $\rho_{\rm NX}$ between the ground truth image and the same image blurred with a Gaussian beam of FWHM equivalent to the maximum fringe spacing of the \sgra observations, 24\ \uas. This value of $\rho_{\rm NX}$ quantifies the potential loss of the image fidelity due to the limited angular resolution.  
Figure \ref{fig:Nxcorr-cut} shows examples of the $\rho_{\rm NX}$ curves between unblurred and blurred ground truth images as a function of the blurring size; the critical value corresponds to those at $\alpha=24$ \uas. Note that this value depends on the true source structure.
Unlike in \citetalias{M87PaperIV}, we find that a relaxation of the $\rho_{\rm NX}$ threshold is required to account for the fact that we reconstruct static images from time variable data sets. 
Hence the critical $\rho_{\rm NX}$ values for all training data sets are multiplied by the relaxation factor of 0.95.
In other words, the threshold for each synthetic data set is set to $0.95 \times \rho_{\rm NX}$ for $\rho_{\rm NX}$ evaluated at $\alpha=24 \uas$.
This relaxed threshold allows for a large enough number of Top Set parameters to be identified for all epochs and imaging pipelines. 
We ensure that the relaxed threshold still reconstructs all representative ground truth morphologies; in \autoref{appendix:topset_selection}, the worst $\rho_{\rm NX}$ images are shown for Top Set reconstructions of each model with each imaging pipeline to demonstrate that the representative features are recovered even in the worst-fidelity Top Set images.

In Tables \ref{table:topset-difmap}, \ref{table:topset-ehtim} and \ref{table:topset-smili}, we summarize the parameters and surveyed values in each pipeline's survey. These tables indicate the fraction of images corresponding to that parameter in each pipeline's Top Set for April 7 data. 
The results on April 6 data are summarized in \autoref{appendix:topset_selection}. 
The tables also provide the total number of surveyed parameter combinations as well as the number of combinations selected for each Top Set.
As seen in each table, there are more than 1000 parameter combinations in each pipeline's Top Set for April 7, which we find is sizable enough for downstream analysis. In contrast, we find that the April 6 Top Set sizes are much smaller, likely due to the poorer \uv-coverage.

\subsection{Image Posteriors with \themis}
\label{subsec:themis_survey}

\begin{deluxetable}{lcc}
\tablecaption{Bayesian Imaging Priors \label{tab:themis_priors}}
\tablehead{
\colhead{Parameter} & 
\colhead{Units} &
\colhead{Prior\tablenotemark{\dag}}
}
\startdata
Control points & Jy~$\mu{\rm as}^{-2}$ & $\mathcal{L}(10^{-3},0.1)$\\
FoV$_x$ & $\mu{\rm as}$ & $\mathcal{U}(50,500)$\\
FoV$_y$ & $\mu{\rm as}$ & $\mathcal{U}(50,500)$\\
Raster rotation & rad & $\mathcal{U}(-0.25\pi,0.25\pi)$\\
Shift in $x$ & $\mu{\rm as}$ & $\mathcal{U}(-100,100)$\\
Shift in $y$ & $\mu{\rm as}$ & $\mathcal{U}(-100,100)$\\
$\ln(\sigma_{\rm ref})$ & -- & $\mathcal{N}(-5.5,1.0)$\\
$\ln(f)$ & -- & $\mathcal{N}(-4.6,1.0)$\\
$a$ & -- & $\mathcal{L}(a_{25\%},a_{75\%})$\tablenotemark{\ddag}\\
$u_0$ & -- & $\mathcal{L}(u_{0,25\%},u_{0,75\%})$\tablenotemark{\ddag}\\
$b$ & -- & $\mathcal{L}(b_{25\%},b_{75\%})$\tablenotemark{\ddag}\\
$c$ & -- & $\mathcal{L}(1.5,2.5)$\\
\enddata
\tablenotetext{\dag}{Linear priors from $a$ to $b$ are represented by $\mathcal{U}(a,b)$, logarithmic priors from $a$ to $b$ are represented by $\mathcal{L}(a,b)$, normal priors with mean $\mu$ and standard deviation $\sigma$ are represented by $\mathcal{N}(\mu,\sigma)$.}
\tablenotetext{\ddag}{$x_{25\%}$ and $x_{75\%}$ are the bottom and top quartile values of the quantity $x$ given in \autoref{tab:premodeling}.}
\end{deluxetable}

The Bayesian imaging method employed in \themis differs from those described above in a number of respects.  Most significantly, apart from sampler tuning -- which affects only the efficiency with which the posterior is explored -- the splined raster model has only two free parameters.  Image resolution, raster orientation, the brightness at each control point, and noise model parameters are determined self-consistently by the fitting process (see \autoref{sec:themisdescription}).  This is achieved by replacing the hyperparameters associated with field of view, scattering threshold in the {\tt Const} prescription, systematic error budget, and those that define the variability mitigation noise with fit parameters, precluding the need to survey over them.  Priors for each quantity are listed in \autoref{tab:themis_priors}, and make use of the pre-imaging constraints described in \autoref{sec::PreImaging_Considerations} and listed explicitly in \autoref{tab:premodeling}.

Importantly, eliminating the hyperparameters associated with the additional contributions to the visibility uncertainties eliminates the noise-related data preparation steps described in \autoref{subsec:pre-im_pipeline}; we do not add any additional uncertainty prior to the \themis analysis.  However, to reduce the data volume (and thus computational expense of the posterior sampling) we incoherently average the flux-normalized data over scans. 
To prevent significant coherence losses, prior to averaging we calibrate the phase gains of the JCMT using the intra-site baseline, JCMT-SMA, and assume that the source is unresolved at the corresponding spatial scales probed by that baseline.

The remaining two unspecified hyperparameters are the raster dimensions, $N_x$ and $N_y$. Initial guesses for these are provided by the diffraction limit; for a typical source size of $80~\mu{\rm as}$, $N_x=N_y=5$ is sufficient to marginally super-resolve the source.  This may then be refined via a modest survey over potential values, with the final values selected by maximizing the Bayesian evidence, computed in \themis via thermodynamic integration \citep{Lartillot_06}.  
In practice, due to the expense of such a survey, we restrict ourselves to $N_x=N_y=5$ for the validation with synthetic data sets with a sole exception. We performed a raster dimension survey for the GRMHD data set presented in Section~\ref{subsec:synthetic_data_images_grmhd}, finding that $N_x=N_y=6$ is preferred.

For application to \sgra, we perform raster dimension surveys independently for April 6 and 7 as described in \autoref{sec:themis_survey_details}, finding preferred dimensions of $N_x=N_y=6$ and $N_x=N_y=7$, respectively.

\subsection{Synthetic Data Images}
\label{subsec:synthetic_data_images}

We first present Top Set and posterior images recovered from geometric model datasets in Section~\ref{subsec:synthetic_data_images_geometric}, followed by images recovered from the GRMHD dataset in Section~\ref{subsec:synthetic_data_images_grmhd}.

\subsubsection{Geometric Model Images}
\label{subsec:synthetic_data_images_geometric}
\begin{figure*}[!t]
    \centering
    \includegraphics[height=0.4\textheight]{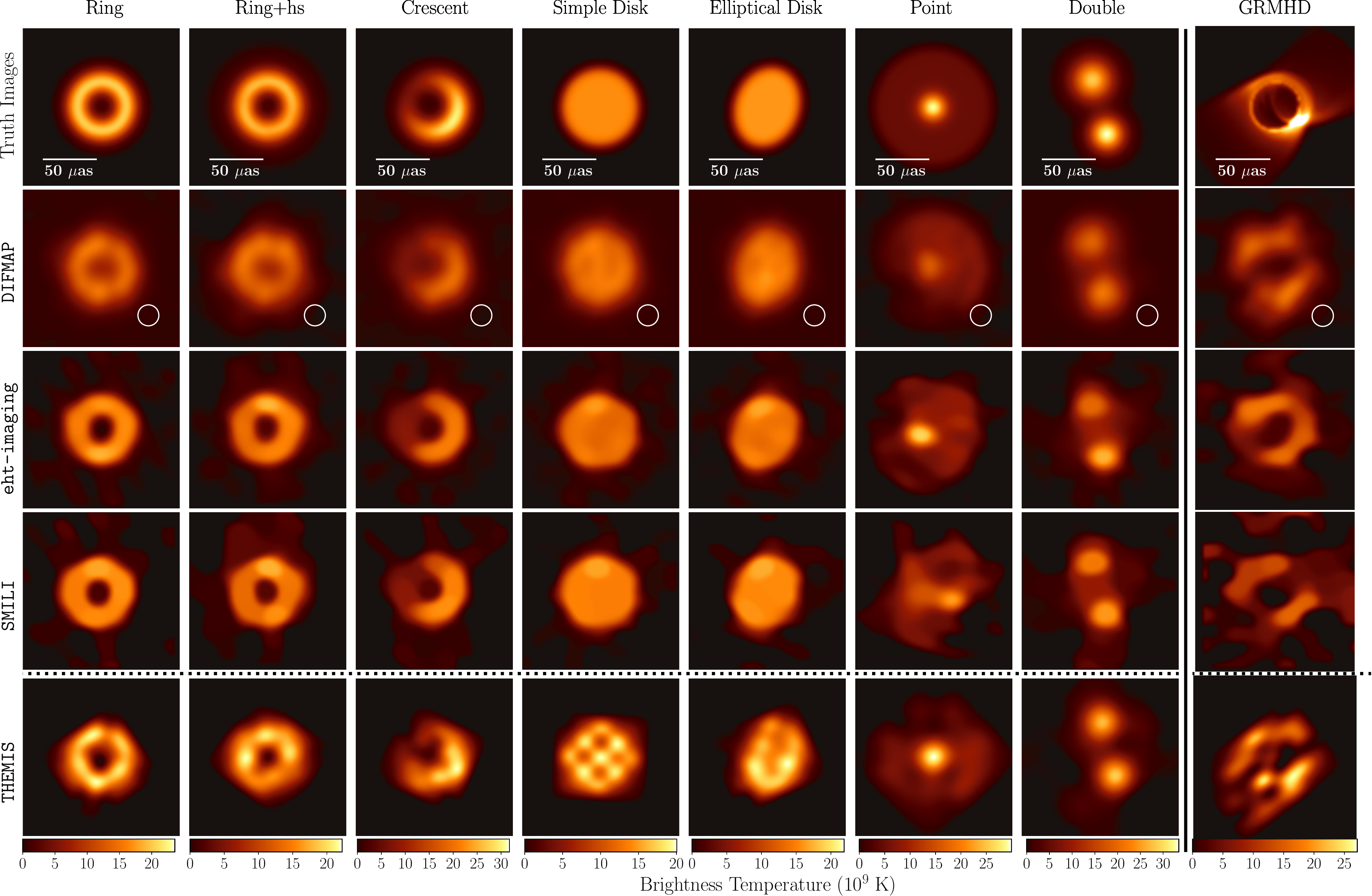} \\
    (a) Descattered Reconstructions \vspace{1em}\\
    \includegraphics[height=0.4\textheight]{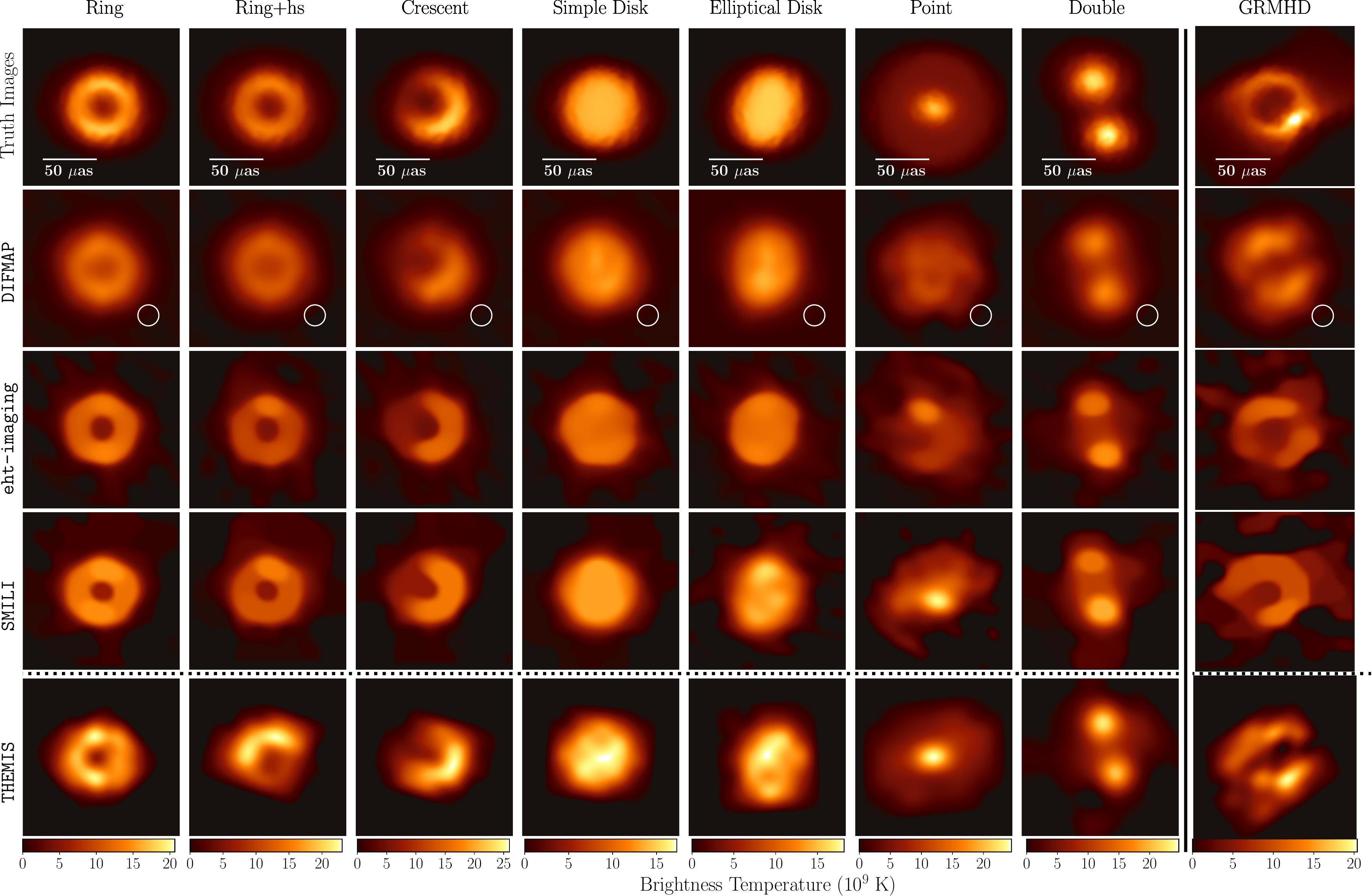}\\
    (b) On-sky Reconstructions \vspace{1em}
    \caption{
    Reconstructed images of synthetic data sets on April 7 (a) with and (b) without scattering mitigation for seven geometric models and the GRMHD model.
    Reconstructions of each geometric model by \difmap, \ehtim and \smili pipelines are made using a parameter combination identified via \textit{cross-validation}: the imaging parameters that result in the the best average $\rho_{\rm NX}$ across all other geometric models. These cross-validation results demonstrate the ability of the selected parameters to correctly reproduce novel source morphologies.
    GRMHD reconstructions for \difmap, \ehtim, and \smili are produced from an imaging parameter combination that performs best on all geometric models. %
    In contrast, for \themis reconstructions the average posterior image is shown for each model; the average posterior image appears to correctly identify the true source morphology in all synthetic data sets tested.
    }
    \label{fig:synthetic-crossvalidation}
\end{figure*}

In \autoref{fig:synthetic-crossvalidation}, we show the time-averaged ground truth images and corresponding image reconstructions for each of the synthetic data sets with April 7 \uv-coverage. 
Images from all pipelines are obtained with and without scattering mitigation (henceforth, \textit{descattered} and \textit{on-sky} reconstructions). 
These descattered and on-sky reconstructions are compared to the time-averaged ground truth images of the intrinsic and scattered structure, respectively.

\difmap, \ehtim and \smili images for each geometric model in \autoref{fig:synthetic-crossvalidation} are obtained using cross-validation: the parameter combination that provides the best mean $\rho_{\rm NX}$ across other geometric models (i.e., except for the model being tested) are selected. 
The cross-validation images in \autoref{fig:synthetic-crossvalidation} (which are contained in each pipeline's Top Set) successfully recover the representative morphology of each geometric model, demonstrating the capability of a single imaging parameter combination to identify various source structures. 
The manifestation of structure significantly different from the ground truth morphology is only seen in a small fraction of the cross-validated Top Set parameters. 
For instance, using the ring classification method described in \autoref{appendix:clustering_method}, only 2\%, $<$1\%, and $<$1\% of the cross-validated Top Set reconstructions for the \texttt{Ring} model are identified as not having a ring feature for \difmap, \ehtim and \smili, respectively; the reconstruction of a non-ring morphology is also found to be limited to small fraction of 3\%, 5\%, and 1\% for the \texttt{Crescent} model. 
Similarly, only a small fraction of the reconstructions from non-ring models are reconstructed with a ring morphology --- in particular, for the \difmap, \ehtim, and \smili pipelines, 7\%, 11\%, and 3\% and $<$1\%, $<$1\%, and $<$1\% of the cross-validated Top Sets reconstructed a ring feature from the \texttt{Point} and \texttt{Double} source models, respectively.

For \themis reconstructions, the mean posterior image is shown for each model. %
\autoref{fig:synthetic-crossvalidation} shows that the posterior images from \themis reconstructions identify the general morphology of each geometric model.
Most of \themis posterior images satisfy the criteria based on a minimum threshold of $\rho _{\rm NX}$ used in Top Set selections for \difmap, \ehtim and \smili pipelines (see \autoref{subsubsec:Top Sets}), with the exception of a few models discussed below. 
For April 7 images, all posterior images show higher $\rho _{\rm NX}$ than the threshold except for 1\% and 7\% of descattered reconstructions for \texttt{Double} source and \texttt{Ring+hs} models, respectfully.
For April 6, the images below the threshold are limited to 1\% and 22\% of descattered reconstructions for the \texttt{Point} and \texttt{Double} source models, respectively. 
However, 95\% of on-sky reconstructions of \texttt{Point} source model are above the $\rho _{\rm NX}$ threshold on April 6 for \themis. 
The high fractions of images beyond the $\rho _{\rm NX}$ threshold for all models on April 7 demonstrate the capability of \themis pipeline to recover various representative morphologies at an acceptable fidelity.

\autoref{fig:synthetic-crossvalidation} also shows the resiliency of the reconstructed morphology to the scattering prescriptions. While the on-sky reconstructions without scattering mitigation tend to be slightly more blurry than those with descattering, there are not many other notable differences in their appearance. 
In particular, the refractive substructure, which adds spatial distortion of images on scales finer than the angular resolution of the EHT, is not well constrained in our EHT data and therefore does not strongly appear in any reconstructions. We further discuss the effects of scattering prescriptions for \sgra images in \autoref{sec:sgra_images:scattering}.

\subsubsection{GRMHD Reconstructions}
\label{subsec:synthetic_data_images_grmhd}

\begin{figure*}[t]
    \centering
    \includegraphics[width=0.85\linewidth]{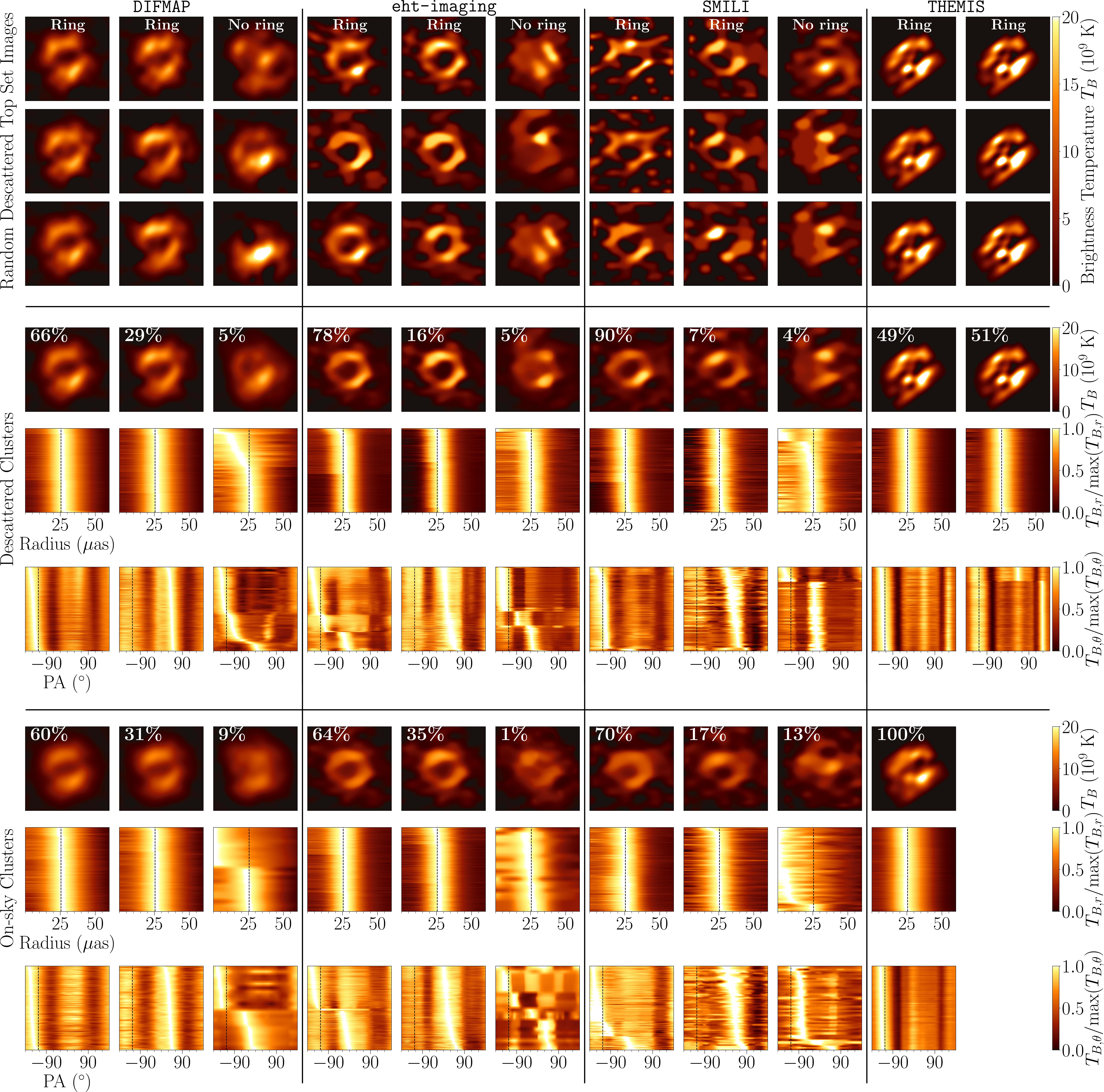}
    \caption{
    The distribution of images obtained from synthetic data provided using a GRMHD movie with April 7 \uv-coverage. 
    From left to right (separated by vertical lines), we show the distributions of Top Set images from the \difmap, \ehtim and \smili pipelines and posterior samples from the \themis pipeline; each vertical panel is further subdivided into clusters identifying common morphologies recovered by each pipeline.
    The figure is comprised of three horizontal panels separated by horizontal lines. 
    The top panel shows individual images randomly sampled from different clusters.
    The middle and bottom panels visualize the distributions of reconstructed descattered and on-sky images for each cluster, respectively. 
    In each panel, from top to bottom, we show the average of each cluster, the distributions of the radial profiles, and the distributions of azimuthal intensity profiles. 
    In the radial profiles, each horizontal slice corresponds to the azimuthally-averaged intensity profile of an image, normalized by its peak value. 
    Similarly, the azimuthal profiles show the azimuthal distribution of the radial peak intensity within a radius of 10 to 40\,\uas, also normalized by its peak value. 
    The dotted lines in the profiles indicate the peak radius or position angle of the ground truth GRMHD movie.
    The vertical order of images in both profiles is independently sorted by the peak radius or position angles; therefore the images are ordered differently in each profile distribution image. These results on synthetic data show that the Top Set and posterior samples from the imaging pipelines are able to recover ring images that resemble the true time-averaged structure of a GRMHD movie (see \autoref{fig:synthetic_summary}). However, the imaging methods sometimes produce non-ring images that still fit the data well.
    }
    \label{fig:grmhd-3599}
\end{figure*}

We show example GRMHD reconstructions on April 7 in the rightmost panels in \autoref{fig:synthetic-crossvalidation}. This GRMHD simulation contains a ring with a diameter of $\sim51$\,\uas. GRMHD images from \difmap, \ehtim, and \smili pipelines are reconstructed with the Top Set parameter combinations that corresponds with the largest average $\rho_{\rm NX}$ value across all seven geometric models.  %
For \themis reconstructions, on the other hand, the expected (i.e., mean) posterior image is shown\footnote{Unlike for \sgra, for this particular GRMHD data set the \themis pipeline used a large scale Gaussian to account for extended emission in the underlying source model.}.%

Unlike with the simple geometric synthetic data sets, the distribution of GRMHD reconstructions show wide variations in the image appearance\footnote{Note that this resembles the varied \sgra reconstructions in \autoref{sec:sgra_images} and \autoref{sec:first_images}}. 
Although GRMHD reconstructions in \autoref{fig:synthetic-crossvalidation} commonly show a ring-like morphology with a diameter of $\sim 50$\,\uas, the azimuthal intensity distribution is not consistent across the Top Sets.
In the top panels of \autoref{fig:grmhd-3599}, we show images from GRMHD April 7 data from each pipeline, which are randomly selected from \difmap, \ehtim and \smili Top Sets and posterior images from \themis. 
Most of the images have clear asymmetric ring features, but a few images show non-ring structures. The diameters of ring features are broadly consistent across different reconstructions, comparable to the ground truth image. 
On the other hand, the azimuthal distributions are not uniquely constrained by the Top Set images --- different position angles are seen in the randomly sampled images from each Top Set. 

To visualize the distribution of images with different morphology, we categorize all images into three major groups: ring images peaked at PAs within (a) the range of $-180^{\circ}\leq {\rm PA}<0^{\circ}$ where the ground truth value of $-124^{\circ}$ is located, (b) the range of $0^{\circ}\leq {\rm PA}<180^{\circ}$, and (c) the remaining images comprising non-ring or other ring-like images with much less consistency. The definition of a ring used in this paper to classify ring versus non-ring images is described in \autoref{appendix:clustering_method}.
Note that the particular definition of a ring will influence the quoted percentages of ring and non-ring images recovered. We find that the particular definition chosen in this paper results in classification that largely aligns with human perception. 
However, the classification of images that are borderline between ring and non-ring classification are sensitive to the exact criteria used. Therefore ring definitions that make use of different criteria can lead to classification that still largely aligns with human perception but varies somewhat in the ring classification percentages quoted in this paper.

In the middle and bottom panels of \autoref{fig:grmhd-3599}, we summarize the distribution of images from each pipeline with and without scattering mitigation, respectively. 
The GRMHD images within each pipeline's Top Set are clustered into image modes. 
The upper sub-panel shows the mean image of each cluster, indicating a common or representative morphology recovered. The middle sub-panel shows the distributions of the azimuthally-averaged radial intensity profiles, where the vertical order of images is sorted by the peak radii of the profiles. The lower sub-panel shows the azimuthal distribution of the radial peak intensity within a radius of 10 to 40\,\uas.

\autoref{fig:grmhd-3599} demonstrates that most of the Top Set or posterior images reconstruct ring images from the GRMHD data set. 
In particular, the radial profiles of the first two ring modes show a broad consistency of the peak radius around $\sim$25\,\uas 
which implies a diameter of $\sim$50\,\uas consistent with that of $\sim$50\,\uas for the ground truth model. %
The capability of identifying a ring with the consistent diameter does not appear to depend on the scattering mitigation. Similarly to geometric synthetic data (see Section \ref{subsec:synthetic_data_images_geometric}), the scattering mitigation does not significantly affect the resulting source morphology in the reconstructed images, except that the images without scattering mitigation tend to be slightly broader. Refractive substructure does not appear in the reconstructed on-sky images, again indicating that the appearance of the refractive substructure is not strongly constrained with EHT 2017 data.

As seen in the azimuthal profiles, each pipeline provides at least one asymmetric ring mode with the peak position angle roughly consistent with that of the mean ground truth image of $-124^{\circ}$. For this particular GRMHD example, this mode is found to be the most popular mode in most pipelines. %
However, \autoref{fig:grmhd-3599} indicates that the ring mode with the correct orientation is not always identified as the most popular mode among image samples --- for instance, the correct orientation is not identified as the most popular ring mode for the \ehtim on-sky pipeline.
Therefore, caution should be taken, as the popularity of a mode in an RML or CLEAN pipeline's Top Set does not necessarily always correspond with the true underlying structure. %

We find that key takeaways from the GRMHD example are consistent with the ``best-bet'' GRMHD models presented in \citetalias{PaperV}, identified based on various criteria using both EHT and non-EHT data. In \autoref{sec:appendix_best_grmhd}, we show example reconstructions of a ``best-bet'' GRMHD model. We find the identification of a ring morphology for the vast majority of the Top Set reconstructions, however multiple ring modes and non-ring images are still reconstructed.
Note also that the peak position angle of the ground truth ``best-bet'' GRMHD model is not necessarily identified as the most popular mode reconstructed in each Top Set. 
These results indicate that the same trends are seen across multiple GRMHD models that are broadly consistent with EHT data.

%% file: topset_images.tex
\section{Horizon-scale Images of \sgra}
\label{sec:sgra_images}
\begin{figure}
    \centering
    \includegraphics[width=\columnwidth]
    {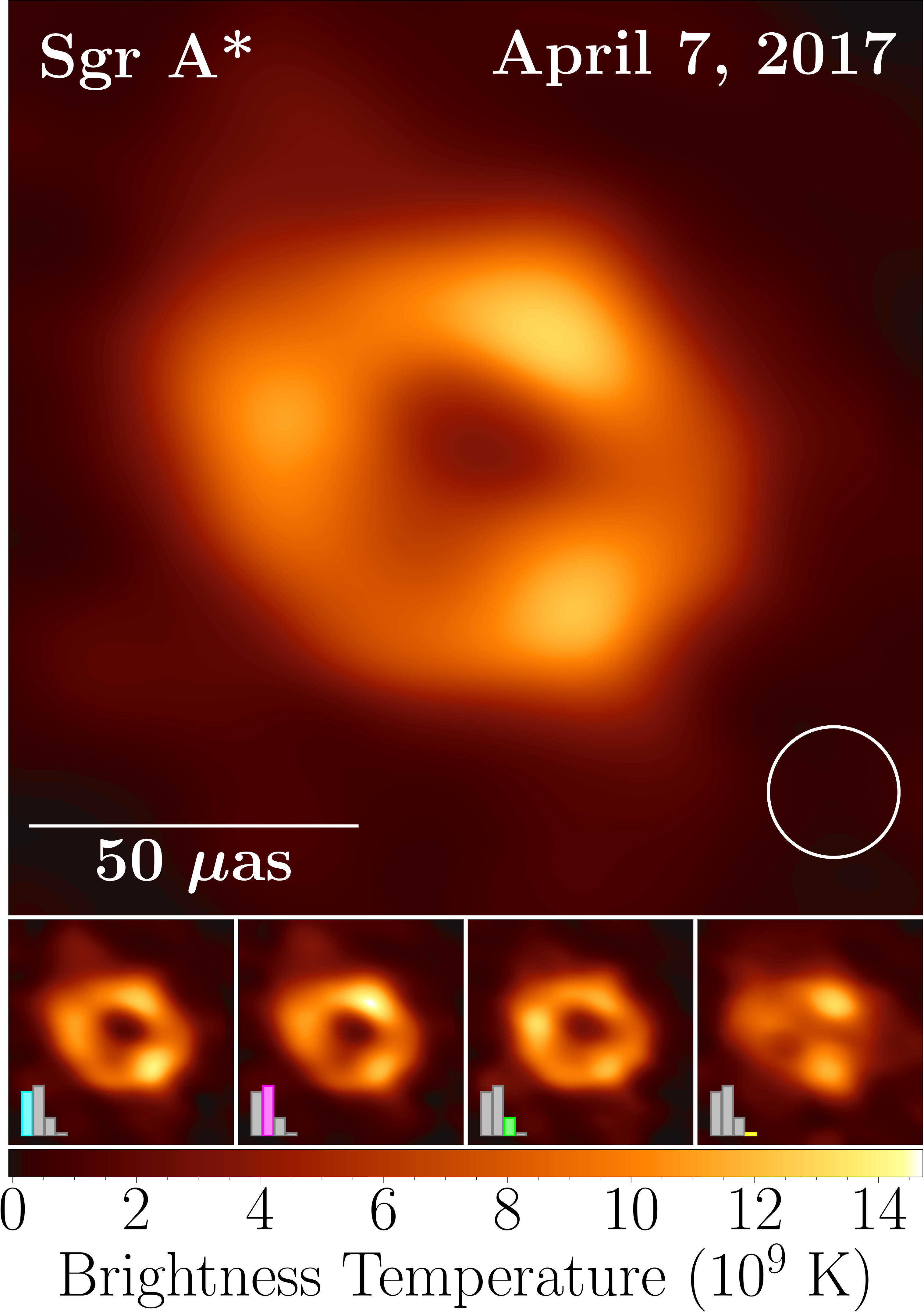}
    \caption{
    In the top main panel we show the representative image of \sgra obtained with the EHT from observations on 2017 April 7. 
    This top image is obtained by averaging the bottom four images. On the bottom from left to right, we show the average images of three prominent ring clusters with different azimuthal structures and a non-ring cluster. 
    The height of the colored bar (bottom left in each panel) represents the relative fraction of images in the Top Sets for each cluster. We note that \themis posterior sample only includes ring images.
    In each cluster, the image is computed through a weighted average over the descattered reconstructions, including all Top Set images from the three imaging methods (\difmap, \ehtim, and \smili) and 1024
    images randomly selected from descattered posteriors from \themis. Images are weighted by the inverse of the total number of Top Set or posterior images used from each pipeline, so that pipelines are represented equally in each image.
    Note that \difmap model images are convolved with 20\,\uas beam (represented by the inset circle), while no blurring is applied to the rest of images.
    }
    \label{fig:rep_sgra_images}
\end{figure}

Having determined Top Set imaging parameters for RML and CLEAN, and validated posterior estimation for \themis via tests on
synthetic data (\autoref{sec:survey}), we now show the result of these methods applied to \sgra data from the 2017 EHT observations. Unlike in the previous EHT imaging of \m87 \citepalias{M87PaperIV,M87PaperVII}, \sgra's recovered structure depends somewhat on the imaging strategy and parameter choices.
Thus, this section presents our main imaging results and analyzes how the image structure is affected by different imaging choices.

We begin in \autoref{sec:sgra_images:overview} by giving an overview of the results, followed by a more detailed discussion of the image structures recovered in each pipeline's Top Set or posterior in \autoref{sec:sgra_images:clustering}. Average images across pipelines, calibrated data sets, and observing days are discussed in  \autoref{sec:sgra_images:meanimages}. In \autoref{subsec:6+7} we present \sgra imaging obtained by combining the data sets for April 6 and 7. In \autoref{sec:sgra_is_a_ring} we address the question of whether \sgra is a ring, and explore how the recovered images are affected by the scattering and temporal variation mitigation strategy.

\subsection{Overview of Recovered \sgra Structure}
\label{sec:sgra_images:overview}
\autoref{fig:rep_sgra_images} summarizes the common morphologies recovered from \sgra data by the four imaging pipelines for April 7. We find that the vast majority
are rings that can be classified into three different clusters with varying azimuthal structures (\autoref{sec:sgra_images:clustering}), shown in the three bottom left panels of \autoref{fig:rep_sgra_images}, and a small fraction 
of non-ring images that also fit the \sgra data well (bottom right panel in \autoref{fig:rep_sgra_images}). Since these non-ring images are not as consistent in structure they largely blur out when averaged together and primarily emphasize a double-structure that is sometimes present. The representative \sgra image obtained by averaging reconstructions from all four clusters is shown in the large top panel of \autoref{fig:rep_sgra_images}, corresponding to that of a ring with a diameter of \sdiam (\autoref{sec:image_analysis}).

\begin{figure*}
    \centering
    \includegraphics[width=\textwidth]{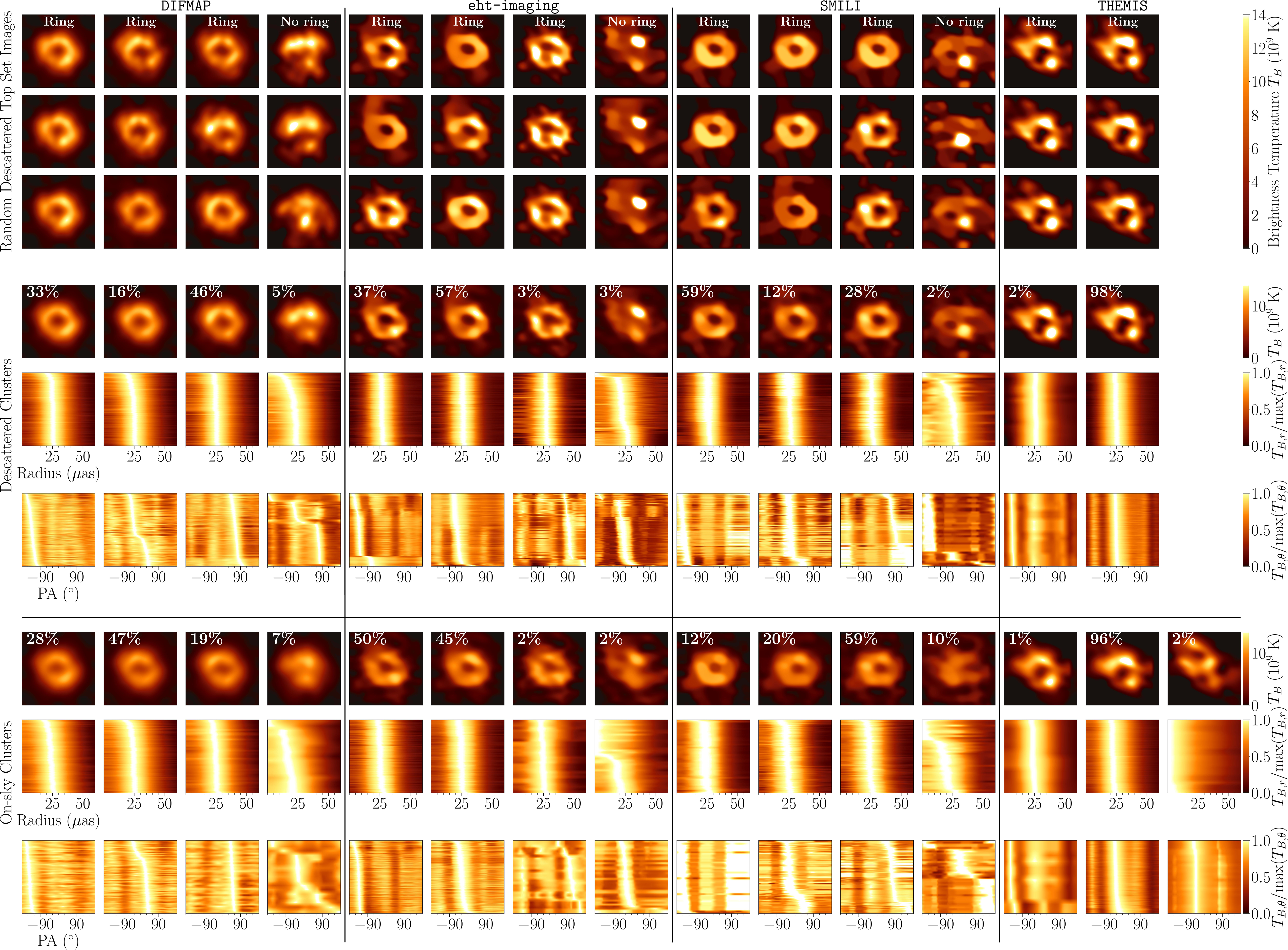}
    \caption{
    The distribution of reconstructed \sgra images on April 7. We show the distribution of images from each pipeline for each cluster with the same convention as \autoref{fig:grmhd-3599}.
    From left to right (separated by vertical lines), we show the distributions of Top Set images from the \difmap, \ehtim, and \smili pipelines and posterior samples from the \themis pipeline; each vertical panel is further subdivided into clusters identifying common morphologies recovered by each pipeline.
    The figure is comprised of three horizontal panels separated by horizontal lines. 
    The top panel shows individual images randomly sampled from different clusters.
    The middle and bottom panels visualize the distributions of reconstructed descattered and on-sky \sgra images for each cluster, respectively. 
    In each panel, from top to bottom, we show the average of each cluster, the distributions of the radial profiles, and the distributions of azimuthal intensity profiles. 
    Note that \themis images have only three clusters for each of descattered and on-sky reconstructions --- their descattered posterior does not contain a non-ring cluster and their on-sky posteriors do not contain an east PA ring cluster.
    }
    \label{fig:SgrA-topset-3599}
\end{figure*}

\begin{figure*}
    \centering
    \includegraphics[width=\textwidth]{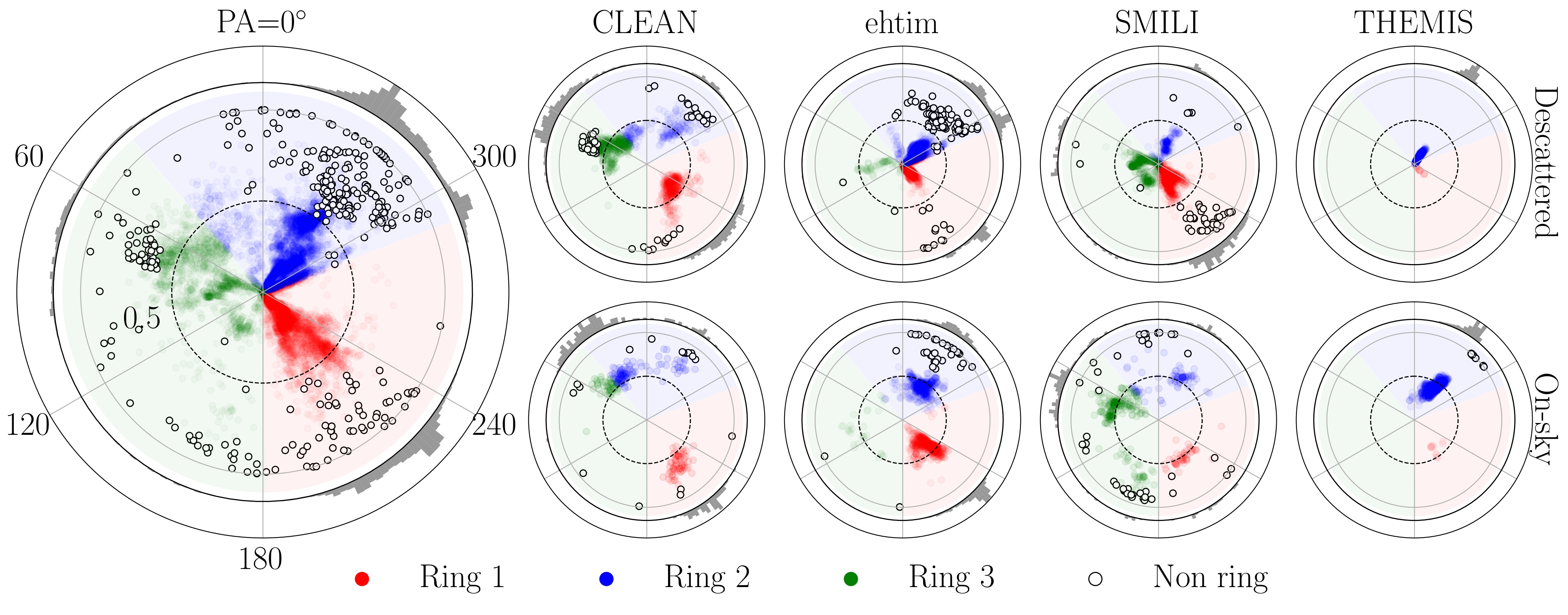}
    \caption{
    Comparison of image characteristics between each cluster. The radial and azimuthal values are the fractional central brightness and azimuthal peak brightness PA, respectively.
    The left panel shows the distribution of all Top Set and posterior images with all imaging pipelines (\difmap, \ehtim, \smili, and \themis) and scattering mitigation (descattered and on-sky), including the three ring clusters with different peak azimuthal brightness [red ($-180^\circ\leq {\rm PA}<-70 ^\circ$), blue ($-70^\circ\leq {\rm PA}<40 ^\circ$), and green ($40^\circ\leq {\rm PA}<180 ^\circ$)], and a non-ring cluster (white).
    The right eight panels show distributions within a single imaging pipeline and scattering mitigation.
    The majority of azimuthal peak brightness among each ring cluster is shown in the outer gray histogram in each panel.
}
    \label{fig:scat_sgra_cluster}
\end{figure*}

\begin{figure*}
    \centering
    \includegraphics[width=\textwidth]{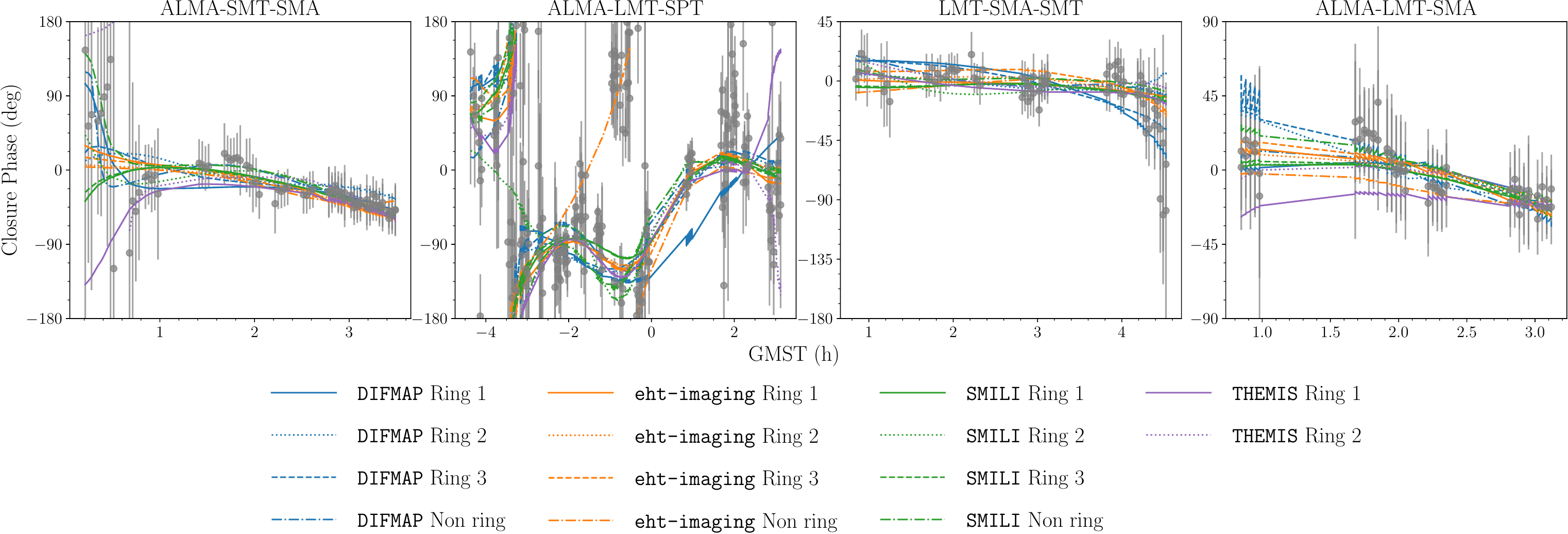}
    \caption{
    Closure phases plotted as function of GMST on four selected triangles from April 7 observations. Each line indicates the corresponding closure phase curves from a single Top Set and posterior image randomly selected from each cluster. The error bars of \sgra data include the fractional 1\,\% noise budget for systematic error and a representative budget for scattering and temporal variability -- in particular, the {\tt J18model1} refractive noise model and a variability model with parameters $a=0.02$, $u_0=2$, $b=2.5$, and $c=2$. These additional noise budgets are all added to the 60 second complex visibility noise budget prior to forming closure quantities. All images show reasonable fits to \sgra data as they all are within two standard deviations of the observed data. %
    }
    \label{fig:SgrA*-cphase-cluster}
\end{figure*}

\subsection{Clustering of Recovered \sgra Images}
\label{sec:sgra_images:clustering}
To effectively visualize the distributions of \sgra images, we cluster all reconstructed Top Set and posterior images using a similar criteria to that used in \autoref{subsec:synthetic_data_images_grmhd} and \autoref{fig:grmhd-3599} (see \autoref{appendix:clustering_method} for details).
Images with a ring feature are grouped by the peak PAs in the southwest ($-180^\circ\leq {\rm PA}<-70^\circ$), northwest ($-70^\circ\leq {\rm PA}<40^\circ$), and east ($40^\circ\leq {\rm PA}<180^\circ$) directions. The \sgra image clustering results for each pipeline are summarized in \autoref{fig:SgrA-topset-3599}; here, we focus on the observed date of April 7 with the HOPS data reduction pipeline (see \autoref{sec:sgra_images:meanimages} for April 6 and CASA-based reconstructions).
The results of the \difmap, \ehtim, \smili, and \themis pipelines are displayed from left to right. Images are separately clustered within each pipeline. 

The middle and bottom panels of \autoref{fig:SgrA-topset-3599} show each cluster's average image for the descattered and on-sky images, respectively. %
To better visualize the properties of individual \sgra images, we present three randomly selected descattered images from each cluster in the columns of \autoref{fig:SgrA-topset-3599}'s top panel; within each cluster images appear to have largely consistent morphologies.

The percentages in each panel of \autoref{fig:SgrA-topset-3599} show the fraction of Top Set or posterior images contained in that particular cluster. These percentages indicate that most of the Top Set and posterior images have ring structures.
For instance, the fraction of non-ring cluster images is $\leq 5$\,\% of the Top Set descattered images from \difmap, \ehtim, and \smili imaging pipelines. Although this is significant, 
we note that in the case of the Top Sets this does not constitute a likelihood and therefore the reported fractions should not be considered as an exact measure of our degree of certainty. In addition, only ring images appear in \themis posterior estimation of descattered images.

The bottom row of both the middle and bottom panel of \autoref{fig:SgrA-topset-3599} shows the azimuthal distribution of the radial peak intensity within a radius of 10 to 40\,\uas.
These profiles are sorted within each cluster by the location of peak brightness to best accentuate variations within a cluster.
By inspecting the profiles in each cluster it can be seen that three primary brightness distributions, with a peak brightness at $\sim -140^\circ$, $-40^\circ$, or $70^\circ$, appear across the pipelines. Thus, even when we restrict our attention to ring-like morphology reconstructions, it is difficult to constrain the azimuthal profile around the ring via imaging. This azimuthal uncertainty could be due to data properties (e.g., sparsity or low \textit{S/N} data) or variation in the intrinsic azimuthal structure of \sgra due to intra-day evolution. 

To visualize the image-domain differences among each cluster, we compare in \autoref{fig:scat_sgra_cluster} the relationship between the fractional central brightness ($f_c$, radial values) and the azimuthal peak brightness (${\rm PA}$, azimuthal values). The polar distribution of Top Sets and posterior images among all imaging pipelines  and with and without scattering mitigation (left panel in \autoref{fig:scat_sgra_cluster}) shows that most of the ring images within each different cluster (indicated by the red, blue, and green points) have a smaller fractional central brightness ($f_c\lesssim 0.5$) than that for the images in the non-ring cluster ($f_c \gtrsim 0.5$). 

The histogram of the azimuthal peak brightness distributions shown in left panel in \autoref{fig:scat_sgra_cluster} provides a clear visualization of the clustering of ring images around position angles of 
$\sim -140^\circ$, $-40^\circ$, or $70^\circ$,
which correspond to the locations of the three knots that commonly appear in ring images. Changes in position angle mostly reflect variations in the relative brightness of these knots. For the case of non-ring images the position angles are, as expected, more randomly distributed. \autoref{fig:scat_sgra_cluster} also confirms that the fractional central brightness are systematically larger for the on-sky images due to the angular broadening produced by the interstellar scattering.

Both the ring and non-ring morphologies identified in Top Set and posterior images show reasonable fits to \sgra data. In \autoref{appendix:topset_selection}, we show $\chi^2$ distributions of Top Set images to \sgra data. After adding the budgets of non-closing systematic errors, representative refractive noise, and time variability, all Top Set images result in a $\chi^2 < 2$ for both closure phases and log closure amplitudes -- we refer the reader to \citepalias{PaperIV} and \citepalias{M87PaperIV} for a discussion of these data products and $\chi^2$ distributions. In \autoref{fig:SgrA*-cphase-cluster} we compare closure phases of \sgra to those of individual images from each cluster and pipeline for four selected closure triangles. As indicated by the $\chi^2$ metrics, multiple ring and non-ring images fit the observed closure phases within the range of deviations anticipated due to temporal variability and refractive scattering effects. Therefore, none of the identified clusters can be excluded from possible \sgra morphologies in terms of the goodness of fit to \sgra data (or through synthetic data tests presented in \autoref{sec:survey}).%

\subsection{Average \sgra Images across Pipelines}
\label{sec:sgra_images:meanimages}

\begin{figure*}[!tb]
    \centering
    \begin{tabular}{ccc}
        \begin{minipage}{0.5\textwidth}
            \centering
            \includegraphics[width=\textwidth]{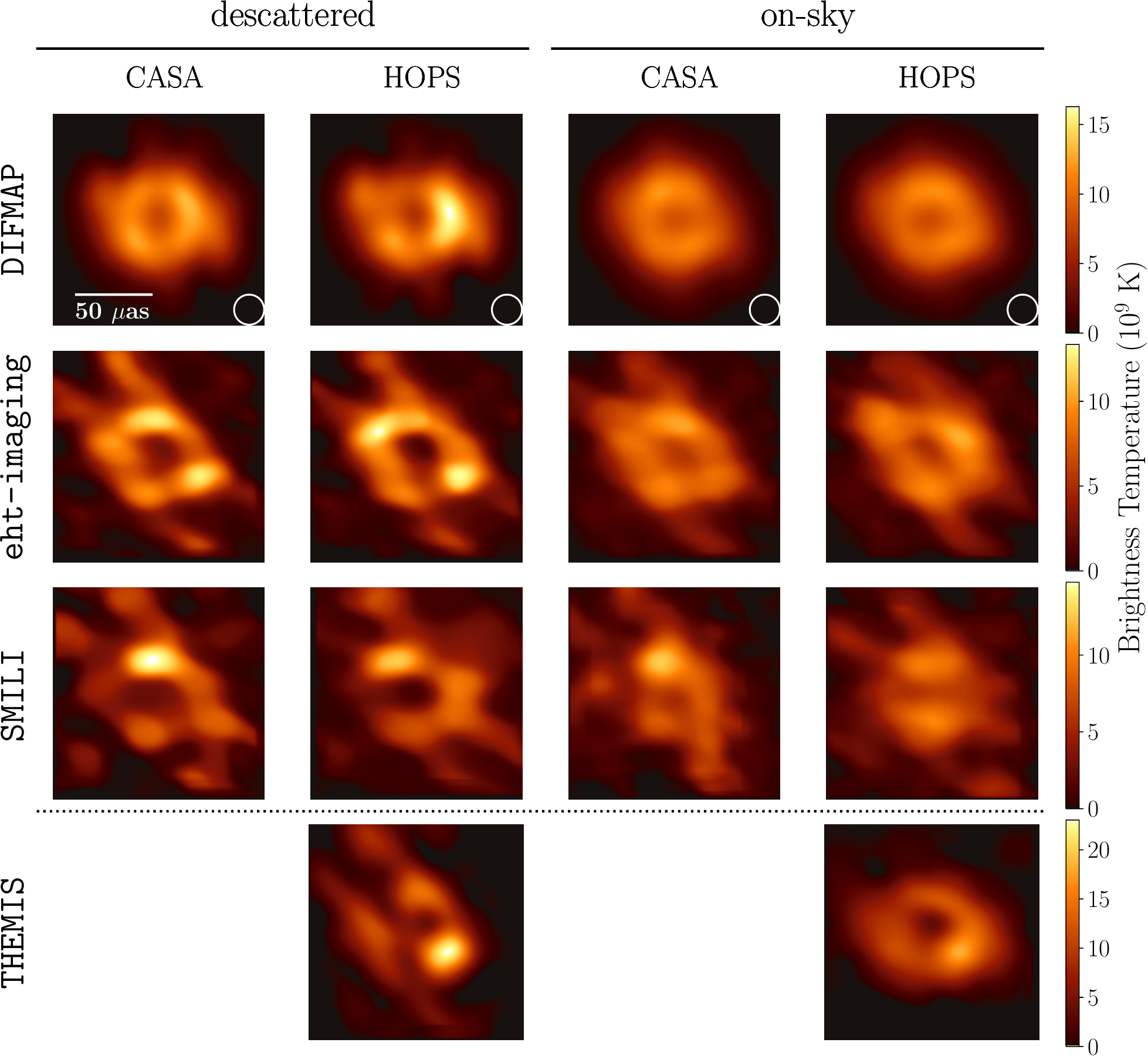}\\
            (a) April 6
        \end{minipage} &
        \begin{minipage}{0.5\textwidth}
            \centering
            \includegraphics[width=\textwidth]{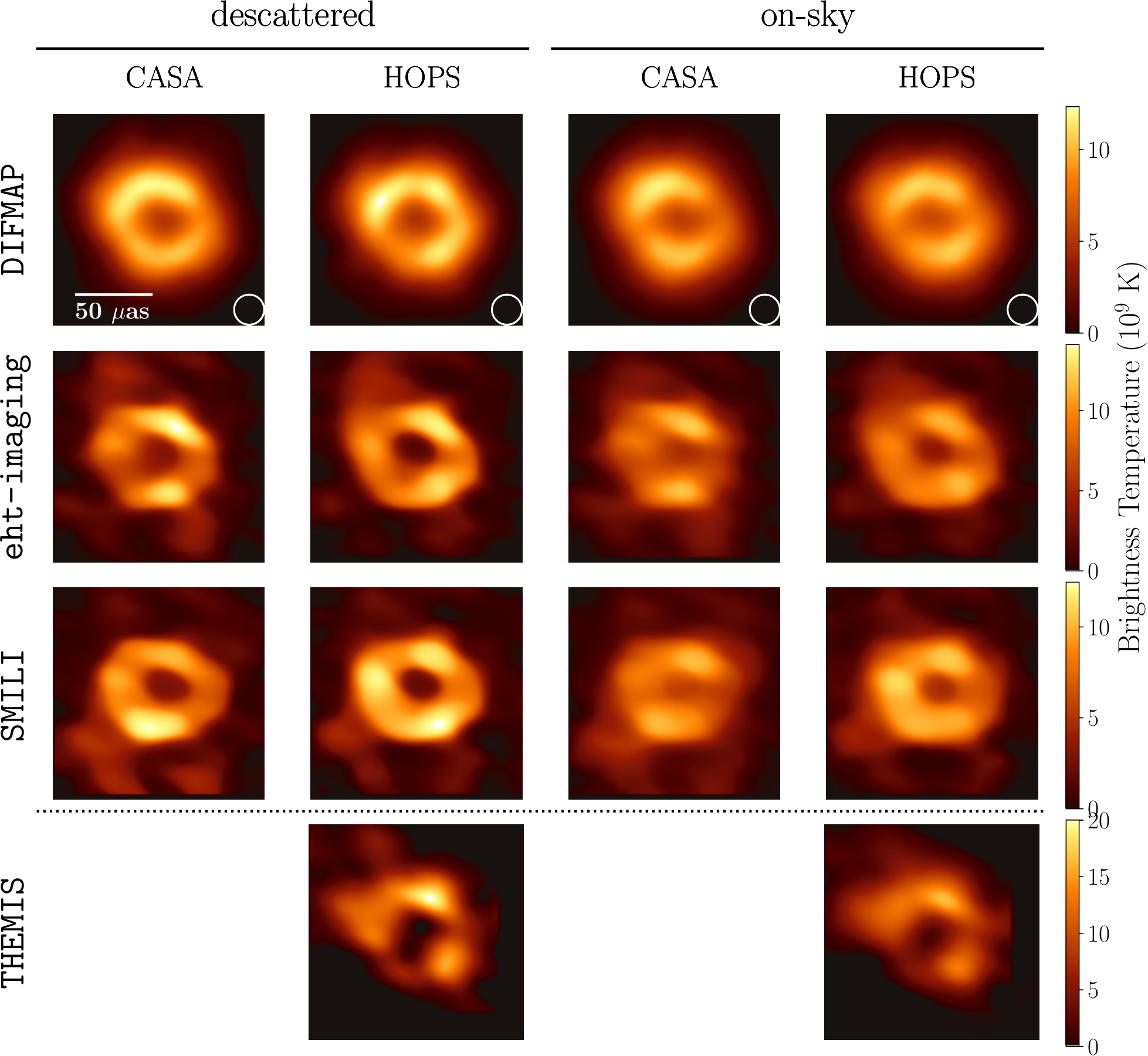}\\
            (b) April 7
        \end{minipage}
    \end{tabular}

    \caption{
    The dominant recovered morphology in \sgra descattered and on-sky reconstructions identified from two VLBI data products (CASA and HOPS data) with all four imaging pipelines (\difmap, \ehtim, \smili and \themis) for two observing days (April 6 and 7). Each panel shows the average image of the corresponding Top Set images for \difmap, \ehtim and \smili pipelines, and the average posterior image for \themis pipeline. Only the HOPS data has been imaged using the \themis pipeline.
    }
    \label{fig:SgrA*-average}
\end{figure*}

\autoref{fig:SgrA*-average} shows the average of \sgra images reconstructed by each of the four imaging pipelines (\difmap, \ehtim, \smili, and \themis) from each calibrated data set (CASA and HOPS) on each observing day (April 6 and 7). Only images from the HOPS data product have been reconstructed using the \themis pipeline. The images from \difmap, \ehtim and \smili are obtained by averaging their respective Top Set images; these average images show the dominant features identified across different combinations of the selected imaging parameters in the Top Sets (refer to \autoref{sec:survey}). 
For the \themis reconstructions we instead show the mean of each posterior obtained by averaging all the posterior samples.
\autoref{fig:SgrA*-average} shows that the majority of images contain a ring-like structure. This ring morphology is common among all imaging pipelines, and is resilient to the scattering mitigation strategy employed (as discussed in \autoref{sec:sgra_images:scattering}). Additionally, we recover largely consistent images between data calibrated by the HOPS and CASA calibration pipelines. Although we recover images with a ring morphology in the majority for all pipelines on April 7, these average images also highlight that the azimuthal brightness distribution is sensitive to small changes in the data and imaging strategy.

\subsection{Imaging Combining April 6 and 7 Data Sets}
\label{subsec:6+7}

\begin{figure}[tb]
\centering 
\includegraphics[width=\columnwidth]{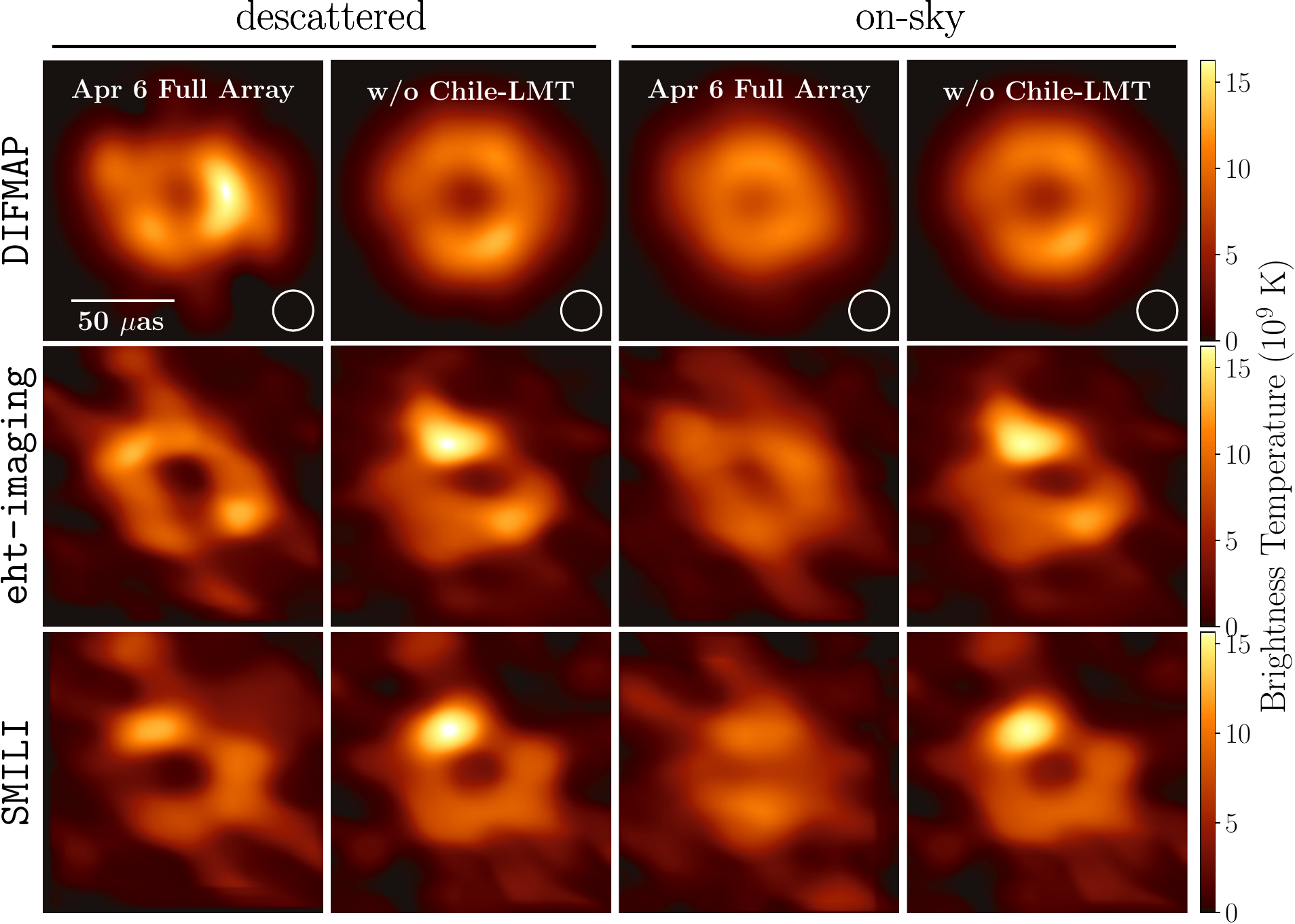}
\caption{
The effect of removing the Chile-LMT baseline from April 6 data reconstructions. Each panel shows the average image of the Top Set images for the \difmap, \ehtim and \smili pipelines from April 6 and the HOPS data product for the descattered and on-sky reconstructions. For comparison we show the average images obtained from full data sets as well as image obtained from data without the Chile-LMT baselines. This Chile-LMT baseline appears near the visibility null, and appears to exhibit significant intra-day variations on April 6 that is likely not captured by the variability noise model presented in \autoref{sec:PSD_noise}.
}
\label{fig:avg_apr6_flag_chile_lmt}
\end{figure}

\begin{figure}[tb]
\centering 
\includegraphics[width=1.0\columnwidth]{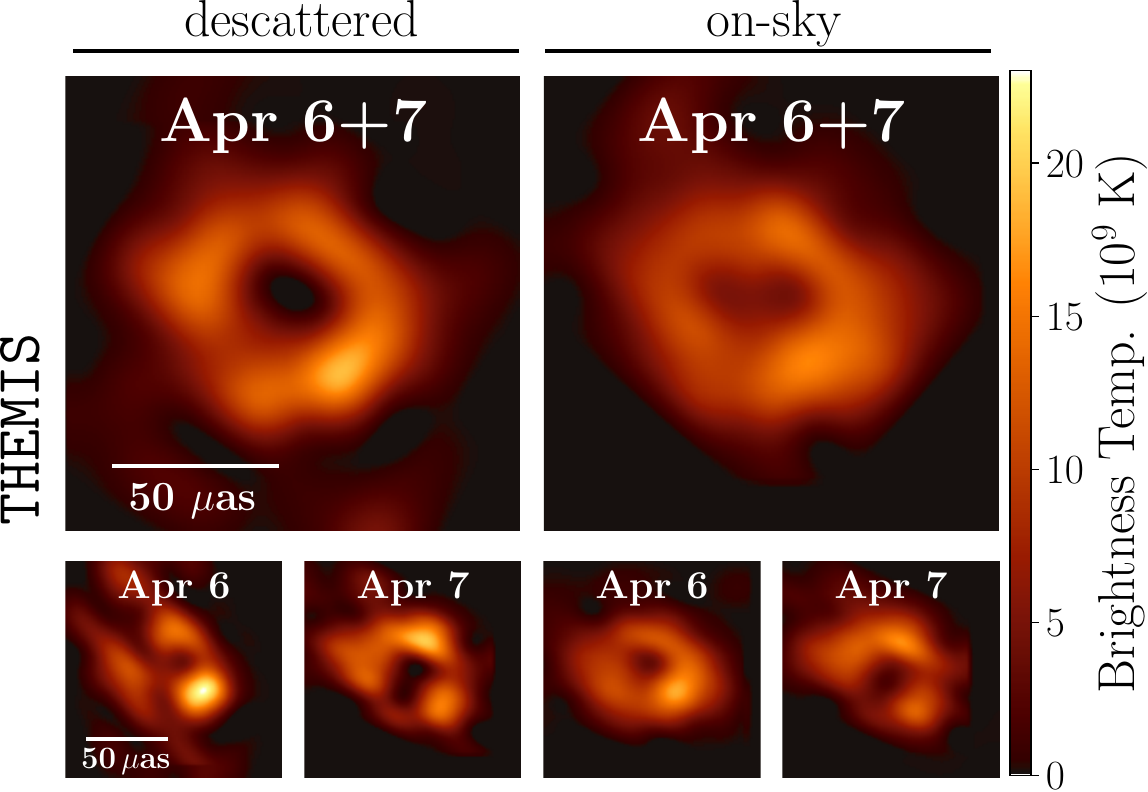}
\caption{Mitigation of the apparent strong intra-day variations seen in April 6 through \themis imaging of the combined April 6 and 7 data. For comparison we show the average posterior images from April 6 and 7 independently and that obtained by combining the data sets for April 6 and 7, exhibiting a clear ring for the descattered and on-sky reconstructions.
}
\label{fig:avg_apr6_themis}
\end{figure}

In \autoref{fig:SgrA*-average} we show results from April 6 data, reconstructed using the same imaging procedure as April 7. We note that although a ring feature appears in most of these reconstructions, it is less prominent. The images also contain a diagonal rail-like feature going from northeast to southwest that corrupts the ring. This feature is especially prominent in the \themis April 6 descattered reconstruction. 
This corrupted or non-ring mode is likely  emphasized in \themis imaging compared to RML and CLEAN pipelines due to the goal of \themis image samples to characterize the probability of an image rather than represent the variety of possible images that can fit the data. %

Through an in-depth inspection of the April 6 data, documented in \citetalias{PaperIV}, the Chile-LMT baselines (i.e., ALMA-LMT and APEX-LMT baselines) were identified as having large coherent visibility swings that are not effectively mitigated by the noise model on a single day, causing this particular feature to arise. \autoref{fig:avg_apr6_flag_chile_lmt} shows the reconstructions using the \difmap, \ehtim\, and \smili imaging pipelines when these particular baselines have been flagged, resulting in a cleaner ring structure.%

The large visibility swings on the Chile-LMT baselines could be the cause of variability around \sgra that exceeds expectations set by the incorporated stationary noise model presented in \autoref{sec:PSD_noise}. In particular, we believe that the variability noise model should capture \sgra's stochastic evolution in expectation, but a single night may contain non-stochastic short-lived variability that can bias a single day's reconstruction. Correlated variability may be mitigated via multi-day fits, which combine statistically independent structural fluctuations; this better matches the assumption within our variability mitigation scheme of an underlying stochastic process, though does carry with it the additional assumption that \sgra is statistically stationary over the multiple days combined. Therefore, in addition to the single day analyses performed by all imaging methods, the \themis image model was also fit to the combined April 6 and 7 scan-averaged data from high and low bands for the HOPS data product. The resulting static image shown in \autoref{fig:avg_apr6_themis} represents an image recovered from the combined data sets. All images within the multi-day \themis posterior exhibit a clear ring-like structure.%

\subsection{Is \sgra a Ring?}
\label{sec:sgra_is_a_ring}

Our primary imaging goal is to answer the question: ``Is \sgra a ring?''. Although our reconstructions are overwhelmingly dominated by ring images, there is a small number of non-ring images that fit the data well and cannot easily be excluded through additional tests. 
There are at least three possible reasons for the recovery of non-ring reconstructions of \sgra: 1) \sgra's intrinsic structure is not ring shaped, 2) scattering causes a distortion of a ring morphology, resulting in a non-ring image, and 3) imaging algorithms recover an incorrect source structure, aggravated by challenges of sparse \uv-coverage and \sgra's intra-day variation. In this section we will explore these three possible origins of non-ring structure. Based on this exploration we conclude that there is evidence that the non-ring reconstructions are caused by our imaging algorithms resulting from the limited $(u,v)$-coverage and rapid variability, rather than being intrinsic to \sgra.

\subsubsection{Manifestation of Rings from Intrinsic Non-rings}%
\label{sec:reconofnonring}

The first possible explanation for the small percentage of non-ring reconstructions of \sgra is simply that \sgra does not possess a ring morphology. Although this possibility cannot be ruled out completely, we explore the possibility of a potential bias in our imaging approach towards recovering ring images from underlying non-ring sources. 

There is always a possibility that the parameters initially explored by the RML and CLEAN parameter surveys were inadvertently biased to produce mostly ring images. 
This would result in an artificially high percentage of ring reconstructions that could give overconfidence in an incorrect ring solution.
To explore this hypothesis we inspect the percentage of ring and non-ring images of \sgra that are in the initial parameter survey versus the restricted Top Set on April 7. We find that although ring images make up a large percentage of the initial parameter surveys, 91\%, 84\%, 60\% for \difmap, \ehtim, and \smili respectively for the descattered reconstructions, this number becomes significantly larger after Top Set selection. In particular, the percentage of ring images raises to values of 95\%, 97\%, 98\%, respectively. Therefore,
when we restrict to those parameters that can best disambiguate between the different ring and non-ring source morphologies contained in our synthetic data sets (\autoref{sec:synthetic_data:geometric}), the number of ring images increases. We also note that \themis posterior samples for April 7 only contains ring images.

To further test the possibility that we are biased to recover primarily rings from underlying non-ring sources, we have explored the performance of our imaging methods on non-ring synthetic data sets to see how they compare to \sgra results. In particular, we performed a Top Set analysis on the \texttt{Point} and \texttt{Double} source models. Two ``cross-validation" Top Sets were identified by excluding the performance of either the \texttt{Point} or \texttt{Double} source model and only using the remaining six geometric models in selection of the imaging parameters. These two cross-validation Top Set parameter sets were then applied to the \texttt{Point} and \texttt{Double} synthetic data sets. Although each cross-validation Top Set does incorrectly reconstruct some ring images, each image set primarily contains reconstructions that look similar to the ground truth \texttt{Point} ad \texttt{Double} source structure. In particular for descattered reconstructions, only 9\% and $<$1\% percentage of the cross-validation Top Set reconstructions possess a ring morphology, compared to 97\% in \sgra reconstructions of April 7. 
The \themis pipeline produces no ring images in the posterior samples for both the \texttt{Point} and \texttt{Double} data sets.
Thus, in the non-ring synthetic data sets we have explored we find that the number of incorrect ring reconstructions is much less than the number that we recover for \sgra. In summary, our imaging pipelines do not appear prone to artificially create a majority of ring structures in sources that do not possess an intrinsic ring morphology.

\subsubsection{Scattering's Effect on Image Reconstruction  }
\label{sec:sgra_images:scattering}

The second possible explanation for the non-ring reconstructions of \sgra is that interstellar scattering causes a non-ring morphology. 
This could take either two forms: a) \sgra's ring morphology has been distorted to a non-ring morphology by an interstellar scattering screen, or b) our imaging algorithms have incorporated an incorrect scattering model that reconstructs a corrupted non-ring morphology.

To address the possibility that interstellar scattering is causing \sgra's intrinsic structure to be distorted to a non-ring, we inspect differences in the descattered versus on-sky reconstructions. 
As outlined in \autoref{sec:pre_scatter}, descattered reconstructions attempt to mitigate two primary effects of interstellar scattering --- diffractive scattering that causes angular broadening and refractive scattering that introduces small-scale structure to the on-sky image. \sgra reconstructions in \autoref{fig:SgrA-topset-3599} show that on-sky images are systematically blurrier than descattered images, as expected due to deblurring that is performed before recovering a descattered image (refer to \autoref{sec:pre_scatter_diff}). This systematic difference between on-sky and descattered images is also seen in synthetic data reconstructions presented in \autoref{subsec:synthetic_data_images}.
This angular broadening causes the central dip of a ring to be less prominent.
Nonetheless, we note that the vast majority of the on-sky (i.e., without any scattering mitigation prescription) images are still rings, with percentages of 93\%, 98\%, 90\%, and 98\% for CLEAN, \ehtim, \smili, and \themis, respectively. 
We note that it is highly unlikely that a non-ring morphology was distorted into an on-sky ring morphology by interstellar scattering.

\begin{figure}
    \centering
    \includegraphics[width=.95\linewidth]{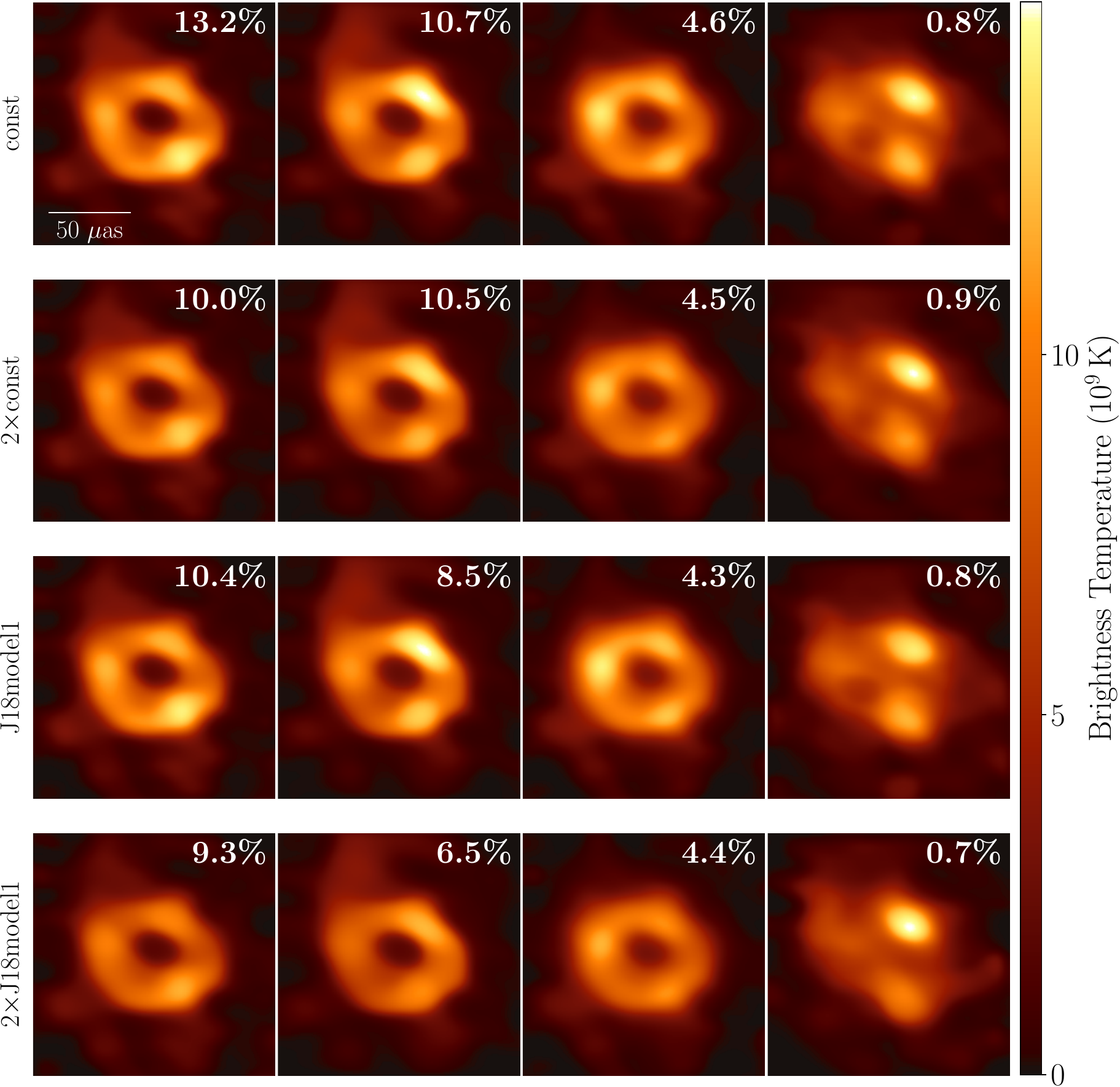}
    \caption{The range of descattered \sgra images recovered using different refractive scattering noise models. From left to right, images are shown clustered in the same order as \autoref{fig:scat_sgra_cluster}. Each panel shows an average of all descattered Top Set images of \sgra on April 7 from the \difmap, \ehtim, and \smili pipelines that were generated using the specified refractive noise model (from top to bottom: \texttt{Const}, 2$\times$\texttt{Const}, \texttt{J18model1}, and 2$\times$\texttt{J18model1}). The number on each panel shows the percentage of  descattered Top Set images that were reconstructed using the specified refractive noise model. These percentages indicate that there is not a clear preference for a particular refractive noise model in Top Set selection. Additionally, the recovered image structure does not appear to correlate with the refractive noise model used.}
    \label{fig:topset_scatt_dep_cluster}
\end{figure}

To address the possibility that our assumed model of the interstellar scattering is reconstructing a corrupted image of the descattered source, we explore the use of multiple refractive noise models to mitigate effects of refractive substructure before deblurring (\autoref{subsec:rml_clean_imaging_surveys}).
The contribution of the included refractive noise budget used to mitigate scattering (for descattered reconstructions) is non-negligible for long baseline data (see \autoref{sec:pre_scatter_ref} and \autoref{fig:scatt_kernel}). 
Nontheless, the RML and CLEAN imaging surveys show that the choice of the refractive noise model does not significantly affect the resulting distributions of image structures. \autoref{fig:topset_scatt_dep_cluster} shows the average descattered \sgra images (clustered into the same four morphologies as is presented in \autoref{fig:rep_sgra_images}) for each refractive noise prescription explored. The comparable fractions of images in each cluster across the different descattering strategies indicates the resiliency of the recovered image structures to different refractive noise models. 
Another piece of evidence suggesting that the scattering prescription does not strongly affect results can be seen in the \themis results; in \themis a constant refractive noise floor is able vary in posterior sampling of descattered images. Although an essentially unlimited refractive noise component is permitted by the \themis model, the posterior estimation instead typically chooses a noise level of $<25$ mJy, only 1\% of the total flux of the source.

In summary, results indicate that for EHT measurements of \sgra, interstellar scattering does not significantly affect the recovered morphology. 
Both on-sky and descattered images contain a majority of ring morphologies. 
In addition, we find the particular choice of scattering mitigation strategy only minimally affects the recovered image structures.

\subsubsection{Variability's Effect on Image Reconstruction}
\label{sec:sgra_images:psdnoise}

The third possible origin for non-ring \sgra images is poor reconstruction quality of the imaging methods. Imaging is solving an ill-posed inverse problem due to sparse \uv-coverage, which always will have the possibility of recovering an incorrect image. This is further exasperated in imaging \sgra by the challenges that come with recovering an evolving source. 
In this section we explore this possibility and find that our imaging methods often reconstruct non-ring sources for variable ring sources with a comparable small percentage as found for \sgra.

The first natural question is how our imaging methods perform on reconstructing ring sources with similar data properties as \sgra. 
We find that our methods produce non-ring images, even when the underlying source structure is ring-like.
As an example, recall that for the variable GRMHD ring-like sources analyzed in \autoref{sec:survey} and \autoref{sec:appendix_best_grmhd}, our imaging methods produced mostly rings, but also a very small fraction of non-ring images. Therefore, we expect that for a variable ring source we would recover some non-rings that fit the data well. 
However, note that the number of non-rings was still fairly small in the case of the GRMHDs: 5\% for the GRMHD presented in \autoref{fig:grmhd-3599} and 4\% for the ``best-bet'' GRMHD presented in \autoref{fig:bestbet-grmhd-3599}.
These values are comparable to the 3\% of non-rings reconstructed for \sgra descattered images across all pipelines. 
We also find that cross-validated Top Set images reconstructed of the \texttt{Crescent} and \texttt{Ring} geometric sources produce a small fraction of non-ring images.
Note that these non-ring percentages for the GRMHD, \texttt{Crescent}, and \texttt{Ring} sources are much less than the percentage of non-rings reconstructed for the variable \texttt{Point} and \texttt{Double} source, as discussed in \autoref{sec:reconofnonring}.

Due to \sgra's intra-day evolution, mitigation of temporal variability in the data is important for reconstructing a single static image of \sgra. 
We next explore how the variability mitigation approach affects the proportion of ring versus non-ring images reconstructed. 
\autoref{sec:PSD_noise} introduced an approach to model the temporal variability as an additional noise budget that could be added in quadrature to the visibility thermal noise budget. 
The dependence of the temporal variability noise model parameters on resulting \sgra reconstructions for the RML and CLEAN pipelines was investigated. 
Although a variability noise budget generally helps with imaging (as evidenced by a higher percentage of Top Set parameters selected -- 89-95\% of the Top Set parameter combinations include a variability noise model on April 7), similar to scattering, we find that there are no significant differences between the images recovered under different variability noise parameters. We suspect that this is partly caused by the complex interplay between regularizers and different noise model parameters (e.g., variability, scattering, and systematic), although no significant trends were identified.

In \autoref{sec:smili_dynam_survey}, we demonstrate how our results are insensitive to a different method of variability mitigation. In particular, we show time-averaged morphologies identified by a large survey over full-track RML dynamical imaging parameters (refer to \autoref{sec:rmldynamicimaging}) that does not rely on the variability noise model presented in \autoref{sec:stratvariabiledata}. We again find the ring modes to be dominantly reconstructed, while a small fraction of non-ring structures are also identified. The broad consistency indicates that our results are resilient to at least two different methods to recover time-averaged morphology.

In summary, we find that our imaging methods do reconstruct a small percentage of non-ring images from ring sources with comparable variability as that seen in \sgra data. Similar behavior is observed across a variety of different imaging approaches to mitigate variability. Therefore, we conclude that we would expect our methods to produce a small fraction of non-ring images from an underlying variable ring morphology.

%% file: image_analysis.tex
\section{Image Analysis}
\label{sec:image_analysis}

\subsection{Ring Parameter Fitting}
To analyze the \sgra images we use two tools, \texttt{REx} \citep{chael_thesis} and \vida\footnote{\url{https://github.com/ptiede/VIDA.jl}} \citep{Tiede_2020}, both of which are able to extract quantitative and pertinent information from the Top Set images. 
Here we briefly review the two image extraction techniques. 

\texttt{REx} attempts to extract ring-like features by directly characterizing the features of the Top Set images. This is the same tool used in 2017 \m87 analysis \citepalias{M87PaperIV}. The detailed definition of \texttt{REx} ring parameters follow those of the \m87 analysis \citepalias{M87PaperIV}. 

In \texttt{REx}, the ring center ($x_0$, $y_0$) is determined so that the dispersion of intensity peak radii is minimized. Around the center, a polar intensity map $I(r,\theta | x_0, y_0)$ is constructed. The ring radius $r_0$ (or diameter $d=2r_0$) is defined as the radius where  azimuthally-averaged  radial intensity peaks. The ring width $w$ is $\left<{\rm FWHM}[I(r, \theta|x_0, y_0) - I_{\rm floor}]\right>_\theta$, where  $I_{\rm{floor}}=\left<I\left(r_{\rm max} = 60\ {\rm\mu as},\theta | x_0, y_0 \right) \right>_{\theta}$\footnote{We set $r_{\rm max} = 60\ \rm \mu as$ instead of the $50\ \rm\mu as$ used in \citetalias{M87PaperIV} due to the larger ring size.}.
To characterize the azimuthal structure, we define the normalized first circular moment at radius $r$ as
\begin{equation}
    m_1(r) = \frac{\int^{2\pi}_{0}I(r,\theta|x_0,y_0)\cos(\theta)\mathrm{d}\theta}{\int^{2\pi}_{0}I(r,\theta|x_0,y_0)\mathrm{d}\theta}.
\end{equation}
The ring position angle $\eta$ and asymmetry $A$ are given by the radially averaged (from $r_0 - w/2$ to $r_0 + w/2$) argument and amplitude of $m_1(r)$ respectively. 
Finally, a central fractional brightness $f_c$ is defined as a ratio of the mean brightness within 5 $\mu$as from the center to the azimuthally-averaged brightness along the ring ($r=r_0$).

\vida takes a forward modeling or template matching approach for image analysis \citep{Tiede_2020}. That is, we approximate the images with parametric families or templates $f_\Theta$, such as as rings, crescents, or Gaussians. The template used depends on both the observed image structure and the features of interest. \vida's approach is therefore similar to geometric modeling presented in \citepalias{PaperIV}, except it is applied to the image reconstructions rather than in the visibility domain. The image features, such as diameter, are then given by the parameters of the optimal template. This differs from \texttt{REx} which defines its quantities directly on the image. 

To find the optimal template we first re-normalize the Top Set image to form a unit flux image $\hat{I}(x,y)$. We then take the L2 norm as our objective function:
\begin{equation}
    J(\Theta) = \mathrm{LS}(f_\Theta||\hat{I}) = \int |f_\Theta(x,y) - \hat{I}(x,y)|^2\mathrm{d} x \mathrm{d} y,
\end{equation}
where $\Theta$ denotes the template parameters. 
 
For analyzing \sgra Top Set images, we use \vida's SymCosineRingwFloor template. 
This template is characterized by a ring center ($x_0$, $y_0$), diameter $d$, FHWM width $w$, and a cosine expansion to describe azimuthal brightness distribution $S$:
\begin{align}
    S(\theta) = 1 - 2\sum_{m=1}^M A_m \cos\left[m(\theta - \eta_m)\right].
\end{align}
For this paper we take $m=4$. Note we also restrict $A_m < 0.5$ to restrict negative intensity in the template\footnote{It is still possible for negative intensity with this restriction. To prevent negative intensity we further take ${\rm max}(0, S(\theta))$ when computing the template.}. The position angle of the image is taken as the phase of the first order cosine expansion, i.e., $\eta_1 = \eta$. Similarly we define the asymmetry $A = A_1$ to match \texttt{REx}'s definition above.
Note, this azimuthal structure is very similar to the $m$-ring model described in \autoref{sec::dynamic}.

In addition, we add a central disk to constrain the central brightness depression of the ring. This disk is forced to have the same radius as the ring, and a Gaussian taper is included that matches the width of the ring. We then compute the central fractional brightness, $f_c$, of the optimal template, following the definition above.

In the next subsection, we provide the results of both \texttt{REx} and \vida for five ring parameters, namely ring diameter $d$, position angle $\eta$, ring width $w$, fractional central brightness $f_c$, and asymmetry $A$. Note that while each methods parameter definition are similar to each other \texttt{REx} produces estimates on the image directly, while \vida parameter values are defined from the optimal template. If the template provides a ``good'' approximation to the on-sky image these estimates should be similar. Note, the negative pixels of \themis images are treated as zero in the \texttt{REx} and \texttt{VIDA} analyses.

\subsection{Ring Fitting Results}
\begin{figure*}[t]
    \centering
    \includegraphics[width=\linewidth]{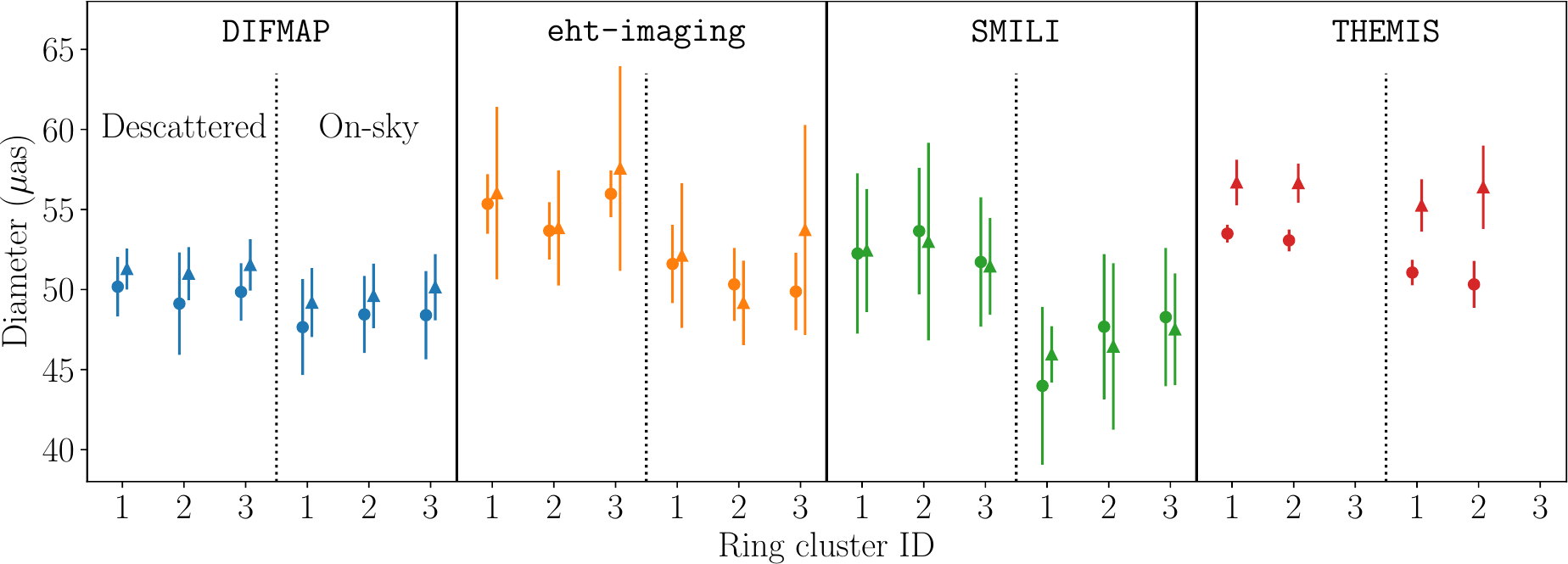}
    \caption{The ring diameters measured from April 7 images, shown separately for each cluster with a ring morphology, each pipeline, and descattered or on-sky reconstructions. Circle and triangle points and associated error bars indicate the means and standard deviations of diameters measured with \texttt{REx} and \vida, respectively.
    Note that \themis error bars appear significantly smaller than the error bars on \difmap, \ehtim, and \smili. This is partly due to the fact that \themis is primarily quantifying aleatoric (e.g., statistical) uncertainty whereas the goal of the other imaging surveys is to also characterize epistemic (e.g., systematic) uncertainty. 
    }
    \label{fig:SgrA_diameter}
\end{figure*}

In this subsection, we present ring parameter results only for images from ring morphology clusters  (\autoref{fig:rep_sgra_images}). 
In \autoref{fig:SgrA_diameter}, we show the diameters measured for on-sky and descattered ring images for each pipeline.
We find that the ring images are consistent with a diameter of \sdiam, as shown for each ring morphology cluster separately in \autoref{sec:sgra_images:clustering} and \autoref{fig:SgrA-topset-3599}. 

\begin{figure*}[t]
    \centering
    \includegraphics[width=\textwidth]{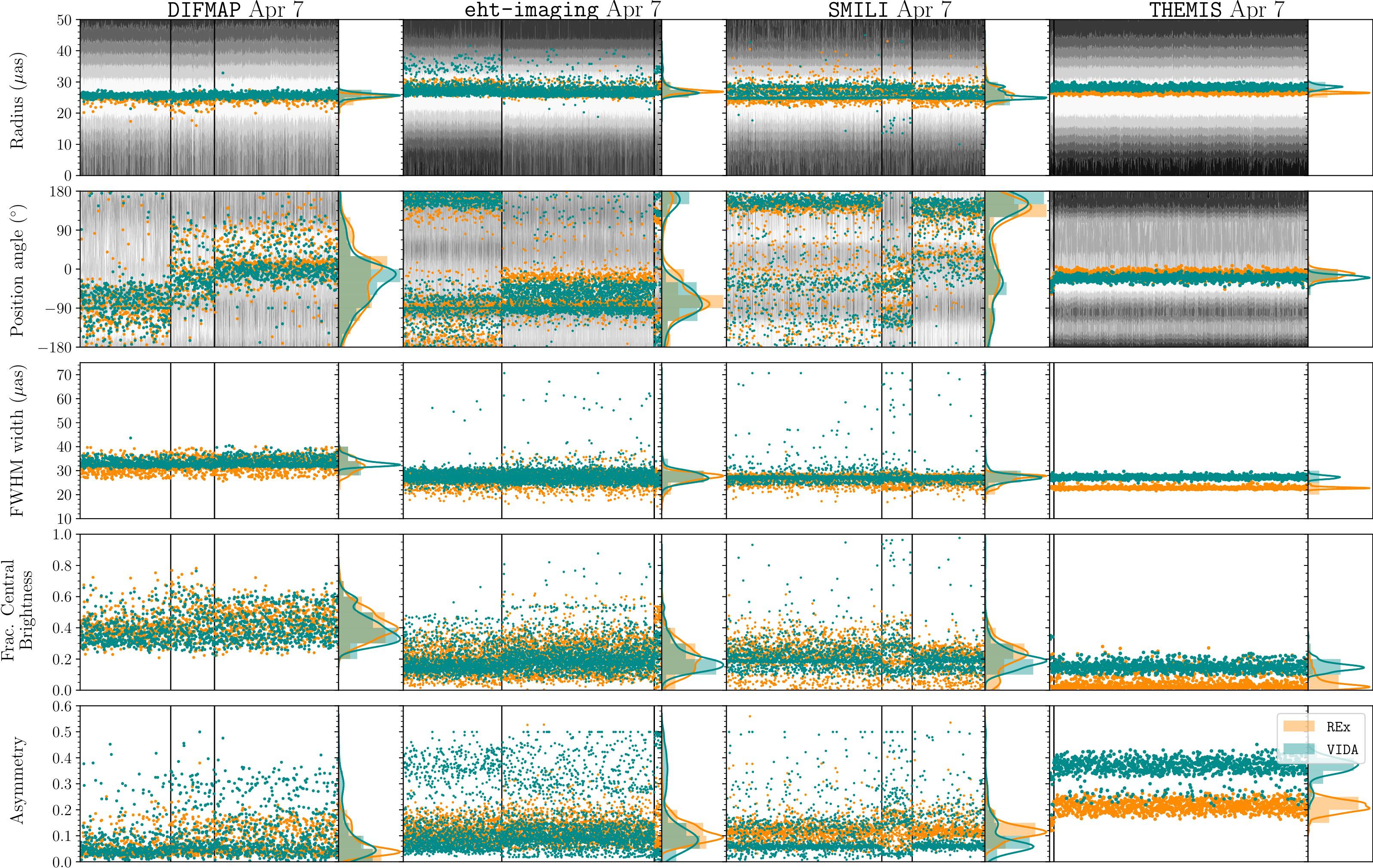}
    \caption{Ring fit results for Sgr A* Top Set descattered images in the ring clusters reconstructed from April 7 data. Each panel shows the distribution of ring parameters corresponding to the images resulting from a single pipeline. In each panel, the scatter plot on the left shows a ring parameter extracted from each image using \texttt{REx} (orange) and \vida (green); the vertical histogram on the right shows the resulting kernel density estimation from the collection of extracted image parameters. Images are ordered by clusters described in \autoref{sec:sgra_images:clustering}, of which boundaries are shown with vertical solid lines. From left to right, the panels show the results for \difmap, \ehtim, \smili and \themis. For the radius and position angle, we also show the radial/azimuthal brightness distribution of each image in the background.
    }
    \label{fig:SgrA-ringfit_deblur}
\end{figure*}

\begin{figure*}[t]
    \centering
    \includegraphics[width=\textwidth]{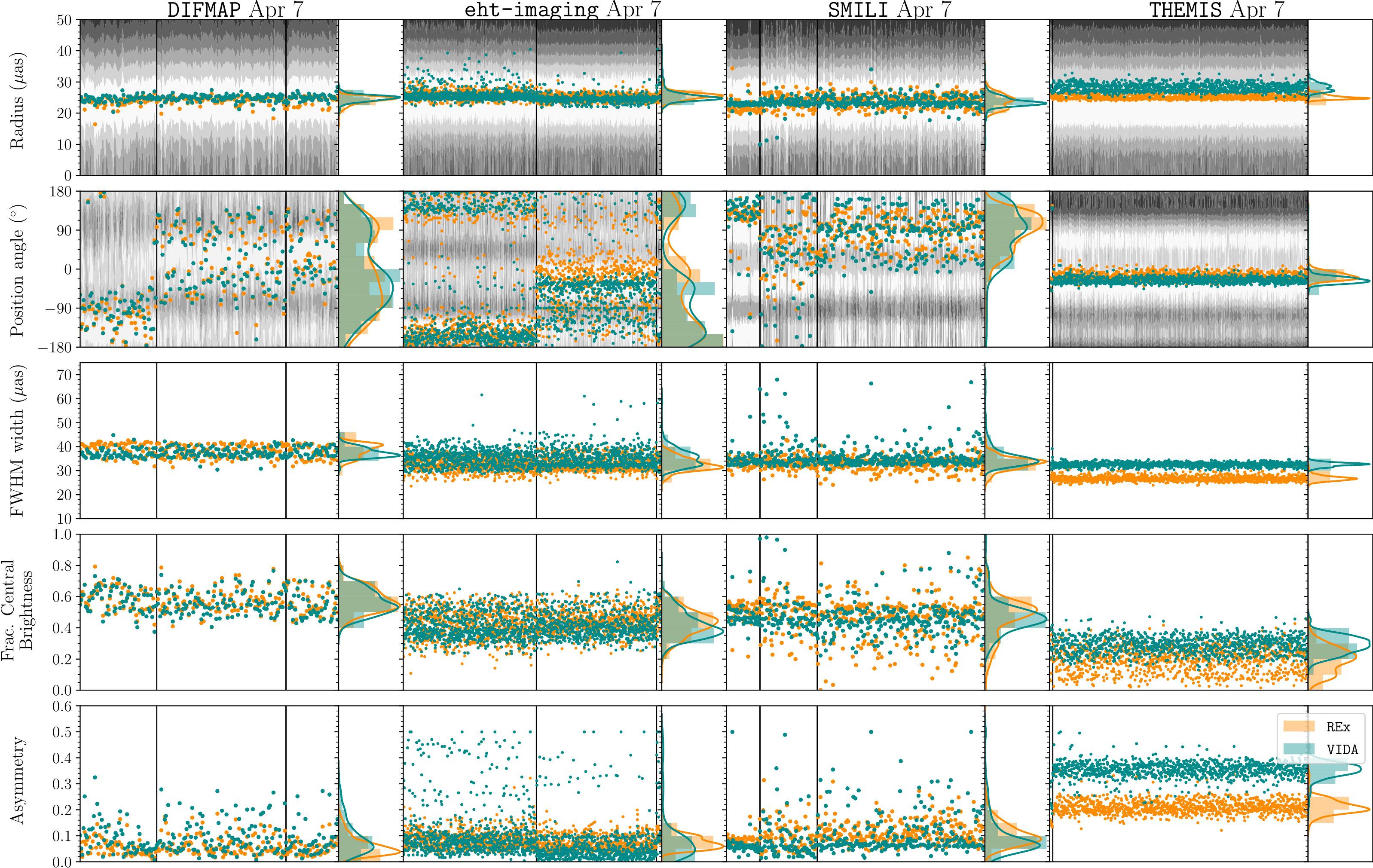}
    \caption{Ring fit results for Sgr A* Top Set on-sky images reconstructed from April 7 data. Refer to the caption of \autoref{fig:SgrA-ringfit_deblur} for more detail.}
    \label{fig:SgrA-ringfit_nodeblur}
\end{figure*}

For more detailed distributions of ring parameters, in \autoref{fig:SgrA-ringfit_deblur} we show the ring parameter fitting results of \sgra descattered Top Set or posterior images for April 7. 
From top to bottom, distributions of ring radius, position angle, ring width, fractional central brightness, and asymmetry are presented for four pipelines. Results from both \texttt{REx} and \vida are shown, and for comparison azimuthally or radially averaged brightness distributions are shown in the background of diameter $d$ and position angle $\eta$ plots.

Radius distributions are clearly peaked at $\sim$ 25 $\mu$as and mostly concentrated between 25 and 30 $\mu$as, being consistent across the three pipelines.
Note that in some occasions, \texttt{REx} and \vida show discrepancy in the radius (and some other parameters), which is mostly due to the difference of the responses to an image with a salient feature.
However, as seen in the kernel density distributions of the radius, contributions of such outliers are negligible for determinations of the mean ring radius.
Meanwhile, the PA value is by far less consistent across the pipelines or even within a single pipeline.
Multiple modes clearly appear in the Top Set images with various PA values, as already seen in \autoref{sec:sgra_images}.
Note that the scatter of PAs tends to be larger when the ring has multiple bright spots that affect the resulting PA, or when the azimuthal profile is close to uniform without a clear peak.
Ring width values are $\sim 30 \ \rm{\mu as}$, and this may come from the angular resolution of the observation. 
Fractional central brightness is mostly $\sim 0-0.3$ for RML and \themis while minor non-ring modes (right-most clusters in each panel) tend to show somewhat higher values. 
These values confirm that majority of the images show ring-like structures with clear central depression. 
CLEAN images tend to have a slightly larger value of width and central brightness than RML and \themis, as expected due to the beam-convolution effect for CLEAN imaging. 
Asymmetry values are $\sim 0.1$, indicating that most of \sgra images have a nearly symmetric azimuthal intensity distribution on ring. 
Apart from time variability, \sgra's apparent symmetry could be one of possible reason for difficulty in constraining the PA.

\autoref{fig:SgrA-ringfit_nodeblur} shows the ring parameter fitting results of the on-sky Top Set or posterior images. Comparing the ring parameters with and without scattering mitigation, PA values are slightly different and FWHM and fractional central brightness values are larger for the on-sky images. 
These differences are reasonable when one considers the angular broadening effect due to scattering that remains in on-sky images. 
On the other hand, radius and asymmetry show similar distributions between descattered and on-sky images --  we obtain a ring diameter of \sdiam for on-sky images, indicating that the ring size is robust with little dependence on the scattering correction.

In \autoref{tab:SgrA_ringfit}, we summarize the mean diameters and their standard deviations over the ring images reconstructed by each pipeline for both descattered and on-sky images and on both April 6 and 7. 
As seen in the table, the mean ring radii for on-sky images are slightly smaller than those of the descattered cases by a few $\mu$as.
This small reduction of ring radii is mainly due to the difference of effective resolutions with/without scattering mitigation.
Nevertheless, the ring radii are within their standard deviations and thus broadly consistent regardless of scattering mitigation.

Comparing the results of April 6 and 7 in \autoref{tab:SgrA_ringfit}, diameters derived from ring reconstructions are consistent within the computed standard deviations, regardless of the pipelines.
The other parameters (except for PA) also give consistent values for both days. 
Again, PA values have a large scatter in April 6 images; the existence of large scatter in PA indicates that it is difficult to constrain the azimuthal brightness distributions along the ring.
In general, most of the Top Set images show a ring morphology with a consistent diameter around \sdiam.

In \autoref{appendix:ring_params} we list the fitted diameter, width, position angle, asymmetry, and fractional central brightness measured for each one of the identified imaging clusters and different pipelines shown in \autoref{fig:SgrA-ringfit_deblur} and \autoref{fig:SgrA-ringfit_nodeblur}.

\begin{table*}[t]
    \centering
    \caption{Mean and standard deviation of diameter $d$ and width $w$ measured from Top Set or posterior \sgra images for each pipeline.}
\label{tab:SgrA_ringfit}
    \begin{tabular}{cc}
        \begin{minipage}{0.45\textwidth}
        \begin{center}
        \begin{tabular}{cccc} \hline \hline
        Descattered &     &              &            \\ \hline
        & & $d\ \rm{(\mu as)}$ & $w\ \rm{(\mu as)}$ \\ \hline
       \difmap &     &              &            \\
        April 6 Ring & \texttt{REx}  &  $46 \pm   4.1 $  &   $33 \pm   3.5 $
        \\
         & \vida  &  $51 \pm   3.1 $  &   $33 \pm   3.1 $ 
        \\
        April 7 Ring &   &  $49 \pm   2.1 $  &   $32 \pm   2.9 $
        \\
         &   &  $51 \pm   1.5 $  &   $33 \pm   1.7 $ 
        \\
        \hline
        \ehtim &     &              &            \\
        April 6 Ring &   &  $56 \pm   4.5 $  &   $24 \pm   2.4 $
        \\
         &   &  $59 \pm   11.3 $  &   $30 \pm   10.4 $ 
        \\
        April 7 Ring &   &  $54 \pm   2.0 $  &   $26 \pm   2.6 $
        \\
         &   &  $54 \pm   4.6 $  &   $27 \pm   3.5 $ 
        \\
        \hline
        \smili &     &              &            \\
        April 6 Ring &   &  $57 \pm   3.4 $  &   $24 \pm   1.9 $
        \\
         &   &  $46 \pm   12.0 $  &   $50 \pm   16.6 $ 
        \\
        April 7 Ring &   &  $52 \pm   4.7 $  &   $26 \pm   2.1 $
        \\
         &   &  $52 \pm   4.0 $  &   $27 \pm   4.7 $ 
        \\
        \hline
        	\themis &     &              &            \\
        April 6 Ring &   &  $51 \pm   3.9 $  &   $25 \pm   1.2 $
        \\
         &   &  $54 \pm   0.9 $  &   $24 \pm   0.9 $ 
        \\
        April 7 Ring &   &  $53 \pm   0.7 $  &   $22 \pm   0.5 $
        \\
         &   &  $56 \pm   1.2 $  &   $27 \pm   0.7 $ 
        \\
        \hline \hline 
        \end{tabular}
        \end{center}
        \end{minipage} &
        \begin{minipage}{0.45\textwidth}
        \begin{center}
        \begin{tabular}{cccc} \hline \hline
        On-sky &     &              &            \\ \hline
        & & $d\ \rm{(\mu as)}$ & $w\ \rm{(\mu as)}$ \\ \hline
        \difmap &     &              &            \\
        April 6 Ring & \texttt{REx}  &  $46 \pm   3.0 $  &   $34 \pm   4.3 $
        \\
         & \vida  &  $47 \pm   1.9 $  &   $39 \pm   3.4 $ 
        \\
        April 7 Ring &   &  $48 \pm   2.7 $  &   $38 \pm   2.7 $
        \\
         &   &  $49 \pm   2.1 $  &   $37 \pm   2.0 $ 
        \\
        \hline
        \ehtim &     &              &            \\
        April 6 Ring &   &  $49 \pm   3.9 $  &   $28 \pm   3.6 $
        \\
         &   &  $50 \pm   5.6 $  &   $41 \pm   6.7 $ 
        \\
        April 7 Ring &   &  $50 \pm   2.5 $  &   $32 \pm   2.7 $
        \\
         &   &  $50 \pm   4.1 $  &   $35 \pm   3.8 $ 
        \\
        \hline
        \smili &     &              &            \\
        April 6 Ring &   &  $43 \pm   0.4 $  &   $28 \pm   3.1 $
        \\
         &   &  $39 \pm   3.9 $  &   $46 \pm   8.8 $ 
        \\
        April 7 Ring &   &  $47 \pm   4.7 $  &   $33 \pm   2.7 $
        \\
         &   &  $47 \pm   3.8 $  &   $35 \pm   5.2 $ 
        \\
        \hline
        	\themis &     &              &            \\
        April 6 Ring &   &  $46 \pm   2.1 $  &   $30 \pm   1.6 $
        \\
         &   &  $47 \pm   3.1 $  &   $33 \pm   2.0 $ 
        \\
        April 7 Ring &   &  $50 \pm   1.5 $  &   $26 \pm   1.1 $
        \\
         &   &  $56 \pm   2.6 $  &   $32 \pm   0.9 $ 
        \\
        \hline \hline 
        \end{tabular}
        \end{center}   
        \end{minipage}
    \end{tabular}
\end{table*}

%% file: best_times.tex
\section{Short-timescale Dynamic Properties on Select Observation Window}
\label{sec::dynamic}

The dynamical timescale at the location of the innermost stable circular orbit for \sgra, $t_{\rm g}=12\pi\sqrt{6} G M /c^3$ for zero spin, is approximately 30 min and can be smaller by a factor of $\sim 10$ if the black hole is spinning rapidly. Variability at these timescales across the electromagnetic spectrum, including at 230 GHz, is one of Sgr A* salient features -- see~\cite{Wielgus_2021} and references therein. As discussed in \autoref{sec:PSD_noise} and \citetalias{PaperII}, a few EHT closure phase triangles show measurable variability across the 2017 observing campaign that can be attributed to intrinsic source variability. In this section, we explore the level and characteristics of structural changes in the \sgra image that are consistent with the observed variability.

Recovering time-resolved structures on these short timescales is especially challenging due to the sparse snapshot \uv-coverage for the EHT array. Indeed, without additional constraints, any observed change in the visibility domain can be interpreted as caused either by intrinsic variability or simply by the rotation of the baselines with the Earth and their probing of different spatial structures -- though fitting fast fluctuations in the visibilities with static emission requires larger fields-of-view.
This is especially true for baselines that probe regions of the \uv-space in which the visibility amplitudes show deep minima (or nulls), across which the complex visibilities change by order unity over infinitesimal changes in baseline length.

In attempts to describe the EHT observations with a static image, we assign any observed variation to spatial structures and mitigate potential effects of variability by inflating the error budget. %
In this section, we instead attempt to fit the time-evolving data directly to produce spatially- and temporally-resolved images of \sgra on minute timescales. Our analysis of dynamic properties presuppose that the 230~GHz emission from \sgra is compact (see  \autoref{sub:data-prop}) and ring-like such that the short-timescale variability we see can be attributed to changes in the image with time. We combine two independent analysis methods – dynamic imaging with temporal regularization between frames and snapshot geometric modeling – to identify trends in the spatial evolution of \sgra.

Our analysis shows that significant uncertainty exists in any attempt to characterize the spatially-resolved dynamics of \sgra using EHT 2017 data.  We expect that future observations with an expanded EHT array will yield significantly improved time- and spatially-resolved movies of \sgra.

\subsection{Selecting an Observation Window}
The rotation of the Earth causes the EHT's snapshot \uv-coverage to change over time. Static imaging and modeling approaches assume the source is unchanging in time, which allows these approaches to combine data from a full night of observations. However, recovery of short-timescale evolution requires that we only consider coverage synthesized on the variability timescale. This ``snapshot" coverage is extremely sparse and introduces artifacts into image reconstructions. To minimize these artifacts, we constructed and evaluated metrics to assess the performance of the snapshot coverage and identify the most promising time windows for dynamic analysis. These metrics rely purely on the \uv-coverage rather than the properties of the underlying \sgra visibilities. The construction and validation of a suite of these metrics
are reviewed in \cite{Farah_2021}. 

We consider metrics that assess several attributes of the \uv-coverage, including the largest gap in coverage (\citealt{github_LCG}), the fraction of the \uv plane covered (\citealt{Palumbo_2019}), and the geometric properties of the coverage (\citealt{Farah_2021}). We summarize the application of these metrics to the 2017 April 7 EHT \sgra data set in \autoref{sec:appendixdynamics}. These three metrics identify a period from approximately 1.5-3.2 GMST on April 6 and 7 that maximally mitigates the EHT's snapshot coverage  limitations. During this time window, all sites participate in observing \sgra except PV, though there is a notable dropout of the LMT between approximately 2.4-2.9 GMST on both days.
All dynamic analyses discussed in the remainder of \autoref{sec::dynamic} are performed \emph{only in this selected time window}.

\begin{figure}
    \centering
    \includegraphics[width=\columnwidth]{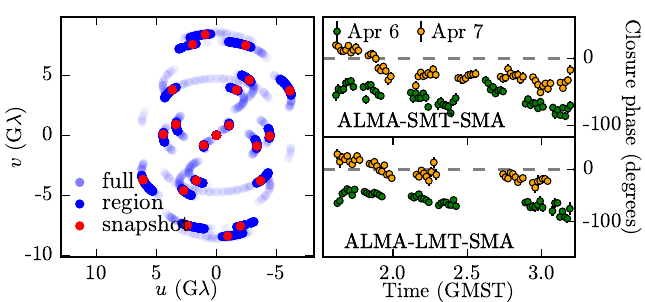}
    \caption{(Left) \uv-coverage for the selected time window for dynamic imaging and modeling. The light blue points show the coverage of the full night of observation, while the dark blue points and the red points represent the coverage for the selected dynamic imaging region and a single 60s snapshot from that region, respectively. 
    (Right) Closure phases for \sgra (green and yellow) on two representative triangles during the selected time region.}
    \label{fig:best_times_coverage_and_closure}
\end{figure}

\autoref{fig:best_times_coverage_and_closure} shows the \uv-coverage for April 7 during the roughly 100 min observation window selected for dynamic analysis, along with the coverage for a single 60 second ``snapshot" integration.
Closure phases from two informative triangles are overlaid for April 6 and 7 during this time window. These closure phases show distinct evolution in the resolved structure of \sgra during the same 100 min window on April 6 and 7. 

\subsection{Dynamic Imaging and Modeling Methods}
\label{sec:dynamapproach}

\begin{figure*}
    \centering
    \includegraphics[width=\textwidth]{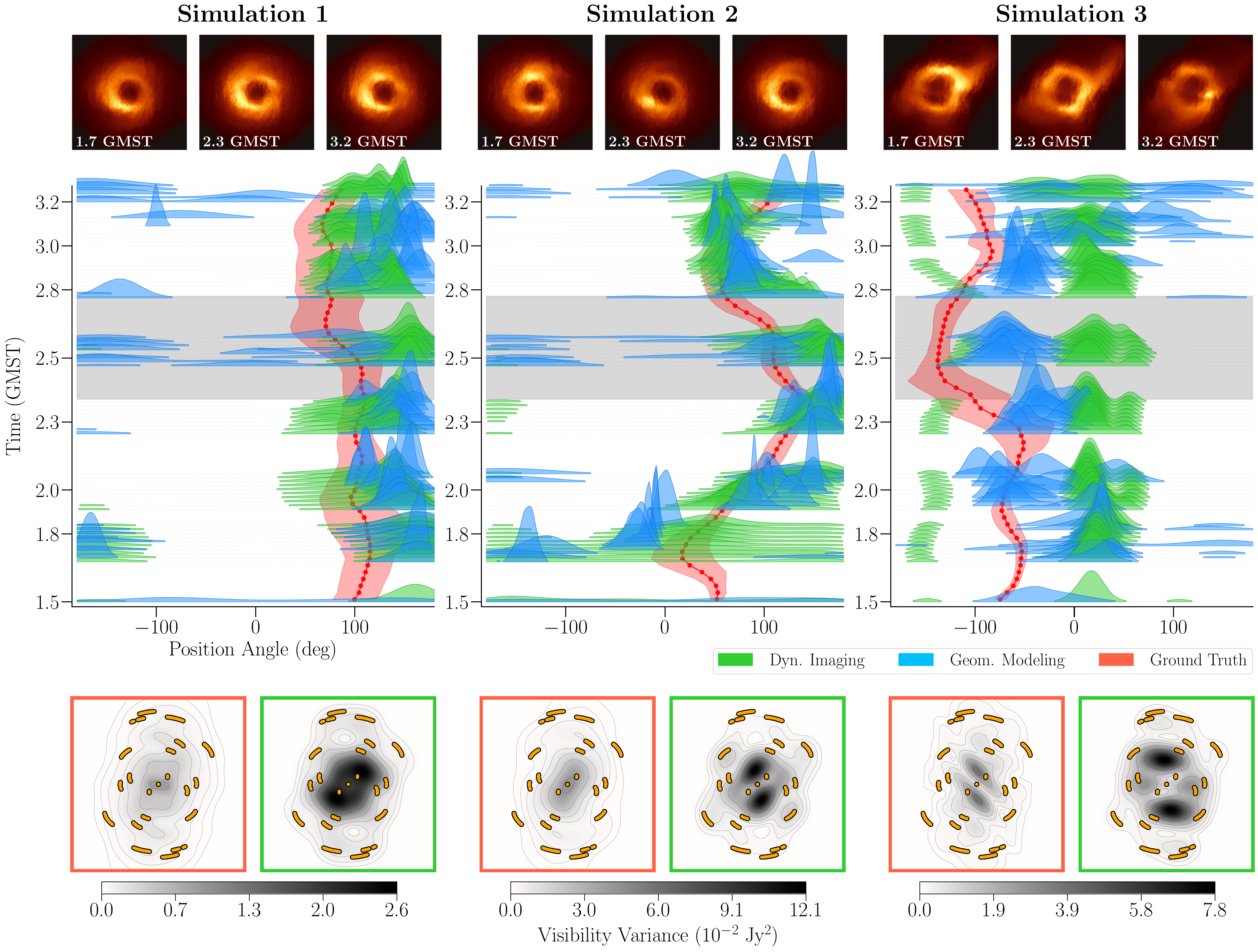}
    \caption{ Position angle (PA) recovered from synthetic data from three different GRMHD simulations on April 7 EHT coverage during the dynamic analysis window using both \texttt{StarWarps} dynamic imaging and \texttt{DPI} snapshot geometric modeling techniques. (Top row) Ground truth GRMHD movie snapshots (including interstellar scattering) from each of the three simulations at 1.7, 2.3, and 3.2 GMST. (Middle row) Plots of PA vs time for the reconstructions compared with the simulation ground truth (in red). The shaded red region indicates the circular standard deviation of the ground truth PA computed using \texttt{REx} (refer to \autoref{sec:image_analysis} and \cite{chael_thesis}). Modeling histograms (blue) correspond to actual marginal posterior distributions, whereas for \texttt{StarWarps} imaging the histograms represent the distribution of PAs and their associated uncertainties for a collection of movies reconstructed under different parameter settings. The gray band at roughly 2.6 GMST indicates the time period where the LMT dropped out of the observation.
    (Bottom row) Visibility variance in the \uv plane over the selected time window for the ground truth simulation movies (left, red) and the reconstruction (right, green).
    In Simulations 1 \& 2, both dynamic imaging and snapshot geometric modeling methods are often able to correctly identify the PA of evolving GRMHD movies during this time window, but they show significant offsets from the correct PA in Simulation 3. From left to right, the maximum variance of the ground-truth (reconstructed) movie is 0.85 (2.62), 4.45 (12.07), and 3.94 (7.79) $\times 10^{-2}$ Jy$^2$.
    Contours start at 90\% of the peak variance and decrease by successive factors of 2 until they reach 0.7\%. 
    }
    \label{fig:grmhd_pa}
\end{figure*}

\begin{figure*}
    \centering
    \includegraphics[width=\textwidth]{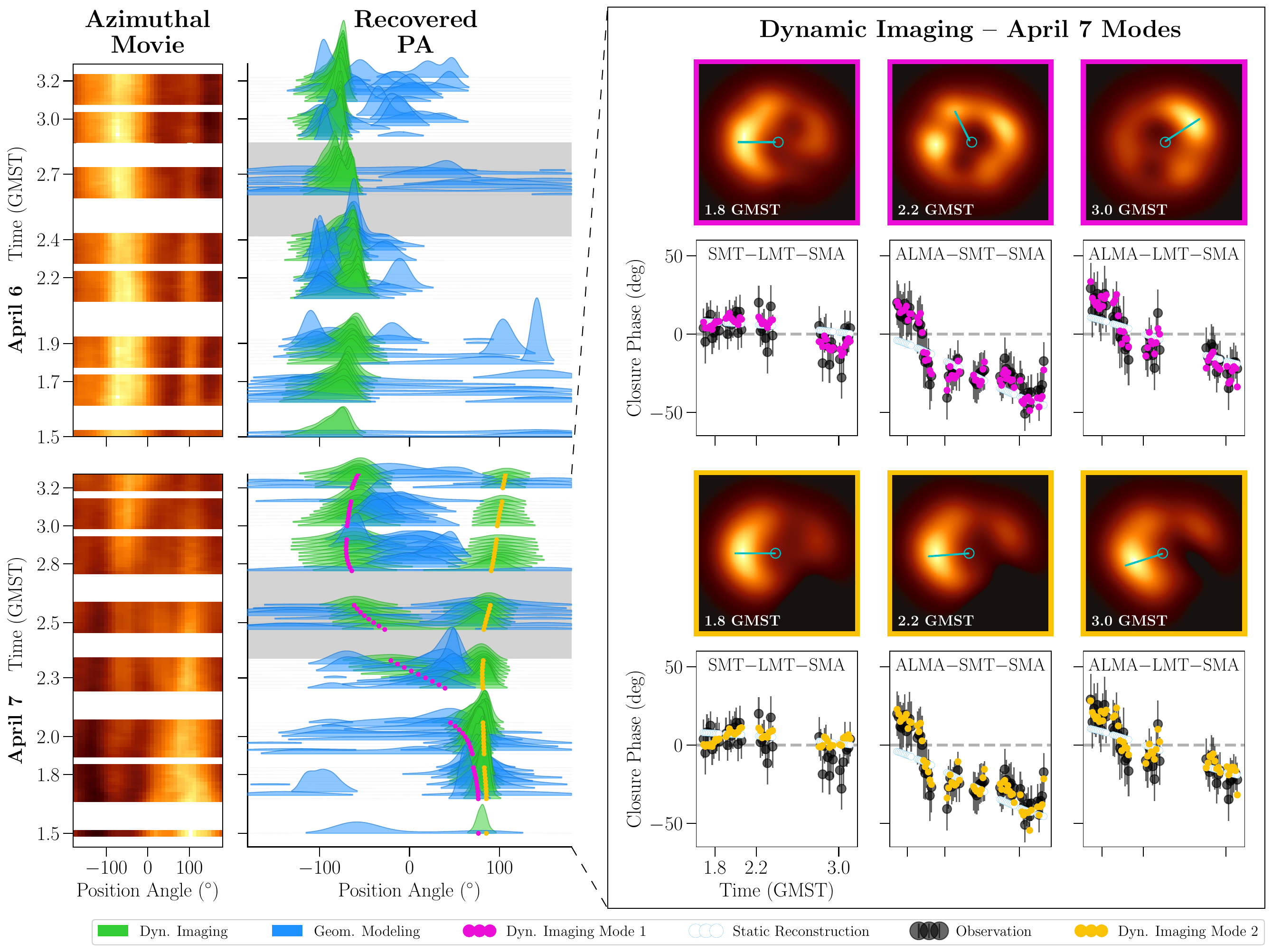}
    \caption{
    (Left) Mean azimuthal brightness profiles from the \texttt{StarWarps} movie reconstructions, unwrapped around the ring as a function of time, and PA distributions obtained from dynamic imaging and snapshot geometric modeling reconstructions of EHT \sgra data on April 6 (top) and 7 (bottom) in the selected time window. Geometric modeling distributions are marginal PA posteriors from \texttt{DPI}. Imaging histograms represent the distribution of PAs and their associated uncertainties from a collection of \texttt{StarWarps} movies reconstructed under different parameter settings with the spatial prior mean $\mu$ and temporal regularization $\beta_Q^{-1}$ held fixed (refer to \autoref{eq:starwarpsreg}). Blank spaces indicate time regions without any data. The gray band at roughly 2.6 GMST indicates the region where the LMT dropped out and data coverage is poor. Both dynamic imaging and modeling appear to identify a nearly constant PA on April 6 but a variable PA over the same time window on April 7. In the reconstructions in this figure, both dynamic imaging and modeling make a prior assumption that the source morphology is ring-like; \texttt{StarWarps} imaging uses a prior/initialization image of a uniform ring, while geometric modeling uses a second order $m$-ring ($m=2$) model. Both dynamic imaging and modeling recover ``descattered" movies using the \texttt{J18model1} refractive noise model. \newline (Right) Focus on the two modes reconstructed by \texttt{StarWarps} on April 7. For each mode, the top panels show three reconstructed snapshots at different times, and bottom panels compare the fitting of the corresponding reconstructed movie (magenta or yellow) and a representative static reconstruction from the \texttt{eht-imaging} static imaging pipeline (white) to the closure phase data measured by three key triangles. The dynamic reconstruction on the top (magenta) shows an evolving PA shift over the observation window. In contrast, the reconstruction on the bottom has a nearly constant PA of $\sim100^\circ$ (yellow). In the selected closure phase plots (bottom rows), the measured data is averaged in 60 s snapshots and the error bars do not include a variability noise model. The static image visibilities (white) capture the general trend of the data, but they do not well fit variability in the closure phases. In contrast, both selected \texttt{StarWarps} dynamic reconstructions better fit the data on minute timescales. We find that the fit's behavior on the SMT-LMT-SMA triangle has a large influence on the resulting PA of the movie on April 7. A positive SMT-LMT-SMA closure value tends to result in a southeast PA ($\sim100^\circ$) whereas a negative value results in a more northwest PA ($\sim-80$$^\circ$).
    }
    \label{fig:days_comparison}
\end{figure*}

To analyze \sgra's spatially-resolved dynamics during the 100 min selected region of time on April 6 and 7, we use 
two methods: dynamic imaging and snapshot geometric modeling (also simply referred to as dynamic modeling).
Both methods fit EHT 2017 data on 60s snapshot integrations within the selected observation window but make different prior assumptions about the structure of the source in space and time. Note that unlike in static imaging, we do not flux normalize the data before dynamic analysis.
Because the \uv-coverage is sparse even in the best available time window, both methods need to make strong prior assumptions about the spatial and/or temporal structure of the source to constrain the fits to the data.
Note that when performing dynamic imaging/modeling fits we {\it do not} include a variability noise budget as is done in static imaging (\autoref{sec:stratvariabiledata}). 

\paragraph{Dynamic Imaging}
Dynamic imaging methods reconstruct a time-evolving image that best fits the observed evolution of \sgra. Our dynamic imaging approach is based on \texttt{StarWarps} \citep{Bouman_2018}, which enforces continuity across an image and in time by means of spatial and temporal regularization (\autoref{subsec:starwarps}). 
Temporal regularization is set by a parameter $\beta_Q^{-1}$ which corresponds to the allowed variance between pixels in snapshots that are typically 60s apart\footnote{Frames are sometimes separated by more than 60s due to the interval between scans.}; smaller values of $\beta_Q^{-1}$ correspond to stronger continuity in time.  
Spatial regularization is imposed
by a multivariate Gaussian prior on snapshots with a mean $\mu$ and covariance $\Lambda$ that encourages spatial smoothness (see \autoref{eq:starwarpsreg} and \citealt{Bouman_2018}).
We examine the sensitivity of time-variable image features (e.g., position angle) to different settings in the \texttt{StarWarps} imaging algorithm by running surveys over different values of the spatial
regularization covariance $\Lambda$ and data weights of the visibility amplitude and log closure amplitude; we typically keep $\beta_Q^{-1}$ and the mean image $\mu$ fixed and examine the sensitivity of our results to these parameters separately across different surveys. 
Unless specified otherwise, in this paper we set $\beta_Q^{-1}$ to $5 \times 10^{-6}$ (Jy/pixel)$^2$ and the prior mean $\mu$ and initialization image to an image of a uniform ring blurred by a 25 $\mu$as beam. These surveys result in distributions of the image features at each snapshot in time. For each of the measurements obtained from 54 parameter combinations, we draw 100 random samples from a normal distribution characterized by the image feature measurement and its associated error. These survey results are \emph{not} posterior probability distributions, but they do provide a sense of the robustness of the sensitivity of movie reconstructions to changes in the algorithm parameters.

\paragraph{Geometric modeling}
In our geometric modeling approach, 
a simple geometric model is fit to each 60 s snapshot \emph{independently}, with no enforced correlations in time.\footnote{In the language of the temporal regularization parameter defined above, for geometric modeling $\beta_Q^{-1}\rightarrow\infty$.}
We consider several different $m$-ring models \citep{Johnson_2020},
described by infinitesimally thin rings with azimuthal brightness variations decomposed into Fourier modes, which are subsequently blurred with a circular Gaussian kernel. 
The complexity of an $m$-ring model depends on the maximum number of Fourier modes, $m$, that are added (e.g. $m=1$ corresponds to a simple crescent).
To model a central floor we include a Gaussian that is located at the center of the ring; the size and brightness of the Gaussian are additional model parameters.

For each $m$-ring model considered, we produce a multi-dimensional posterior using two modeling approaches. First, we consider a variational inference based approach, \texttt{DPI}, that fits to the log closure amplitudes and closure phases~\citep{sun2021, sun2021alpha}.
Second, we consider a sampling based method, \texttt{Comrade}~\citep{comrade}\footnote{\url{https://github.com/ptiede/Comrade.jl}}, which fits to visibility amplitudes and closure phases. \texttt{Comrade} uses the nested sampling package \texttt{dynesty} \citep{dynesty} and the probabilistic programming language \texttt{Soss} \citep{scherrer2020soss}.
The different data products used by \texttt{DPI} and \texttt{Comrade} imply different assumptions made about the telescope amplitude gains -- they are unconstrained in \texttt{DPI}, while in \texttt{Comrade} the gain amplitudes are more constrained and are included as model parameters during fitting (see \citetalias{PaperIV}). 
As a result of these different data products, \texttt{DPI} and \texttt{Comrade} produce slightly different posteriors. 
Details of the geometric modeling approaches are further explained in \citetalias{PaperIV}\footnote{Note that model fits in \citetalias{PaperIV} uses 120~second snapshots, while in this section, we use 60~second snapshots.}.

\paragraph{Comparing dynamic imaging \& modeling}
Both dynamic imaging and modeling share a common goal of extracting time- and spatially-resolved structure in \sgra but there are key differences between the methods in how prior assumptions about the spatial and temporal variability are incorporated. \texttt{StarWarps} imaging allows for more freedom in the recovered spatial structure but assumes strong temporal regularization between frames. In contrast, snapshot geometric modeling is restricted to a parameterized set of spatial structures, but makes no assumptions on image correlations in time. 
Although snapshot geometric modeling cannot recover spatial structures outside of the $m$-ring model specifications, it allows for quantifying the uncertainty in $m$-ring model features (and their temporal variability) as it estimates full posterior distributions for the particular geometric model used. 

\paragraph{Diagnostics}
To characterize our dynamic reconstructions, we mostly investigate trends of the brightness position angle (PA; see Eq. 21 in \citetalias{M87PaperIV}) with time. The PA is a simple and easily characterizable feature of the brightness distribution around an asymmetric ring. 
For \texttt{Starwarps} reconstructed movies, we extract the ring PA on the different snapshots using \texttt{REx}; in $m-$ring model fitting results, the PA is obtained directly from the fitted model as the argument of the first azimuthal Fourier mode.

\paragraph{Ring assumption}
Many of the results in this section apply strong prior assumptions that \sgra's underlying structure is ring-like, motivated by the 
ring morphology recovered in static image reconstructions using the full \uv-coverage (\autoref{sec:sgra_images}).
\texttt{StarWarps} reconstructions enforce a ring constraint by setting the mean prior image $\mu$ to a ring with $\approx 50\,\mu$as diameter and a width set by a circular Gaussian blurring kernel. 
In geometric modeling, the ring assumption is intrinsically imposed by the structure of the $m$-ring model. 
In \autoref{sec:dynamicmodelingchoices} and in \autoref{sec:appendixdynamics} we
explore the sensitivity of our results to the choice of mean image $\mu$ in \texttt{StarWarps}. 

\subsection{Synthetic Data Tests}
Sparse snapshot \uv-coverage can lead to artifacts in both imaging and geometric modeling results. These artifacts appear in static imaging, but are further amplified in dynamic imaging due to the far-sparser coverage
~\citep{Farah_2021}. Thus, it is important to assess how the sparse \uv-coverage during the selected time window may affect the recovered results and whether it may introduce biases in recovered image features, particularly the position angle (PA) of ring-like images. 

\subsubsection{Static Crescents}

In \autoref{sec:appendixdynamics_uvcov} 
we present synthetic data tests conducted to characterize the effect of the sparse snapshot EHT2017 \uv-coverage on PA recovery from static crescent images. 
These tests show that there are significant biases in the recovered PA from 60s snapshots when the brightness asymmetry of the ground truth ring image is low. When there is a strong asymmetry in the brightness distribution around the ring, however, the PA is accurately recovered even with 60s snapshot \uv-coverage. 

\subsubsection{GRMHD Simulations}
We explored how our methods perform in recovering time-varying position angles from three selected GRMHD simulation movies. We used three representative GRMHD movies from the GRMHD library presented in \citetalias{PaperV}  \footnote{ Simulation 1 \& 2 is using a MAD GRMHD model with parameters $a_{*}=0$, $i=10$, $R_{\rm high}=10$. Simulation 3 is using a SANE GRMHD model with $a_{*}=-0.94$, $i=50$, $R_{\rm high}=160$.} (see \autoref{sec:synthetic_data:grmhd}).
We generated visibility data from the three movies over the 100 min dynamic analysis window on April 7 using the same procedure described in \autoref{sec:synthetic_data}, including atmospheric noise, telescope gain errors, and polarimetric leakage. 

\autoref{fig:grmhd_pa} presents results obtained from both dynamic imaging and modeling reconstructions of these three synthetic data sets.
The ground truth simulation PA evolution is recovered (with $\sim$30$^\circ$) for the first two models (Simulation 1 \& 2). However, there are several localized deviations in the recovered PA distributions from the ground truth in these models, especially when the the instantaneous \uv-coverage worsens (e.g. during the LMT dropout time region). For the third model (Simulation 3) both the dynamic imaging and modeling methods recover significant offset from the true PA. 
One potential cause of this offset is the prominent extended jet structure to the northwest of the ring in the SANE simulation. This extended structure cannot be captured in either the dynamic imaging or modeling methods due to their strong prior assumptions of a ring-like morphology. 

In the bottom row of \autoref{fig:grmhd_pa}, we investigate the complex variance of the Fourier transform of the reconstructed images across the selected time window for both the ground truth (scattered) movies and the reconstructions. 
We find that in all three cases the reconstructions tend to introduce more variability than is present in the ground truth. 
In Simulation 1 and 2 this excess variability does not prevent the reconstructions from qualitatively recovering the correct PA trend, but the PA results are not correctly recovered in Simulation 3.

These results indicate that although we often recover the PA accurately from some realistic synthetic GRMHD data sets, we should remain cautious when interpreting \sgra dynamic results. Our methods sometimes incorrectly recover the PA, especially if extended structure is present that is not captured by the prior assumptions on the source structure made by the dynamic imaging and modeling methods. In addition to effects from extended structure, there may be additional uncharacterized systematic uncertainty from prior assumptions in the reconstructions in these results derived from extremely sparse snapshot \sgra coverage. 

\subsection{\sgra Spatiotemporal Characterization and Uncertainty}
\label{sec::dynamic_results}

\begin{figure*}
    \centering
    \includegraphics[width=\textwidth]{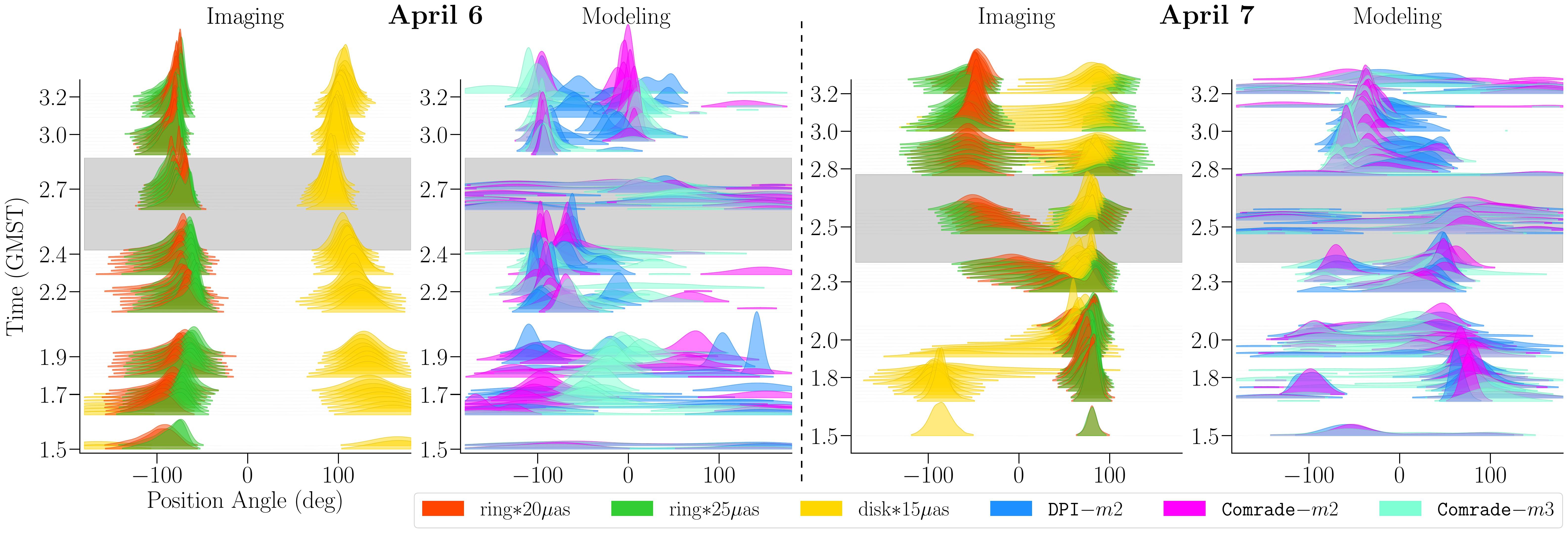}
    \caption{The PA for the 2017 \sgra data recovered using dynamic imaging and geometric modeling techniques under different assumptions. \texttt{StarWarps} imaging results were obtained using a spatial prior image set to either a uniform ring convolved with a circular Gaussian kernel with FWHM of 20 or 25 $\mu$as  (see \autoref{fig:synthetic_summary}), or a uniform disk blurred with a kernel with FWHM of 15 $\mu$as. Descattered modeling results were obtained from geometric models with increasing complexity ($m$-ring 2 vs $m$-ring 3) and different fitting software packages (\texttt{DPI} vs \texttt{Comrade}). All results were obtained from low-band data on April 6 and 7 that has been descattered with a \texttt{J18model1} refractive noise floor. 
    The gray band at roughly 2.6 GMST indicates the region where the LMT has dropped out and data coverage is poor.}
    \label{fig:prior_comparison_descattered}
\end{figure*}

Here we present results of our analysis on \sgra's spatially resolved temporal variability on minute timescales on April 6 and 7, using both dynamic imaging and snapshot geometric modeling methods. 
In \autoref{fig:days_comparison}, we show detailed results for the \sgra PA evolution and data fits reconstructed using a restricted range of dynamic imaging and modeling parameters. In \autoref{fig:prior_comparison_descattered} we show PA results obtained under a broader range of parameter settings. 

\begin{figure*}
    \centering
    \includegraphics[width=0.49\textwidth]{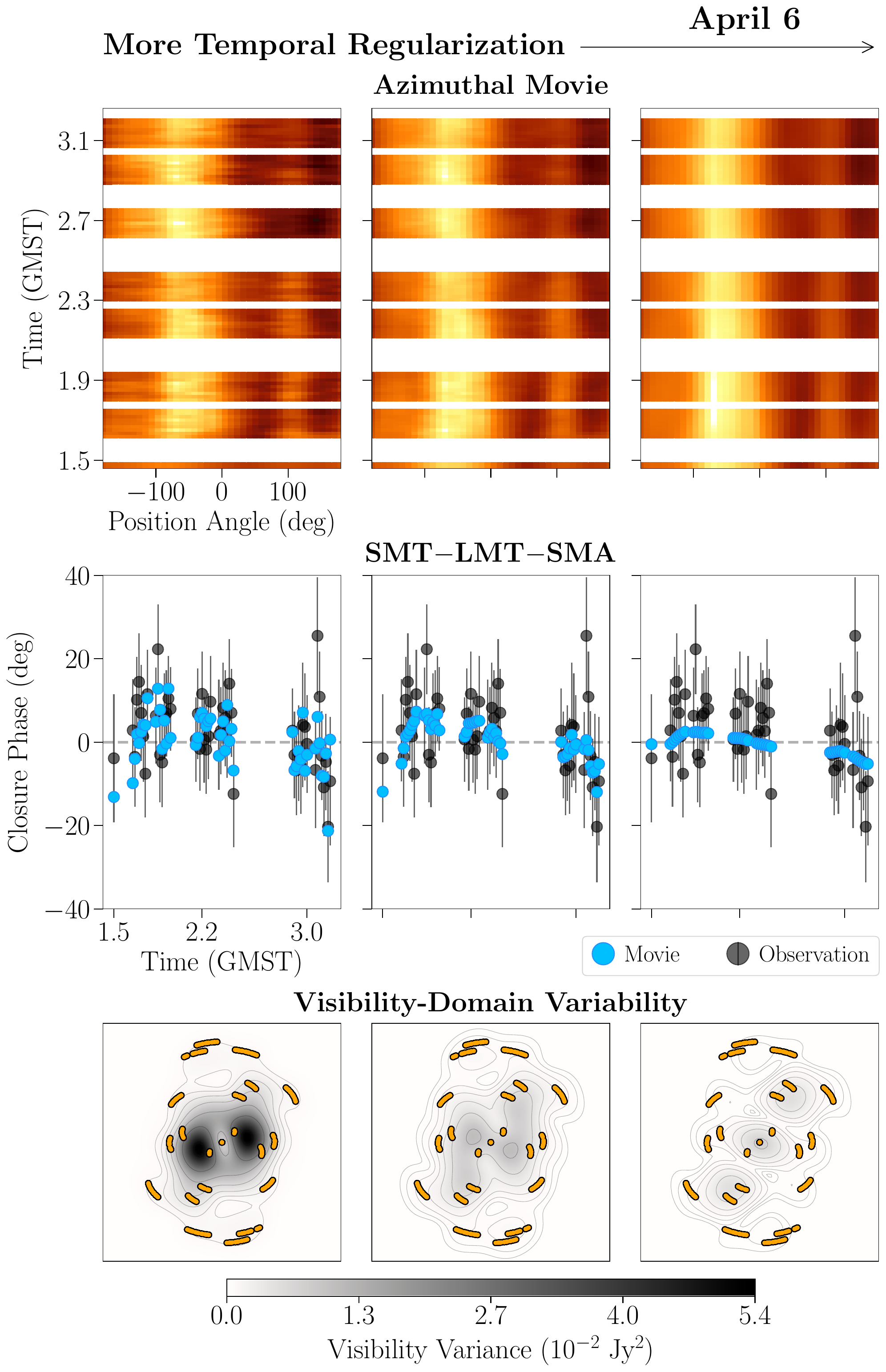}
    \includegraphics[width=0.49\textwidth]{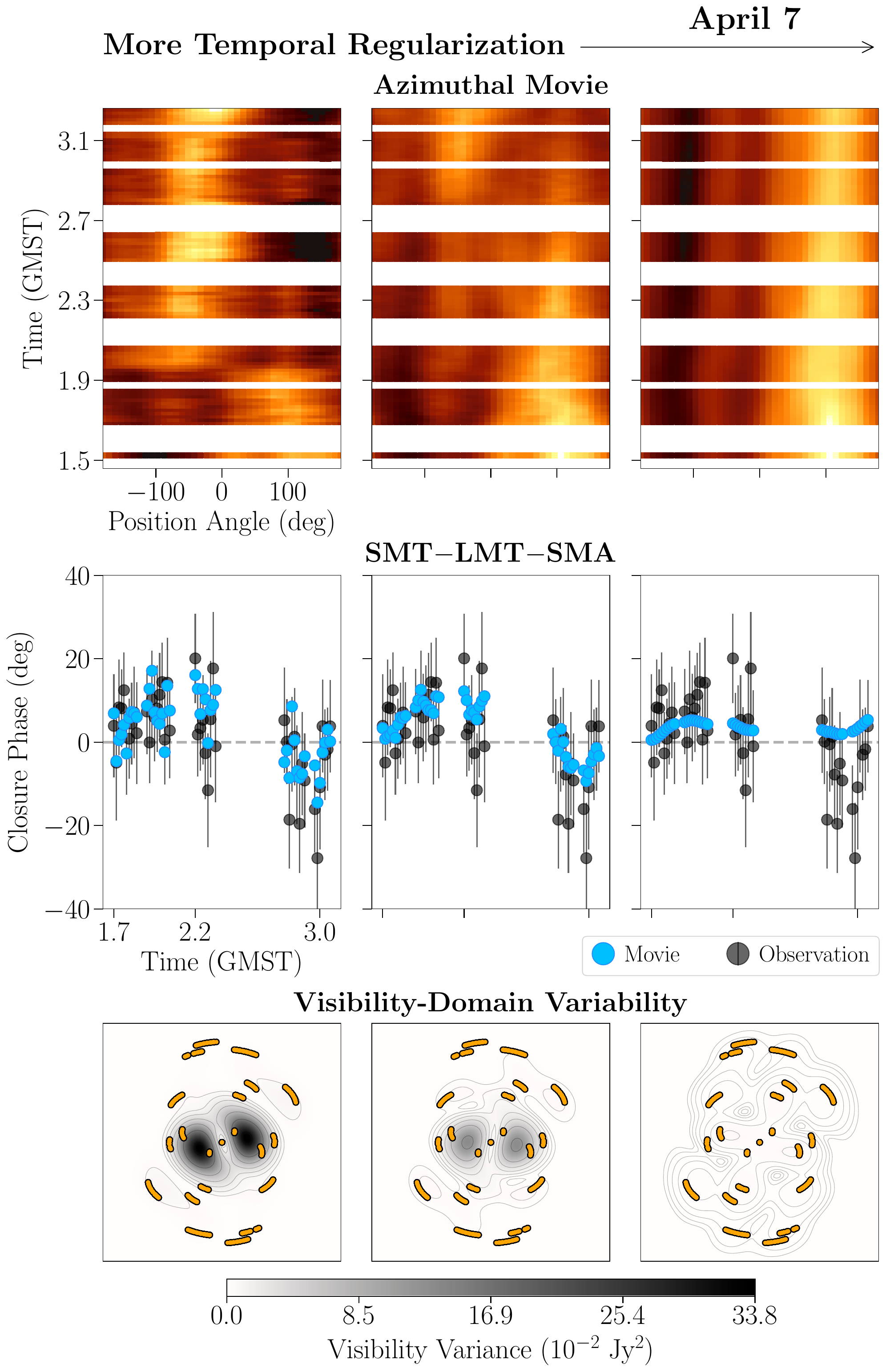}
    \caption{Comparing the effect of temporal regularization on the reconstructed \texttt{StarWarps} movies for April 6 and 7. The temporal regularization strength is increased from left to right ($\beta_Q^{-1}=5\times10^{-4}$, $5\times10^{-6}$, and $5\times10^{-8}$ (Jy/pixel)$^2$). For each value of $\beta_Q^{-1}$, we show the mean unwrapped movie around the ring (top), the mean closure phase values on the triangle SMT-LMT-SMA (middle), and the variance of the complex visibilities across the \uv plane (bottom). As temporal regularization is increased the recovered movies become more static, and the degree of \uv plane variability decreases. 
    From weak to strong regularization, the maximum variance of the reconstructed movie is 5.37, 1.11, and 0.73 $ \times 10^{-2}$ Jy$^2$ on April 6, and 33.81, 13.24, and 0.50 $\times 10^{-2}$ Jy$^2$ on April 7. Contours start at 90\% of the peak variance and decrease by successive factors of 2 until they reach 0.7\%. For comparison, the variance in the light curve over this interval is ${\sim} 0.5 \times 10^{-2} {\rm Jy}^2$. Thus, the leftmost reconstruction with the weakest temporal regularization produces a movie with visibility variance substantially exceeding the light curve variance due to overfitting to the thermal noise. 
    }
    \label{fig:temporal_reg}
\end{figure*}

In general, we find that snapshot geometric modeling
results performed under different $m$-ring orders, scattering mitigation strategies, and modeling codes produce fairly consistent results.
The modeling results show broad posteriors of PA at each 60\,s snapshot but still indicate significant differences between April 6 and 7 and between the first and second halves of the 100 min window on April 7. 
In rough terms, the PA on April 6 is centered around $\sim-50^\circ$, with some scatter around this value over the time window. 
In contrast, on April 7 the modeling results show PA posteriors that are initially centered around $\sim90^\circ$ and then shift to values centered around $\sim-50^\circ$ in the second half of the time window.

Compared to snapshot geometric modeling, dynamic imaging allows for more freedom in spatial and temporal regularization, and as a result is more sensitive to parameter choices. 
Dynamic imaging results can produce movies that reproduce the PA trends on April 6 and 7 recovered by snapshot modeling. These PA trends -- a stable PA on April 6 and a shifting PA on April 7 -- are predominantly seen in dynamic imaging reconstructions with low temporal regularization and ring-like spatial priors. Geometric modeling makes similar assumptions, imposing no temporal regularization and enforcing ring structure in the form of the model. 
In \autoref{fig:days_comparison}, we directly compare dynamic modeling and imaging results under these strong assumptions. 

Even in the limited space of dynamic imaging reconstructions conducted with weak temporal regularization and ring-like mean prior images, the imaging results are sensitive to other hyperparameters. In particular \autoref{fig:days_comparison} indicates that we recover two modes of position angle evolution on April 7 even with the mean prior image and temporal regularization level fixed. We present representative snapshots and fit to the closure phase data from these two modes in the right panels of \autoref{fig:days_comparison}.

When the ring-like mean image prior is changed or the weak temporal regularization is increased in \texttt{StarWarps} dynamic imaging, significantly different PA variations can be recovered from the same data. 
In \autoref{fig:prior_comparison_descattered}, we show that when the ring assumption is relaxed and a disk prior is used in reconstruction, \texttt{StarWarps} results show drastically different PA trend over time. In particular, in reconstructions initialized with a disk prior, the PA curves on April 6 and 7 appear to be flipped by $180^\circ$ (i.e., on April 7 the PA transitions from $\sim90^\circ$ to $\sim-50^\circ$). 
We further show in \autoref{fig:temporal_reg} that when using stronger temporal regularization in the \texttt{StarWarps} dynamic imaging, the PA stays constant, eliminating the shift from positive to negative PA trend on April 7. We discuss these tests further in the \autoref{sec:dynamicmodelingchoices} and \autoref{sec:app_model_imaging_choices}.

Note that April 6 and 7 have nearly identical \uv-coverage during the selected 100 min region of time. We can thus compare the results obtained over these two days to help disentangle effects of \uv-coverage from any effects due to intrinsic evolution. If the PA trends that we recover are due primarily to biases from the \uv-coverage,  we would expect to recover the same PA trends on both days. However, we consistently see different PA trends with time using the same parameter settings in both dynamic imaging and modeling methods. This implies that differences in the visibilities, not the \uv-coverage, drive differences in the PA evolution we see on the two nights in some reconstructions, but it does not help select among any of the different reconstruction modes on either day.

\subsubsection{Effect of Model and Imaging Choices}
\label{sec:dynamicmodelingchoices}

The PA evolution recovered with dynamic imaging and modeling methods is sensitive to choices made in the imaging and modeling procedures that enforce constraints on the spatial and temporal structure of the reconstructions.
Enforcing strong spatial or temporal priors will suppress any structural variability
while adding too much flexibility in a model with sparse data constraints will lead to overfitting or uninformative posteriors. In \autoref{sec:app_model_imaging_choices} we present in detail several tests of these choices for both imaging and modeling methods; here, we highlight the most important results.

\paragraph{Spatial Priors}
Constraints on the spatial structure are enforced via the $m$-ring order in geometric model fitting
and via the choice of mean image prior in dynamic imaging. 
To test the effects of different mean prior images in \texttt{StarWarps}, 
we produced reconstructions using uniform ring priors with increasing widths (from convolution of the ring described in \autoref{sec:synthetic_data} with circular beams of 11, 15, 20, and 25 $\mu$as FWHM; henceforth ring$\textasteriskcentered$11$\mu$as, ring$\textasteriskcentered$15$\mu$as, ring$\textasteriskcentered$20$\mu$as, and ring$\textasteriskcentered$25$\mu$as priors, respectively). We also used a tapered disk with diameter $\sim$ 74$\mu$as (see \autoref{fig:synthetic_summary}) as a prior that does not feature any central dip. The disk is tapered via convolution with a circular beam of 15 $\mu$as (henceforth, disk$\textasteriskcentered$15$\mu$as prior). We discuss the details of these prior choices in \autoref{sec:app_imaging_choices}.
For geometric modeling, we tested ring-like models of increasing complexity in their azimuthal brightness distribution, from crescent models ($m=1$) to higher order $m$-rings. When fitting $m$-rings to \sgra snapshot data, we explored $m$ from 1 to 4 and selected the order to use based on the Bayesian evidence across all data sets -- settling on $m=2$ (see \autoref{sec:app_modeling_choices}).

\autoref{fig:prior_comparison_descattered} shows 
histograms of the dynamic imaging PA results made using different mean prior images and PA posterior distributions from geometric modeling results from different $m-$ring orders. 
We also compare modeling results from two different modeling codes in \autoref{fig:prior_comparison_descattered}.
The reconstructed PA trends are fairly consistent on both days among the different $m$-ring orders in geometric model fitting. In \texttt{StarWarps} imaging, reconstructions using ring-like mean prior images of several different thicknesses produce similar trends, with a stable PA on April 6 and a PA transition from positive to negative values on April 7. However, when a disk prior is used in \texttt{StarWarps} the PA trends of both April 6 and 7 change drastically and appear to be flipped by $180^\circ$. \autoref{fig:dynamicvsstatic_alldays2} in \autoref{sec:appendixdynamics} shows image snapshots and data fits for \texttt{StarWarps} reconstructions with both disk and ring mean prior images.
Note that although the position angle evolution is different than the movie reconstructed using a ring prior, the movie reconstructed with a disk prior still results in a ring-like structure, though with a less prominent central brightness depression. 

\paragraph{Temporal Regularization}

Geometric modeling enforces no correlations in between temporal snapshots, while dynamic imaging can enforce correlations via temporal regularization. \autoref{fig:temporal_reg} shows that when using stronger temporal regularization in \texttt{StarWarps} (lower values of $\beta^{-1}_Q$) the PA becomes constant in time on both April 6 and 7 -- a result of the method enforcing strong continuity between frames.
Note that in the case of high temporal regularization, the SMT-LMT-SMA closure triangle fits in the second half of the time window on April 7 appear offset with respect to the data, although still within two standard deviations of most data points.

We also show that reconstructions with low levels of temporal regularization produce prominent variance in the model visibilities plane at \uv points that are not sampled by our coverage during this time window\footnote{As shown in the GRMHD synthetic data tests, it is not necessary that the peaks in the \uv-plane variability map align with measured data points to correctly identify PA evolution.}. In contrast, reconstructions with more temporal regularization lower the overall variance of model visibilities everywhere in the Fourier plane and place the peaks in the variance maps at \uv points sampled by the EHT. We discuss the interpretation of the different temporal regularization values $\beta^{-1}_Q$ used here in \autoref{sec:app_sw_reg_norm} and further tests of the  \texttt{StarWarps} temporal regularization further in \autoref{sec:app_imaging_choices}.

\paragraph{Scattering}
Another choice made in both dynamic imaging and modeling procedures common to both static and dynamic reconstruction methods is the strategy for mitigating the effects of interstellar scattering in \sgra data. We investigate the effects of the same five prescriptions for scattering mitigation we use in static imaging on the dynamic reconstructions in \autoref{sec:app_dyn_scat}. In general, we find that choices made in the scattering mitigation procedure contribute less to our overall uncertainty than choices related to the spatial prior or temporal regularization.

\subsection{\sgra Dynamic Property Conclusions}
Our aim in this section has been to use the 2017 EHT data to explore spatially-resolved dynamics of \sgra on minute timescales.  First, we identified the time windows with the best \uv-coverage during the observation run -- a roughly 100 min window on April 6 and 7.  
We identify a significant difference between the closure phases on April 6 and 7, signifying that the underlying structure is different on the two days.
We reconstruct movies from this small slice of the EHT data using dynamic imaging and geometric snapshot modeling methods. %
We track the average position angle (PA) in our dynamic imaging and modeling results as a way of following a specific, dynamic, and measurable aspect of the source over time. 
We find that we are able to recover the PA in synthetic EHT data from some GRMHD simulation movies; however, there are prominent cases when this is not the case and both geometric modeling and dynamic imaging methods recover biased results. 

On April 6, most dynamic imaging and modeling results show a stable PA in the \sgra images over the selected window.
In contrast, the recovered PA evolution on April 7 is more dependent on prior assumptions on the spatial structure and temporal regularization.
On April 7, when using a ring image as a spatial prior and weak temporal regularization, dynamic imaging results largely align with geometric modeling results and show an evolution in the PA of $\sim140^\circ$ over the $\sim 100$ min window. 
However, we also see several other PA trends in the dynamic imaging results, including a PA evolution in the opposite direction and modes where the PA is static on both days.

These results, along with our synthetic data tests, show that
while the 230 GHz image of \sgra may exhibit interesting and  measurable dynamics,
our current methods cannot conclusively determine the PA evolution of \sgra. 
Dynamic reconstructions of \sgra with EHT2017 coverage should thus be interpreted with caution.
This analysis provides a promising starting point for further studies of future evolution seen in \sgra EHT observations with denser \uv-coverage.

%% file: summary.tex
\section{Summary and Conclusions}
\label{sec::summary}

We present \sgra static and dynamic imaging results for data taken with the EHT in April 2017. \sgra was observed with eight EHT stations at six geographic sites over five observing days, out of which the highly sensitive phased ALMA array participated in April 6, 7, and 11. 
April 7 is the only day in which the easternmost station PV participated, providing the best $(u,v)$-coverage that probes a null at $\sim$3.0 G$\lambda$. 
On April 11 \sgra exhibits the highest variability in the light curve during the 2017 campaign \citep{Wielgus_2021}, possibly related to an X-ray flare observed shortly before the start of the EHT observations \citepalias{PaperII}, rendering static imaging for this day particularly challenging. For these reasons, April 7 has been used as the primary dataset, while April 6 is considered as a secondary data set. Results from the remaining 2017 EHT observations of \sgra will be presented in future publications.

Similar to the 2017 EHT observations of \m87 \citepalias{M87PaperI,M87PaperII,M87PaperIII,M87PaperIV,M87PaperV,M87PaperVI}, the data sets analyzed in this paper are the first with sufficient sensitivity and $(u,v)$-coverage to reconstruct images of \sgra on event-horizon scales with an angular resolution of $\sim$20\,\uas. Imaging \sgra with the EHT is however significantly more challenging than \m87 due to the interstellar scattering towards the Galactic Center, and most importantly the rapid intra-day variability that characterizes \sgra, with timescales much shorter than the duration of our typical EHT observing runs.

To mitigate the scattering effects in our \sgra reconstructions we have developed a strategy based on \citet{Fish_2014}, \citet{Psaltis_2018} and \citet{Johnson_2018} to account for the angular broadening and substructure induced by diffractive (\autoref{sec:pre_scatter_diff}) and refractive (\autoref{sec:pre_scatter_ref}) scattering, respectively. Images of \sgra have been obtained with and without these scattering mitigation prescriptions to asses their impact in our reconstructions.

It is however the rapid intra-day intrinsic variability of \sgra, coupled with the sparse $(u,v)$-coverage as compared with regular VLBI observations, that poses the strongest challenge for reconstructing horizon-scale images of \sgra with the EHT. With a typical variability timescale of a few minutes, the horizon-scale brightness distribution can change significantly during our typical multi-hour observing runs, which violates the fundamental assumption for Earth rotation VLBI aperture synthesis. To overcome this extraordinary challenge we have included a ``variability noise budget" in the observed visibilities (\autoref{sec:stratvariabiledata}) that facilitates the reconstruction of static full-track images capturing the time-averaged structure.

Static full-track imaging of \sgra has been conducted through surveys over a wide range of imaging assumptions using the classical CLEAN algorithm (implemented in \difmap), regularized maximum likelihood methods (\ehtim and \smili), and a Bayesian posterior sampling method (\themis), as described in \autoref{sec::Background}. Imaging surveys, exploring $\sim$10$^{3-5}$ parameter combinations, were first performed on synthetic data sets designed to be qualitatively consistent with \sgra measurements, including its characteristic temporal variability (\autoref{sec:synthetic_data}). The use of synthetic data sets allows us to asses the capability to accurately reproduce different morphologies with our imaging methods, and to select the ``Top Sets": imaging parameter combinations for RML and CLEAN methods that successfully reproduce the known ground truth movies (\autoref{subsubsec:Top Sets}). %

Unlike for \m87, where a persistent ring structure is observed across imaging pipelines \citepalias{M87PaperIV}, our static reconstructions of \sgra show structural changes within and across the different imaging methods used. The variety of images can be classified into four main clusters of images corresponding to ring images with three different azimuthal brightness distributions, and a cluster that contains a small number of non-ring images with multiple morphologies. This classification is resilient to the scattering mitigation prescription used, including no mitigation.

Although the relative fraction of non-ring images in the Top Set reconstructions from RML and CLEAN is very small ($\le$5\% in the April 7 descattered images), we note that the Top Sets do not sample from the Bayesian posterior likelihood and are instead meant to characterize the range of possible images due to epistemic uncertainty.
On the other hand, the full Bayesian imaging approach implemented in the \themis pipeline does provide a Bayesian posterior exploration that characterizes aleatoric uncertainty. Scattering mitigated \sgra reconstructions from \themis for April 7 only contain ring images with a very similar structure as that found for RML and CLEAN methods, although it also identifies a small number (2\%) of on-sky non-ring images (\autoref{sec:sgra_images}). 

The April 6 data set suffers from poorer \uv-coverage and likely more unusual intrinsic variability. This results in April 6 reconstructions that contain a less prominent ring structure in the RML and CLEAN pipelines. Nonetheless, the diameter of the ring structures recovered for April 6 are consistent with that of the April 7 ring images.
In addition, although descattered \themis samples of April 6 primarily show a corrupted ring or non-ring structure, \themis posterior samples for the combined April 6 and 7 data sets exhibits only clear ring-like images for both, the on-sky and scattered mitigated reconstructions.

Imaging of a synthetic GRMHD data set with data properties similar to those of \sgra and characterized by a ring image of $\sim$50\,$\mu$as (\autoref{subsec:synthetic_data_images_grmhd}) shows similar imaging results to those found in \sgra: RML and CLEAN methods recover ring images with different azimuthal orientations and a small ($\le5\%$ ) Top Set fraction of non-ring images, while \themis posterior sample only recovers ring images.

We conclude that the Event Horizon Telescope \sgra data show compelling evidence for an image that is dominated by a bright ring of emission. This conclusion is based on our extensive analysis of the effects of sparse $(u,v)$-coverage, source variability and interstellar scattering, as well as studies of simulated visibility data, which find that non-ring images are recovered in the minority by our imaging pipelines for variable sources with an intrinsic ring morphology (\autoref{sec:sgra_is_a_ring}).
Representative first event horizon-scale images of \sgra are shown in \autoref{fig:rep_sgra_images}, obtained from the April 7 data set by averaging similar Top Set and posterior images together.
Despite the different azimuthal brightness distributions observed, all ring images have a ring diameter of \sdiam on both April 6 and 7, consistent with the expected ``shadow" of a $4\times10^6 M_\odot$ black hole in the Galactic Center located at a distance of 8\,kpc.

We also present preliminary dynamic imaging and modeling analysis of \sgra on horizon scales in an attempt to characterize the azimuthal variations on minute timescales and their uncertainties. We applied dynamical analysis methods to a 100-minute interval of the 2017 \sgra data with the best coverage on April 6 and 7. On April 6, most dynamic imaging and modeling methods recover a stable position angle. In contrast, on April 7 the recovered PA evolution is more dependent on spatial and temporal regularization; April 7 frequently shows an evolving position angle over the 100 minute window when we impose strong priors on the spatial structure but weak temporal regularization. However, when expanding the parameter space available to the imaging and modeling methods, we recover disparate modes in the position angle behavior, including some reconstructions that are nearly static. Our initial results show that significant uncertainty exists in any attempt to characterize the spatially-resolved dynamics of \sgra using sparse EHT 2017 data, but this analysis provides a promising starting point for further dynamical studies of \sgra with future EHT observations.

%% file: Acks_PaperIII_may9.tex
We thank the anonymous referee for helpful suggestions that improved this paper. 

The Event Horizon Telescope Collaboration thanks the following
organizations and programs: the Academia Sinica; the Academy
of Finland (projects 274477, 284495, 312496, 315721); the Agencia Nacional de Investigaci\'{o}n 
y Desarrollo (ANID), Chile via NCN$19\_058$ (TITANs) and Fondecyt 1221421, the Alexander
von Humboldt Stiftung; an Alfred P. Sloan Research Fellowship;
Allegro, the European ALMA Regional Centre node in the Netherlands, the NL astronomy
research network NOVA and the astronomy institutes of the University of Amsterdam, Leiden University and Radboud University;
the ALMA North America Development Fund; the Black Hole Initiative, which is funded by grants from the John Templeton Foundation and the Gordon 
and Betty Moore Foundation (although the opinions expressed in this work are those of the author(s) 
and do not necessarily reflect the views of these Foundations);
Chandra DD7-18089X and TM6-17006X; the China Scholarship
Council; China Postdoctoral Science Foundation fellowship (2020M671266); Consejo Nacional de Ciencia y Tecnolog\'{\i}a (CONACYT,
Mexico, projects  U0004-246083, U0004-259839, F0003-272050, M0037-279006, F0003-281692,
104497, 275201, 263356);
the Consejer\'{i}a de Econom\'{i}a, Conocimiento, 
Empresas y Universidad 
of the Junta de Andaluc\'{i}a (grant P18-FR-1769), the Consejo Superior de Investigaciones 
Cient\'{i}ficas (grant 2019AEP112);
the Delaney Family via the Delaney Family John A.
Wheeler Chair at Perimeter Institute; Direcci\'{o}n General
de Asuntos del Personal Acad\'{e}mico-Universidad
Nacional Aut\'{o}noma de M\'{e}xico (DGAPA-UNAM,
projects IN112417 and IN112820); 
the Dutch Organization for Scientific Research (NWO) VICI award
(grant 639.043.513) and grant OCENW.KLEIN.113; the Dutch National Supercomputers, Cartesius and Snellius  
(NWO Grant 2021.013); 
the EACOA Fellowship awarded by the East Asia Core
Observatories Association, which consists of the Academia Sinica Institute of Astronomy and
Astrophysics, the National Astronomical Observatory of Japan, Center for Astronomical Mega-Science,
Chinese Academy of Sciences, and the Korea Astronomy and Space Science Institute; 
the European Research Council (ERC) Synergy
Grant ``BlackHoleCam: Imaging the Event Horizon
of Black Holes" (grant 610058); 
the European Union Horizon 2020
research and innovation programme under grant agreements
RadioNet (No 730562) and 
M2FINDERS (No 101018682);
the Generalitat
Valenciana postdoctoral grant APOSTD/2018/177 and
GenT Program (project CIDEGENT/2018/021); MICINN Research Project PID2019-108995GB-C22;
the European Research Council for advanced grant `JETSET: Launching, propagation and 
emission of relativistic jets from binary mergers and across mass scales' (Grant No. 884631); 
the Institute for Advanced Study; the Istituto Nazionale di Fisica
Nucleare (INFN) sezione di Napoli, iniziative specifiche
TEONGRAV; 
the International Max Planck Research
School for Astronomy and Astrophysics at the
Universities of Bonn and Cologne; 
DFG research grant ``Jet physics on horizon scales and beyond'' (Grant No. FR 4069/2-1);
Joint Princeton/Flatiron and Joint Columbia/Flatiron Postdoctoral Fellowships, 
research at the Flatiron Institute is supported by the Simons Foundation; 
the Japan Ministry of Education, Culture, Sports, Science and Technology (MEXT; grant JPMXP1020200109); the Japanese Government (Monbukagakusho:
MEXT) Scholarship; 
the Japan Society for the Promotion of Science (JSPS) Grant-in-Aid for JSPS
Research Fellowship (JP17J08829); the Joint Institute for Computational Fundamental Science, Japan; the Key Research
Program of Frontier Sciences, Chinese Academy of
Sciences (CAS, grants QYZDJ-SSW-SLH057, QYZDJSSW-SYS008, ZDBS-LY-SLH011); 
the Leverhulme Trust Early Career Research
Fellowship; the Max-Planck-Gesellschaft (MPG);
the Max Planck Partner Group of the MPG and the
CAS; the MEXT/JSPS KAKENHI (grants 18KK0090, JP21H01137,
JP18H03721, JP18K13594, 18K03709, JP19K14761, 18H01245, 25120007); the Malaysian Fundamental Research Grant Scheme (FRGS) FRGS/1/2019/STG02/UM/02/6; the MIT International Science
and Technology Initiatives (MISTI) Funds; 
the Ministry of Science and Technology (MOST) of Taiwan (103-2119-M-001-010-MY2, 105-2112-M-001-025-MY3, 105-2119-M-001-042, 106-2112-M-001-011, 106-2119-M-001-013, 106-2119-M-001-027, 106-2923-M-001-005, 107-2119-M-001-017, 107-2119-M-001-020, 107-2119-M-001-041, 107-2119-M-110-005, 107-2923-M-001-009, 108-2112-M-001-048, 108-2112-M-001-051, 108-2923-M-001-002, 109-2112-M-001-025, 109-2124-M-001-005, 109-2923-M-001-001, 110-2112-M-003-007-MY2, 110-2112-M-001-033, 110-2124-M-001-007, and 110-2923-M-001-001);
the Ministry of Education (MoE) of Taiwan Yushan Young Scholar Program;
the Physics Division, National Center for Theoretical Sciences of Taiwan;
the National Aeronautics and
Space Administration (NASA, Fermi Guest Investigator
grant 80NSSC20K1567, NASA Astrophysics Theory Program grant 80NSSC20K0527, NASA NuSTAR award 
80NSSC20K0645); 
NASA Hubble Fellowship 
grant HST-HF2-51431.001-A awarded 
by the Space Telescope Science Institute, which is operated by the Association of Universities for 
Research in Astronomy, Inc., for NASA, under contract NAS5-26555; 
the National Institute of Natural Sciences (NINS) of Japan; the National
Key Research and Development Program of China
(grant 2016YFA0400704, 2017YFA0402703, 2016YFA0400702); the National
Science Foundation (NSF, grants AST-0096454,
AST-0352953, AST-0521233, AST-0705062, AST-0905844, AST-0922984, AST-1126433, AST-1140030,
DGE-1144085, AST-1207704, AST-1207730, AST-1207752, MRI-1228509, OPP-1248097, AST-1310896, AST-1440254, 
AST-1555365, AST-1614868, AST-1615796, AST-1715061, AST-1716327,  AST-1716536, OISE-1743747, AST-1816420, AST-1935980, AST-2034306); 
NSF Astronomy and Astrophysics Postdoctoral Fellowship (AST-1903847); 
the Natural Science Foundation of China (grants 11650110427, 10625314, 11721303, 11725312, 11873028, 11933007, 11991052, 11991053, 12192220, 12192223); 
the Natural Sciences and Engineering Research Council of
Canada (NSERC, including a Discovery Grant and
the NSERC Alexander Graham Bell Canada Graduate
Scholarships-Doctoral Program); the National Youth
Thousand Talents Program of China; the National Research
Foundation of Korea (the Global PhD Fellowship
Grant: grants NRF-2015H1A2A1033752, the Korea Research Fellowship Program:
NRF-2015H1D3A1066561, Brain Pool Program: 2019H1D3A1A01102564, 
Basic Research Support Grant 2019R1F1A1059721, 2021R1A6A3A01086420, 2022R1C1C1005255); 
Netherlands Research School for Astronomy (NOVA) Virtual Institute of Accretion (VIA) postdoctoral fellowships; 
Onsala Space Observatory (OSO) national infrastructure, for the provisioning
of its facilities/observational support (OSO receives
funding through the Swedish Research Council under
grant 2017-00648);  the Perimeter Institute for Theoretical
Physics (research at Perimeter Institute is supported
by the Government of Canada through the Department
of Innovation, Science and Economic Development
and by the Province of Ontario through the
Ministry of Research, Innovation and Science); the Spanish Ministerio de Ciencia e Innovaci\'{o}n (grants PGC2018-098915-B-C21, AYA2016-80889-P,
PID2019-108995GB-C21, PID2020-117404GB-C21); 
the University of Pretoria for financial aid in the provision of the new 
Cluster Server nodes and SuperMicro (USA) for a SEEDING GRANT approved towards these 
nodes in 2020;
the Shanghai Pilot Program for Basic Research, Chinese Academy of Science, 
Shanghai Branch (JCYJ-SHFY-2021-013);
the State Agency for Research of the Spanish MCIU through
the ``Center of Excellence Severo Ochoa'' award for
the Instituto de Astrof\'{i}sica de Andaluc\'{i}a (SEV-2017-
0709); the Spinoza Prize SPI 78-409; the South African Research Chairs Initiative, through the 
South African Radio Astronomy Observatory (SARAO, grant ID 77948),  which is a facility of the National 
Research Foundation (NRF), an agency of the Department of Science and Innovation (DSI) of South Africa; 
the Toray Science Foundation; Swedish Research Council (VR); 
the US Department
of Energy (USDOE) through the Los Alamos National
Laboratory (operated by Triad National Security,
LLC, for the National Nuclear Security Administration
of the USDOE (Contract 89233218CNA000001); and the YCAA Prize Postdoctoral Fellowship.

We thank
the staff at the participating observatories, correlation
centers, and institutions for their enthusiastic support.
This paper makes use of the following ALMA data:
ADS/JAO.ALMA\#2016.1.01154.V. ALMA is a partnership
of the European Southern Observatory (ESO;
Europe, representing its member states), NSF, and
National Institutes of Natural Sciences of Japan, together
with National Research Council (Canada), Ministry
of Science and Technology (MOST; Taiwan),
Academia Sinica Institute of Astronomy and Astrophysics
(ASIAA; Taiwan), and Korea Astronomy and
Space Science Institute (KASI; Republic of Korea), in
cooperation with the Republic of Chile. The Joint
ALMA Observatory is operated by ESO, Associated
Universities, Inc. (AUI)/NRAO, and the National Astronomical
Observatory of Japan (NAOJ). The NRAO
is a facility of the NSF operated under cooperative agreement
by AUI.
This research used resources of the Oak Ridge Leadership Computing Facility at the Oak Ridge National
Laboratory, which is supported by the Office of Science of the U.S. Department of Energy under Contract
No. DE-AC05-00OR22725. We also thank the Center for Computational Astrophysics, National Astronomical Observatory of Japan.
The computing cluster of Shanghai VLBI correlator supported by the Special Fund 
for Astronomy from the Ministry of Finance in China is acknowledged.

APEX is a collaboration between the
Max-Planck-Institut f{\"u}r Radioastronomie (Germany),
ESO, and the Onsala Space Observatory (Sweden). The
SMA is a joint project between the SAO and ASIAA
and is funded by the Smithsonian Institution and the
Academia Sinica. The JCMT is operated by the East
Asian Observatory on behalf of the NAOJ, ASIAA, and
KASI, as well as the Ministry of Finance of China, Chinese
Academy of Sciences, and the National Key Research and Development
Program (No. 2017YFA0402700) of China
and Natural Science Foundation of China grant 11873028.
Additional funding support for the JCMT is provided by the Science
and Technologies Facility Council (UK) and participating
universities in the UK and Canada. 
The LMT is a project operated by the Instituto Nacional
de Astr\'{o}fisica, \'{O}ptica, y Electr\'{o}nica (Mexico) and the
University of Massachusetts at Amherst (USA). The
IRAM 30-m telescope on Pico Veleta, Spain is operated
by IRAM and supported by CNRS (Centre National de
la Recherche Scientifique, France), MPG (Max-Planck-Gesellschaft, Germany) 
and IGN (Instituto Geogr\'{a}fico
Nacional, Spain). The SMT is operated by the Arizona
Radio Observatory, a part of the Steward Observatory
of the University of Arizona, with financial support of
operations from the State of Arizona and financial support
for instrumentation development from the NSF.
Support for SPT participation in the EHT is provided by the National Science Foundation through award OPP-1852617 
to the University of Chicago. Partial support is also 
provided by the Kavli Institute of Cosmological Physics at the University of Chicago. The SPT hydrogen maser was 
provided on loan from the GLT, courtesy of ASIAA.

This work used the
Extreme Science and Engineering Discovery Environment
(XSEDE), supported by NSF grant ACI-1548562,
and CyVerse, supported by NSF grants DBI-0735191,
DBI-1265383, and DBI-1743442. XSEDE Stampede2 resource
at TACC was allocated through TG-AST170024
and TG-AST080026N. XSEDE JetStream resource at
PTI and TACC was allocated through AST170028.
This research is part of the Frontera computing project at the Texas Advanced 
Computing Center through the Frontera Large-Scale Community Partnerships allocation
AST20023. Frontera is made possible by National Science Foundation award OAC-1818253.
This research was carried out using resources provided by the Open Science Grid, 
which is supported by the National Science Foundation and the U.S. Department of 
Energy Office of Science. 
Additional work used ABACUS2.0, which is part of the eScience center at Southern Denmark University. 
Simulations were also performed on the SuperMUC cluster at the LRZ in Garching, 
on the LOEWE cluster in CSC in Frankfurt, on the HazelHen cluster at the HLRS in Stuttgart, 
and on the Pi2.0 and Siyuan Mark-I at Shanghai Jiao Tong University.
The computer resources of the Finnish IT Center for Science (CSC) and the Finnish Computing 
Competence Infrastructure (FCCI) project are acknowledged. This
research was enabled in part by support provided
by Compute Ontario (http://computeontario.ca), Calcul
Quebec (http://www.calculquebec.ca) and Compute
Canada (http://www.computecanada.ca).

The EHTC has
received generous donations of FPGA chips from Xilinx
Inc., under the Xilinx University Program. The EHTC
has benefited from technology shared under open-source
license by the Collaboration for Astronomy Signal Processing
and Electronics Research (CASPER). The EHT
project is grateful to T4Science and Microsemi for their
assistance with Hydrogen Masers. This research has
made use of NASA's Astrophysics Data System. We
gratefully acknowledge the support provided by the extended
staff of the ALMA, both from the inception of
the ALMA Phasing Project through the observational
campaigns of 2017 and 2018. We would like to thank
A. Deller and W. Brisken for EHT-specific support with
the use of DiFX. We thank Martin Shepherd for the addition of extra features in the Difmap software 
that were used for the CLEAN imaging results presented in this paper.
We acknowledge the significance that
Maunakea, where the SMA and JCMT EHT stations
are located, has for the indigenous Hawaiian people.


%% file: appendix_imaging_methods.tex
\section{Estimating Static Image Posteriors with \themis}
\label{sec:themisdescription}

\subsection{The \themis Image Model}
\themis is a general sampling-based parameter estimation framework developed for comparing parameterized models with the VLBI data produced by the EHT \citep{Broderick_2020a}.  Implemented within \themis is a wide variety of image models, ranging from phenomenological geometric models (e.g., Gaussians, rings, etc.) to physically motivated ray-traced radiative transfer models (e.g., semi-analytic accretion flow models).  Of relevance here is a set of splined raster models, described in detail in \citet{Broderick_2020b}, and applied to polarized reconstructions in \citetalias{M87PaperVII}.  In these, the image reconstruction process is replaced with a model reconstruction problem in which the intensity at control points on a rectilinear raster grid are varied, with the final image produced at arbitrary locations via a cubic spline.

The image model presented in \citet{Broderick_2020b} has a fixed field of view in the two cardinal directions.
As described in \citetalias{M87PaperVII}, this restriction has been subsequently relaxed, with a more general set of models, adaptive splined rasters, that permit a rotation of the raster and a resizing of the field of view among its principle axes.  In this way, not only can the brightness at each control point vary, but the locations of the control points themselves can evolve.  This results in a very flexible image model even when the dimension of the raster is small. The splined raster and adaptive splined raster models may be combined with any other \themis models, and typically large-scale features will be absorbed by a large-scale Gaussian component.  The dimension of the underlying image model is, then, the sum of the number of control points, $N_x\times N_y$, two fields of view and the orientation of the raster, and the number of parameters associated with a potential geometric addition.

The \themis model has been fit to a variety of EHT data types.  The bulk of the \themis-based analyses presented here employed the light-curve-normalized, LMT-calibrated complex visibilities directly.  The residual complex station gains are reconstructed during each model evaluation and marginalized over via the Laplace approximation \citep[see Section 6.8 of][]{Broderick_2020a}.  Log-normal priors are imposed on the station gain amplitudes with uncertainties that depend on the a priori calibration estimates (see \autoref{sub:data-red}).  Network calibrated sites (ALMA, APEX, JCMT, SMA) are assumed to have a log-normal 1$\sigma$ uncertainty of 1\%; following the LMT gain amplitude correction it is assumed to have a residual log-normal 1$\sigma$ uncertainty of 20\%; and the remaining sites (PV, SMT, SPT) are assigned a log-normal 1$\sigma$ uncertainty of 10\%.
In this way \themis fully explores the potential station gain space during its exploration of the underlying image space.
Some early analyses made use of visibility amplitudes and closure phases.  For these the station gain amplitudes were reconstructed in a similar fashion as described above, though with a larger log-normal prior on LMT gain of 100\% and 20\% on all other stations.

\subsection{Scattering and Variability Mitigation}
Application of \themis-based imaging methods to \sgra is complicated by the presence of the confounding factors described in \autoref{sec::PreImaging_Considerations}.  Unless otherwise stated, diffractive scattering is mitigated via the scattering model from \citet{Johnson_2018} directly to the model visibilities, i.e., instead of deblurring the data, the model is blurred. 
Refractive scattering and variability are mitigated via a scheme similar to that described in \autoref{sec:pre_scatter_ref}.  Unlike the description there, for the \themis analyses, the additional noise terms are treated as parameterized contributions to the visibility uncertainties as described in detail in \citet{Georgiev_2021}.

Explicitly, the \themis analyses assume that the visibility uncertainties are described by
\begin{multline}
\sigma_j^2(\sigma_{\rm ref},f,a,b,c,u_0)
=\\
\sigma_{j,\rm th}^2
+
\sigma_{\rm ref}^2
+
f^2 |V|^2
+
\sigma_{\rm var}^2(a,b,c,u_0),
\label{eq:themis_noise_model}
\end{multline}
where $\sigma_{j,\rm th}$ is the intrinsic thermal uncertainty,
$|V|$ is the measured visibility amplitude, and $\sigma_{\rm var}$ is given in \autoref{eq:PSD_noise}.  Within this prescription, the $\sigma_{\rm ref}$ is conceptually identical to the {\tt Const} approach to refractive scattering mitigation described in \autoref{sec:pre_scatter_ref}, with the exception that it is not fixed.  Similarly, the fractional contribution, $f|V|$ is conceptually identical to the systematic noise term, again with the sole distinction being to treat $f$ as a free parameter.

\subsection{Likelihood and Parameter Priors}
For analyses that fit the visibility amplitudes and closure phases, the \themis log-likelihood is given by the appropriate combination of log-likelihoods from Sections 6.3 and 6.8 from \citet{Broderick_2020a}.  For analyses that fit the full complex visiblities with the noise modeling, the log-likelihood is
\begin{multline}
\mathcal{L}({\bf p},{\bf q})=\\
- \sum_{\rm data} \bigg\{
\frac{|\hat{V}_j-\hat{V}({\bf u}_j;{\bf p})|^2}{2\sigma_j^2({\bf q})}
+
\frac{1}{2}\ln\left[2\pi
\sigma_j^2({\bf q})
\right]
\bigg\},
\end{multline}
where $\{\hat{V}_j\}$ is the light-curve-normalized visibility data, ${\bf p}$ are the image-model parameters and ${\bf q}=\{\sigma_{\rm ref},f,a,b,c,u_0\}$ are the noise-model parameters.  The novel terms on the right are simply the $\sigma$-dependent normalization; while normally constant, in this instance they induce a penalty term that constrains the ${\bf q}$.  When descattered, the diffractive scattering kernel is applied {\em to the model}, as opposed to deblurring the data, which maintains a consistent treatment of the ``noise'' model among the descattered and on-sky analyses.

Priors on each of the image model parameters are listed in \autoref{tab:themis_priors}.  In summary, uninformative logarithmic priors are assumed on the brightness at the control points, the raster size and orientation, and a potential shift from the correlation phase center.  Similarly uninformative priors are imposed on the refractive scattering mitigation and systematic uncertainty terms of the noise model.  In contrast, informed priors are imposed on the remaining parameters of the noise model, set by the pre-modeling considerations from \citet{Georgiev_2021} and applied to \sgra in \citetalias{PaperIV}.  The relevant interquartile ranges are data-set specific, and are listed for the \themis analyses reported here in \autoref{tab:premodeling} for \sgra and the various synthetic data sets in \autoref{sec:synthetic_data}. 

\subsection{Reconstructing the Posterior with \themis}
\themis provides a number of sampling schemes with which to explore the likelihood surface.  The bulk of the \themis-based analyses presented here employ a set of Markov Chain Monte Carlo (MCMC) samplers to explore the large-dimensional parameter space, $\{{\bf p},{\bf q}\}$. To efficiently traverse the prior, identify multiple modes, and rapidly locate the global maximum, \themis makes use of parallel tempering, in which multiple, tempered copies of the likelihood, $\mathcal{L}_\beta = \beta \mathcal{L}$ are simultaneously explored; higher ``temperature'' copies ($\beta\rightarrow0$) become the reference distribution which we take to be uniform in the prior bounds, while the bottom ``temperature'' ($\beta=1$) is the posterior to be sampled.  These are coupled via the deterministic even–odd swap tempering scheme \citep{DEO:2019}, which efficiently moves information through the multiple, tempered levels. Periodically, the $\beta$ are optimized to efficiently move information from the prior to the posterior following the scheme from \citep{DEO:2019}. Each individual tempering level is sampled via the Hamiltonian Monte Carlo sampling kernel from the Stan package \citep{Stan:2017}.

Convergence is assessed via chain statistics, such as, integrated autocorrelation time, approximate split-$\hat{R}$ tests, parameter rank distributions, and visual inspection. Reconstruction quality is determined by inspection of the best-fit sample's residuals, and ``reasonableness'' of the associated image and gain reconstructions.  Pathological solutions are uncommon, and are typically indicative of poor sampler initialization and/or errors in the input. The number of tempering levels is chosen to exceed the global communication barrier $\Lambda$ \citep{DEO:2019}, typically near 20, and is run for $5\times 10^4$ to $10^5$ MCMC steps per tempering level.

%% file: appendix_scattering.tex
\section{Refractive Noise Levels Anticipated for EHT Data\label{appendix:ref_scatt}}
 
\begin{figure}[t]
    \centering
    \includegraphics[width=1.0\linewidth]{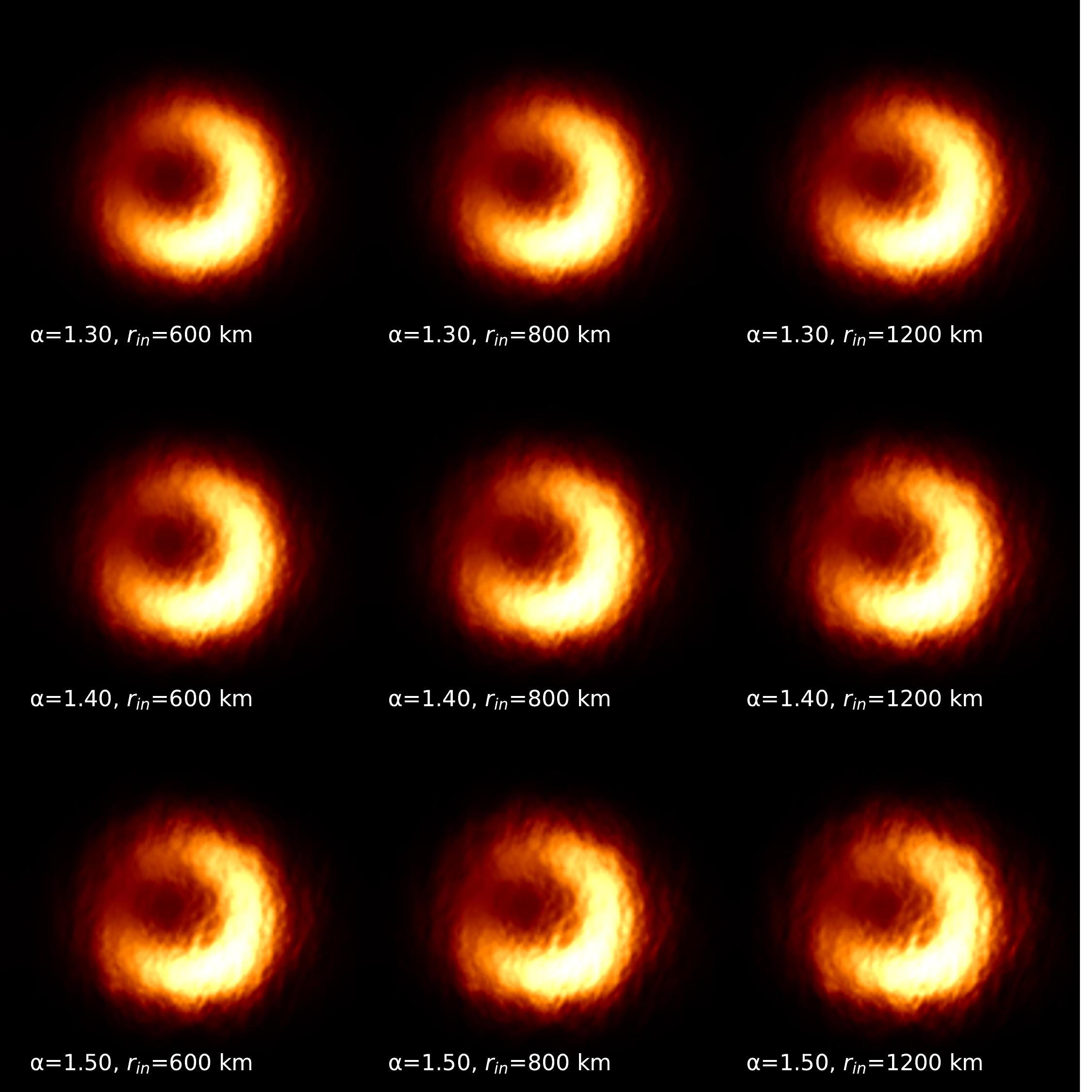}\\
    (a) On-sky images\\
    \includegraphics[trim={0.0cm 0.0cm 0.0cm 0.cm},clip,width=1.0\linewidth]{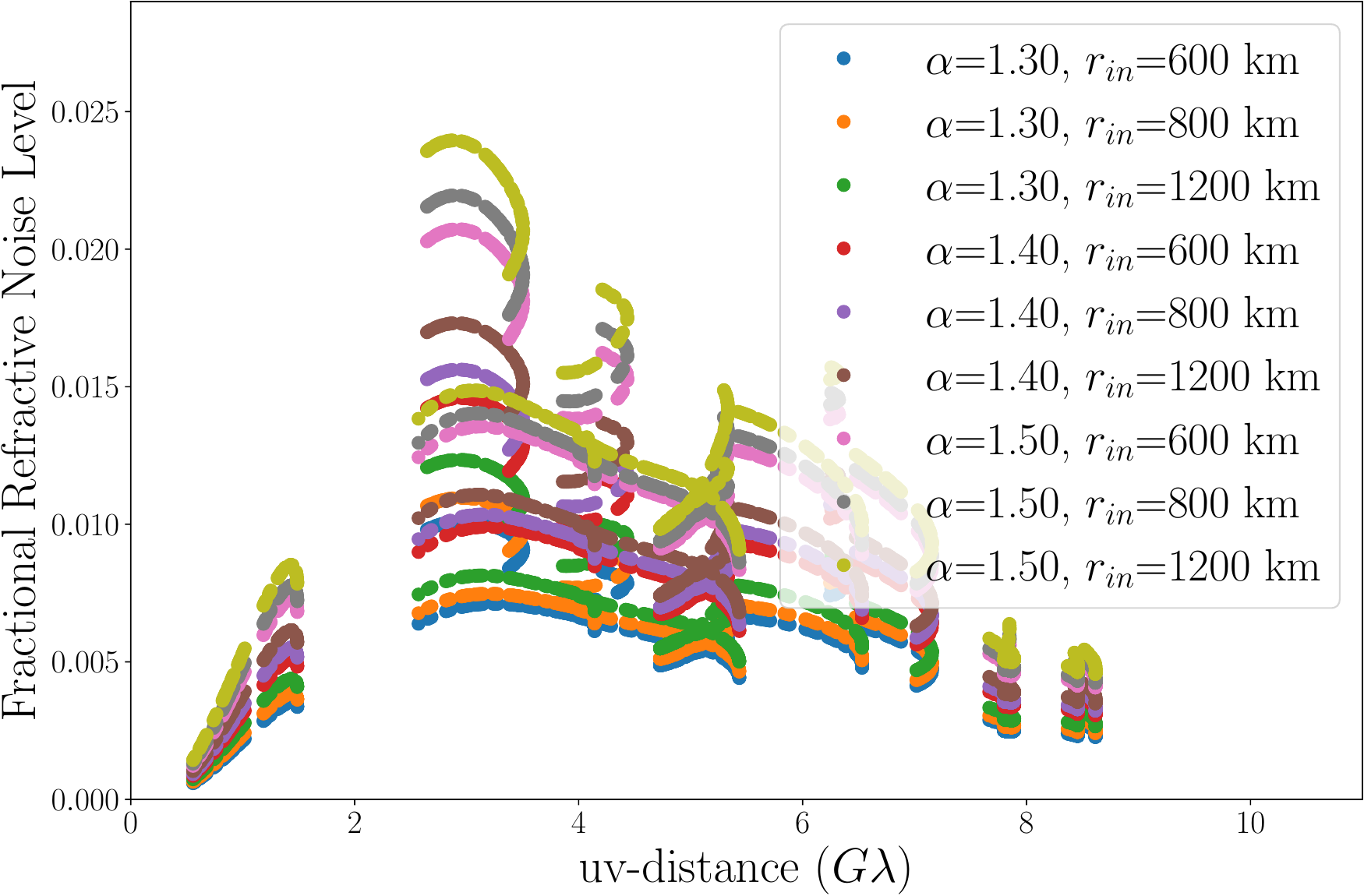}\\
    (b) Refractive noise levels
    \caption{Dependence of refractive substructures on the power-law index $\alpha$ and inner scale of the turbulence $r_\mathrm{in}$.
    (a) Example realizations of the on-sky (i.e., scattered) images of the crescent model in \autoref{sec:synthetic_data}.
    Each row shows a realization for a different power-law index $\alpha$ (1.3, 1.4 and 1.5), while each column shows a different $r_{in}$ (600, 800 and 1200\,km).
    (b) The corresponding levels of the flux-normalized refractive noise anticipated for \uv-coverage of April 7 \sgra data.
    The corresponding normalized visibility amplitudes are shown in \autoref{fig:two_days_amps}. 
    }
    \label{fig:ref_noise_alpha_rin}
\end{figure}

To assess the effects of refractive scattering on our EHT measurements, we use the \texttt{stochastic optics} module implemented in \texttt{eht-imaging} library~\citep{Johnson_2016, Chael_2016}. The \texttt{stochastic optics} allows the simulation of scattering realizations for a given intrinsic image and generation of the corresponding synthetic data sets. 
Here, we focus only on the refractive noise that would affect our imaging, i.e., the substructures on the scattered image. The other effects --- centroid shift due to the position wander or total flux modulation--- do not affect the image reconstruction of data calibrated by self-calibration and obtained from a single on-sky realization of scattering~\citep{Johnson_2018}. 

The refractive noise level at each \uv-coordinate is determined by the standard deviation of the flux-normalized visibilities for all realizations, where the centroid of each realization is shifted to the phase center to mitigate the effects of position wander. 
We adopt a dipole model for the magnetic field wandering in the scattering screen~\citep{Johnson_2018}. 
Note that the refractive noise does not depend strongly on the choice of the model for field wander~\citep{Psaltis_2018} for the source size anticipated from EHT data and scattering parameters constrained by \citet{Johnson_2018}. %

We first compare the refractive effects for different combinations of the power-law spectral index $\alpha$ of the density fluctuation of the ionized plasma (see \autoref{sec:pre_scatter_diff}) and the inner scale of turbulence $r_{in}$. 
Our refractive noise prescriptions {\tt J18model1} and {\tt J18model2} are derived with $\alpha=1.38$ and $r_{in}=800\,{\rm km}$, which are the best-fit values in 
\citet{Johnson_2018}. These measurements have the uncertainties of $1.3 < \alpha < 1.5$ and $600\,{\rm km}
< r_{in} < 1500\,{\rm km}$~\citep{Johnson_2018}.
\autoref{fig:ref_noise_alpha_rin} shows an example gallery of scattered images and corresponding noise levels in the visibility domain for each combination of $\alpha$ and $r_{in}$ in these ranges.
The level of refractive substructure as well as the kernel size of diffractive angular broadening increase with $\alpha$ and $r_{in}$ \citep[][]{Psaltis_2018}.
Based on the results shown in \autoref{fig:ref_noise_alpha_rin}, we adopt a scaling factor of up to 2 to cover the cases of highest refractive noise level in the possible ranges of these two parameters.

\begin{figure}[t]
    \centering
    \includegraphics[width=1.0\columnwidth]{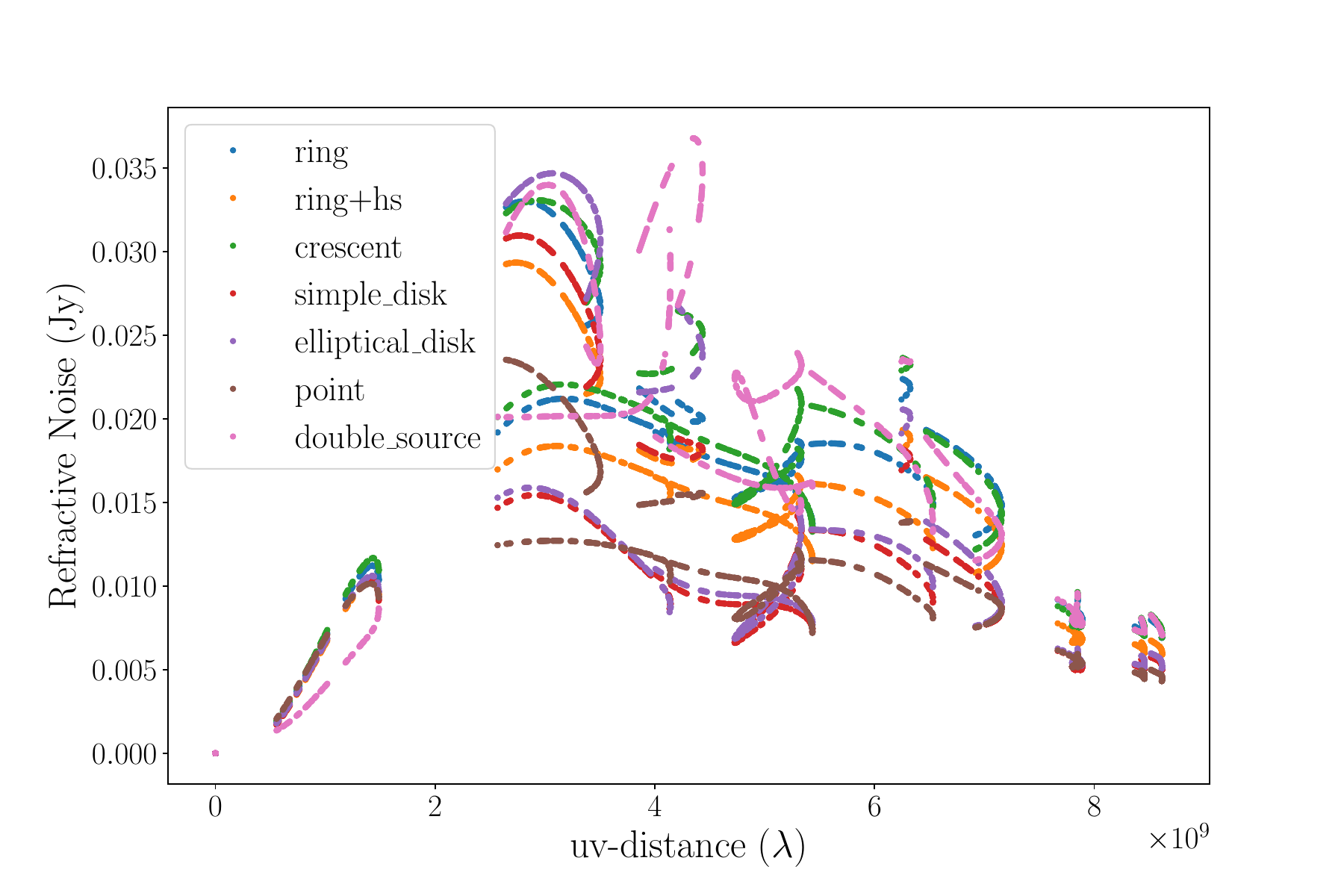}
    \caption{The levels of the flux-normalized refractive noises as a function of \uv-distance for different intrinsic structures. The refractive noise level is computed on \uv-coverage of April 7 \sgra data.
    The corresponding normalized visibility amplitudes are shown in \autoref{fig:two_days_amps}.}
    \label{fig:refnoise_uv}
\end{figure}

Refractive noise is expected to also be dependent on the intrinsic source structure~\citep[e.g.][]{Johnson_Narayan_2016}.
To assess this dependence, we estimate the refractive noise levels based on a circular Gaussian and the time-averaged images of the seven geometric models introduced in \autoref{sec:synthetic_data:geometric}, for $\alpha=1.38$ and $r_{in}=800\,{\rm km}$.
As described in \autoref{sec:synthetic_data:geometric}, these geometric models provide a broad consistency with the visibility amplitudes profile of \sgra data.
The size of the circular Gaussian is adjusted to have the FWHM of 45\,\uas, consistent with the effective FWHMs of other seven geometric models estimated from their second moments.
In \autoref{fig:refnoise_uv}, we show the refractive noise levels for the different intrinsic source models. 
The refractive noise level is only a few percents of the total flux density regardless of the intrinsic source structures.
The standard deviations across different geometric models are typically $\lesssim$0.4\% of the total flux density.

{\tt J18model2} was derived by taking the mean of the refractive noise levels of seven geometric models.
{\tt J18model1}, on the other hand, is based on the refractive noise level of the the circular Gaussian model. However, the circular Gaussian model has the lowest refractive noise level in the most of \uv-coverage of \sgra data, indicating that the refractive noise level from the circular Gaussian model, if not scaled properly, may underestimate that of \sgra data. 
We derived a scaling factor to make the median of the $\chi^{2}$ of all seven geometric models unity to minimize the effects of its systematically low refractive noise level.

%% file: appendix_first_imaging.tex
\section{First \sgra Images from Blind Imaging}
\label{sec:first_images}
Independently performed comparison between synthetic and real observational data can help to identify which image features are likely intrinsic and which are most likely spurious (see this approach in \citet{Bouman_Thesis} for synthetic data and \citetalias{M87PaperIV} for observed EHT data of M87$^{*}$). 

To initiate the exploration of possible \sgra images while minimizing influence from collective bias, in our first stage of analysis we reconstructed images of \sgra in five independent teams. 
The teams were blind to each others’ work and prohibited from discussing their imaging results.
This procedure was similar to that done in \citetalias{M87PaperIV} for first images of \m87, with two notable differences. First, since the participants in all teams were already aware of problematic data from previous EHT imaging work, the team activities were not blind to previously-identified problems in the data that could be affecting results. Second, as teams approached the date to submit results it was made known to the organizers that teams would not be comfortable submitting only a single image or movie, and the instructions were changed to submit three representative images or movies per team. 

No restrictions were imposed on the data pre-processing or imaging procedures used by each team. Teams 1 and 2 focused primarily on RML methods. Teams 3 and 4 primarily on CLEAN methods. Team 5 used a Bayesian method. All teams used an early-release engineering data set. 
While the reduction procedures and metadata corresponded to a preliminary version of the calibration pipeline, the data products was deemed mature enough for the blind imaging consistency test.
The April 7 data set was selected for the first comparison, as it had the best $(u,v)$-coverage and a very stable light curve over the full observation \citep{Wielgus_2021}. 

\begin{figure}[t]
\centering
\includegraphics[width=\columnwidth]{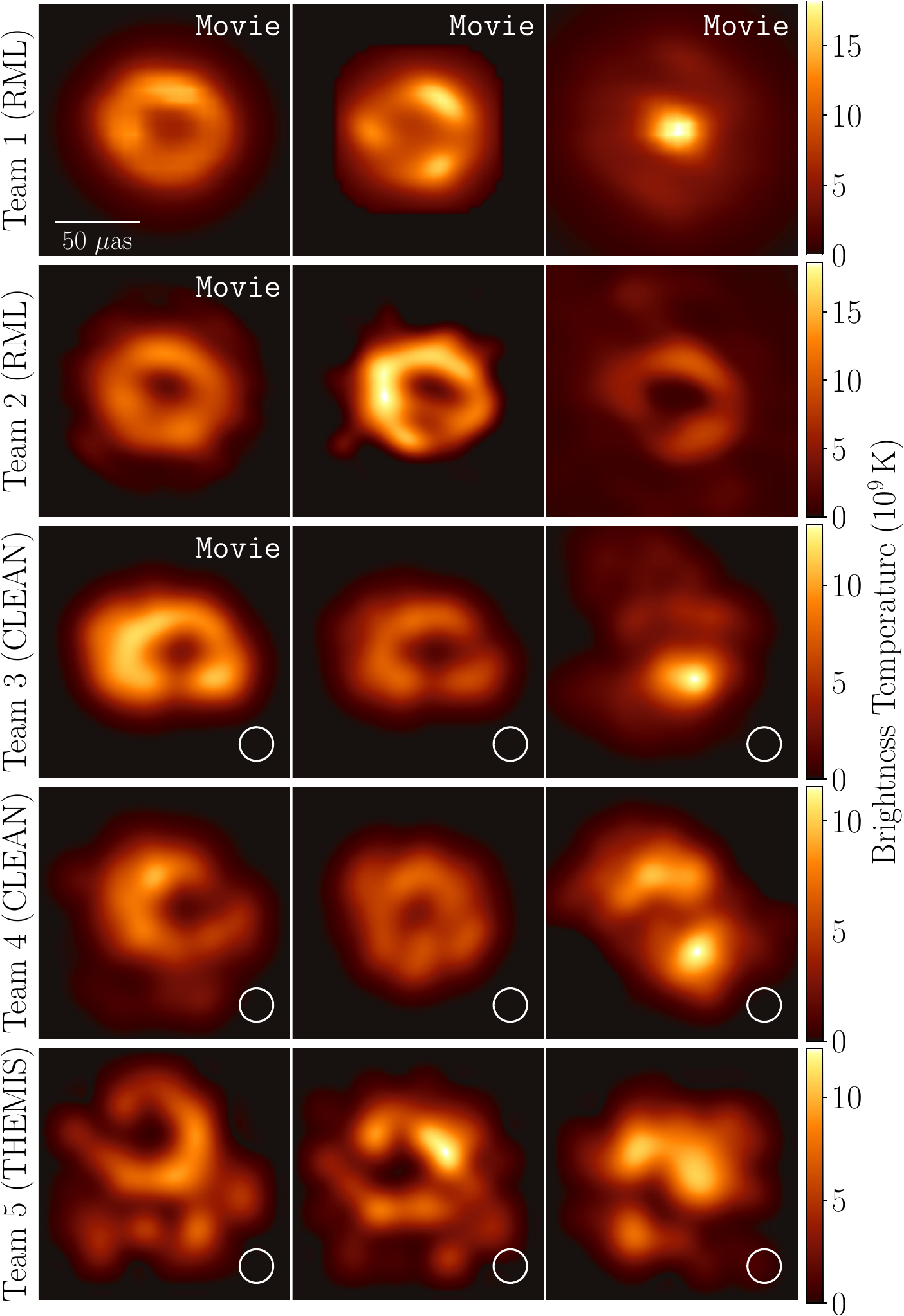}
\caption{The first EHT images of \sgra, blindly reconstructed by five independent imaging teams using an early, engineering release of Stokes I data from the April 7 observations. Images from Teams 1 and 2 used RML methods (no restoring beam); images from Teams 3 and 4 used CLEAN (restored with a circular 20\,\uas beam, shown in the lower right); images from Team 5 used \themis. Each Team shows three representative images from their reconstructions. Some images with the label ``Movie'' are time-averaged images of movie reconstructions. Many images show ring-like morphology with the diameter of $\sim 50$\,\uas, while some reconstructions show non ring-like morphology. The significant differences in brightness temperature between images are caused primarily by different assumptions regarding the total compact flux density (see \autoref{tab:first_images}) and also restoring beams applied only to CLEAN and \themis images.} 
\label{fig:first_images}
\end{figure}

\begin{table*}[t]
    \centering
    \caption{Image Properties and Data Consistency Metrics for the First Sgr A* Images}
    \label{tab:first_images}
    \begin{tabular}{@{}lcccccc}
        \toprule
         & & Team 1 & Team 2 & Team 3  & Team 4 & Team 5  \\ \midrule
        \multicolumn{6}{@{}l}{\noindent Image Properties}\\
        \quad Method 				& Image ID & RML & RML  & CLEAN & CLEAN & THEMIS\\
        \quad $F_{\rm cpct}$ (Jy)	
        & 1 & 2.37 & 2.07 & 1.90 & 1.64 & 1.59\\
	& 2 & 2.38 & 2.24 & 1.29 & 1.18 & 1.61\\
	& 3 & 2.38 & 1.99 & 1.25 & 0.69 & 1.73\\[0.5ex]
        \multicolumn{6}{@{}l}{\noindent Engineering Data (10-sec avg., stokes I, 0\% sys.~error)}\\
        \quad $\chi^2_{\text{CP}}$ 
& 1 & 2.27 (1.85) & 2.35 (1.75) & 4.52 (4.20) & 2.21 & 7.78\\
& 2 & 2.30 (1.81) & 2.08 & 2.51 & 3.31 & 6.79\\
& 3 & 2.81 (1.80) & 2.08 & 2.30 & 13.80 & 13.58\\[0.5ex]
        \quad $\chi^2_{\text{log\,CA}}$ 
& 1 & 3.44 (1.77) & 4.67 (2.19) & 2.76 (2.98) & 2.06 & 5.90\\
& 2 & 2.44 (2.07) & 2.00 & 2.49 & 2.84 & 4.88\\
& 3 & 4.73 (2.08) & 6.44 & 2.93 & 1.91 & 4.95\\[0.5ex]
        \multicolumn{6}{@{}l}{\noindent Science Release (60-sec avg., Stokes $I$, 0\% sys.~error)}\\
        \quad $\chi^2_{\text{CP}}$  
& 1 & 5.55 (2.53) & 6.09 (1.96) & 25.05 (21.77) & 4.58 & 54.11\\
& 2 & 6.08 (2.32) & 3.89 & 7.27 & 12.11 & 45.09\\
& 3 & 8.78 (2.31) & 4.11 & 5.73 & 108.88 & 109.10\\[0.5ex]
        \quad $\chi^2_{\text{log\,CA}}$ 
& 1 & 7.12 (1.52) & 10.75 (1.69) & 8.31 (5.92) & 2.79 & 24.22\\
& 2 & 4.02 (1.46) & 2.81 & 5.43 & 9.66 & 19.94\\
& 3 & 6.47 (1.79) & 16.95 & 8.21 & 7.45 & 10.01\\[0.5ex]
        \multicolumn{6}{@{}l}{\noindent Science Release (60-sec avg., Stokes $I$, 10\% sys.~error)}\\
        \quad $\chi^2_{\text{CP}}$   
& 1 & 2.42 (1.33) & 2.62 (1.18) & 7.80 (6.92) & 2.32 & 10.24\\
& 2 & 2.41 (1.25) & 2.09 & 2.85 & 5.93 & 8.02\\
& 3 & 4.24 (1.17) & 2.07 & 2.56 & 13.59 & 16.69\\[0.5ex]
        \quad $\chi^2_{\text{log\,CA}}$ 
& 1 & 3.36 (0.74) & 4.73 (0.89) & 3.98 (2.86) & 1.32 & 10.70\\
& 2 & 1.89 (0.76) & 1.26 & 2.49 & 3.88 & 7.55\\
& 3 & 3.57 (0.88) & 6.88 & 3.72 & 3.59 & 4.66\\[0.5ex]
         \bottomrule
    \end{tabular}\\
    \begin{minipage}{0.7\textwidth}
    {\footnotesize $^\dagger$ $F_{\rm cpct}$, $\chi^2_{\text{CP}}$, and $\chi^2_{\text{log\,CA}}$ mean the compact total flux density of the VLBI scale images, $\chi ^2$ to closure phases and log closure amplitudes, respectively. For movies, the values in the parentheses show $\chi ^2$ to the original movies, while the values outside of them show $\chi ^2$ to the time-averaged images. }
    \end{minipage}
\end{table*}

\autoref{fig:first_images} shows images of the three representative submissions made by each team for \sgra. In cases where dynamic movie reconstructions of \sgra were submitted, we show the time-averaged image and indicate that the original submission was a movie. 
Although most submissions contain flux with a $\sim$50\,\uas separation, not all submissions contain a clear ring feature. Additionally, of the submissions that do contain a ring feature, the azimuthal flux distribution varies drastically. For example, Team 1 submitted three movies: one containing a ring with a position angle of $\sim 30^\circ$, another containing a ring with position angle of $\sim -70^\circ$, and the third containing no ring at all. 

This initial imaging stage indicated that a $\sim$50\,\uas feature was likely to exist in the image, but there is significant uncertainty in the presence of a ring structure, as well as the flux distribution around a possible ring structure.
In \autoref{tab:first_images}, we show $\chi ^2$ values to closure phases ($\chi ^2_{\rm CP}$) and log closure amplitudes ($\chi ^2_{\rm \log CA}$) for both engineering data used for imaging and science release data described in \autoref{sec::Observations}, as well as the compact total flux densities ($F_{\rm cpct}$).
There are no significant differences seen in $\chi ^2$ values of images regardless of the presence of a ring structure and between different azimuthal intensity distributions of the ring structure.  
These ambiguous results motivated a series of systematic tests to identify if source evolution (similar to that expected to be present in \sgra) could explain the multiple solutions recovered.

%% file: appendix_pipelines.tex
\section{Imaging Pipelines}
\subsection{\difmap Pipeline}
\label{appendix:difmap-pipeline}

For this second stage of the imaging process we developed a scripted version of CLEAN in \texttt{DIFMAP} (version higher than \texttt{2.5k}), together with a python wrapper used over the script, for carrying out an extensive parameter search. The initial manual CLEAN analysis (see \autoref{sec:first_images}) together with the pre-imaging considerations (see \autoref{sec::PreImaging_Considerations}) provided the set and range of parameters to be explored on both the synthetic data sets (described in \autoref{sec:synthetic_data}) and the actual April 6 and 7 \sgra data, processed with both fringe fitting pipelines HOPS and CASA, with high and low bands combined.

Prior to CLEAN imaging, the EHT data was pre-processed using the pre-imaging pipeline discussed previously. This includes intra-site normalization to the \sgra light curve, coherent averaging of the visibilities using 10 and 60 seconds, the inclusion of an additional systematic error by a factor of 0, 2, and 5\%, and the scattering and noise mitigation prescriptions with the common parameters listed in \autoref{subsec:pre-im_pipeline} (see also, \autoref{table:topset-difmap}).

The first step in the \texttt{DIFMAP} imaging process is the gridding of the visibilities in the $(u,v)$-plane, that was done by using two different approaches: uniform and natural weighting. %
The next step is to down-weight the data on all baselines to phased-ALMA in the self-calibration process. As found for the case of M87 \citepalias{M87PaperIV}, this prevents the ALMA baselines to dominate the phase and amplitude self-calibration process due to their significantly higher \textit{S/N}. We have tested scaling factors for ALMA baselines weights during self-calibration of 0.1 and 0.5.

Atmospheric fluctuations at the short wavelength of 1.3\,mm 
severely limit the \textit{S/N} and the capability to reliably measure the visibility phases. We therefore rely on the closure phase measurements and the reconstruction of station phases using the Cornwell Wilkinson hybrid phase self-calibration approach. 
However the self-calibration with CLEAN model 
solely relies on the visibility phases and amplitudes, which forces an initial self-calibration of the phases (visibility amplitudes are much better constrained; \citetalias{PaperII}), a process which is anchored on the more robustly measured closure phases.

To mitigate any possible bias in our choice for the initial model to self-calibrate the phases, and therefore to minimize the chances of reaching a local minimum in our manifold of image models, we have explored different initial models for the first phase self-calibration. 
The ``fiducial'' initial model has been chosen, based on the reduced-$\chi^2$ of closure phases with the first reconstructed CLEAN model after the phase self-calibration with the initial model among three different types: a Gaussian with a FWHM of 15 $\mu$as (i.e., unresolved symmetric model), a uniform disk with a size ranging between 56 and 84 $\mu$as (in steps of 4 $\mu$as), and a uniform ring with sizes ranging between 36 and 68 $\mu$as (also in steps of 4 $\mu$as, no width).

A distinctive feature of CLEAN imaging is the use of masks (also known as cleaning windows) to define the area of the image to be cleaned. %
For this purpose we used a set of centered circular-disk-shaped CLEAN windows (i.e., with no hole in the middle to avoid any biasing).

The size of the cleaning windows should match the expected emission extend of the imaged source. To avoid introducing any prior bias we have surveyed for a relatively large range of mask sizes, from 80 to 110\,$\mu$as, that covers our prior \sgra image size constrains \citepalias{PaperII}. We note that due to the limited \uv-coverage larger masks ($\geq$ 100 $\mu$as) may pick up some emission from the main side-lobe of the interferometer, which may be particularly problematic for the point source model given its diffuse emission, more difficult to recover with the CLEAN algorithm.

Contrary to the case of M87 \citepalias{M87PaperIV}, we do not expect a large extended missing flux in \sgra \citepalias{PaperII}, hence our CLEAN stopping criterion is based on setting a minimum threshold for the relative decrease in the rms of the residual image over the image noise estimated from the visibilities.

As a result, the total number of surveyed parameters has been 1680 for the on-sky data set, and 6720 for descattered data. Note that the descattered results are based on a 4 times larger number of parameters due to the scattering mitigation (i.e., refractive noise floor).

\subsection{\ehtim Pipeline}
\label{appendix:ehtim-pipeline}

The \ehtim software library implements the RML imaging technique described in \autoref{subsec:rml_static}. As a RML method, a successful image reconstruction is achieved when the proper hyperparameters balance the contribution of the selected data products and regularizers to the minimization of \autoref{eq::objfunc}. Thus, we designed a framework around our imaging pipeline to explore an extensive range of imaging and data pre-calibration parameters.

The pipeline pre-process the input data following the procedure described in \autoref{subsec:pre-im_pipeline} as a first step. Only low- and high-band light-curve normalized data sets were used for static imaging to minimize the effects of the source variability. We opted for an integration time of 60\, s and explored all the non-closing systematic error, scattering and intra-day variability noise budgets tabulated in \autoref{table:topset-ehtim}. The algorithm is then initialized using a symmetric flat disk image contained in a square-shaped grid of 80$\times$80 pixels, which corresponds to a FOV of 150 \uas. The disk diameter is surveyed as a parameter with values of 70, 80, and 90 \uas. Note that, among the different imaging regularizers used in the pipeline, the maximum entropy regularizer (MEM) rewards similarity to a prior image, which in this case is assumed to be equal to the initialization image.

Full closure quantities and visibility amplitudes are used as data products. Contrary to M87$^*$, the relative weighting of the visibility amplitudes adopted range from a less to an equally important contribution as closure quantities, since complex station gains were derived from calibrator sources beforehand (see Section 5.1.3 in \citetalias{PaperII}). Several imaging regularizers and their relative weighting are employed to constraint on the image properties, such as MEM, total variation (TV), and total squared variation (TSV), which enforce similarity to a prior image and/or smoothness over neighboring pixels, respectively.

Convergence to an optimal image which minimizes \autoref{eq::objfunc} is then carried out in an iterative process in which the reconstructed image is blurred using a Gaussian kernel with the FWHM of the array nominal resolution ($\sim$24\,\uas), and then used as initialization for the next one to prevent the algorithm of being caught in local minima. In contrast to the M87$^*$ pipeline, no self-calibration was performed to the data, since the $\chi^2$ statistics obtained for the output images of just one cycle of this iterative procedure were already good enough.

\subsection{\smili Pipeline}
\label{appendix:smili-pipeline}

The \smili imaging pipeline was designed to reconstruct images with RML imaging techniques utilizing three imaging regularizers: weighted-$\ell_1$ ($\ell_1^w$), TV, and TSV, similar to \citetalias{M87PaperIV}.
Before the imaging process, both low- and high-band EHT data are pre-processed as described in \autoref{subsec:pre-im_pipeline}. First, data are normalized by the time-dependent intra-site flux density, and coherently time-averaged at an integration time of 120 sec. The intra-scan fluctuations in LMT baselines are corrected by self-calibrating the shortest VLBI baseline between SMT and LMT to a circular Gaussian with the total flux density of 1\,Jy and the FWHM size of 60\,\uas. Finally, the pipeline applies the parameterized mitigation schemes for the non-closing errors, scattering effects and intra-day variations as described in \autoref{subsec:pre-im_pipeline}.

The pipeline reconstructs an image from pre-processed data at both low and high bands jointly with the field of view of $150\ \mu\text{as}$ discretized by 2\,\uas pixels. Throughout the imaging, the pipeline adopts a prior image of a circular Gaussian with the FWHM size of 140, 160 or 180\,\uas for $\ell_1$ prior. The imaging process consists of in total nine imaging cycles, including a single self-calibration of complex visibilities. Each imaging cycle performs 5000 iterations of the L-BFGS-B algorithm used for the gradient-descent optimization in the \smili's image solver.  

In the first eight cycles of imaging, the images are reconstructed from visibility amplitudes, closure phases and log closure amplitudes. To account for residual errors in the amplitude calibration, the fractional 5 \% and  30 \% uncertainties are added in quadrature to the errors of visibility amplitudes on non-LMT and LMT baselines, respectively. The first cycle starts with a circular disk image with a $70\ \mu\text{as}$ diameter with the total flux of 1 Jy, while the later cycles subsequently adopt the image from the previous cycle as the initial image after re-centering its center of the mass and blurring with a $5\ \mu\text{as}$ circular Gaussian. 

After the eight cycles of imaging, complex visibilities are self-calibrated with its output image. Since the budgets of scattering and intra-day variations are not fractional to visibility amplitudes, these two budgets will not be properly scaled with gains solved with a self-calibration. Therefore, to reflect these two budgets accurately on self-calibrated data, the \smili pipeline updates these two budgets added to complex visibilities and closure quantities after self-calibration. Finally, the final image is reconstructed by a single imaging cycle using visibility amplitudes and closure quantities.

\subsection{\themis Raster Dimension Survey}
\label{sec:themis_survey_details}

The sole hyperparameters in the \themis analyses not self-consistently explored during the construction of the posterior are the raster dimensions, $N_x$ and $N_y$.  While reasonable initial guesses may be made based on the diffraction limit, in practice we employ data-driven method: comparing the Bayesian evidence (or appropriate proxies) of different dimensions \citep{Broderick_2020b}.

Due to the computational expense of performing \themis posterior reconstructions, we limit the number of raster surveys performed to three: for the \sgra April 6 data, the \sgra April 7 data, and the GRMHD synthetic data set presented in \autoref{subsec:synthetic_data_images_grmhd}.  Of these, we present only that on the \sgra April 7 data here for brevity; surveys on the other data sets were conducted similarly.  Furthermore, while we did experiment with $N_x\ne N_y$, given the apparent symmetry of the \sgra image, and not wishing to introduce any potential biases away from symmetry in the model specification, we restrict ourselves here to square rasters, i.e., $N_x=N_y$.

\begin{figure*}[!t]
    \centering
    \includegraphics[width=\textwidth]{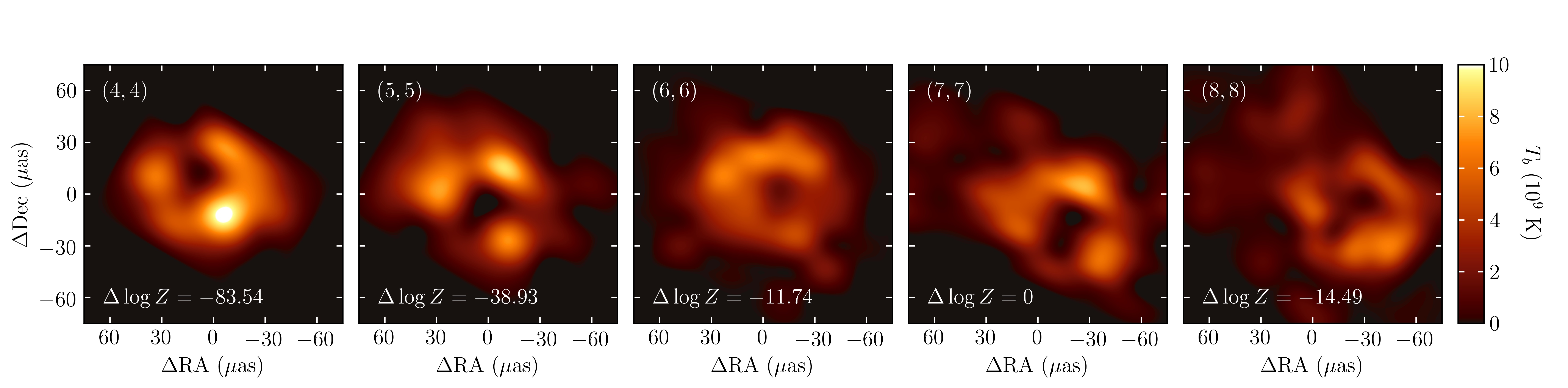}
    \caption{Mean images from the \themis posteriors for $(N_x,N_y)=(4,4)$, $(5,5)$, $(6,6)$, $(7,7)$ and $(8,8)$ from left to right.  In each the difference in the logarithm of the Bayesian evidence, $\Delta\log Z$, relative to the $(7,7)$ model is listed in the lower left.}
    \label{fig:themis_raster_survey}
\end{figure*}

Five individual analyses were performed on the April 7, combined high- and low-band complex visibility data\footnote{A similar study was performed using only the low-band data with similar results.} as described in \autoref{sec:themisdescription}, differing in the raster dimensions: $(N_x,N_y)=(4,4)$, $(5,5)$, $(6,6)$, $(7,7)$ and $(8,8)$.  For each, after convergence the Bayesian evidence, $Z$, was computed via thermodynamic integration across tempering levels \citep{Lartillot_06}. Mean images from each analysis and the relative Bayesian evidence are shown in Figure~\ref{fig:themis_raster_survey}.

For raster resolutions that are too small, the fit quality is poor.  For raster resolutions that are too large, the added model complexity is not justified by the fit improvement.  Importantly, by selecting on $\log Z$, we avoid potential complications associated with the modifications of the likelihood due to noise modeling, correlations between the high- and low-band gains, and non-Gaussianity of the posterior.

As roughly anticipated by the diffraction limit, the preferred raster size is $(7,7)$, which we adopt henceforth.  This is accompanied by a convergence of the image structure for raster dimensions of $(5,5)$ and larger.  Smaller and larger raster dimensions are overwhelmingly disfavored.  Nevertheless, by $(4,4)$ the locations of the knots are identified, though the model misspecification prevents a faithful recovery of their relative brightness.  

An identical procedure applied to the \sgra April 6 data finds that a smaller $(6,6)$ raster is preferred, ostensibly due to the smaller data volume.  Similarly, when applied to the GRMHD synthetic data set in \autoref{subsec:synthetic_data_images_grmhd}, a raster dimension of $(6,6)$ is again preferred, possibly due to the competition between the simplicity of the ring structure.

%% file: appendix_topsetselection.tex
\section{Top Set Selection \label{appendix:topset_selection}}
\subsection{The Effect of Relaxation in $\rho_{NX}$ Criteria}
As discussed in \autoref{subsec:rml_clean_imaging_surveys}, we relaxed the normalized cross-correlation criteria for each image by multiplying the value of $\rho_{NX}$ obtained for $\alpha=24$ \uas beam by a relaxation factor of 0.95. This relaxation factor allows us to recover a large enough number of Top Set parameters for evolving synthetic data with \sgra \uv-coverage.
This contrasts with the Top Set selection for M87 \citepalias{M87PaperIV}, where no relaxation factor was considered.
The necessity of the relaxation factor originates from the fact that temporal variations of the source structure cause a loss in the image fidelity of the reconstructions.%

\begin{figure*}
    \centering
    \includegraphics[width=0.85\linewidth]{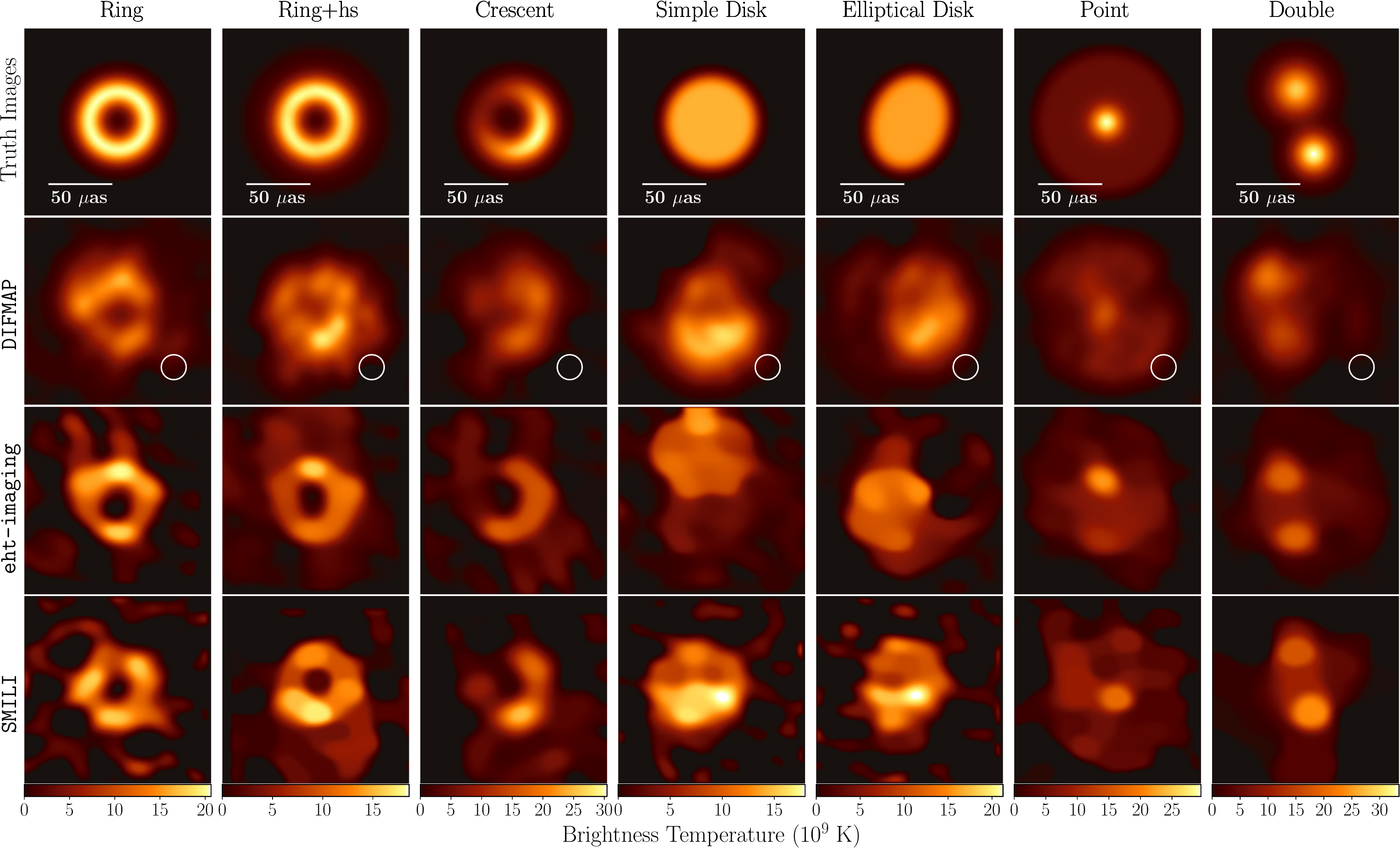}\\
    (a) Descattered reconstructions\vspace{1em}\\
    \includegraphics[width=0.85\linewidth]{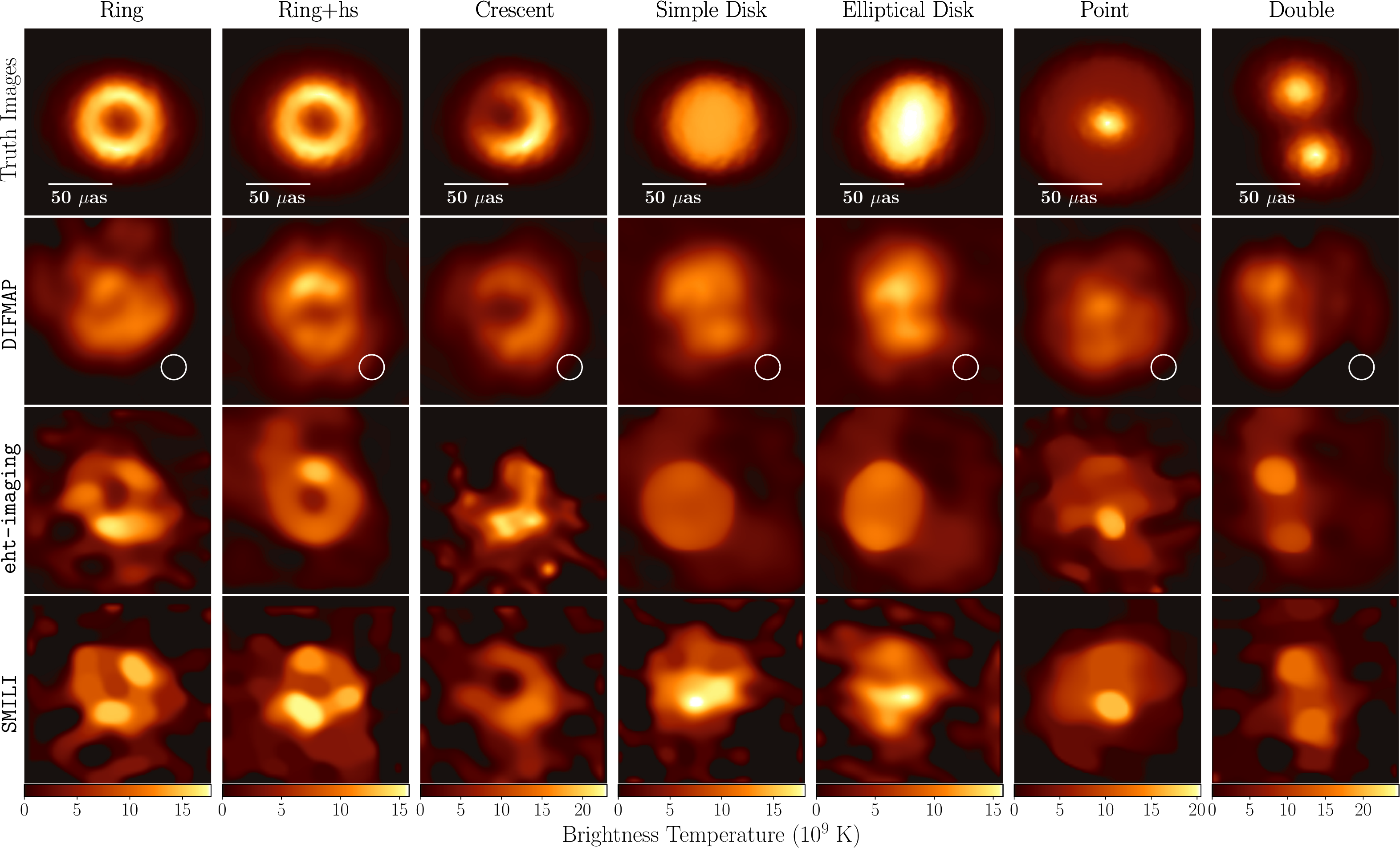}\\    (b) On-sky reconstructions \vspace{1em}
    \caption{The worst (i.e., lowest) $\rho_{\rm NX}$ images of synthetic geometric models among those reconstructed from the Top Sets of imaging parameters (a) with and (b) without scattering mitigation.}    
    \label{fig:synthetic_deblur-worst_0.95}
\end{figure*}

To ensure that this relaxed criteria is still able to distinguish different morphologies, we show the Top Set images with the worst (i.e. lowest) $\rho_{NX}$ for each geometric model in \autoref{fig:synthetic_deblur-worst_0.95}.
Each descattered or on-sky reconstruction in \autoref{fig:synthetic_deblur-worst_0.95} shows the image that has the worst $\rho_{NX}$ among the Top Set images for April 7.
These images still retain resemblance to the ground truth morphology, demonstrating that Top Set selection with a relaxation factor of 0.95 is still capable of reconstructing the basic structures of synthetic images with varied morphologies.

\subsection{$\chi^2$ Distribution of \sgra Images}
\begin{figure}
    \centering
    \includegraphics[width=\columnwidth]{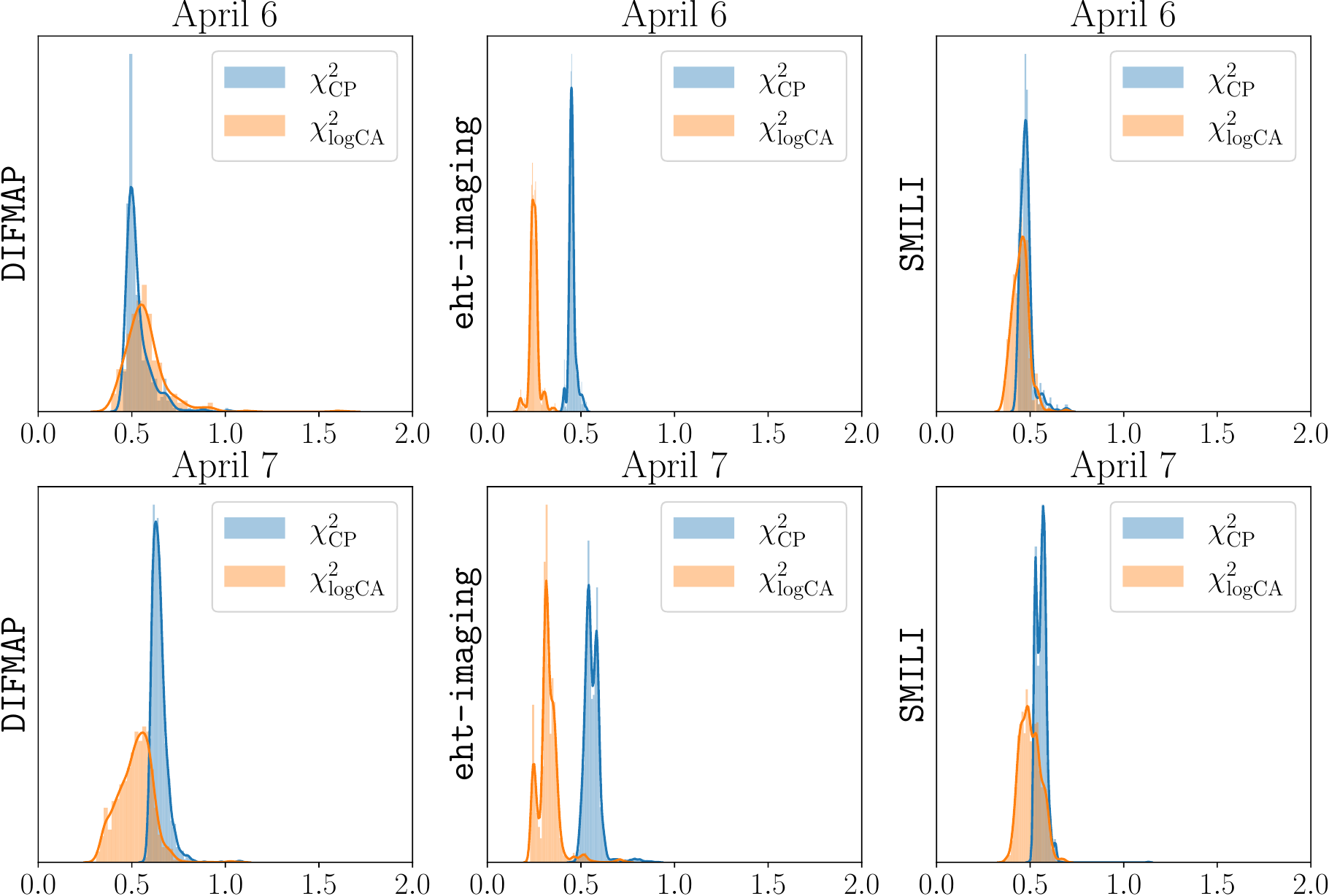}
    \caption{$\chi^2$ distributions of the images from each pipeline and epoch which passed the $\rho_{\rm NX}$ criteria. $\chi^2$ of closure phases and log closure amplitudes are showed in blue and orange respectively. We assumed $1 \ \rm{\%}$ systematic noise and representative variability noise for scattered images and we also added refractive noise when calculating $\chi^2$ of deblurred images.}
    \label{fig:chi2}
\end{figure}
Although the Top Set selection described in \autoref{subsec:synthetic_data_images_geometric} is sorely based on the image fidelity of the reconstructions from the synthetic data sets, Top Set images of \sgra provide reasonable fits to EHT measurements. 
In \autoref{fig:chi2}, we show $\chi^2$ distributions of all the \sgra images passing the $\rho_{NX}$ criterion using synthetic data reconstructions. 
$\chi^2$ values were computed for closure phases and log closure amplitudes formed from data after averaged at 1\,min, based on the same definition with \citetalias{M87PaperIV}. To account for non-closing effects, refractive scattering and time-variability, we include 1\,\% of the fractional errors and the representative budgets for scattering and temporal variability (see \autoref{sec:pre_scatter} and \autoref{sec:PSD_noise}), respectively 
-- in particular, the {\tt J18model1} refractive noise model and a variability model with parameters $a=0.02$, $u_0=2$, $b=2.5$, and $c=2$ are used. 
As shown in \autoref{fig:chi2}, most of the images have $\chi^2$ less than unity, and all the images provide $\chi^2<2$, demonstrating that all Top Set images provide reasonable fits to the \sgra data within the anticipated deviations of time-variable on-sky images from its intrinsic time-averaged structure.
Therefore, we adopt all the parameter sets satisfying the $\rho_{NX}$ criteria as the Top Sets without any cut by $\chi^2$.

\subsection{Top Sets of Imaging Parameters for April 6}

\begin{table}[t]
\footnotesize
\tabcolsep=0.05cm
\centering
\caption{Parameters in the \difmap Pipeline Top Set on April\,6
}
\label{table:topset-difmap-appendix}
\begin{tabularx}{\columnwidth}{lccccccc} 
\hline
\hline
\multicolumn{8}{c}{April~6 (8400 Param.\ Combinations; 365 in Top Set) } \\
\hline
\textbf{Systematic} & \multicolumn{2}{c}{\textbf{0}} & \multicolumn{2}{c}{\textbf{0.02}} & \multicolumn{3}{c}{\textbf{0.05}} \\
\textbf{error} & \multicolumn{2}{c}{15.6\%} & \multicolumn{2}{c}{17.3\%} & \multicolumn{3}{c}{67.1\%} \\
\hline
\textbf{$\bf Ref\ Type$} & \textbf{No} & \textbf{Const} & \multicolumn{2}{c}{\textbf{2$\times$Const}} & \textbf{J18} & \multicolumn{2}{c}{\textbf{2$\times$J18}} \\
& 18.1\% & 23.0\% & \multicolumn{2}{c}{14.2\%} & 24.1\% & \multicolumn{2}{c}{20.5\%} \\
\hline
\textbf{$\bf a_{\bf \rm psd}$} & \textbf{No} & \multicolumn{2}{c}{\textbf{0.015}} & \multicolumn{2}{c}{\textbf{0.02}} & \multicolumn{2}{c}{\textbf{0.025}} \\
& 21.9\% & \multicolumn{2}{c}{26.8\%} & \multicolumn{2}{c}{25.8\%} & \multicolumn{2}{c}{25.5\%} \\
\hline
\textbf{$\bf b_{\bf \rm psd}$} & \textbf{No} & \multicolumn{2}{c}{\textbf{1}} & \multicolumn{2}{c}{\textbf{3}} & \multicolumn{2}{c}{\textbf{5}} \\
& 21.9\% & \multicolumn{2}{c}{17.3\%} & \multicolumn{2}{c}{24.4\%} & \multicolumn{2}{c}{36.4\%} \\
\hline
\textbf{$|u|_{0}$} & \multicolumn{3}{c}{\textbf{No}} & \multicolumn{4}{c}{\textbf{2}} \\
& \multicolumn{3}{c}{21.9\%} & \multicolumn{4}{c}{78.1\%} \\
\hline
\textbf{Time average} & \multicolumn{3}{c}{\textbf{10}} & \multicolumn{4}{c}{\textbf{60}}\\
(sec) & \multicolumn{3}{c}{49.9\%} & \multicolumn{4}{c}{50.1\%} \\
\hline
\textbf{ALMA weight} & \multicolumn{3}{c}{\textbf{0.1}} & \multicolumn{4}{c}{\textbf{0.5}} \\
& \multicolumn{3}{c}{39.5\%} & \multicolumn{4}{c}{60.5\%} \\
\hline
\textbf{UV weight} & \multicolumn{3}{c}{\textbf{0}} & \multicolumn{4}{c}{\textbf{2}} \\
& \multicolumn{3}{c}{72.6\%} & \multicolumn{4}{c}{27.4\%} \\
\hline
\textbf{Mask Diameter} & \textbf{80} & \textbf{85} & \textbf{90} & \textbf{95} & \textbf{100} & \textbf{105} & \textbf{110}\\
(\uas) & 10.7\% & 19.2\% & 36.2\% & 17.0\% & 6.0\% & 7.9\% & 3.0\%\\
\hline
\end{tabularx}
\\ \vspace{0.3cm}
\raggedright{\textbf{Note. }
In each row, the upper line with bold text shows the surveyed parameter value corresponding to the parameter of left column, while the lower line shows the number fraction of each value in Top Set. The total number of surveyed parameter combinations and Top Set are shown in the first row. } 
\end{table}

\begin{table}[t]
\footnotesize
\tabcolsep=0.2cm
\centering
\caption{
Parameters in the \ehtim Pipeline Top Set on April\,6
}
\label{table:topset-ehtim-appendix}
\begin{tabularx}{\columnwidth}{lccccc} 
\hline
\hline
\multicolumn{6}{c}{April~6 (112320 Param.\ Combinations; 1415 in Top Set) } \\
\hline
\textbf{Systematic} & \multicolumn{2}{c}{\textbf{0}} & \textbf{0.02} & \multicolumn{2}{c}{\textbf{0.05}} \\
\textbf{error} & \multicolumn{2}{c}{19.0\%} & 39.7\% & \multicolumn{2}{c}{41.3\%} \\
\hline
\textbf{$\bf Ref\ Type$} & \textbf{No} & \textbf{Const} & \textbf{2$\times$Const} & \textbf{J18} & \textbf{2$\times$J18} \\
& 7.8\% & 29.2\% & 26.4\% & 17.0\% & 19.6\%\\
\hline
\textbf{$\bf a_{\bf \rm psd}$} & \textbf{No} & \multicolumn{2}{c}{\textbf{0.015}} & \textbf{0.02} & \textbf{0.025}\\
& 13.0\% & \multicolumn{2}{c}{34.2\%} & 29.4\% & 23.4\%\\
\hline
\textbf{$\bf b_{\bf \rm psd}$} & \textbf{No} & \textbf{1} & \textbf{2} & \textbf{3} & \textbf{5}\\
& 13.0\% & 16.3\% & 25.4\% & 22.1\% & 23.1\%\\
\hline
\textbf{$|u|_{0}$} & \multicolumn{3}{c}{\textbf{No}} & \multicolumn{2}{c}{\textbf{2}} \\
& \multicolumn{3}{c}{13.0\%} & \multicolumn{2}{c}{87.0\%} \\
\hline
\textbf{TV} & $\bf 0$ & \multicolumn{2}{c}{\textbf{0.01}} & \textbf{0.1} & \textbf{1}\\
& 12.4\% & \multicolumn{2}{c}{27.8\%} & 51.8\% & 8.0\%\\
\hline
\textbf{TSV} & $\bf 0$ & \multicolumn{2}{c}{\textbf{0.01}} & \textbf{0.1} & \textbf{1}\\
& 29.0\% & \multicolumn{2}{c}{34.8\%} & 35.8\% & 0.4\%\\
\hline
\textbf{Prior size} & \multicolumn{2}{c}{\textbf{70}} & \textbf{80} & \multicolumn{2}{c}{\textbf{90}} \\
(\uas) & \multicolumn{2}{c}{15.3\%} & 27.1\% & \multicolumn{2}{c}{57.6\%} \\
\hline
\textbf{MEM} & $\bf 0$ & \multicolumn{2}{c}{\textbf{0.01}} & \textbf{0.1} & \textbf{1}\\
& 3.4\% & \multicolumn{2}{c}{15.3\%} & 77.9\% & 3.5\%\\
\hline
\textbf{Amplitude} & \multicolumn{2}{c}{\textbf{0}} & \textbf{0.1} & \multicolumn{2}{c}{\textbf{1}} \\
\textbf{weight} & \multicolumn{2}{c}{0\%} & 4.2\% & \multicolumn{2}{c}{95.8\%} \\
\hline
\end{tabularx}
\\ \vspace{0.3cm}
\raggedright{\textbf{Note. }
Same as \autoref{table:topset-difmap-appendix}} 
\end{table}

\begin{table}[t]
\footnotesize
\tabcolsep=0.2cm
\centering
\caption{Parameters in the \smili Pipeline Top Set on April\,6
}
\label{table:topset-smili-appendix}
\begin{tabularx}{\columnwidth}{lccccc} 
\hline
\hline
\multicolumn{6}{c}{April~6 (54000 Param.\ Combinations; 292 in Top Set) } \\
\hline
\textbf{Systematic} & \multicolumn{2}{c}{\textbf{0}} & \textbf{0.02} & \multicolumn{2}{c}{\textbf{0.05}} \\
\textbf{error} & \multicolumn{2}{c}{34.2\%} & 31.8\% & \multicolumn{2}{c}{33.9\%} \\
\hline
\textbf{$\bf Ref\ Type$} & \textbf{No} & \textbf{Const} & \textbf{2$\times$Const} & \textbf{J18} & \textbf{2$\times$J18} \\
& 7.9\% & 14.0\% & 18.8\% & 24.3\% & 34.9\% \\
\hline
\textbf{$\bf a_{\bf \rm psd}$} & \textbf{No} & \multicolumn{2}{c}{\textbf{0.015}} & \textbf{0.02} & \textbf{0.025}\\
& 0.3\% & \multicolumn{2}{c}{23.3\%} & 46.2\% & 30.1\%\\
\hline
\textbf{$\bf b_{\bf \rm psd}$} & \textbf{No} & \textbf{1} & \textbf{2} & \textbf{3} & \textbf{5}\\
& 0.3\% & 28.8\% & 37.0\% & 28.1\% & 5.8\%\\
\hline
\textbf{$|u|_{0}$} & \multicolumn{2}{c}{\textbf{No}} & \textbf{1} & \multicolumn{2}{c}{\textbf{2}} \\
& \multicolumn{2}{c}{0.3\%} & 44.5\% & \multicolumn{2}{c}{55.1\%} \\
\hline
\textbf{TV} & $\bf 10^2$ & \multicolumn{2}{c}{$\bf 10^3$} & $\bf 10^4$ & $\bf 10^5$\\
& 7.2\% & \multicolumn{2}{c}{92.5\%} & 0.3\% & 0.0\%\\
\hline
\textbf{TSV} & $\bf 10^2$ & \multicolumn{2}{c}{$\bf 10^3$} & $\bf 10^4$ & $\bf 10^5$ \\
& 41.4\% & \multicolumn{2}{c}{58.6\%} & 0.0\% & 0.0\% \\
\hline
\textbf{Prior size} & \multicolumn{2}{c}{\textbf{140}} & \textbf{160} & \multicolumn{2}{c}{\textbf{180}} \\
(\uas) & \multicolumn{2}{c}{31.5\%} & 41.1\% & \multicolumn{2}{c}{27.4\%} \\
\hline
\textbf{$\bf \ell_1$} & \multicolumn{2}{c}{\textbf{0.1}} & \textbf{1} & \multicolumn{2}{c}{$\bf 10$} \\
& \multicolumn{2}{c}{99.7\%} & 0.3\% & \multicolumn{2}{c}{0.0\%} \\
\hline
\end{tabularx}
\\ \vspace{0.3cm}
\raggedright{\textbf{Note. }
Same as \autoref{table:topset-difmap-appendix}} 
\end{table}

In \autoref{subsubsec:Top Sets}, we show the results of the Top Set selection for April 7 data.
Here, we show the summary of Top Set parameters of each pipeline for April 6 data in \autoref{table:topset-difmap-appendix}, \autoref{table:topset-ehtim-appendix} and \autoref{table:topset-smili-appendix}, which are selected based on the method identical to that of April 7 data.
Although the number of Top Set parameters for April 6 exceeds 100, this number is significantly smaller than the Top Set size identified for April 7 data. In particular, the number of Top Set parameters is reduced to only $22\ \%$ (\difmap), $25\ \%$ (\ehtim) and $11\ \%$ (\smili) of the April 7 Top Set size, as shown in \autoref{table:topset-difmap}, \autoref{table:topset-ehtim} and \autoref{table:topset-smili}. This significant reduction is most likely due to poorer \uv-coverage on April 6. 

%% file: appendix_clustering.tex
\section{Classification of Ring Images}
\label{appendix:clustering_method}
As described in \autoref{sec:synthetic_data:grmhd} and \autoref{sec:sgra_images:clustering}, we categorize GRMHD and \sgra images into clusters of images that share similar morphologies. Clustering of images is performed separately for on-sky and descattered reconstructions from each pipeline.

Prior to clustering, the images are aligned with an iterative method described in below.
We first derive the averaged image of all images. Each image is then aligned to maximize the cross-correlation with the averaged image. After aligning all images, the averaged image is recomputed with the aligned ones, and used to align each image again. We repeat this procedure to align each image for a total of three times, providing the convergence of the alignment.
After the above relative alignments between images, all images are centered using the average image; the images are uniformly shifted with the same amount of positional offset to maximize the cross-correlation between the average image and the time-averaged image of \texttt{Simple Disk} model.

The clustering of images has two major steps: the identification of ring images and non-ring images, and the clustering of the ring images by the peak position angles. The ranges of the peak position angles are described in \autoref{sec:synthetic_data:grmhd} for the GRMHD model and \autoref{sec:sgra_images:clustering} for the \sgra reconstructions.
The separation of ring and non-ring images is based on two morphological criteria as described in below.

First, we identify non-ring images by the degree of the central depression seen in the image.
We measure a typical brightness of the central region by taking the median of the intensity within a radius of 10\,\uas, defined by
\begin{equation}
    I_{\rm c} \equiv \median _{r < 10,\,\theta \in [-\pi, \pi)} I(r, \theta),
\end{equation}
where $I(r, \theta)$ denotes the intensity at the radius of $r$ in \uas and the position angle of $\theta$ in radians.
To measure the degree of the central depression, we compare the measured central brightness $I_{\rm c}$ with the maximum of the azimuthally-averaged intensity at the outer area within a radius of 10 to 40\,\uas, given by
\begin{equation}
    I_{\rm o} \equiv \max _{r \in [10,40]} \left( \frac{1}{2\pi}\int^{\pi}_{-\pi}I(r,\theta)\,d\theta \right).
\end{equation}
We define the degree of the central depression by $f_{\rm cd} \equiv 1-I_c/I_o$, and classify as non-ring images those with $f_{\rm cd} < 0.2$ and $f_{\rm cd} < 0.15$ for descattered and on-sky reconstructions, respectively. The slightly lower threshold for on-sky images takes account of the angular broadening effects due to scattering, causing a systematic decrease in the central depression of a ring emission. This criterion can effectively distinguish point-like images from ring images. 

The second criterion identifies non-ring images by the smoothness of the azimuthal intensity distributions.
To extract the azimuthal profile of the ridge intensity for the outer emission, we take the maximum intensity of the outer area within the radius of 10-40 \uas for each position angle, given by
\begin{equation}
    I_p(\theta) \equiv \max_{r \in [10,40]} I(r,\theta).
\end{equation}
For each position angle $\theta$, we evaluate the difference between the ridge brightness $I_p(\theta)$ to the central brightness $I_c$ normalized by the maximum intensity of the outer area defined by
\begin{equation}
    f_p(\theta) \equiv \frac{I_p(\theta) - I_c}{\max_{\theta \in [-\pi,\pi)} I_p(\theta)}.
\end{equation}
$f_p(\theta)$ allows to assess whether the azimuthal distribution of the outer area has a dark gap area comparable or lower than the central depression. We identify a continuous dark area with $f_p(\theta) < 0.2$ and $f_p(\theta) < 0.15$ over a range of position angles broader than $70^{\circ}$ as a gap in the azimuthal intensity distribution of descattered and on-sky reconstructions, respectively.
If an image has more than two distinct gaps separated by $>70^{\circ}$, we classify it as a non-ring image with two or more distinct blobs not smoothly connected with each other. We also classify an image as a non-ring one with an incomplete ring if it does not have a continuous bright area without any gaps over the range of position angles less than $180^{\circ}$ (i.e. not completing more than a half circle).
The above criteria effectively exclude multiple blob or highly corrupted ring-like images.

The morphological criteria discussed above provides a classification of ring and non-ring images that is broadly consistent with human perception. However, as noted in Section~\ref{subsec:synthetic_data_images_grmhd}, the classification of images that are borderline between ring and non-ring classification are sensitive to the exact criteria used. Therefore ring definitions that make use of slightly different criteria can lead to classification that still largely aligns with human perception but varies in the ring classification percentages quoted in this paper. In this work, motivated by method interpretability, we chose to make use of the simplest classification criteria that still largely aligned with human perception.

%% file: appendix_smili_dynamical.tex
\section{Validation of Static Imaging Results with Full-track Dynamic Imaging}
\label{sec:smili_dynam_survey}

\begin{table}[t]
\footnotesize
\tabcolsep=0.4cm
\centering
\caption{Parameters in the \smili dynamic Pipeline Top Set on April\,7
}
\label{table:topset-dynamic-appendix}
\begin{tabularx}{\columnwidth}{lcccc} 
\hline
\hline
\multicolumn{5}{c}{April~7 (7776 Param.\ Combinations; 345 in Top Set) } \\
\hline
\textbf{Systematic} & \textbf{0} & \multicolumn{2}{c}{\textbf{0.02}} & \textbf{0.05} \\
\textbf{error} & 32.8\% & \multicolumn{2}{c}{35.4\%} & 31.9\% \\
\hline
\textbf{$\bf Ref\ Type$} & \multicolumn{2}{c}{\textbf{No}} & \multicolumn{2}{c}{\textbf{J18model1}} \\
& \multicolumn{2}{c}{18.8\%} & \multicolumn{2}{c}{81.2\%} \\
\hline
\textbf{TV} & $\bf 10^2$ & $\bf 10^3$ & $\bf 10^4$ & $\bf 10^5$\\
& 9.0\% & 22.0\% & 69.0\% & 0\%\\
\hline
\textbf{TSV} & $\bf 10^2$ & $\bf 10^3$ & $\bf 10^4$ & $\bf 10^5$ \\
& 32.2\% & 44.6\% & 23.2\% & 0\%\\
\hline
\textbf{$\bf \ell_1$} & \textbf{0.1} & \multicolumn{2}{c}{\textbf{1}} & $\bf 10$ \\
& 11.9\% & \multicolumn{2}{c}{88.1\%} & 0\% \\
\hline
\textbf{Prior size} & \textbf{140} & \multicolumn{2}{c}{\textbf{160}} & \textbf{180} \\
(\uas) & 29.6\% & \multicolumn{2}{c}{36.5\%} & 33.9\% \\
\hline
\textbf{$\bf R_t$} & $\bf 10^4$ & \multicolumn{2}{c}{$\bf 10^5$} & $\bf 10^6$ \\
& 39.7\% & \multicolumn{2}{c}{24.1\%} & 36.2\% \\
\hline
\textbf{$\bf R_i$} & $\bf 10^4$ & \multicolumn{2}{c}{$\bf 10^5$} & $\bf 10^6$ \\
& 14.2\% & \multicolumn{2}{c}{40.9\%} & 44.9\% \\
\hline
\end{tabularx}
\\ \vspace{0.3cm}
\raggedright{\textbf{Note. }
In each row, the upper line with bold text shows the surveyed parameter value corresponding to the parameter of left column, while the lower line shows the number fraction of each value in Top Set. The total number of surveyed parameter combinations and Top Set are shown in the first row. } 
\end{table}

\begin{figure}[t]
    \centering
    \includegraphics[width=0.85\linewidth]{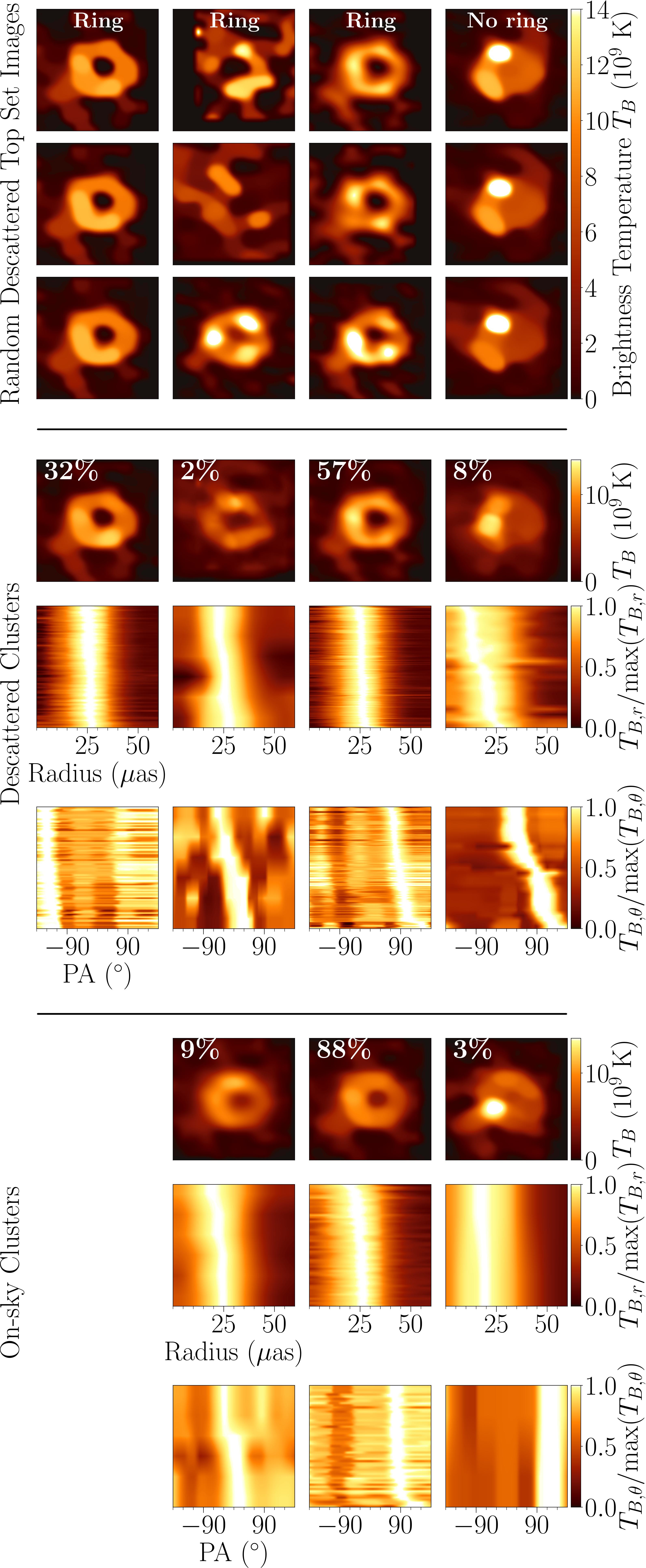}
    \caption{
The distribution of \sgra Top Set images on April 7 reconstructed with the \smili dynamic imaging pipeline. We show the distribution of images for each cluster in the same convention with \autoref{fig:SgrA-topset-3599} with three horizontal panels separated by horizontal lines. 
The top panel shows individual images randomly sampled from different clusters.
The middle and bottom panels visualize the distributions of reconstructed descattered and on-sky \sgra images for each cluster, respectively. 
In each panel, from top to bottom, we show the average of each cluster, the distributions of the radial profiles, and the distributions of azimuthal intensity profiles. 
}
    \label{fig:sgra_smili_dynam}
\end{figure}
As described in \autoref{sec:PSD_noise}, the temporal variation of \sgra is anticipated to cause significant deviations of visibilities from those of the time-averaged morphology, of which levels can be well characterized by a broken power-law model. 
As described in \autoref{sec:sgra_images:psdnoise}, the variability noise model allows us to enhance the overall fidelity of synthetic data reconstructions, and enable more sets of the imaging parameter combinations to be selected as Top Sets.
However, the prescription for the temporal variability in \autoref{sec:PSD_noise} include some simplifications of its characteristic properties: for instance, the circular symmetry assumed for the levels of the variability amplitudes in Fourier space and no inclusion of the correlated variations considered in data metrics as covariant components.
Here, to assess the dependence of the mitigation scheme for temporal variability, we show the time-averaged images identified by full-track dynamic imaging not relying on the variability noise model. We emphasize that the goal of this section is not to characterize the dynamic evolution of \sgra on short timescales, which is the primary focus of \autoref{sec::dynamic}.

As shown in \autoref{fig:sgra_smili_dynam}, we find the primary three ring morphologies identified in \autoref{sec:sgra_images} with similar azimuthal variations in the most of Top Set images, while non-ring structures are also reconstructed in a small fraction of Top Sets. 
The results are broadly consistent with those from the imaging survey presented in \autoref{sec:sgra_images}, and strongly indicate that our results described in \autoref{sec:sgra_images} are resilient to the methods to recover the time-averaged morphology.
We briefly describe the imaging process in \autoref{sec:smili_dynam_survey:methods} and the results in  \autoref{sec:smili_dynam_survey:results}.

\subsection{\smili Dynamic Imaging Pipeline and Top Set Selections}
\label{sec:smili_dynam_survey:methods}
Similar to the RML and CLEAN parameter surveys described in \autoref{subsec:rml_clean_imaging_surveys}, we conducted a large imaging survey with RML dynamic imaging methods (see \autoref{sec:dynamicimagingmethods}) implemented in a scripted pipeline using \smili. The survey was performed on all seven geometric models and \sgra data.

After completing the common pre-imaging process of data (\autoref{subsec:pre-im_pipeline}), the pipeline reconstructs the time-averaged images on $\sim 8000$ sets of the imaging parameter combinations across a broad parameter space, as outlined in \autoref{table:topset-dynamic-appendix}.
With the exactly same criteria described in \autoref{subsec:rml_clean_imaging_surveys}, Top Sets of the imaging parameters are then selected based on the fidelity of the synthetic data reconstructions.

The \smili dynamic imaging pipeline shares the same procedures with the \smili static imaging pipeline described in \autoref{appendix:smili-pipeline}, except for a major difference. 
Instead of the variability noise model used in \autoref{subsec:rml_clean_imaging_surveys}, we allow temporal variations on time scales of typical scan intervals --- for each set of parameters, the pipeline reconstructs a movie with the frame interval of 1\,h, and then time-averages the movie to obtain the resultant reconstruction of the time-averaged morphology. 
We utilize two temporal regularizers, denoted by $R_t$ and $R_i$, enforcing the continuity of the frame-to-frame intensity variations and the continuity between each frame and time-averaged intensity distributions, respectively, based on the Euclidian distance between images \citep[see][for details]{johnson2017dynamical}.
We note that here we only explore \texttt{J18model1} for the scattering mitigation, given consistency among the different scattering mitigation schemes (\autoref{sec:sgra_images:scattering}).

\subsection{Results}
\label{sec:smili_dynam_survey:results}
As shown in \autoref{fig:sgra_smili_dynam}, Top Set images from \smili dynamic imaging pipeline can be categorized into four clusters shown in \autoref{sec:sgra_images:clustering}. Similar to the RML, CLEAN and \themis static imaging pipelines, the ring morphology with the diameter of \sdiam was found as the dominant feature, while the non-ring images were found in a small ($\sim 5$\%) fraction of Top Sets. 
The differences in the image appearance between on-sky and descattered reconstructions are also consistent with \autoref{sec:sgra_images:scattering}; the on-sky reconstructions are more blurry than the descattered ones. 
We note that similar to Top Sets presented in \autoref{sec:synthetic_data} and \autoref{sec:sgra_images}, here Top Set images from \smili dynamic imaging pipeline do not constitute a likelihood and therefore the fractions should not represent our degree of certainty. 

%% file: appendix_bestbetMAD_from_theory.tex
\section{Reconstructions of a Best-bet GRMHD Model in Paper V}
\label{sec:appendix_best_grmhd}

In \autoref{sec:survey}, we utilize synthetic data based on a GRMHD model selected from a library of time-dependent GRMHD models presented in \citetalias{PaperV} (see \autoref{sec:synthetic_data:grmhd}) to assess the performance of our imaging procedures on a physically motivated evolving source.
While this GRMHD model is broadly consistent with our criteria in \autoref{sec:synthetic_data} based on 1.3\,mm EHT data and light curves, it is not identified as a ``best-bet" model in \citetalias{PaperV}; ``best-bet" models satisfy heterogeneous constraints derived from 1.3\,mm EHT data, 86 GHz VLBI observations with the GMVA, 2.2 ${\rm \mu m}$ flux density and x-ray luminosity.
\citetalias{PaperV} identifies three GRMHD models that lay in a ``best-bet'' region  of strongly magnetized (MAD) models at low inclination with prograde spin. 

\begin{figure}[t]
\centering
\includegraphics[width=\linewidth]{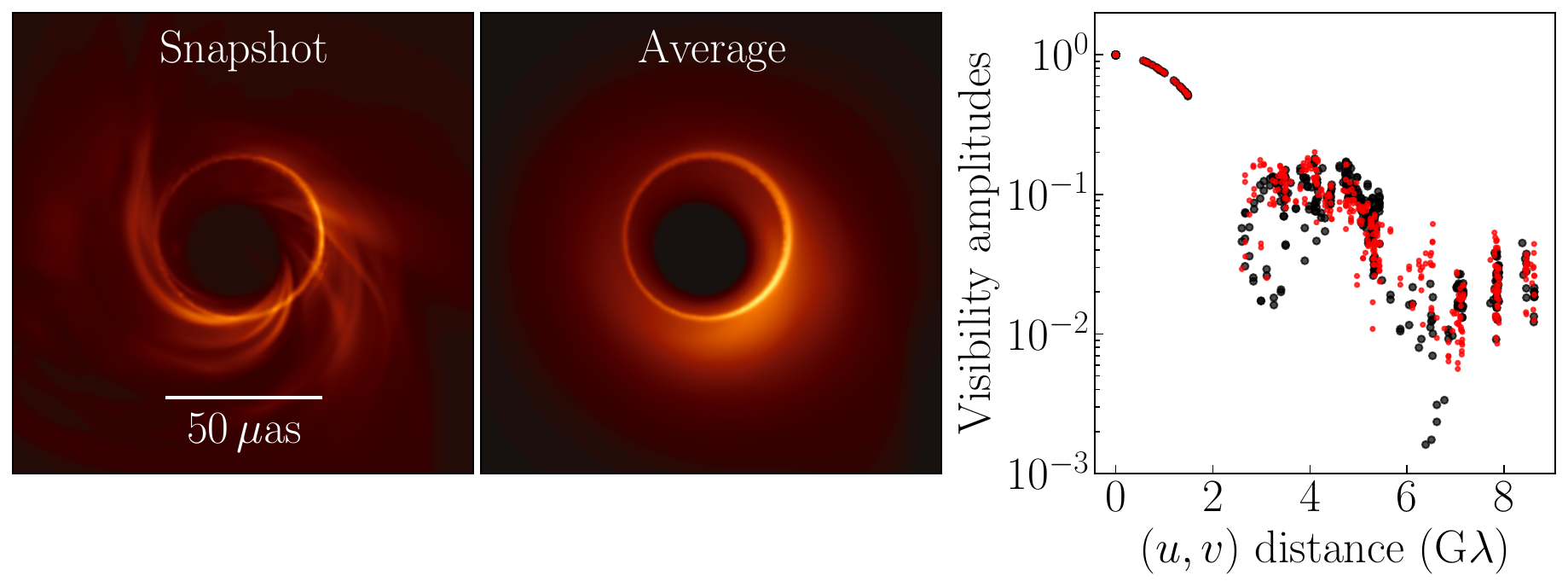}
    \caption{
    Image and visibility characteristics of the best-bet GRMHD models in \citetalias{PaperV} (MAD, $a_{*}=0.5, i=30^\circ, R_{\rm high}=160$). Left two panels are the snapshot and averaged images of the ground-truth movie. Right panel shows the simulated visibility amplitudes (red) and real \sgra measurements (black) as a function of projected baseline length.
    }
    \label{fig:bestbet-grmhd-groundtruth}
\end{figure}

\begin{figure*}[t]
    \centering
    \includegraphics[width=0.85\linewidth]{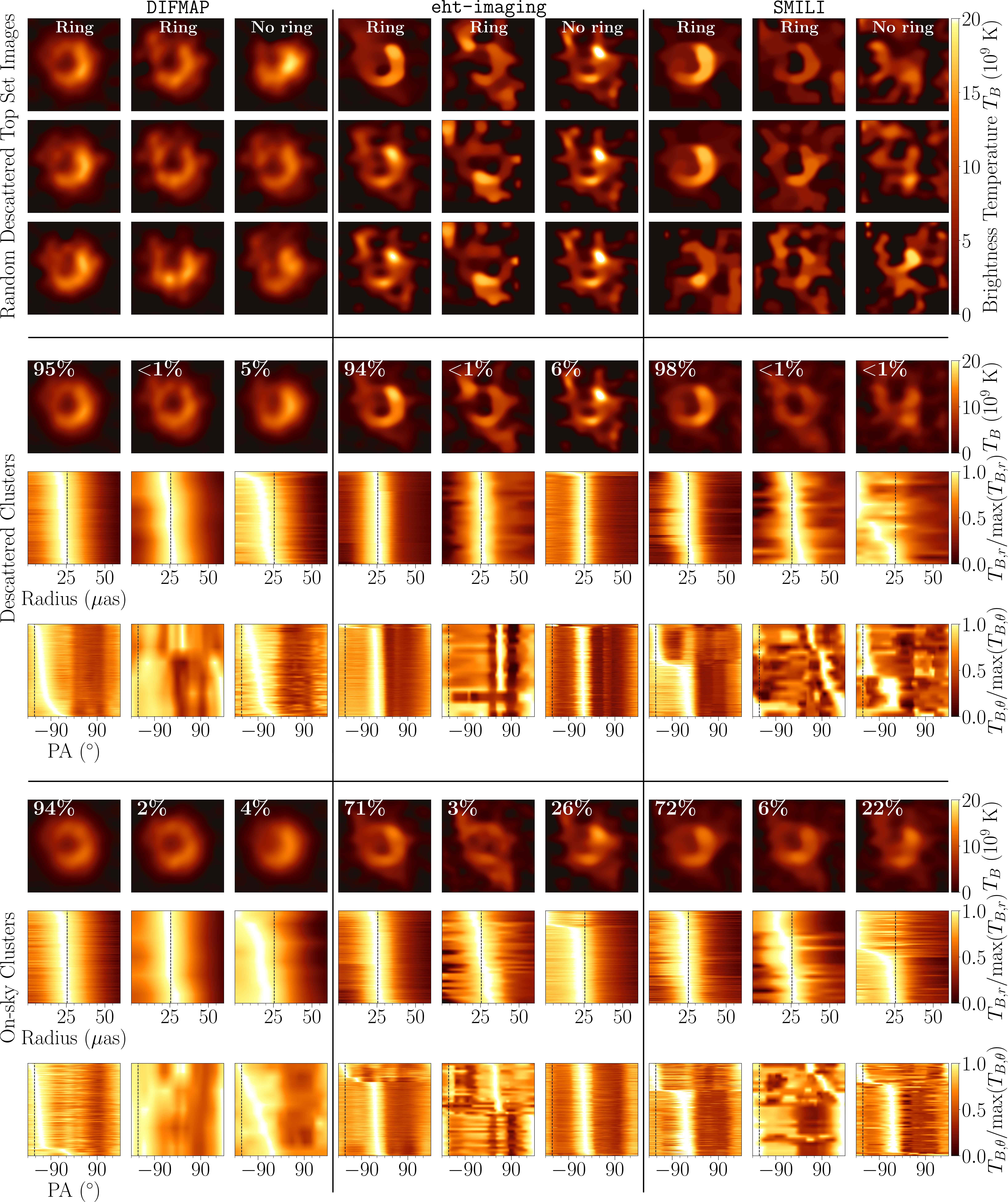}
    \caption{
    The distribution of reconstructed images with a best-bet GRMHD model identified in paper V. The distribution of images from each pipeline for each cluster is shown with the same convention as \autoref{fig:grmhd-3599}.
    }
    \label{fig:bestbet-grmhd-3599}
\end{figure*}

Here we present example reconstructions of synthetic data derived from the ``best-bet" model shown in Figure 16 of \citetalias{PaperV}, with positive spin of $a_{*}=0.5$ and electron temperature of $R_{\rm high}=160$ viewed at $i=30^\circ$. 
In \autoref{fig:bestbet-grmhd-groundtruth}, we show a snapshot and time-averaged images of this GRMHD model, as well as the amplitude versus \uv-distance corresponding to synthetic data generated with April 7 coverage (generated in the same manner with \autoref{sec:synthetic_data}). As shown in \autoref{fig:bestbet-grmhd-groundtruth}, visibility amplitudes are broadly consistent with EHT \sgra data. 
\autoref{fig:bestbet-grmhd-3599} shows descattered and on-sky images, which are reconstructed with \difmap, \ehtim, and \smili pipelines using their Top Set parameters and then further categorized by the same clustering method presented in \autoref{appendix:clustering_method}.

As shown in \autoref{fig:bestbet-grmhd-3599}, the distributions of the Top Set images share key results from those of the GRMHD reconstructions described in \autoref{subsec:synthetic_data_images_grmhd}  (see also \autoref{fig:grmhd-3599}). 
The vast majority of the reconstructions identify a ring morphology with a diameter of \sdiam, consistent with the ground-truth model. However, a small fraction of images have a non-ring morphology. 
Furthermore, ring reconstructions have multiple azimuthal intensity modes.
In particular, ring images that have a peak PA of $\sim -154^{\circ}$ consistent with the ground-truth image do not appear as the most popular modes in \ehtim and \smili reconstructions.
The broad consistency with the results in \autoref{subsec:synthetic_data_images_grmhd} suggest that our main results (e.g., that for the RML and CLEAN pipelines our methods produce a small fraction of non-ring modes for an underlying ring model, and that the most popular mode reconstructed is not always that true mode) likely generalize across GRMHD models that are in a broad agreement with 1.3\,mm EHT data.

%% file: appendix_ringfits.tex
\section{Ring fitting parameters}
\label{appendix:ring_params}
In \autoref{tab:ring_fulfits_descattered} and \autoref{tab:ring_fulfits_on-sky} we list the diameter $d$, width $w$, position angle $\eta$, asymmetry $A$, and fractional central brightness $f_c$ measured from Top Set \sgra images for each identified cluster (see \autoref{sec:image_analysis}) corresponding to the descattered and on-sky images, respectively.

\begin{table*}[t]
\begin{center}
\caption{Mean and standard deviation of ring parameters, diameter $d$, width $w$, position angle $\eta$, asymmetry $A$, and fractional central brightness $f_c$ measured from Top Set or posterior descattered \sgra images for each cluster.}
\label{tab:ring_fulfits_descattered}
\begin{tabular}{ccccccc} \hline \hline
& & $d\ \rm{(\mu as)}$ & $w\ \rm{(\mu as)}$ & $\eta\ \rm{(^\circ)}$ & $A$ & $f_c$ \\ \hline
\difmap &     &              &                   &                      &     &       \\
April 6 Ring & \texttt{REx}  &  $46 \pm   4.1 $  &   $33 \pm   3.5 $                     &  $-100.6 \pm   73.4 $ & $0.15 \pm   0.08 $   &    $0.47 \pm   0.14 $
\\
 & \vida  &  $51 \pm   3.1 $  &   $33 \pm   3.1 $                     &  $-108.3 \pm   79.2 $ & $0.26 \pm   0.12 $   &    $0.46 \pm   0.11 $
\\
April 7 Ring 1&   &  $50 \pm   1.9 $  &   $31 \pm   2.7 $                         &  $-92.0 \pm   42.0 $ & $0.06 \pm   0.04 $   &    $0.40 \pm   0.08 $
\\
 &   &  $51 \pm   1.3 $  &   $33 \pm   1.4 $                         &  $-82.7 \pm   40.5 $ & $0.07 \pm   0.07 $   &    $0.37 \pm   0.08 $
\\
April 7 Ring 2&   &  $49 \pm   3.2 $  &   $32 \pm   2.9 $                         &  $-22.5 \pm   54.4 $ & $0.08 \pm   0.06 $   &    $0.45 \pm   0.12 $
\\
 &   &  $50 \pm   1.7 $  &   $32 \pm   1.4 $                         &  $-27.0 \pm   47.4 $ & $0.12 \pm   0.11 $   &    $0.42 \pm   0.11 $
\\
April 7 Ring 3&   &  $49 \pm   1.8 $  &   $32 \pm   2.9 $                         &  $20.5 \pm   47.2 $ & $0.08 \pm   0.05 $   &    $0.44 \pm   0.09 $
\\
 &   &  $51 \pm   1.6 $  &   $33 \pm   1.8 $                         &  $5.3 \pm   38.9 $ & $0.12 \pm   0.10 $   &    $0.41 \pm   0.10 $
\\
\hline
\ehtim &     &              &                   &                      &     &       \\
April 6 Ring & \texttt{REx}  &  $56 \pm   4.5 $  &   $24 \pm   2.4 $                     &  $32.3 \pm   86.4 $ & $0.17 \pm   0.05 $   &    $0.16 \pm   0.14 $
\\
 & \vida  &  $59 \pm   11.3 $  &   $30 \pm   10.4 $                     &  $70.0 \pm   88.3 $ & $0.24 \pm   0.14 $   &    $0.23 \pm   0.19 $
\\
April 7 Ring 1&   &  $55 \pm   1.9 $  &   $26 \pm   2.3 $                         &  $-140.6 \pm   62.9 $ & $0.11 \pm   0.04 $   &    $0.18 \pm   0.08 $
\\
 &   &  $56 \pm   5.4 $  &   $27 \pm   2.5 $                         &  $-156.0 \pm   60.6 $ & $0.15 \pm   0.13 $   &    $0.18 \pm   0.09 $
\\
April 7 Ring 2&   &  $53 \pm   1.8 $  &   $27 \pm   2.7 $                         &  $-60.1 \pm   35.8 $ & $0.12 \pm   0.05 $   &    $0.23 \pm   0.10 $
\\
 &   &  $53 \pm   3.6 $  &   $27 \pm   4.0 $                         &  $-71.7 \pm   33.1 $ & $0.15 \pm   0.11 $   &    $0.22 \pm   0.10 $
\\
April 7 Ring 3&   &  $55 \pm   1.5 $  &   $27 \pm   3.0 $                         &  $170.4 \pm   101.4 $ & $0.12 \pm   0.07 $   &    $0.31 \pm   0.17 $
\\
 &   &  $57 \pm   6.4 $  &   $26 \pm   4.5 $                         &  $179.7 \pm   67.1 $ & $0.18 \pm   0.21 $   &    $0.27 \pm   0.11 $
\\
\hline
\smili &     &              &                   &                      &     &       \\
April 6 Ring & \texttt{REx}  &  $57 \pm   3.4 $  &   $24 \pm   1.9 $                     &  $-27.7 \pm   43.8 $ & $0.23 \pm   0.09 $   &    $0.03 \pm   0.10 $
\\
 & \vida  &  $46 \pm   12.0 $  &   $50 \pm   16.6 $                     &  $-71.6 \pm   123.8 $ & $0.18 \pm   0.12 $   &    $0.59 \pm   0.33 $
\\
April 7 Ring 1&   &  $52 \pm   5.0 $  &   $26 \pm   2.0 $                         &  $151.9 \pm   75.8 $ & $0.12 \pm   0.04 $   &    $0.22 \pm   0.10 $
\\
 &   &  $52 \pm   3.8 $  &   $27 \pm   3.9 $                         &  $175.9 \pm   77.6 $ & $0.10 \pm   0.08 $   &    $0.22 \pm   0.09 $
\\
April 7 Ring 2&   &  $53 \pm   4.0 $  &   $25 \pm   2.9 $                         &  $-39.3 \pm   51.2 $ & $0.13 \pm   0.06 $   &    $0.23 \pm   0.11 $
\\
 &   &  $52 \pm   6.2 $  &   $29 \pm   8.4 $                         &  $-59.2 \pm   64.9 $ & $0.15 \pm   0.12 $   &    $0.27 \pm   0.18 $
\\
April 7 Ring 3&   &  $51 \pm   4.0 $  &   $26 \pm   1.9 $                         &  $109.1 \pm   55.2 $ & $0.13 \pm   0.04 $   &    $0.19 \pm   0.09 $
\\
 &   &  $51 \pm   3.0 $  &   $27 \pm   3.8 $                         &  $116.2 \pm   65.0 $ & $0.10 \pm   0.08 $   &    $0.19 \pm   0.08 $
\\
\hline
	\themis &     &              &                   &                      &     &       \\
April 6 Ring & \texttt{REx}  &  $51 \pm   3.9 $  &   $25 \pm   1.2 $                     &  $-128.6 \pm   10.0 $ & $0.20 \pm   0.04 $   &    $0.27 \pm   0.10 $
\\
 & \vida  &  $54 \pm   0.9 $  &   $24 \pm   0.9 $                     &  $-121.0 \pm   21.7 $ & $0.27 \pm   0.06 $   &    $0.34 \pm   0.08 $
\\
April 7 Ring 1&   &  $53 \pm   0.5 $  &   $23 \pm   1.4 $                         &  $-37.5 \pm   11.3 $ & $0.14 \pm   0.01 $   &    $0.09 \pm   0.07 $
\\
 &   &  $56 \pm   1.4 $  &   $27 \pm   0.9 $                         &  $-37.6 \pm   7.6 $ & $0.30 \pm   0.05 $   &    $0.19 \pm   0.07 $
\\
April 7 Ring 2&   &  $53 \pm   0.7 $  &   $22 \pm   0.5 $                         &  $-12.5 \pm   8.3 $ & $0.21 \pm   0.02 $   &    $0.05 \pm   0.05 $
\\
 &   &  $56 \pm   1.2 $  &   $27 \pm   0.7 $                         &  $-20.6 \pm   6.1 $ & $0.36 \pm   0.04 $   &    $0.15 \pm   0.03 $
\\
April 7 Ring 3& & -   &   - &  -  & -    &- 
\\
 &  & -   &   - &  -  & -    &- 
\\
\hline
\end{tabular}
\end{center}
\end{table*}

\begin{table*}[t]
\begin{center}
\caption{Mean and standard deviation of ring parameters, diameter $d$, width $w$, position angle $\eta$, asymmetry $A$, and fractional central brightness $f_c$ measured from Top Set and posterior on-sky \sgra images for each cluster.}
\label{tab:ring_fulfits_on-sky}
\begin{tabular}{ccccccc} \hline \hline
& & $d\ \rm{(\mu as)}$ & $w\ \rm{(\mu as)}$ & $\eta\ \rm{(^\circ)}$ & $A$ & $f_c$ \\ \hline
\difmap &     &              &                   &                      &     &       \\
April 6 Ring & \texttt{REx}  &  $46 \pm   3.0 $  &   $34 \pm   4.3 $                     &  $97.9 \pm   88.9 $ & $0.11 \pm   0.05 $   &    $0.50 \pm   0.15 $
\\
 & \vida  &  $47 \pm   1.9 $  &   $39 \pm   3.4 $                     &  $94.8 \pm   87.0 $ & $0.14 \pm   0.07 $   &    $0.56 \pm   0.10 $
\\
April 7 Ring 1&   &  $47 \pm   3.0 $  &   $38 \pm   2.7 $                         &  $-111.6 \pm   39.7 $ & $0.06 \pm   0.03 $   &    $0.59 \pm   0.07 $
\\
 &   &  $49 \pm   2.2 $  &   $37 \pm   1.9 $                         &  $-99.9 \pm   40.2 $ & $0.09 \pm   0.06 $   &    $0.58 \pm   0.08 $
\\
April 7 Ring 2&   &  $48 \pm   2.4 $  &   $38 \pm   2.6 $                         &  $31.3 \pm   81.2 $ & $0.06 \pm   0.04 $   &    $0.58 \pm   0.07 $
\\
 &   &  $49 \pm   2.0 $  &   $37 \pm   2.1 $                         &  $16.0 \pm   75.7 $ & $0.10 \pm   0.06 $   &    $0.56 \pm   0.07 $
\\
April 7 Ring 3&   &  $48 \pm   2.7 $  &   $37 \pm   2.9 $                         &  $40.4 \pm   63.5 $ & $0.05 \pm   0.03 $   &    $0.57 \pm   0.08 $
\\
 &   &  $50 \pm   2.1 $  &   $38 \pm   1.9 $                         &  $27.6 \pm   60.9 $ & $0.08 \pm   0.06 $   &    $0.55 \pm   0.08 $
\\
\hline
\ehtim &     &              &                   &                      &     &       \\
April 6 Ring & \texttt{REx}  &  $49 \pm   3.9 $  &   $28 \pm   3.6 $                     &  $13.8 \pm   123.8 $ & $0.15 \pm   0.05 $   &    $0.33 \pm   0.15 $
\\
 & \vida  &  $50 \pm   5.6 $  &   $41 \pm   6.7 $                     &  $110.8 \pm   133.2 $ & $0.20 \pm   0.10 $   &    $0.55 \pm   0.13 $
\\
April 7 Ring 1&   &  $51 \pm   2.4 $  &   $32 \pm   2.7 $                         &  $-165.9 \pm   45.0 $ & $0.08 \pm   0.03 $   &    $0.42 \pm   0.10 $
\\
 &   &  $52 \pm   4.5 $  &   $35 \pm   3.5 $                         &  $-170.7 \pm   41.1 $ & $0.11 \pm   0.10 $   &    $0.41 \pm   0.08 $
\\
April 7 Ring 2&   &  $50 \pm   2.3 $  &   $32 \pm   2.7 $                         &  $-30.4 \pm   68.1 $ & $0.07 \pm   0.03 $   &    $0.44 \pm   0.07 $
\\
 &   &  $49 \pm   2.6 $  &   $35 \pm   4.0 $                         &  $-71.6 \pm   53.5 $ & $0.06 \pm   0.08 $   &    $0.42 \pm   0.08 $
\\
April 7 Ring 3&   &  $49 \pm   2.4 $  &   $33 \pm   3.3 $                         &  $161.6 \pm   67.3 $ & $0.10 \pm   0.02 $   &    $0.50 \pm   0.08 $
\\
 &   &  $53 \pm   6.6 $  &   $35 \pm   7.5 $                         &  $173.3 \pm   70.8 $ & $0.19 \pm   0.14 $   &    $0.52 \pm   0.10 $
\\
\hline
\smili &     &              &                   &                      &     &       \\
April 6 Ring & \texttt{REx}  &  $43 \pm   0.4 $  &   $28 \pm   3.1 $                     &  $163.1 \pm   10.1 $ & $0.10 \pm   0.04 $   &    $0.57 \pm   0.01 $
\\
 & \vida  &  $39 \pm   3.9 $  &   $46 \pm   8.8 $                     &  $-98.6 \pm   93.3 $ & $0.17 \pm   0.14 $   &    $0.67 \pm   0.16 $
\\
April 7 Ring 1&   &  $43 \pm   4.9 $  &   $33 \pm   1.7 $                         &  $127.7 \pm   29.9 $ & $0.09 \pm   0.08 $   &    $0.51 \pm   0.07 $
\\
 &   &  $45 \pm   1.8 $  &   $34 \pm   3.1 $                         &  $136.8 \pm   36.8 $ & $0.07 \pm   0.06 $   &    $0.48 \pm   0.07 $
\\
April 7 Ring 2&   &  $47 \pm   4.5 $  &   $33 \pm   2.8 $                         &  $58.4 \pm   57.5 $ & $0.10 \pm   0.05 $   &    $0.48 \pm   0.11 $
\\
 &   &  $46 \pm   5.2 $  &   $37 \pm   7.4 $                         &  $59.8 \pm   81.7 $ & $0.08 \pm   0.08 $   &    $0.50 \pm   0.13 $
\\
April 7 Ring 3&   &  $48 \pm   4.3 $  &   $32 \pm   2.9 $                         &  $89.6 \pm   34.9 $ & $0.11 \pm   0.04 $   &    $0.45 \pm   0.14 $
\\
 &   &  $47 \pm   3.5 $  &   $35 \pm   4.3 $                         &  $84.8 \pm   48.2 $ & $0.11 \pm   0.08 $   &    $0.46 \pm   0.11 $
\\
\hline
	\themis &     &              &                   &                      &     &       \\
April 6 Ring & \texttt{REx}  &  $46 \pm   2.1 $  &   $30 \pm   1.6 $                     &  $-139.4 \pm   52.9 $ & $0.15 \pm   0.05 $   &    $0.34 \pm   0.11 $
\\
 & \vida  &  $47 \pm   3.1 $  &   $33 \pm   2.0 $                     &  $-127.4 \pm   53.0 $ & $0.17 \pm   0.08 $   &    $0.42 \pm   0.07 $
\\
April 7 Ring 1&   &  $51 \pm   0.8 $  &   $28 \pm   1.3 $                         &  $-37.8 \pm   53.6 $ & $0.15 \pm   0.01 $   &    $0.21 \pm   0.10 $
\\
 &   &  $55 \pm   1.6 $  &   $32 \pm   2.4 $                         &  $-41.4 \pm   52.6 $ & $0.25 \pm   0.05 $   &    $0.30 \pm   0.07 $
\\
April 7 Ring 2&   &  $50 \pm   1.5 $  &   $26 \pm   1.1 $                         &  $-16.0 \pm   8.9 $ & $0.21 \pm   0.03 $   &    $0.20 \pm   0.08 $
\\
 &   &  $56 \pm   2.6 $  &   $32 \pm   0.9 $                         &  $-25.2 \pm   7.3 $ & $0.35 \pm   0.03 $   &    $0.29 \pm   0.06 $
\\
April 7 Ring 3& & -   &   - &  -  & -    &- 
\\
 &  & -   &   - &  -  & -    &- 
\\
\hline
\end{tabular}
\end{center}
\end{table*}

%% file: appendix_besttimes.tex
\section{Dynamic Imaging and Snapshot Model Fitting tests}
\label{sec:appendixdynamics}

\subsection{Selection of Time Windows with the Best \uv-coverage}

\begin{figure}
    \centering
    \includegraphics[width=\columnwidth]{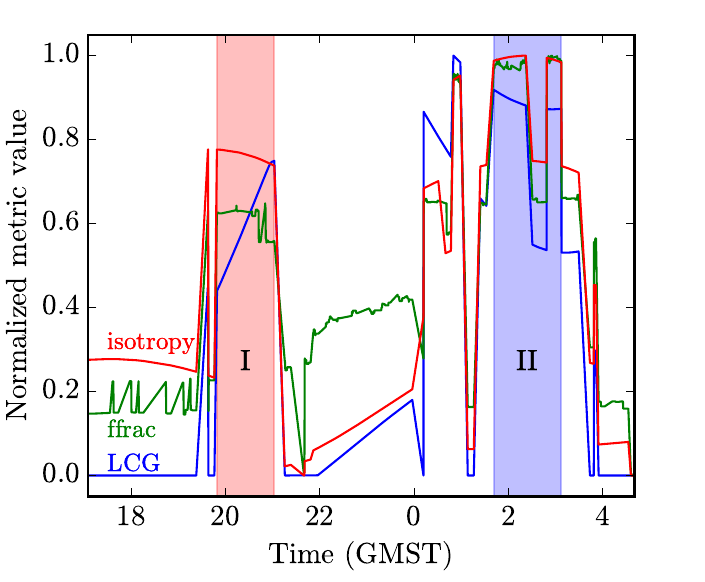}
    \caption{Normalized metric computations for every scan of the April 7, 2017 EHT coverage of \sgra. 0:00 GMST marks the day change from April 7 to April 8. Though the metrics have different considerations, all highlight the region (labelled ``II'') from ${\sim}$1:30 GMST to ${\sim}$3:10 GMST as a candidate region for dynamic imaging.}
    \label{fig:sdi_window_selection}
\end{figure}

\cite{Farah_2021} demonstrates that the changing \uv-coverage created by the Earth's rotation during the aperture synthesis process leads to regions of time that produce dynamic reconstructions of varying quality. The quality of a short-timescale reconstruction is partially determined by the snapshot \uv-coverage geometry, which introduces certain artifacts during the imaging process. 

The scale and severity of these artifacts can be predicted by quantitatively scoring the \uv-coverage as a function of time, which can be done in a number of ways. Some metrics examine how much of the Fourier plane is covered by an interferometer (e.g., \citealt{Palumbo_2019}), while others look at gaps created by the sparse coverage (e.g., \citealt{github_LCG}). In addition to these metrics, \cite{Farah_2021} derives a novel metric that probes both the anisotropy and radial homogeneity of the coverage.

By applying these metrics to the April 6 and 7 EHT \uv-coverage on \sgra, we can assess the scan-by-scan performance and identify regions of time which are likely to produce the best reconstructions, independent of the underlying source structure. The result of such an analysis for April 7 is shown in \autoref{fig:sdi_window_selection}, and two candidate regions are highlighted. The metrics predict dynamic imaging reconstructions will have the highest quality in the region from 1.5-3.2 GMST (Region II); the reconstructions will produce substantially worse results in the region from 19.4-21 GMST (Region I). We validate this prediction by testing on high \textit{S/N} data in \cite{Farah_2021} and show that Region II indeed allows for significantly better recovery of the source variability than Region I.
Therefore, based only on the EHT's \uv-coverage, we focus on dynamic imaging/modeling Region II throughout \autoref{sec::dynamic}.

\subsection{Starwarps Temporal Regularizer Normalization}
\label{sec:app_sw_reg_norm}
\begin{figure}
    \centering
    \includegraphics[width=\columnwidth]{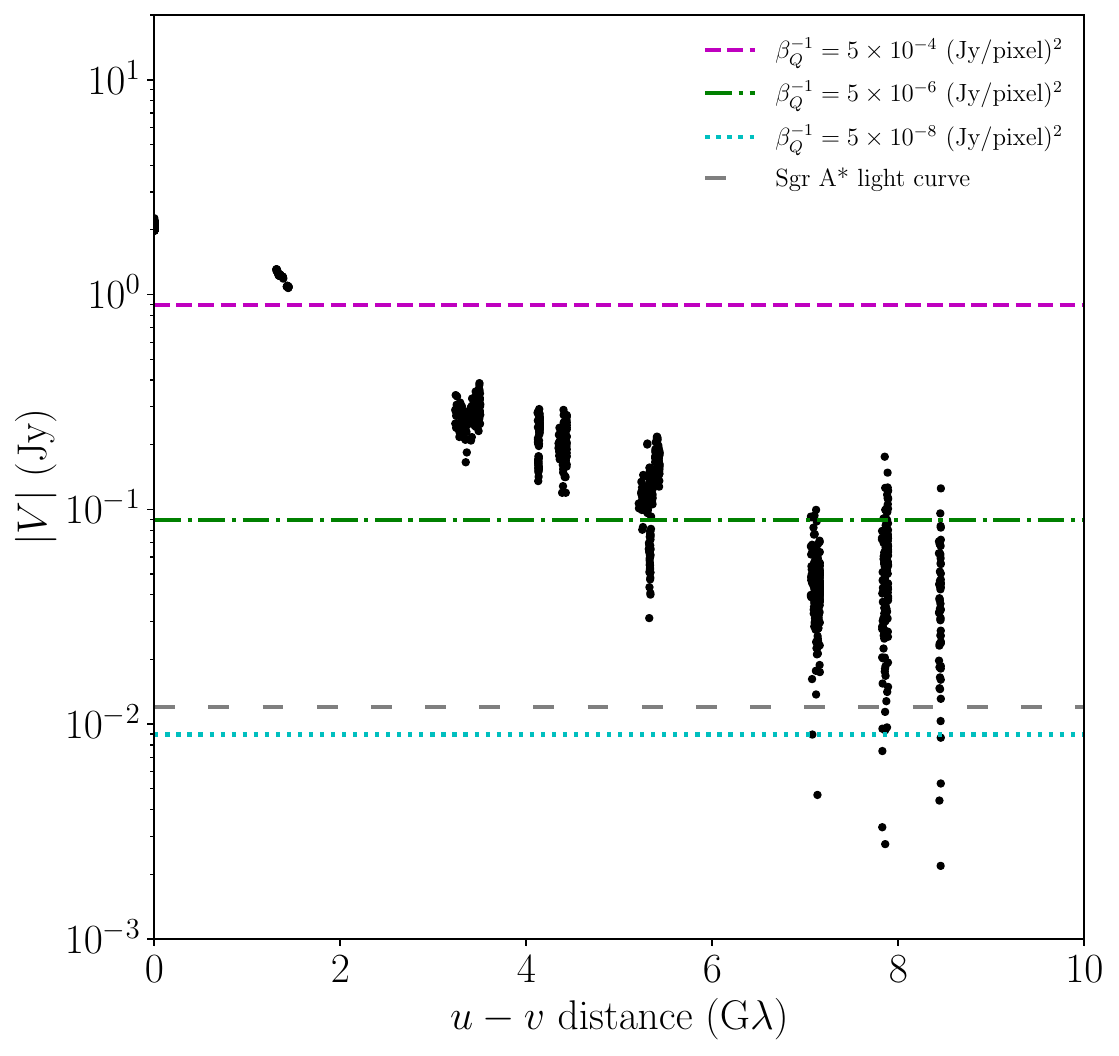}
    \caption{
    April 7 EHT visibility amplitudes over the selected window for dynamic imaging and modeling (black points). The horizontal lines show the expected standard deviation of the visibility amplitudes in  \texttt{StarWarps} reconstructions for different values of $\beta^{-1}_Q$ in units of $(\text{Jy/pixel})^2$: $5\times10^{-4}$ (magenta), $5\times10^{-6}$ (green), and $5\times10^{-8}$ (cyan).
    }
    \label{fig:sw_reg_norm}
\end{figure}

Temporal regularization in \texttt{StarWarps} is controlled by a parameter $\beta_Q^{-1}$. This parameter corresponds to the variance of the conditional distribution  of pixel intensities for a given snapshot holding the previous snapshot fixed: $p(I_k|I_{k-1}) = \mathcal{N}(I_{k-1},\beta^{-1}_Q \mathbbm{1})$. The units of $\beta_Q^{-1}$ are $(\text{Jy/pixel})^2$.  In the main text, we consider values $\beta^{-1}_Q \in \{5\times10^{-4},5\times10^{-6},5\times10^{-8}\}$. Larger values of $\beta^{-1}_Q$ correspond to less temporal regularization, as the conditional distribution $p(I_k|I_{k-1})$ becomes wider. 

We can also interpret the values $\beta^{-1}_Q$ in visibility space. The Fourier transform of the image $I_k$ is given by a $N_{\rm pix}\times N_{\rm pix}$ matrix $\mathbf{F}$:
\begin{equation}
    V_k = \mathbf{F}I_k.
    \label{eq:mov_fourier}
\end{equation}
In our convention, the pixel values in $I_k$ have units Jy/pixel, so the entries of $\mathbf{F}$ are pure phase terms without a $1/\sqrt{N}$ normalization (so that, for instance, the zero-baseline visibility in Jy is just the sum of the pixel intensities).
As a result $\mathbf{F}\mathbf{F}^\dagger = N_{\rm pix} \mathbbm{1}$. Because \autoref{eq:mov_fourier} is a linear transformation, $p(V_k|I_{k-1})$ is also a normal distribution, with a mean $V_{k-1}$ and a covariance:
\begin{equation}
    \mathbf{\Sigma} = \mathbf{F} \left[\beta^{-1}_Q \mathbbm{1} \right] \mathbf{F}^\dagger = \beta^{-1}_Q N_{\rm pix} \mathbbm{1}. 
\end{equation}

Thus,  $\sigma_{\rm vis} \equiv \sqrt{\beta^{-1}_Q N_{\rm pix}}$ is the standard deviation of a snapshot visibility measurement in \texttt{StarWarps}, holding the previous frame fixed.

In \autoref{fig:sw_reg_norm}, we compare this quantity to the measured EHT visibility amplitudes in the selected dynamic imaging window on April 11. The \texttt{StarWarps} movie reconstructions in the main text have $N_{\rm pix} =  40\times40 = 1600$. GRMHD simulations (see \citetalias{PaperIV} and \citealt{Georgiev_2021}) and the light curve of \sgra \citep[see][]{Wielgus_2021} suggest that the variations on minute timescales should have a zero-baseline standard deviation of $\sigma_{\rm vis} \sim 10\,{\rm mJy}$. Thus, reconstructions with $\beta^{-1}_Q \sim 10^{-7} ({\rm Jy/pixel})^2$ are expected to give variability that is consistent with what is measured in \sgra. 

Larger values of $\beta^{-1}_Q$ correspond to lower temporal regularization and allow for larger variations in the visibility amplitudes. For instance,  our reconstructions in \autoref{sec::dynamic} with $\beta^{-1}_Q=5 \times 10^{-6}$ permit somewhat more variability than is seen in simulations and observations of \sgra. Nevertheless, we have also found that allowing excess variability helps to trace evolution in tests on synthetic data from GRMHD simulations.

\subsection{Testing \uv-coverage Effects}
\label{sec:appendixdynamics_uvcov}

As discussed in \autoref{sec:dynamapproach}, the geometry of the \uv-coverage can have an effect on the recovered image structure, especially in cases where the coverage is extremely sparse. To study the effects of \uv-coverage on dynamic fits to \sgra data, we perform a number of tests on synthetic data sets and study the effect of different \uv baselines on fits to the real data. 

\subsubsection{Recovering the Position Angle of a Static Crescent}
\label{sec:dynamic_crescents}

As most of our analysis of the dynamic structure of \sgra revolves around tracking the position angle (PA) of brightness around the ring, it is important to assess our ability to recover the PA accurately in realistic synthetic data. To that end, we constructed synthetic EHT data sets from 4 static crescent models with peak brightness points rotated at 60$^\circ$ increments around the ring. The brightness ratio of each crescent model was chosen to roughly match the 1.5:1 ratio recovered from geometric model fitting to \sgra data. \autoref{fig:crescentws_besttimes} shows the imaging and geometric modeling results obtained by fitting to these synthetic data sets in the selected 1.7 h region. Note that, for both approaches, the true PA is recovered as the primary mode for most of the crescents. 
The imaging methods contain temporal regularization, which likely makes it easier to recover a static underlying structure; however, the geometric modeling results do not assume any temporal regularization.

\subsubsection{Uniform Ring Synthetic Data}

\begin{figure}
    \centering
    \includegraphics[width=\columnwidth]{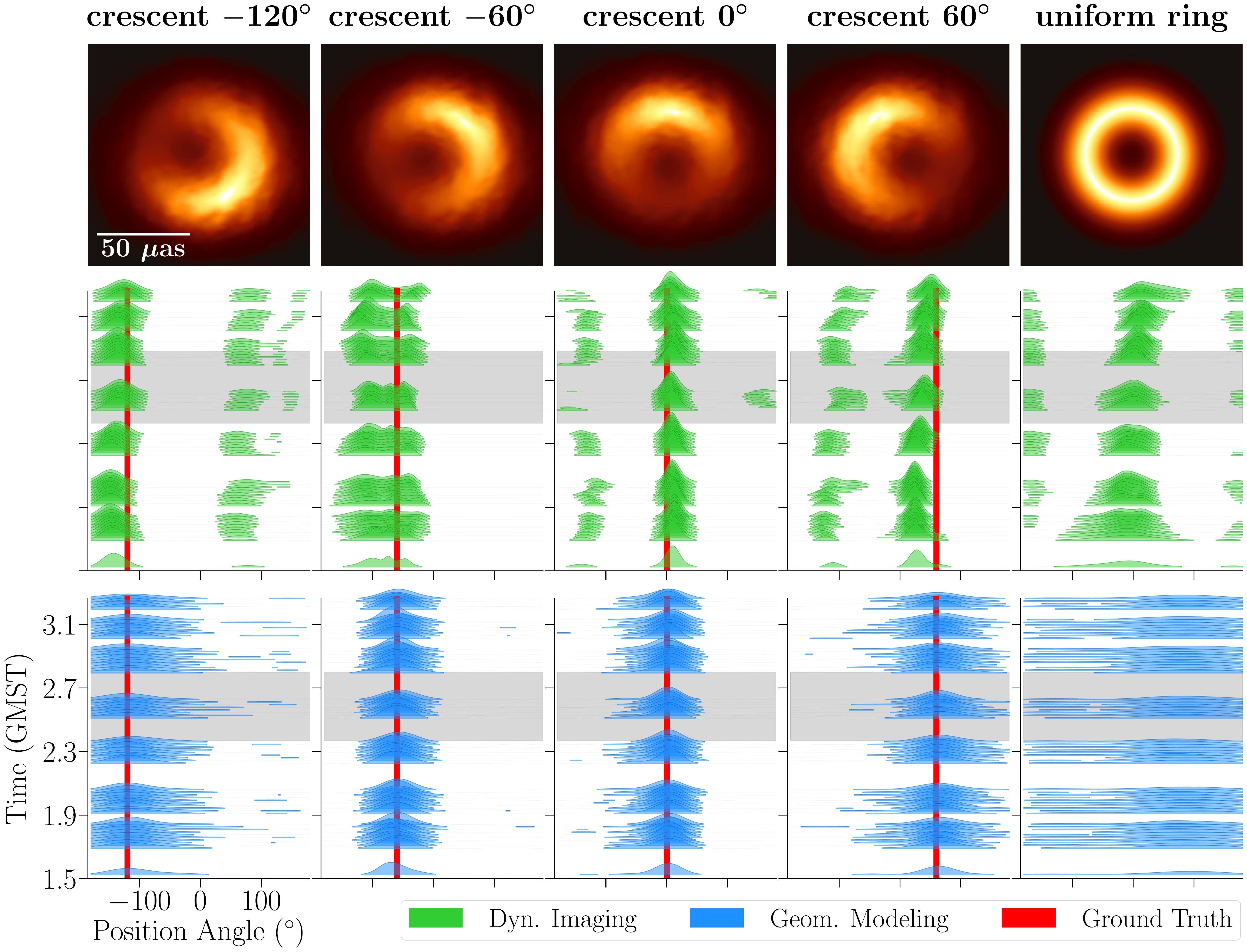}
    \caption{Position angle (PA) recovered from differently oriented static crescents synthetic data sets and a uniform ring synthetic data set, using both dynamic imaging (green) and geometric modeling (blue) techniques. The crescents' ground truth PA is shown as a vertical red line. Imaging uses a prior image $\mu$ of a uniform ring convolved with a 25 $\mu$as beam.}
    \label{fig:crescentws_besttimes}
\end{figure}

The interplay between the source size and sidelobes in the dirty beam pattern from sparse coverage can cause imaging artifacts that appear in the form of bright ``knots" around ring sources.
Computing the dirty image of an underlying uniform ring source reveals the location of these knots when using calibrated visibilities.  
To assess the impact of these knot artifacts on our analysis we performed imaging and geometric model fitting on data generated from a uniform ring (with no brightness changes in azimuth) with diameter of 49 $\mu$as (refer to the uniform ring in \autoref{fig:synthetic_summary}). As can be seen in \autoref{fig:crescentws_besttimes}, both the imaging and modeling results indicate an image structure with a preferred PA -- $\sim 0 ^\circ$ or $\sim90^\circ$ for imaging and $\sim100 ^\circ$ for modeling. However, the associated asymmetry of the recovered uniform ring is very low, a brightness ratio of less than 1.1:1 compared to 1.5:1 for \sgra. Thus, in combination with the results of \autoref{sec:dynamic_crescents}, we conclude that although the \uv-coverage will bias the PA in the limit of low image asymmetry, for the level of image asymmetry recovered in \sgra this bias should have a small effect. 

\subsubsection{Baseline Test}

\begin{figure*}
    \centering
    \includegraphics[width=.9\textwidth]{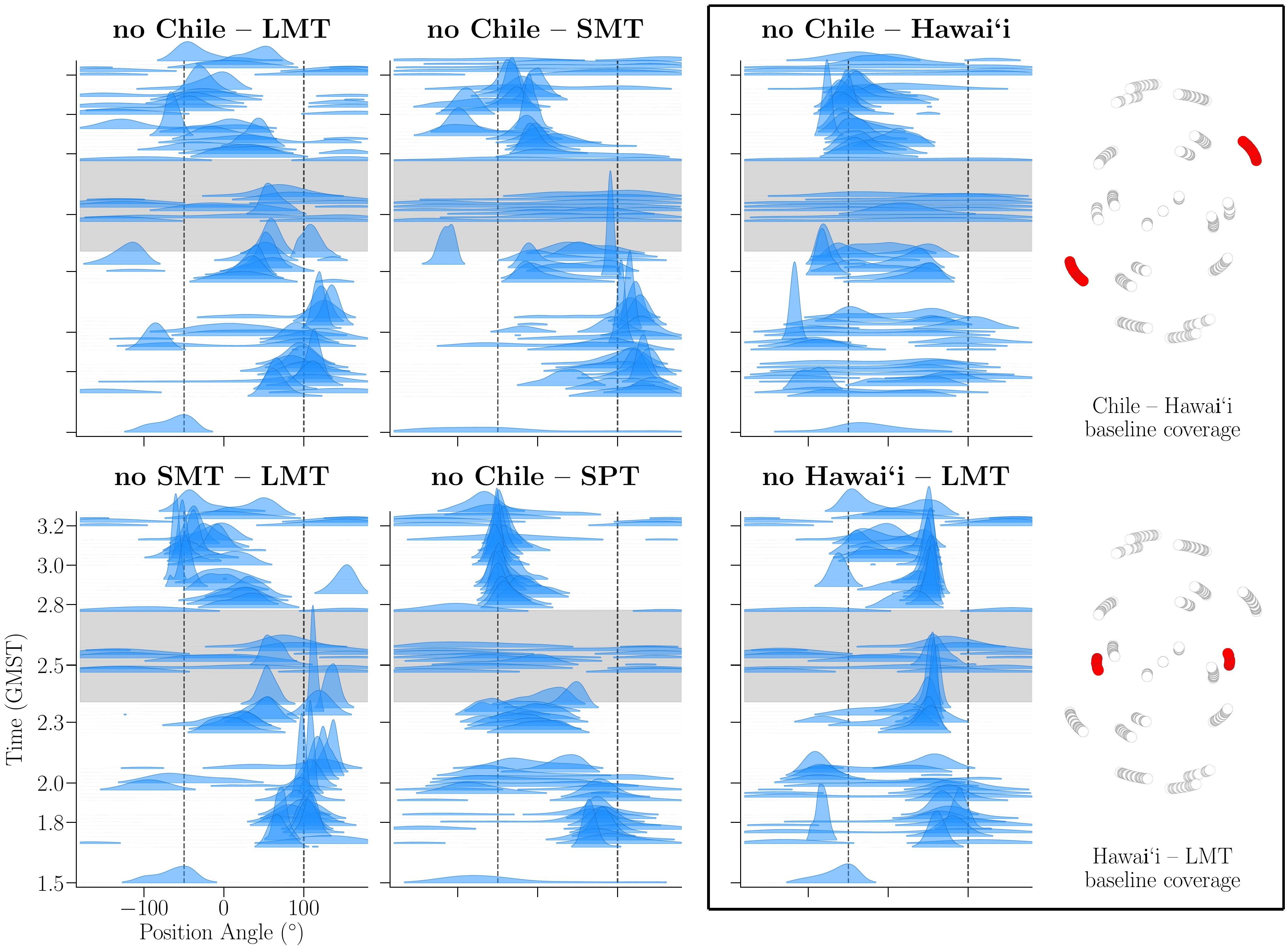}
    \caption{Position angle (PA) recovered using \texttt{DPI} geometric modeling after removal of a particular baseline from \sgra data on April 7. The flagged baseline appears in the title of each panel. The Chile-\hawaii and \hawaii-LMT baselines are highlighted and their location in the \uv plane is shown. These two baselines probe the east--west and southeast--northwest orientations.}
    \label{fig:baseline_dropping}
\end{figure*}

In order to evaluate the contribution of each baseline to the recovered evolution in \sgra, we compared results obtained on data sets modified to remove a particular baseline. In particular, we compared the PA posteriors obtained using geometric modeling on 11 different data sets -- 10 data set each with a single baselines removed
and one complete data set. 
As can be seen in \autoref{fig:baseline_dropping}, we find that most baselines do not heavily affect the trends we see on April 7. 
However, there are two baselines that appear to have a significant effect on the results: Chile--\hawaii and LMT--\hawaii. Without the Chile--\hawaii baseline we are not able to discriminate between the northwest and southeast PA; without the LMT--\hawaii baseline we do not recover as significant of a PA shift. 
Upon inspecting the \uv-coverage of these baselines it becomes apparent that these two baselines probe the northwest to southeast orientations that we are interested in, and thus without them we are unable to properly discriminate between these two PA orientations. 
It is also worth noting that removal of the Chile-SPT baseline appears to ``clean up" the modeling results, suggesting that small-scale features probed by this baseline may not be properly captured in our geometric model fits.  

\subsection{Testing Scattering Mitigation Strategies}
\label{sec:app_dyn_scat}

In producing dynamic reconstructions and model fits of the \sgra data, the choice of scattering mitigation strategy is a potential source of uncertainty.
We have explored the sensitivity of our dynamic imaging and snapshot model fitting results to the same five scattering mitigation strategies we consider in the static image surveys in \autoref{sec:survey} and \autoref{sec:sgra_images:scattering}. Namely, we produce reconstructions and model fits to the unmodified data (i.e., on-sky with no descattering), as well as with visibilities deblurred by the \sgra diffractive scattering kernel and with the thermal noise error bars inflated by four models of the refractive noise: the \texttt{Const} model, the \texttt{J18model1} model, and then double the additional error tolerance from each of these models (\texttt{2$\times$Const}, \texttt{2$\times$J18model1}). Based on the analysis done in \autoref{sec:pre_scatter}, this selection is conservative and spans our  uncertainty in \sgra's refractive noise. 

\begin{figure*}
    \centering
    \includegraphics[width=\textwidth]{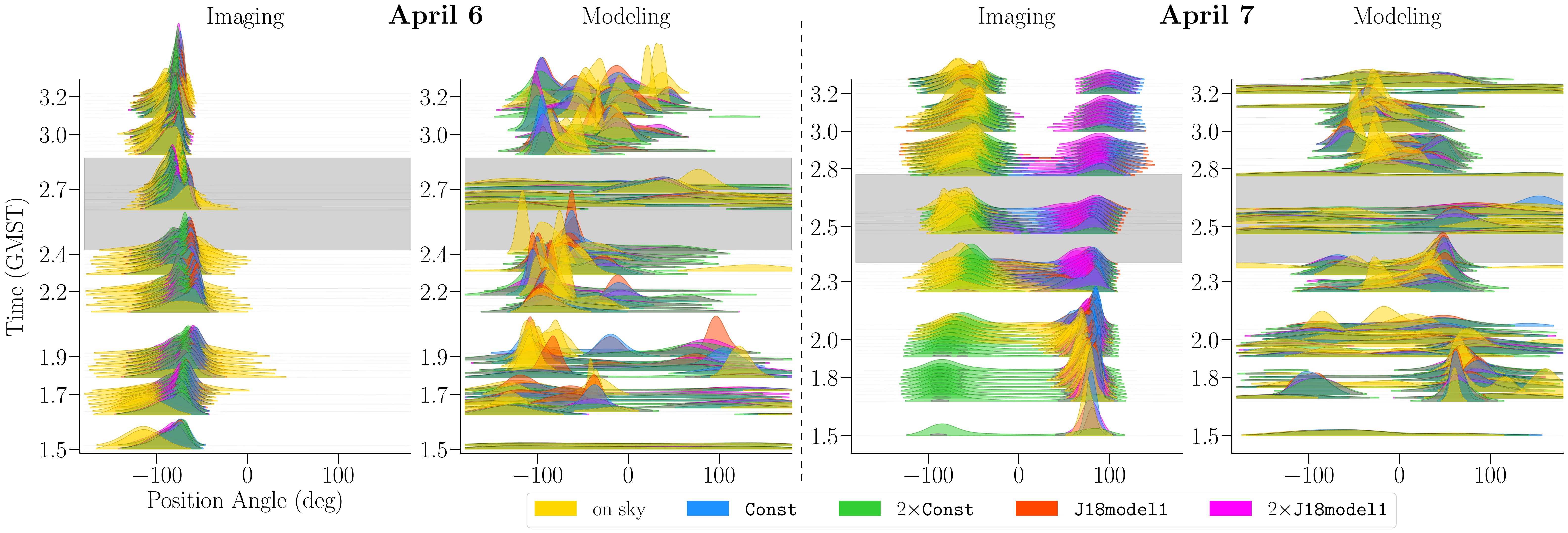}
    \caption{The position angle (PA) recovered using dynamic imaging and geometric modeling techniques under
    different scattering mitigation assumptions. These range from no scattering mitigation whatsoever (yellow), to different amounts of refractive noise added to the deblurred data: a constant noise floor (blue), the refractive noise model \texttt{J18model1} (red), and these two scaled by a factor of two (green and magenta, respectively). Imaging results were obtained using an initialization/prior image of a uniform ring blurred with a Gaussian kernel with FWHM of 25 $\mu$as. Modeling results were obtained from the \texttt{DPI} pipeline using the $m$-ring 2 geometric model. The gray band at roughly 2.6 GMST indicates the region where the LMT has dropped out and data coverage is poorer.}
    \label{fig:dynamic_scattering_comparison}
\end{figure*}

\autoref{fig:dynamic_scattering_comparison} presents comparisons of the \texttt{StarWarps} reconstructions and snapshot model fitting results on the April 6 and 7 \sgra data with all five scattering mitigation strategies. We find that the general trends in the ring position angle we discuss in \autoref{sec::dynamic_results} are not significantly changed by any of the five scattering mitigation strategies we explore for geometric modeling, although the position angle posteriors are significantly broader (sometimes spanning a full 360$^\circ$) when using the larger refractive noise budgets of \texttt{2$\times$Const} and \texttt{2$\times$J18model1}. 
In contrast, for imaging, on April 7, we observe a transition from positive to negative PA to be the most commonly recovered trend with all of the on-sky, \texttt{Const} and \texttt{J18model1} scattering mitigation strategies when using a ring prior. However, when we add a very large amount of refractive noise tolerance to the error bars in the \texttt{2$\times$Const} and \texttt{2$\times$J18model1} models, the PA becomes more stable over the observation window. This is due to the interplay of temporal regularization with an increased flexibility in fitting the data with a static model due to the expanded noise budget.
In this figure, imaging with \texttt{StarWarps} makes use of the ring$\textasteriskcentered$25$\mu$as ring prior/initialization and modeling with \texttt{DPI} uses a second order $m$-ring ($m=2$) model with a parameterized central Gaussian floor.

\subsection{Testing the Effects of Different Imaging Priors and Model Specifications}
\label{sec:app_model_imaging_choices}
As overviewed in \autoref{sec:dynamicmodelingchoices}, the recovered PA of the azimuthal brightness distribution in the ring-like morphology of \sgra is sensitive to the modeling choices made in both imaging and geometric modeling. In this section we go into further detail on some of the effects seen.%

\subsubsection{Imaging}
\label{sec:app_imaging_choices}
\paragraph{Temporal regularization} The level of continuity enforced between recovered frames is controlled by temporal regularization. In particular, \texttt{StarWarps} encourages frames to be similar by probabilistically modeling each frame $I_k$ as being a sample from a normal distribution with mean $I_{k-1}$ and covariance $\beta_{Q}^{-1} \mathbbm{1}$ (see \autoref{eq:starwarpsreg}). Thus, decreasing the multiplier $\beta_{Q}^{-1}$ will increase recovered continuity between frames. This is seen, as expected, in the recovered movies of \sgra visualized in \autoref{fig:temporal_reg}; 
recovered movies with low temporal regularization experience fast and drastic variability on large scale features, while movies with high temporal regularization experience slow yet steady variability on large-scale features and absorb the remaining variability in the data with small scale fluctuations.
Due to the extreme sparsity of data on each snapshot, although these movies contain significantly different levels of recovered variability, they all fit the data in terms of $\chi^2$ fairly well. The only substantial difference between data fits can be seen when inspecting the SMT-LMT-\hawaii closure triangle. %
It is worth noting that the positive to negative flip on the April 7's closure triangle is not reproduced by the recovered video with high temporal regularization. Nonetheless, since all movies still match all remaining baselines indistinguishably well it is difficult to form any solid conclusions on the type of variability in \sgra based on this one closure triangle. 

\paragraph{\texttt{StarWarps} spatial prior images}

\begin{figure}[h!]
    \centering
    \includegraphics[width=\columnwidth]{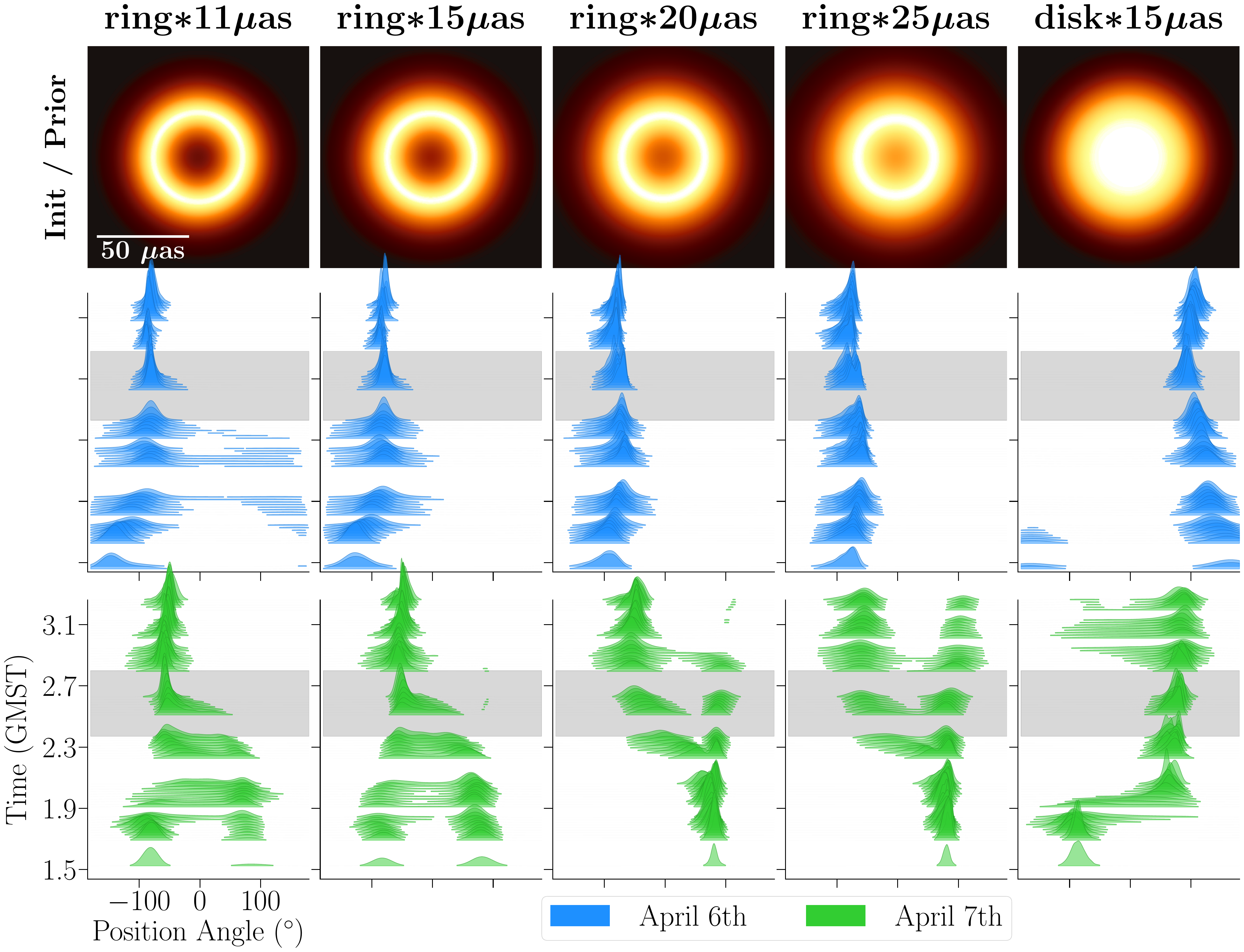}
    \caption{Position angle (PA) reconstructed using \texttt{StarWarps} under different init/prior assumptions for both April 6 (blue) and 7 (green). These include uniform rings with increasing Gaussian blurring and a uniform disk with no central brightness depression.}
    \label{fig:initprior}
\end{figure}

\begin{figure*}
    \centering
    \includegraphics[width=\textwidth]{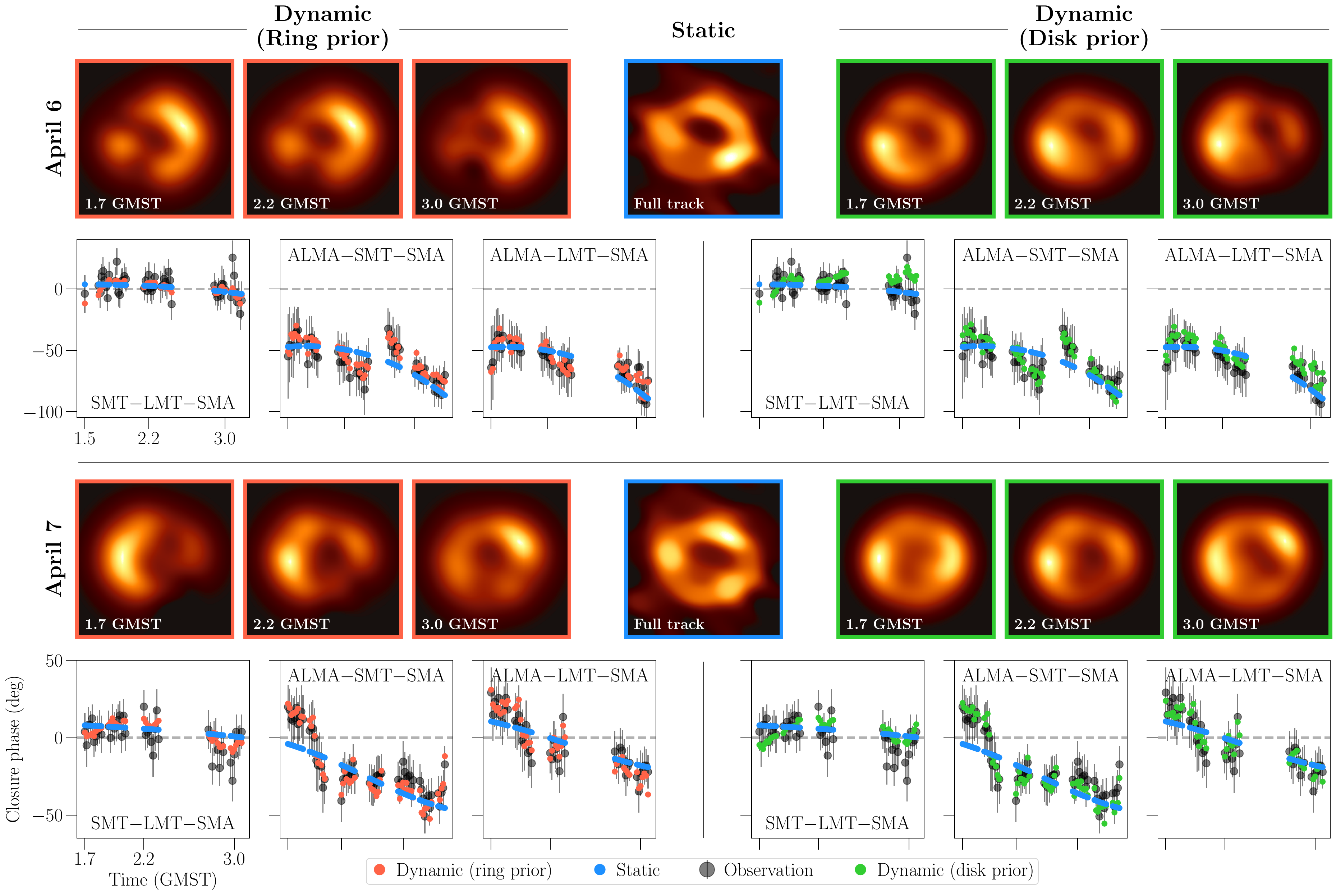}
    \caption{Comparing data fit of descattered dynamic imaging versus static imaging on April 7. A representative descattered image from the \texttt{eht-imaging} static imaging pipeline is shown. In the selected closure plots below the measured data (60 s avg. without a variability noise budget) is shown overlaid with the corresponding closure phases. }
    \label{fig:dynamicvsstatic_alldays2}
\end{figure*}

To explore the sensitivity of results to the ring features encouraged during imaging, we introduce 5 different images used as both the initialization and mean prior image $\mu$ in \texttt{StarWarps}. \autoref{fig:initprior} shows the mean prior images explored for imaging the original data and deblurred data. Note that recovered flux is constrained to evolve within the regions that have flux in the prior image, therefore the puffier rings and the tapered disk with more extended flux are less constraining during imaging. As expected, movies recovered with the less constraining puffier prior images result in recovered movies with a less clear underlying ring structures. Nonetheless, in all these cases (with a ring init/prior), the same general trend in PA is recovered in one of the modes, even when the central indent is very weak. 
When a disk prior is used with no central indent whatsoever the same PA trend is not recovered; instead, the PA trend appears to be reversed in sign (as discussed in \autoref{fig:prior_comparison_descattered}). \autoref{fig:dynamicvsstatic_alldays2} shows more detailed comparisons of \texttt{StarWarps} movies reconstructed with ring and disk mean prior images on April 6 and 7, including data fits to representative closure phases. 

\subsubsection{Geometric Modeling}
\label{sec:app_modeling_choices}
For snapshot geometric modeling, we have two competing effects. One is that we require a geometric model that can adequately explain the on-sky image. However, given the sparseness of the \uv-coverage for each snapshot, the risk of over-fitting the data is considerable and potentially leads to artificially un-informative posteriors. However, under-fitting the data can lead to large biases in the recovered parameters, and artificially narrow posteriors. To find the preferred model we use relative measures. That is, we don't compare the absolute fit quality using a metric like the $\chi^2$ statistic, but rather how well a model does compared to the others considered. For this purpose we use the Bayesian evidence (also referred to as evidence),
\begin{equation}
    Z({\rm data} | M) = \int \mathcal{L}({\rm data} | \theta, M)p(\theta | M)\mathrm{d}\theta.
\end{equation}
The evidence measures the marginal probability of the data, after averaging over all possible parameter values of the model. The preferred model is then the one that maximizes the evidence of the model. For snapshot modeling we select the preferred model by computing the log-evidence in each snapshot, and then sum the log-evidence across all snapshots. Note, that we are only able to estimate the evidence for the \texttt{Comrade} pipeline. \texttt{DPI}/Variational inference cannot estimate the evidence, but instead can compute an effective lower bound to use as a proxy. 

\begin{figure}
    \centering
    \includegraphics[width=\linewidth]{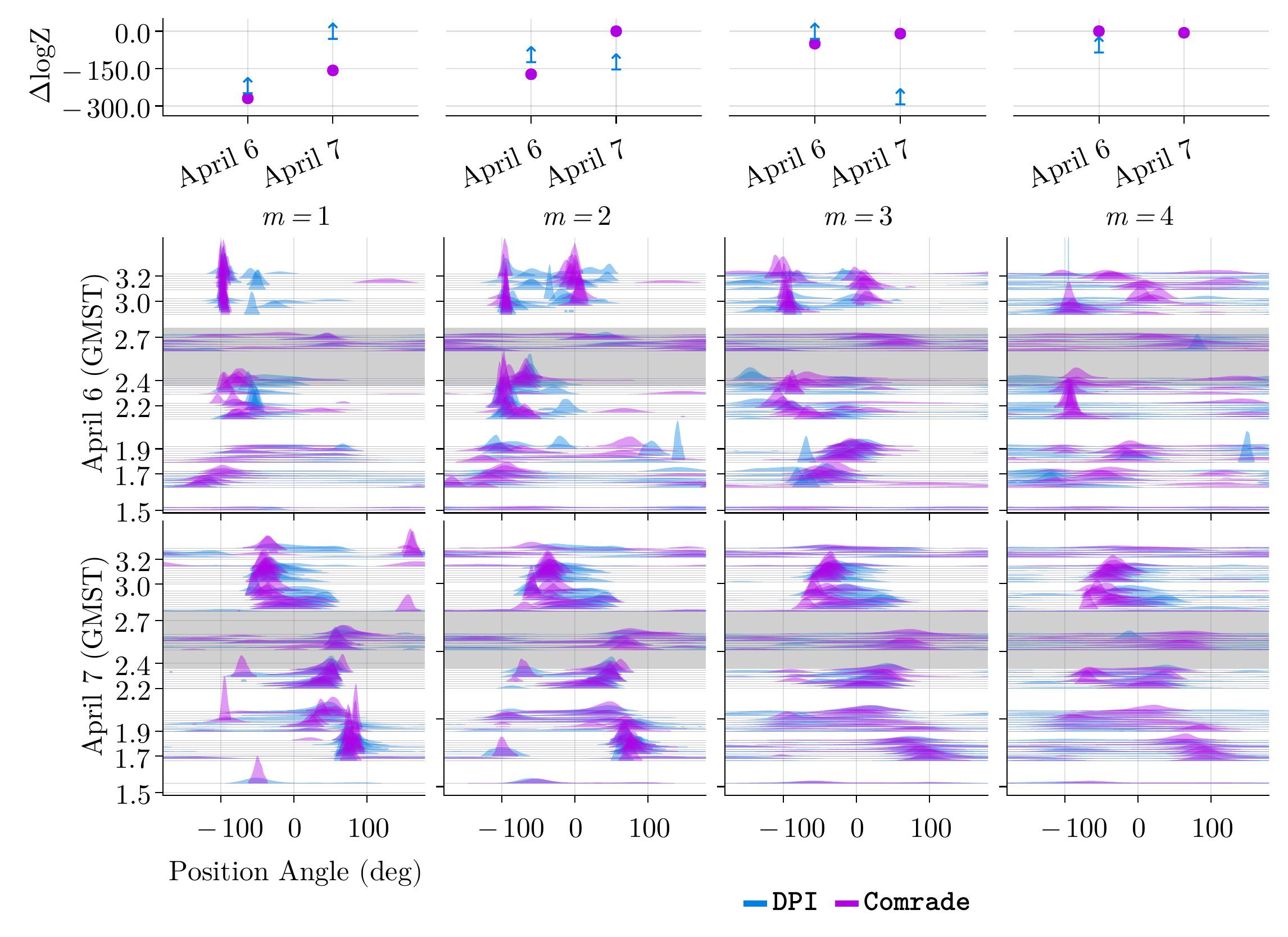}
    \caption{Comparing the $m$-ring results across orders 1, 2, 3, and 4 in both \texttt{Comrade} (magenta) and \texttt{DPI} (blue). The impact of the different $m$-ring orders on the position angle evolution for April 6 and 7 are shown in the middle and bottom rows. The top row is the change in the log-evidence across $m$-ring orders and days. The evidence lower bound produced by \texttt{DPI} is shown as a blue upwards arrow.}
    \label{fig:modeling_order}
\end{figure}

\paragraph{M-ring Order} 
To assess the impact of different model choices on the posterior samples, we considered an $m$-ring model with 1--4 modes.  
The results for the $m$-ring model from the \texttt{Comrade} pipeline are shown in \autoref{fig:modeling_order}. We find that the trend for the dipole moment phase is consistent across model specifications, albeit the posteriors become more uncertain for the higher-order $m$-ring models. The recovered total evidence for each model is shown in \autoref{fig:modeling_order}. For April 6 $m=4$ is the preferred model with a log-evidence of $1499$. On April 7, the $m=2$ $m$-ring is preferred with a log-evidence of $1667$. On both days the overall trend of the position angle maintains stable for $m=1,2,3$, but the distributions become noticeably wider. 

For the \texttt{DPI} pipeline we find similar PA trends for April 6 and 7. Using the evidence-lower-bound (ELBO) that is calculated as part of variational methods, $m=3$ being preferred on April $6$ and $m=1$ on April 7. Furthermore, on April 7 the PA posterior for $m=3,4$ is very broad, becoming essentially unconstrained.

Comparing the \texttt{DPI} results to \texttt{Comrade} we find that $m=1$ is preferred on April 7 and $m=3$ on April 6 according to the evidence lower bound. \texttt{DPI} fits closure products which are equivalent to placing uniform priors on gains, meaning that the data is less constraining. Therefore, it is not surprising that \texttt{DPI} prefers simpler models compared to \texttt{Comrade}. 
To select a fiducial model across days and bands we considered both the evidence, the relative impact on the posterior, and impact on each pipeline. For these reasons we take $m=2$ for \texttt{DPI} and $m=2$ and sometimes $m=3$ for \texttt{Comrade}. The $m=2$ model is preferred for \texttt{Comrade} on April 7 and the $m=2$ distribution for \texttt{DPI} is similar but broader than the $m=1$.

